\documentclass[journal=jacsat,layout=traditional]{achemso}

\usepackage[version=3]{mhchem} 
\usepackage{caption}
\usepackage{subcaption}
\usepackage{array}
\newcolumntype{M}[1]{>{\centering\arraybackslash}m{#1}}
\usepackage{tabularx}
\usepackage[table]{xcolor}
\usepackage{multirow}
\usepackage{textcomp}
\usepackage[normalem]{ulem}
\usepackage{bm,bbm}
\usepackage{hyperref}
\hypersetup{
    colorlinks=true,
    linkcolor=blue,
    citecolor=blue}
\setkeys{acs}{doi = true}
\usepackage{tabularx,multirow,booktabs}
\newcolumntype{L}{>{\raggedright\arraybackslash}X}
\newcolumntype{C}{>{\centering\arraybackslash}X}
\newcolumntype{R}{>{\raggedleft\arraybackslash}X}

\usepackage{tikz}
\usepackage{pgfplots}
\usepackage{pgfplotstable}
\usetikzlibrary{shapes.geometric, arrows}
\usetikzlibrary{patterns}
\usetikzlibrary{positioning}
\usetikzlibrary{matrix,calc,fit}

\pgfplotsset{compat=1.7}

\tikzstyle{startstop} = [rectangle, rounded corners, line width = 0.6pt, minimum width=3cm, minimum height=1cm,text centered, draw=black, fill=red!30]
\tikzstyle{long_startstop} = [rectangle, rounded corners, line width = 0.6pt, minimum width=4cm, minimum height=1cm, text width=4cm, text centered, draw=black, fill=red!30]
\tikzstyle{io} = [trapezium, trapezium left angle=70, trapezium right angle=110, line width = 0.6pt, minimum width=3cm, minimum height=1cm, text centered, draw=black, fill=blue!30]
\tikzstyle{process} = [rectangle, line width = 0.6pt, minimum width=3cm,  text width=3cm, minimum height=1cm, text centered, draw=black, fill=orange!30]
\tikzstyle{short_process} = [rectangle, line width = 0.6pt, minimum width=2.5cm,  text width=2.5cm, minimum height=1cm, text centered, draw=black, fill=orange!30]
\tikzstyle{long_process} = [rectangle, minimum width=4cm, line width = 0.6pt, minimum height=1cm, text centered, text width=4cm, draw=black, fill=orange!30]
\tikzstyle{very_long_process} = [rectangle, minimum width=5cm, line width = 0.6pt, minimum height=1cm, text centered, text width=5cm, draw=black, fill=orange!30]
\tikzstyle{decision} = [diamond, line width = 0.6pt, minimum width=3cm, minimum height=1cm, text centered, aspect=3, draw=black, fill=green!30]
\tikzstyle{arrow} = [thick,->,>=stealth]

\SectionNumbersOn

\author{Shivang Agarwal}
\affiliation[ucla-ece]
{Department of Electrical and Computer Engineering, University of California, Los Angeles}
\author{Daniel R. Kattnig}
\affiliation[uoe]
{Department of Physics and Living Systems Institute, University of Exeter}
\author{Clarice D. Aiello}
\affiliation[ucla-ece]
{Department of Electrical and Computer Engineering, University of California, Los Angeles}
\author{Amartya S. Banerjee}
\email{asbanerjee@ucla.edu}
\affiliation[ucla-mse]
{Department of Materials Science and Engineering, University of California, Los Angeles}

\title[Posner]{The Biological Qubit: Calcium Phosphate Dimers, not Trimers}

\begin{document}








\begin{abstract}
The Posner molecule (calcium phosphate trimer), has been hypothesized to function as a biological quantum information processor due to its supposedly long-lived entangled \ce{^{31}P} nuclear spin states. This hypothesis was challenged by our recent finding that the molecule exists as an asymmetric dynamical ensemble without a well-defined rotational axis of symmetry --- an essential assumption in the proposal for Posner-mediated neural processing. Following up, we investigate here the spin dynamics of the molecule's entangled \ce{^{31}P} nuclear spins within the asymmetric ensemble. Our simulations show that entanglement between two nuclear spins prepared in a Bell state in separate Posner molecules decays on a sub-second timescale --- much faster than previously hypothesized, and not long enough for super-cellular neuronal processing. Calcium phosphate dimers however, are found to be surprisingly resilient to decoherence and are able to preserve entangled nuclear spins for hundreds of seconds, suggesting that neural processing might occur through them instead.
\end{abstract}

The Posner molecule (calcium phosphate trimer, \ce{Ca9(PO4)6}), was first hypothesized to exist in the bone mineral hydroxyapatite \cite{posner1975synthetic}, and has since been identified as the structural unit of amorphous calcium phosphate \cite{treboux2000symmetry}. Its aggregation is thought to underpin bone growth \cite{yin2003biological,du2013structure,dey2010g,wang2012posner}. Although the Posner molecule is yet to be unambiguously experimentally observed in an isolated form, its biochemical relevance has long been recognized. In recent years, Fisher and co-workers have suggested that pairs of isolated Posner molecules could act as ``neural qubits'' by harboring long-lived entangled spin states amongst the twelve \ce{^{31}P} nuclei \cite{fisher2015quantum,swift2018posner,weingarten2016new}. This has been hypothesized to facilitate long-range quantum-correlated release of \ce{Ca^{2+}} ions in pre-synaptic neurons and thus, to potentially give rise to correlated post-synaptic neuron firing \cite{fisher2015quantum}.

Central to the above-mentioned proposal is the supposed \ce{S6}-symmetric arrangement of the molecule, wherein a rotational axis of symmetry allows the binding and unbinding of Posner molecules to act as a ``pseudospin'' entangler for the nuclear spin states \cite{fisher2015quantum,swift2018posner}. However, through an extensive series of first principles simulations, we demonstrated recently  \cite{agarwal2021dynamical} that the Posner molecule does not exhibit the required symmetry and instead exists as a dynamical ensemble of predominantly low-symmetry clusters. Given that all prior works \cite{swift2018posner,player2018posner}, until now, have considered the \ce{S6}-symmetric structure of the Posner molecule to make theoretical predictions on the entanglement times of the nuclear spin states and for examining the viability of the molecule as a potential biomolecular qubit, here, we explore if, and for how long, \ce{^{31}P} nuclear spin coherences can be maintained for the multiple asymmetric configurations found in our  work.

\begin{figure}[ht]
    \centering
    \includegraphics[width=10cm]{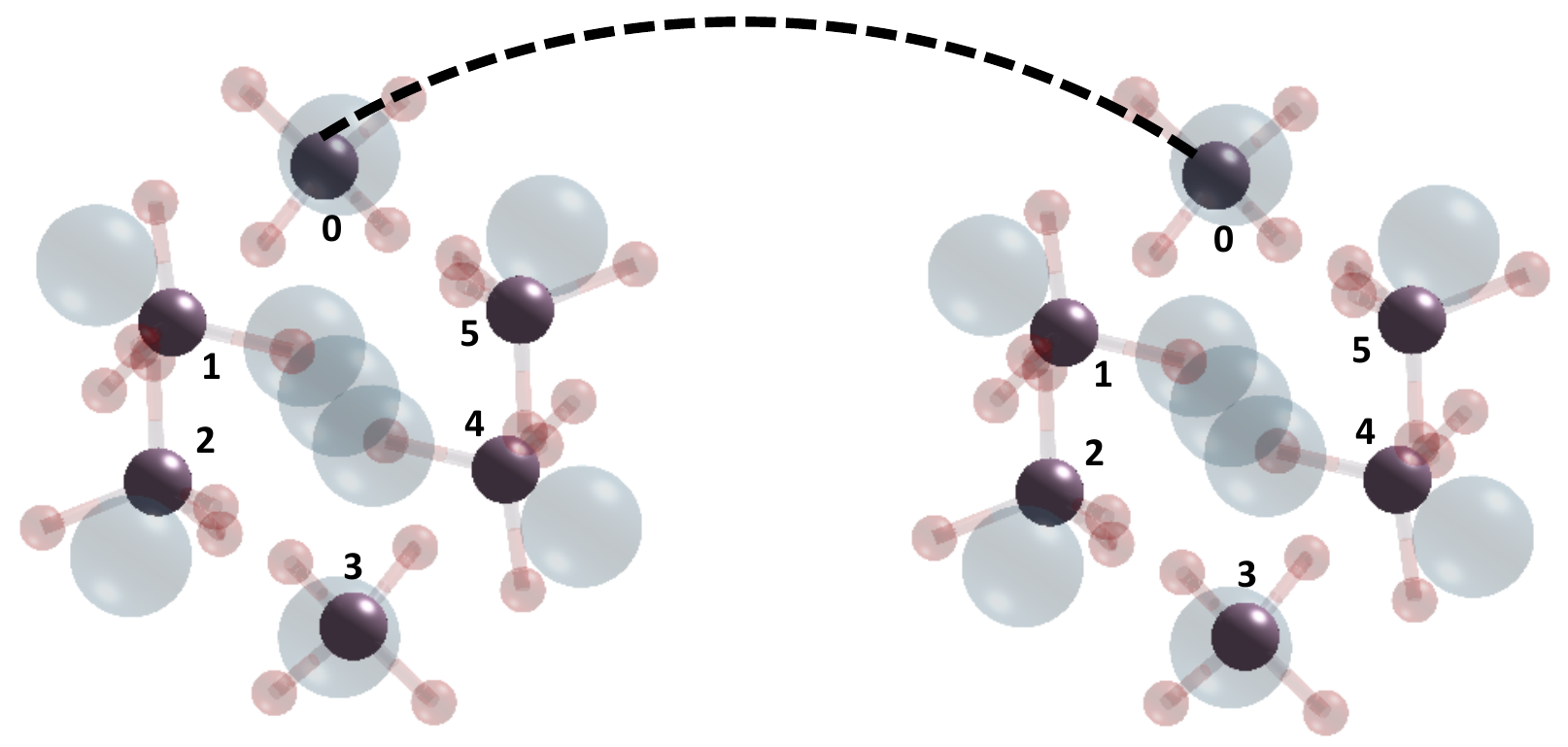}
    \caption{A pair of \ce{^{31}P} nuclei, namely the pair labeled $(0,0)$, entangled and initialized in the singlet state in two separate but identical Posner molecules. The separation between the molecules, although arbitrary, is large enough for intermolecular interactions between the \ce{^{31}P} nuclear spins to be negligible. The entanglement between the nuclear spins has been depicted by a dotted line.}
    \label{fig:entanglement}
\end{figure}

Keeping Fisher's original proposal in mind \cite{fisher2015quantum,swift2018posner}, and closely following the work by Player et al.\ \cite{player2018posner}, one of the simplest measures to assess the pertinent spin coherences is the temporal evolution of the (maximally entangled) singlet state for a pair of \ce{^{31}P} nuclei in identical, spatially-separated Posner molecules, with the remaining ten uncorrelated nuclear spins serving as background (see Fig.~\ref{fig:entanglement}).
In this arrangement, the evolution of the singlet probability over time serves as a measure of conservation of quantum correlation among the Posner molecules, and can be calculated for a variety of structural configurations of the molecule. Eventually, for the thermalized spin system, the singlet probability will settle down to $1/4$ --- the value expected for a maximally mixed state. A value of the singlet probability greater than $1/2$ indicates that entanglement is maintained (\textit{i.e.},\ a mixed state of singlets and unpolarized triplet states in which more than $50\%$ of the spins are in a singlet state, is entangled) \cite{player2018posner}. Additionally, the two-qubit concurrence \cite{wootters1998entanglement, player2018posner} between the \ce{^{31}P} nuclear spin pairs serves as a direct measure of entanglement. A value of one indicates a maximally entangled state; a value of zero indicates total loss of entanglement. A priori, we expect that the absence of symmetry in the Posner molecule will increase the number of unique scalar spin-spin couplings ($J$-couplings) in the coupling network \cite{buckingham1970theory,perras2013symmetry} which, as we explain later, dominate the coherent spin dynamics for this system. A large number of unique coupling constants implies a large number of unique frequencies that characterize the coherent evolution of the singlet probability. In turn, this is expected to accelerate the decay of the singlet probability through destructive interference in our model \cite{player2018posner,annabestani2018dipolar,lidar2012review}. Here, we address if this accelerated decay of spin correlation is realized for all structures of the ensemble, and if this is detrimental to the Fisher's proposal of calcium phosphate-mediated quantum information processing in neurons. Note that our detailed discussion on the link between the symmetry of the Posner molecule and the longevity of the singlet state is motivated by the importance given to the former as a key ingredient for the theory of Posner-mediated neural processing \cite{fisher2015quantum,swift2018posner}. However, as we show later, our calculations on the tricalcium biphosphate dimers and subsequent analysis shown in the Supplementary Information (SI) suggest that while the symmetry of the molecule and its coupling constant values are certainly relevant, the longevity of the singlet state is primarily dictated by the number of coupled nuclear spins, i.e., the number of \ce{^{31}P} atoms.

Using an \ce{S6} molecular point group symmetry for the Posner molecule, which renders the six \ce{^{31}P} nuclei magnetically equivalent, a singlet relaxation time of $37$ minutes has been calculated previously by Player et al.~\cite{player2018posner}. The authors of that work, while refining the original estimate of $21$ days for the entanglement lifetime of nuclear spin states \cite{fisher2015quantum,swift2018posner}, acknowledge that the singlet state may relax even faster due to various interactions present in realistic scenarios. Here, we calculate the singlet probability for a pair of \ce{^{31}P} nuclear spins due to their coherent evolution in the presence of dominant relaxation mechanisms. Our approach closely follows the above reference \citep{player2018posner}, but is extended to explicitly include spin relaxation in the calculation. Unlike the above work, however, which analyzed only a couple of symmetric structures, while using coupling constants from Ref.\ \citenum{swift2018posner} (calculated for a \emph{different} \ce{S6}-symmetric structure), we aim to explore the full ensemble of symmetric and asymmetric structures. To this end, and because the $J$-coupling constants are expected to be different for each structural configuration of the molecule, we estimate the coupling constants for all investigated structures through first principles  calculations. We contend that our calculations and findings are more directly relevant to investigations of the viability of Fisher's proposal, than earlier studies. Details of our Density Functional Theory (DFT) approach used can be found in the SI. 

Lastly, given that the structures of the calcium phosphate monomer, \ce{Ca3(PO4)2}, and the dimer, \ce{Ca6(PO4)4}, are more conclusively known\cite{kanzaki2001calcium,agarwal2021dynamical}, we also performed similar calculations on these structures and found extremely long-lived singlet states, irrespective of the symmetry of the molecule, for the case of the dimer. Their presence \textit{in vivo} has not yet been considered, but is something that invites thorough future investigation. Further details about the monomer and dimer results can be found in the SI.


We consider the coherent evolution of the six \ce{^{31}P} nuclear spins in each Posner molecule subject to the spin Hamiltonian $\hat{H}_0$, where:
\begin{align}
    \hat{H}_0 = \omega_0\sum_{k}\hat{I}_{k,z} + 2\pi\sum_{j<k}\sum_{k} J_{jk}\hat{I}_{j}\cdot\hat{I}_{k} \ .
\label{eq:ref}
\end{align}
Here, $\omega_0=-\gamma_k \left(1-\sigma\right)B$ is the Larmor frequency of the \ce{^{31}P} nucleus; $\sigma$ is the isotropic chemical shielding constant; $\hat{I}_{k}$ is the spin angular momentum operator for nucleus $k$; $\hat{I}_{k,z}$ is the $z$-component of the spin angular momentum operator; and $J_{jk}$ is the scalar coupling constant between nuclear spins indexed $j$ and $k$. The first term accounts for the Zeeman interactions, whereby differences in the chemical shielding will be assumed to be negligible in the low magnetic field considered. The second term corresponds to the intramolecular scalar spin-spin coupling interactions, characterized by the coupling constants $J_{jk}$. Since the evolution of the singlet probability and entanglement of pairs of spins are strongly dependent on the scalar coupling constants, it is critical to obtain accurate values of $J_{jk}$. Moreover, the coupling constants differ for each molecular structure. Here, the $J_{jk}$ values have been derived from DFT calculations for every molecular geometry using the pcJ-n basis set \cite{jensen2006basis}, built specifically for the calculation of these interaction constants, for \ce{P} and \ce{O} atoms. The pcseg-n basis \cite{jensen2015segmented} set, optimized for the calculation of nuclear magnetic shielding, was used for the \ce{Ca} atoms. Further details can be found in the SI.  We remark that the values obtained by us are appreciably different from those reported, and used, in earlier studies \cite{swift2018posner,player2018posner}. Our use of an optimized method and of a more accurate and specialized basis set for each atom gives us confidence in the preciseness of constants employed for the calculations reported here.

Molecular motion modulates the spin Hamiltonian, which induces spin relaxation (details in the SI). Here, we discuss possible relaxation pathways such as intra- and intermolecular dipole-dipole interactions, chemical shielding anisotropy (CSA), dipolar coupling with the solvent, and spin-rotation relaxation. The first of these (i.e.\ intramolecular dipole-dipole interactions) is expected to be the dominant relaxation pathway for rapidly and independently rotating dilute Posner molecules in the geomagnetic field ($\sim 50\;\mu$T). Based on a simple estimate of spin-lattice relaxation times due to the modulation of the dipolar coupling by translational and rotational diffusion, the (concentration-dependent) intermolecular dipolar relaxation is expected to be slower than the intramolecular relaxation for concentrations of up to 7 mol/L (!). The CSA contribution is generally relevant in strong magnetic fields and can be considered negligible in the geomagnetic field, as is the case here (our estimates suggest a marked influence of CSA relaxation only for fields larger than $1$ T). Any dipolar coupling with the solvent is expected to average out to zero in the presence of rapid, independent tumbling motion of the molecule. Finally, spin-rotation relaxation, which is thought to be the dominant relaxation mechanism for pyrophosphates \cite{korenchan202131,korenchan2022limits}, is neglected here under the assumption that the standard deviation in the angular velocity is expected to be much smaller for the larger Posner molecule. Of all these pathways then, the intermolecular contribution is clearly negligible at reasonable concentrations, and the dominant relaxation is induced by the $15$ pair-wise intramolecular dipolar nuclear spin couplings, the detailed form of which can be found in the SI. 

We calculate and evolve the singlet state as follows. Indexing the \ce{^{31}P} nuclear spins in two separate Posner molecules, $A$ and $B$, from $0_{(A,B)}$ through $5_{(A,B)}$, we assume that, without loss of generality, the system is initialized in a state with spins $0_A$ and $0_B$ in the maximally entangled singlet state, while the ten other spins are uncorrelated. We represent this singlet state as $|S_{0_{A},0_{B}}\rangle = \frac{1}{\sqrt{2}}\left(|\alpha_{0_{A}}\beta_{0_{B}}\rangle - |\beta_{0_{A}}\alpha_{0_{B}}\rangle\right)$, where $|\alpha\rangle$ is the spin-up state, and $|\beta\rangle$ is the spin-down state. Thus, the initial state density operator is proportional to the singlet projection operator, given as:
\begin{align}
\begin{split}
    \hat{P}_{0_{A},0_{B}}&=\sum_{\gamma\neq(0_{A},0_{B})}|S_{0_{A},0_{B}}\text{;}\;\gamma\rangle\langle S_{0_{A},0_{B}}\text{;}\;\gamma| \\
    &=|S_{0_{A},0_{B}}\rangle\langle S_{0_{A},0_{B}}|\otimes \mathbbm{1} \,.
\end{split}
\end{align}
Here $|S_{0_{A},0_{B}};\gamma\rangle=\frac{1}{\sqrt{2}}\left(|\alpha_{0_{A}}\beta_{0_{B}}\rangle - |\beta_{0_{A}}\alpha_{0_{B}}\rangle\right)\otimes|\gamma\rangle$, and $|\gamma\rangle$ is any of a set of states such that $\sum_{\gamma\neq(0_{A},0_{B})}|\gamma\rangle\langle\gamma| = \mathbbm{1}$, \textit{i.e.}, it assembles the maximally mixed spin state of the other $10$ nuclear spins. The singlet probability $p_{0_{A},0_{B}}(t)$ is given by $\text{Tr}\left[\hat{\rho}(t)\hat{P}_{0_{A},0_{B}}\right]$ with $\hat{\rho}(0)=\hat{P}_{0_{A},0_{B}}/\text{Tr}(\hat{P}_{0_{A},0_{B}})$. To obtain the evolution of $p_{0_{A},0_{B}}(t)$ through time, we solve the Liouville-von Neumann equation \cite{breuer2002theory}:
\begin{align}
    \frac{d\hat{\rho}(t)}{dt}=-i\hat{\hat{L}}\hat{\rho}(t) \ ,
\end{align}
where the Liouvillian superoperator $\hat{\hat{L}}$ is given by $\hat{\hat{H}}_0 + i\hat{\hat{\Gamma}}$, with the relaxation superoperator $\hat{\hat{\Gamma}}$ given, in the extreme narrowing limit applicable to fast rotational motion, by $-\langle\hat{\hat{H}}_1(t)\hat{\hat{H}}_1(t)\rangle\tau_{c}$ \cite{player2018posner}. Here, double hats denote the commutator superoperator of the operators in question, and $\tau_c$ is the rotational correlation constant of the molecule. 

To validate the form of the relaxation superoperator, we have calculated the rotational correlation constant of the molecule from molecular dynamics (MD) simulations. Closely following the work by Demichelis et al.\ \cite{demichelis2018simulation}, we simulated the Posner molecule in a box of water molecules and followed its trajectory over $34$ ns (i.e., at the simulation limits of computational resources available), using LAMMPS \cite{thompson2022lammps}. Further details about the simulation can be found in the SI. A value of $\tau_c \sim 177$ ps was obtained by analyzing the rotational correlation function in 3 dimensions. Thus, we find a slower rotational dynamics (by roughly a factor of 3) compared to the previous estimate by Player et al.\ \cite{player2018posner}, which uses the approximate Stokes-Einstein-Debye relation. The value derived here is still well within the extreme narrowing limit, thus justifying the form of $\hat{\hat{\Gamma}}$. Details about the calculation of the two-qubit concurrence \cite{wootters1998entanglement,player2018posner}, $\mathcal{C}_{0_A,0_B}$, can be found in the SI.

\begin{figure}[!ht]
    \centering
    \includegraphics[width=15cm]{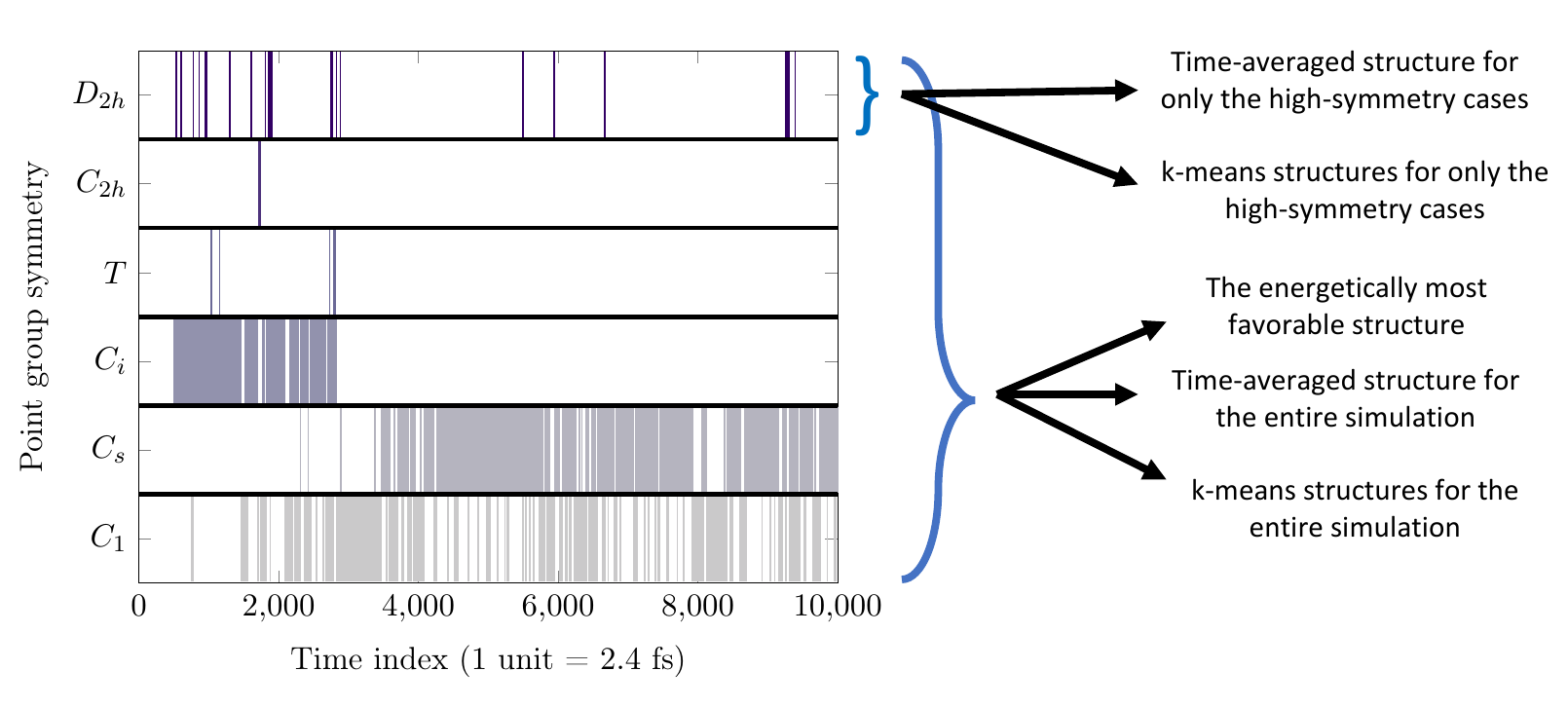}
    \caption{Illustration of the methodology for obtaining different configurations for spin dynamics calculations, starting from the AIMD structures presented in Ref.~\citenum{agarwal2021dynamical}. Here, we only show the type of structures obtained from one AIMD simulation (out of the eight that were performed). Overall, our approach gave us $102$ unique structures. The graph on the left shows the evolution of the molecular point-group symmetry during a dynamic simulation of a transition-state structure of the Posner molecule.}
    \label{fig:AIMD_structs}
\end{figure}

With this computational framework for spin relaxation calculations in hand, we build upon our previous work \cite{agarwal2021dynamical}, wherein we demonstrated that the Posner molecule exists as an asymmetric dynamic ensemble. \textit{Ab initio} molecular dynamics (AIMD) simulations of eight transition state structures of the molecule formed the core of that previous contribution. Here, we examine the singlet probability and concurrence for not only the above transition state structures, but also for the time-averaged structures of each of the eight AIMD simulations. Additionally, to cover a wider range of possibly relevant structures, we also calculate here the singlet probability for the energetically most favorable structures from each simulation, and the time-averaged configurations for structures associated with high-symmetries, as shown in Fig.~\ref{fig:AIMD_structs}. Lastly, we also use \textit{k}-means clustering to further generate statistically relevant structures --- both for the entire simulation and for the high-symmetry cases only --- and calculate the singlet probabilities and concurrence for all of them. This thorough examination gives us over 100 unique structures to analyze.

\begin{figure}[!ht]
    \begin{subfigure}[b]{0.3\textwidth}
    \centering
    \includegraphics[width=\textwidth]{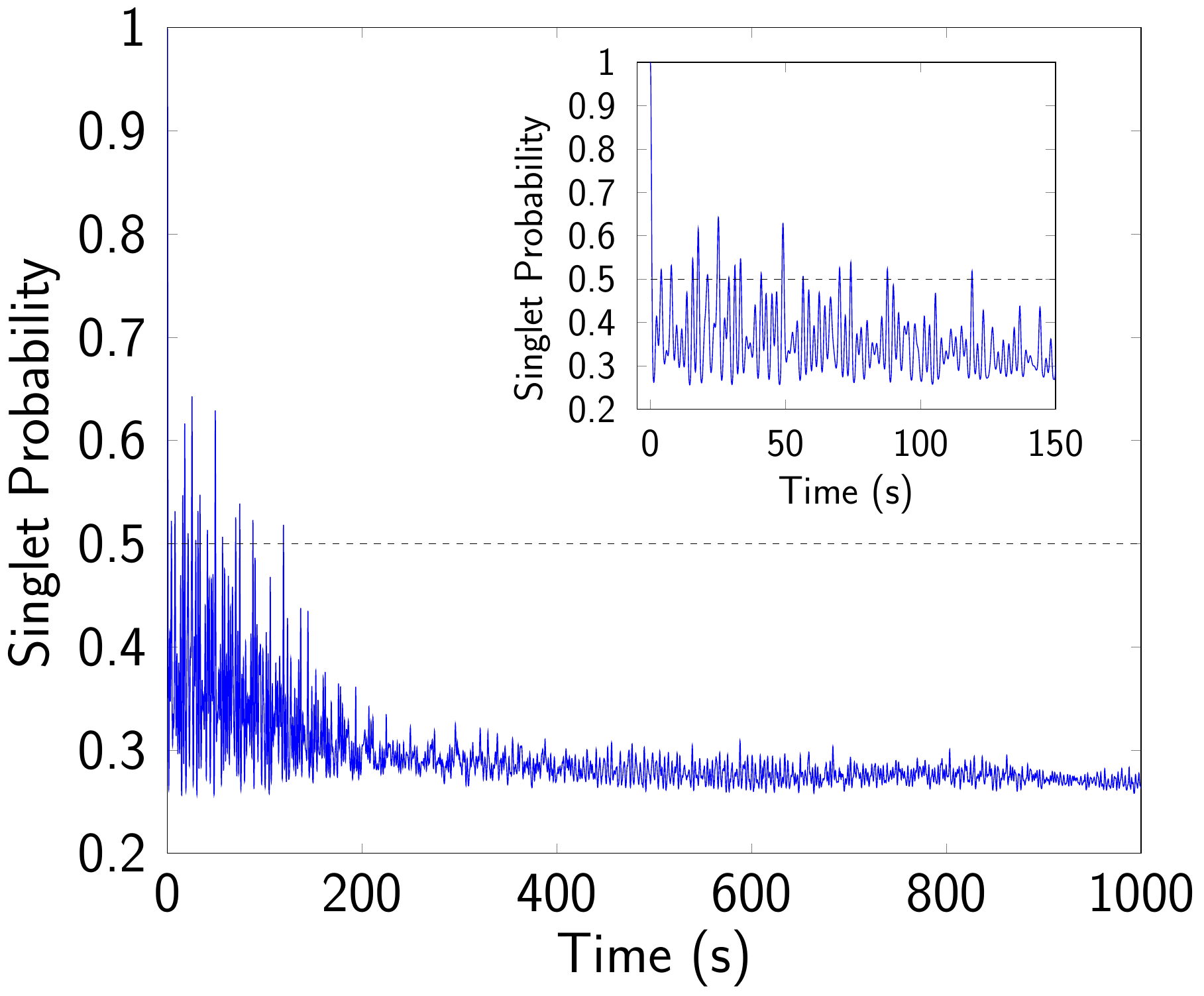}
    \caption{The longest-lived singlet state occurs for a \textit{k}-means structure of high symmetry.}
    \label{fig:max_sing}
    \end{subfigure}
    \hspace{0.25cm}
    \begin{subfigure}[b]{0.3\textwidth}
      \centering
      \includegraphics[width=\textwidth]{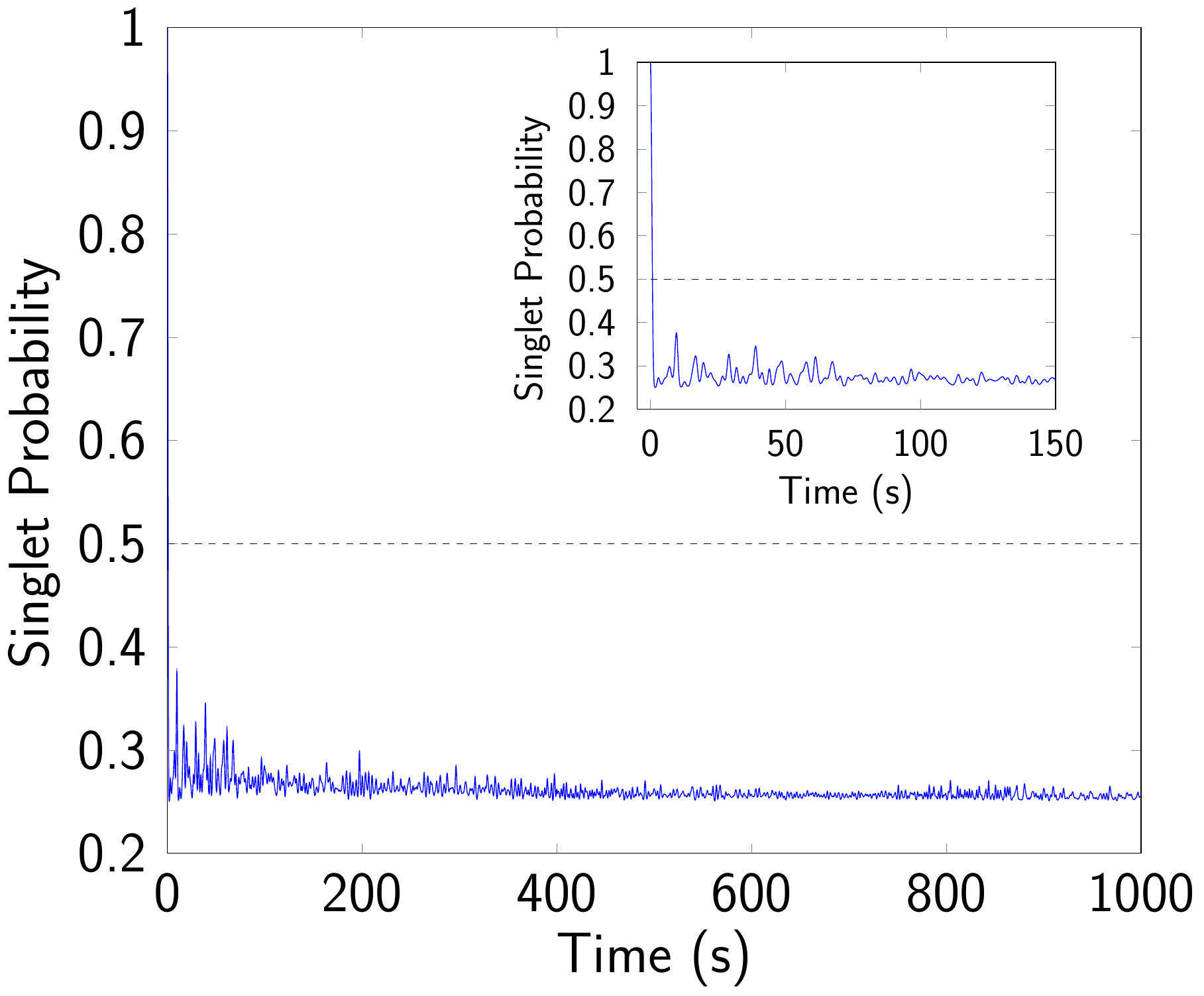}
      \captionsetup{font={small}}
      \caption{The singlet probability for the energetically most stable structure. \cite{agarwal2021dynamical}}
      \label{fig:cs_ms_sp}
    \end{subfigure}
    \hspace{0.25cm}
    \begin{subfigure}[b]{0.3\textwidth}
      \centering
      \includegraphics[width=\textwidth]{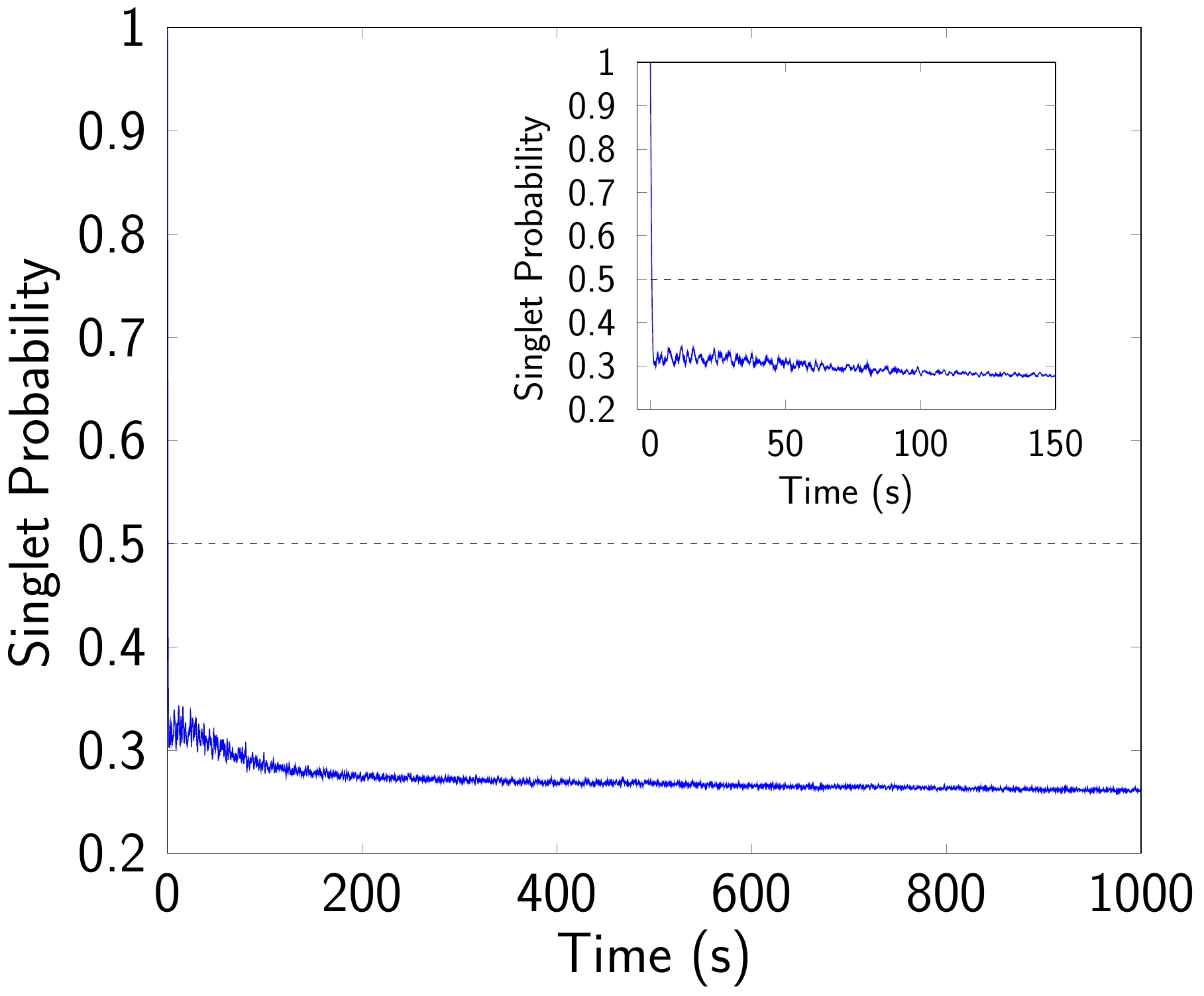}
      \captionsetup{font={small}}
      \caption{Averaged singlet probability for all 102 cases considered in the study.}
      \label{fig:sing_avg}
    \end{subfigure} \\ \vspace{0.3cm}
    \begin{subfigure}[b]{0.3\textwidth}
      \centering
      \includegraphics[width=\textwidth]{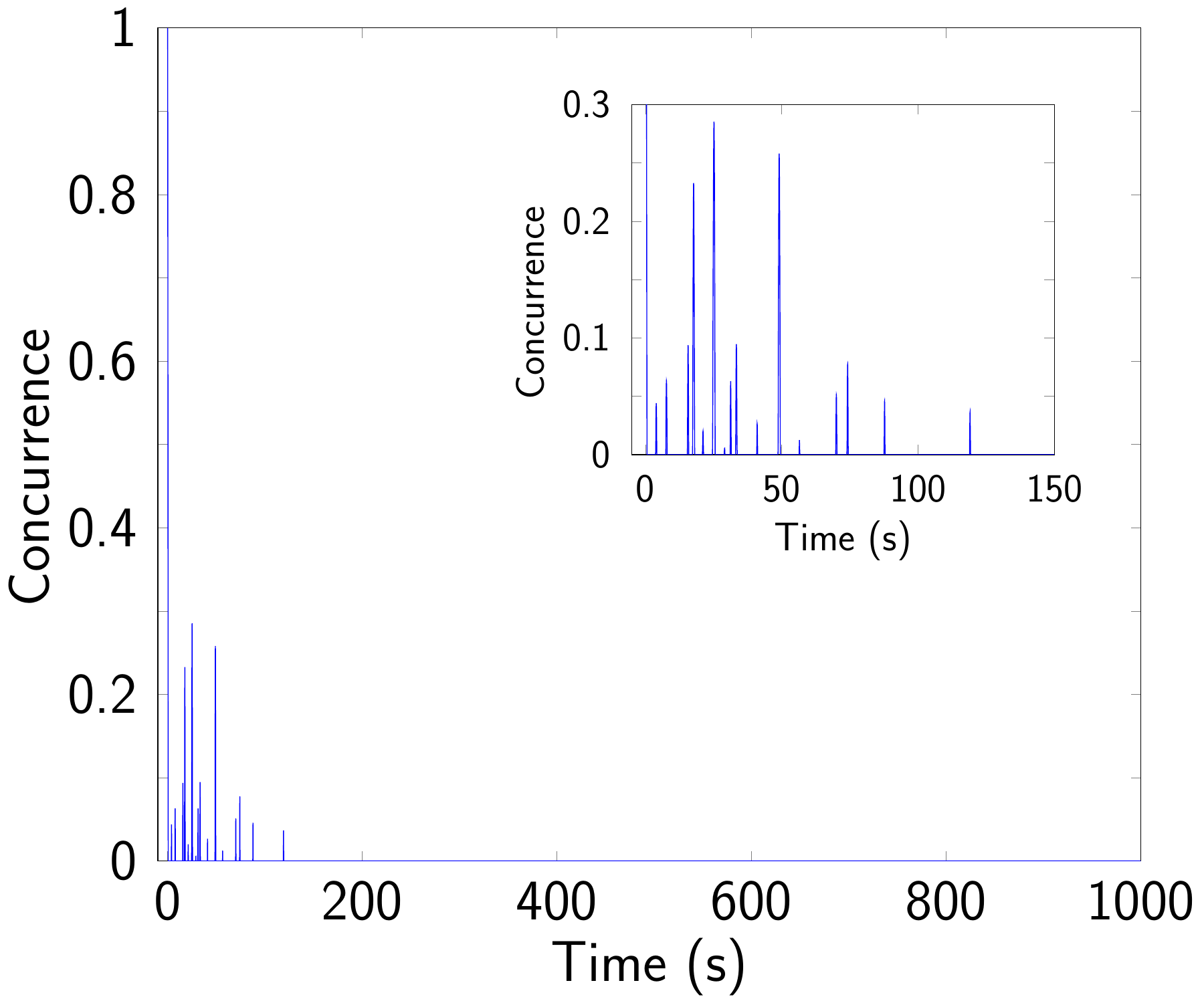}
      \captionsetup{font={small}}
      \caption{Concurrence values for even the longest-lived singlet state are extremely low.}
      \label{fig:max_conc}
    \end{subfigure}
    \hspace{0.25cm}
    \begin{subfigure}[b]{0.3\textwidth}
      \centering
      \includegraphics[width=\textwidth]{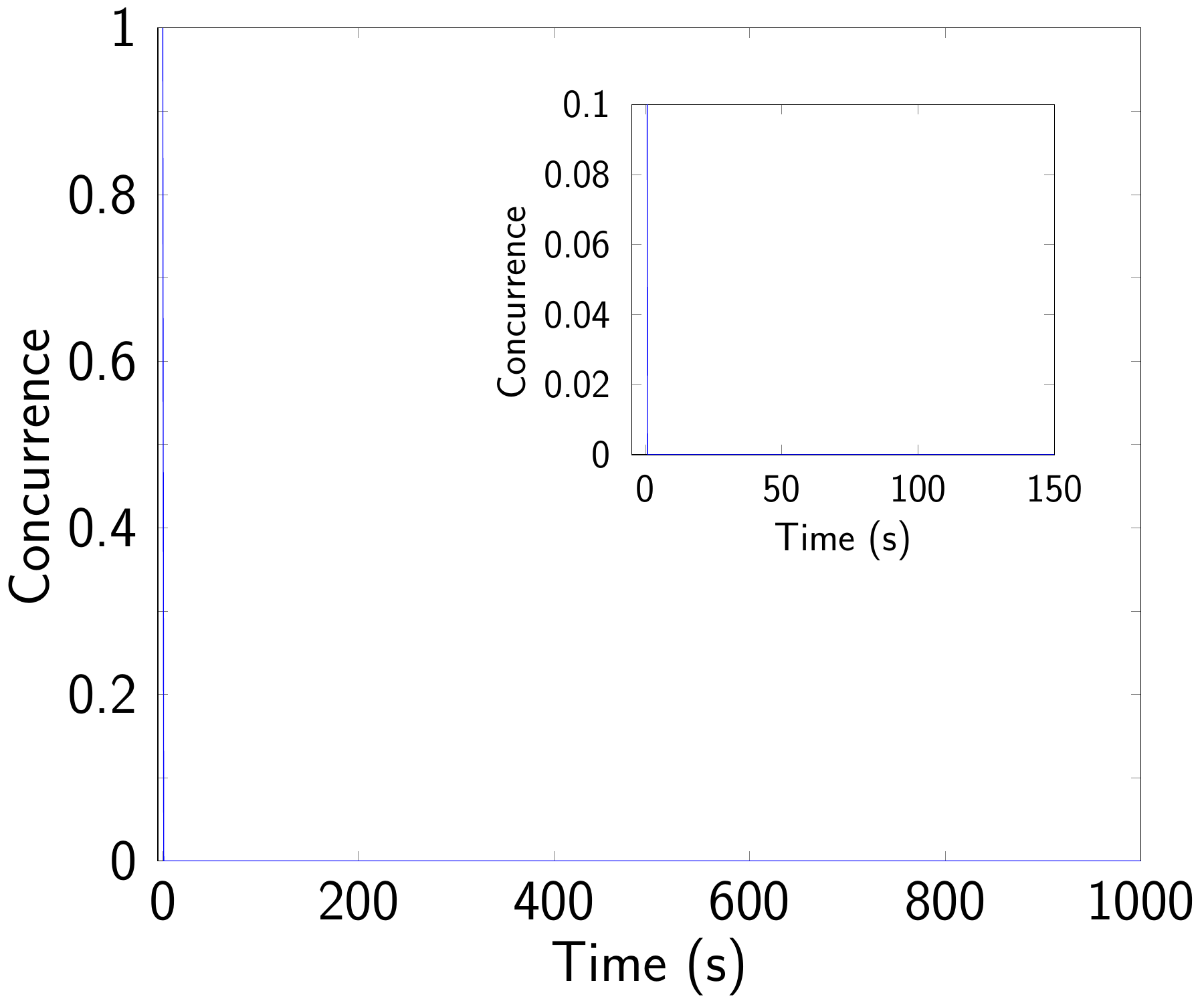}
      \captionsetup{font={small}}
      \caption{Concurrence for the energetically most favorable structure.}
      \label{fig:cs_ms_conc}
    \end{subfigure}
    \hspace{0.25cm}
    \begin{subfigure}[b]{0.3\textwidth}
    \centering
    \includegraphics[width=\textwidth]{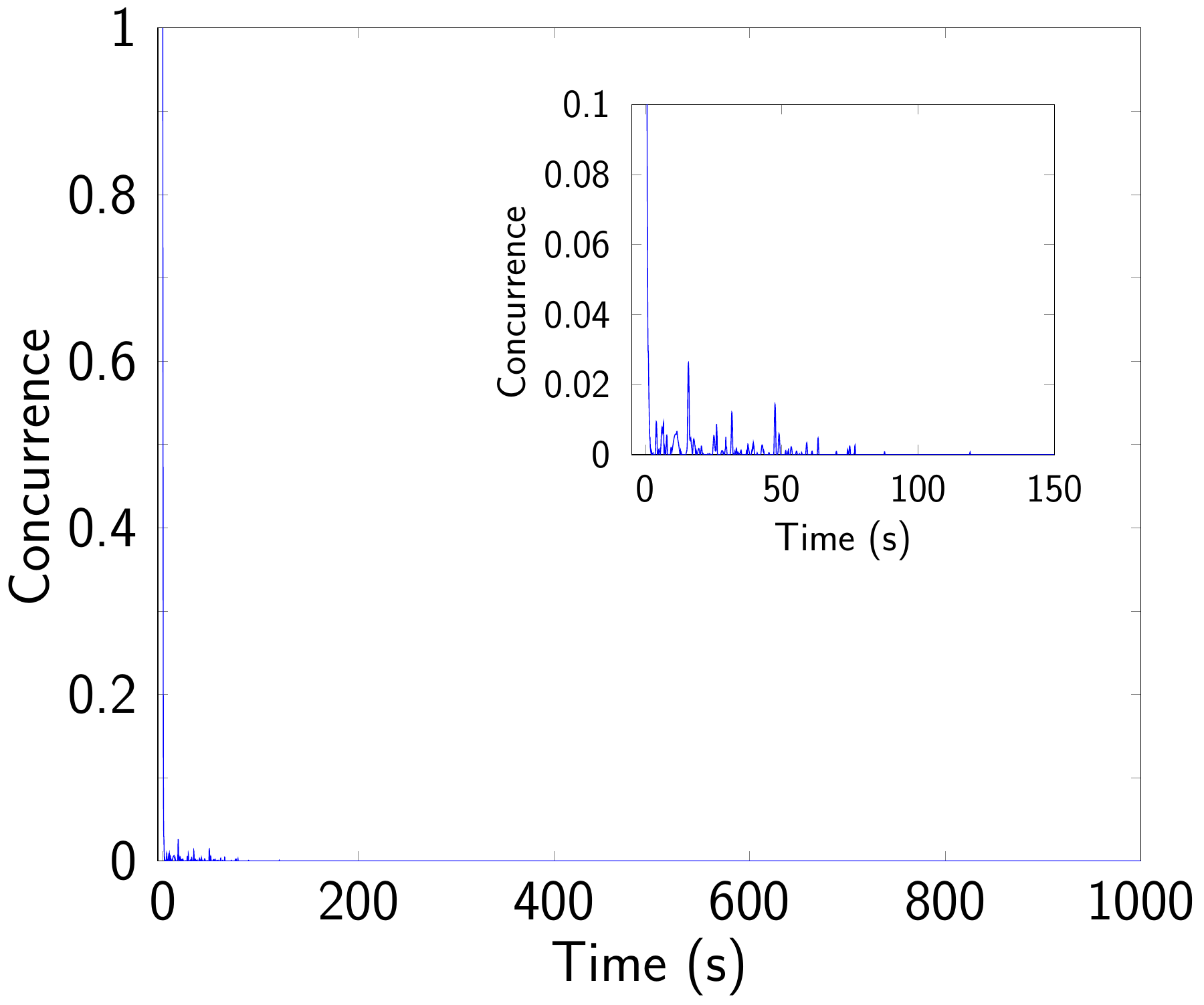}
    \caption{Averaged concurrence for all 102 cases considered in the study.}
    \label{fig:avg_conc}
    \end{subfigure}
    \caption{The short-lived nature of the singlet probability and concurrence for the Posner molecule, shown for three representative structures. Insets show zoomed in regions of the plots. (a) and (d) correspond to the structure with the longest-lived singlet state --- a \textit{k}-means structure with \ce{D_{2h}} symmetry; (b) and (e) correspond to the energetically most stable structure with no symmetry in our dynamical study \cite{agarwal2021dynamical}; (c) and (f) are the average singlet probability and concurrence for all $102$ structures considered. Contrary to previous studies \cite{swift2018posner,fisher2015quantum,player2018posner}, the plots above suggest that the system is not suitable for quantum information processing at biologically relevant time scales.}
    \label{fig:sing_and_conc}
\end{figure}

\begin{figure}[h!]
  \centering
  \begin{tabular}{c}
  \scalebox{0.65}{
  \includegraphics[width=0.77\textwidth]{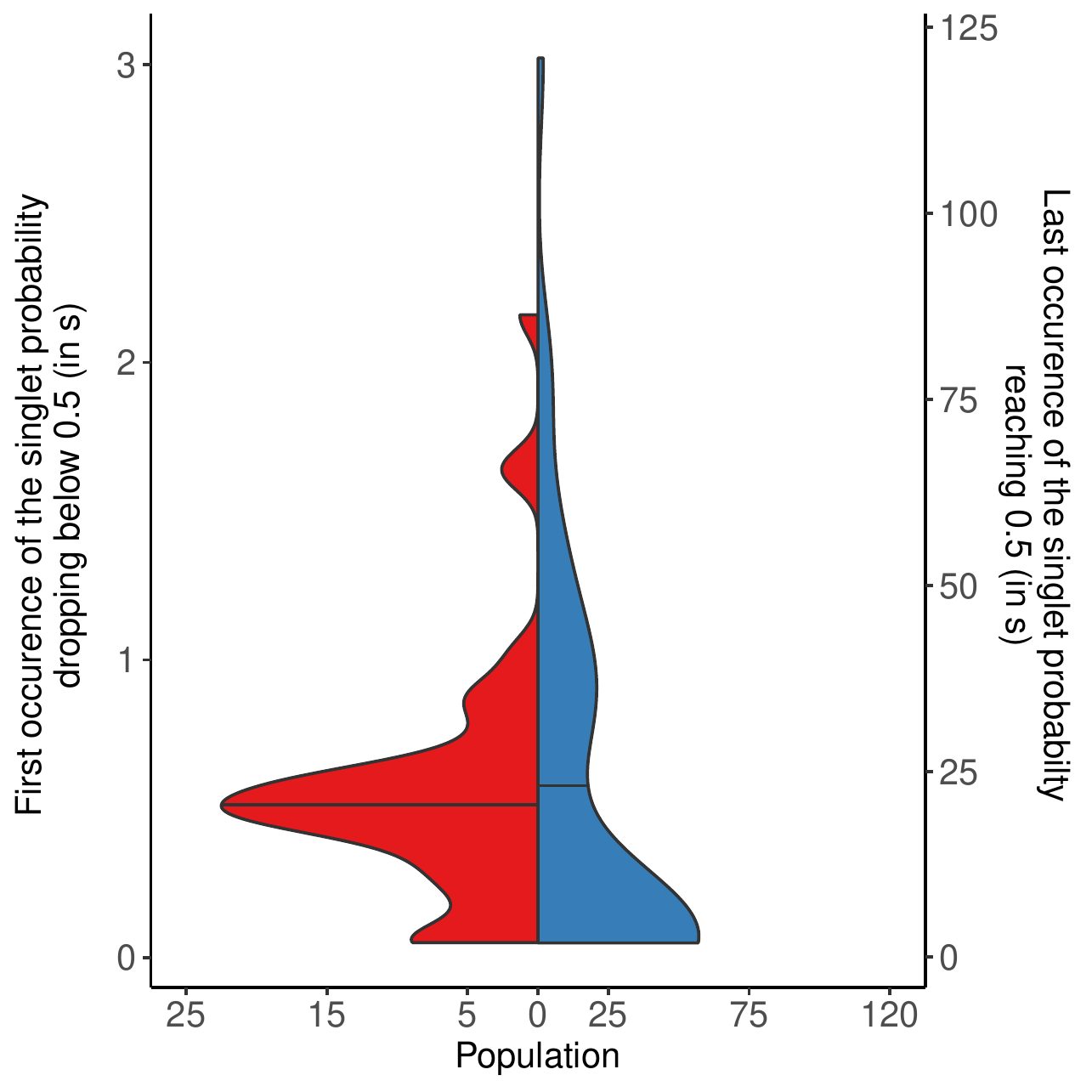}  
  }
  \end{tabular}
  \caption{The first and last instances of the singlet probability crossing the threshold of $1/2$. The number of structures for each case, with different scales, has been represented on the \textit{x}-axis. Note that the \textit{y}-axes have different scales. A total of $102$ structures have been considered, obtained from methods depicted in Fig.~\ref{fig:AIMD_structs}. The $50\%$ quantile line has been shown in each case, and confirms that a majority of the structures are unable to maintain the singlet state for more than a second.}
  \label{fig:sing_hist}
\end{figure}

 We show the time evolution of the singlet probability and concurrence for two representative cases in Fig.~\ref{fig:sing_and_conc} --- namely the longest-lived singlet state (Figs.\ \ref{fig:max_sing} and \ref{fig:max_conc}), and the energetically most stable structure as per our dynamical analysis \cite{agarwal2021dynamical} (Figs.\ \ref{fig:cs_ms_sp} and \ref{fig:cs_ms_conc}). Given that the Posner molecule is expected to exist as a dynamic ensemble, we also show the singlet probability and the concurrence averaged over all $102$ cases described above (Figs.\ \ref{fig:sing_avg} and \ref{fig:avg_conc}). We immediately observe that, in general, the singlet probabilities decay at a rate much faster than what may be necessary for neural processing \cite{fisher2015quantum,swift2018posner}, and fall below the $1/2$ entanglement threshold within a second. Correspondingly, the concurrence plots also suggest that the system loses entanglement within a second. Note that the plots presented in this study are for a model incorporating spin relaxation (in contrast with those presented in Player et al.'s work \cite{player2018posner}), although spin relaxation is a minor contributor to the decay of the singlet probability on the timescales shown. For a majority of the cases, as shown in Fig.~\ref{fig:sing_hist}, the trend of Fig.~\ref{fig:cs_ms_sp} is followed. Overall, it is important to emphasize that the singlet probability rapidly decays below the $1/2$ threshold within a second, regardless of the symmetry of the Posner molecule. Moreover, in cases where the singlet state is longer-lived, it is not entangled throughout, \textit{i.e.}, the singlet probability decays quickly and then refocuses sporadically. This has major implications on the viability of the molecule as a quantum information processor, because any information processing will have to be done only at the instances when the system is entangled, \textit{i.e.}, when the singlet probability is greater than $1/2$. In other words, the molecule may only act as a biological qubit for a very short periods of time, or only at specific time instances.

The longest time for which a singlet state was sustained in the presence of relaxation was a recurrence at $119$ seconds (Fig.~\ref{fig:max_sing}), observed in one of the structures obtained using \textit{k-}means clustering of the high-symmetry phase of a dynamic simulation (\ce{D_{2h}} point group symmetry). Note that this time is defined as the last instance when the singlet probability was above the threshold of $1/2$, and that it does not correspond to the the singlet lifetime. A comprehensive overview of the relevant singlet probability parameters for the structures examined in this study is provided in Fig.~\ref{fig:sing_hist}. It is clear that, for $95$\% of the cases, the singlet state loses entanglement in less than a second. While a few structures do show recurrence of entanglement at time scales of tens of seconds, we reiterate that the molecule may act as a biological qubit at only the instances when the probability breaches the $1/2$ --- threshold, and not throughout. Additionally, we also consider the transfer of entanglement from one pair of \ce{^{31}P} nuclear spins to another, owing to their interactions via the $J-$coupling constants. Unsurprisingly, the singlet probabilities for all spin pairs remains low ($\sim 0.25$) throughout the considered time duration (further details can be found in the SI).

\begin{figure}[h!]
\centering
\subfloat[First instances of the singlet probability dropping below $1/2$ versus the number of symmetry operations in molecular point group.]{\scalebox{0.4}{
\includegraphics[width=1.1\textwidth]{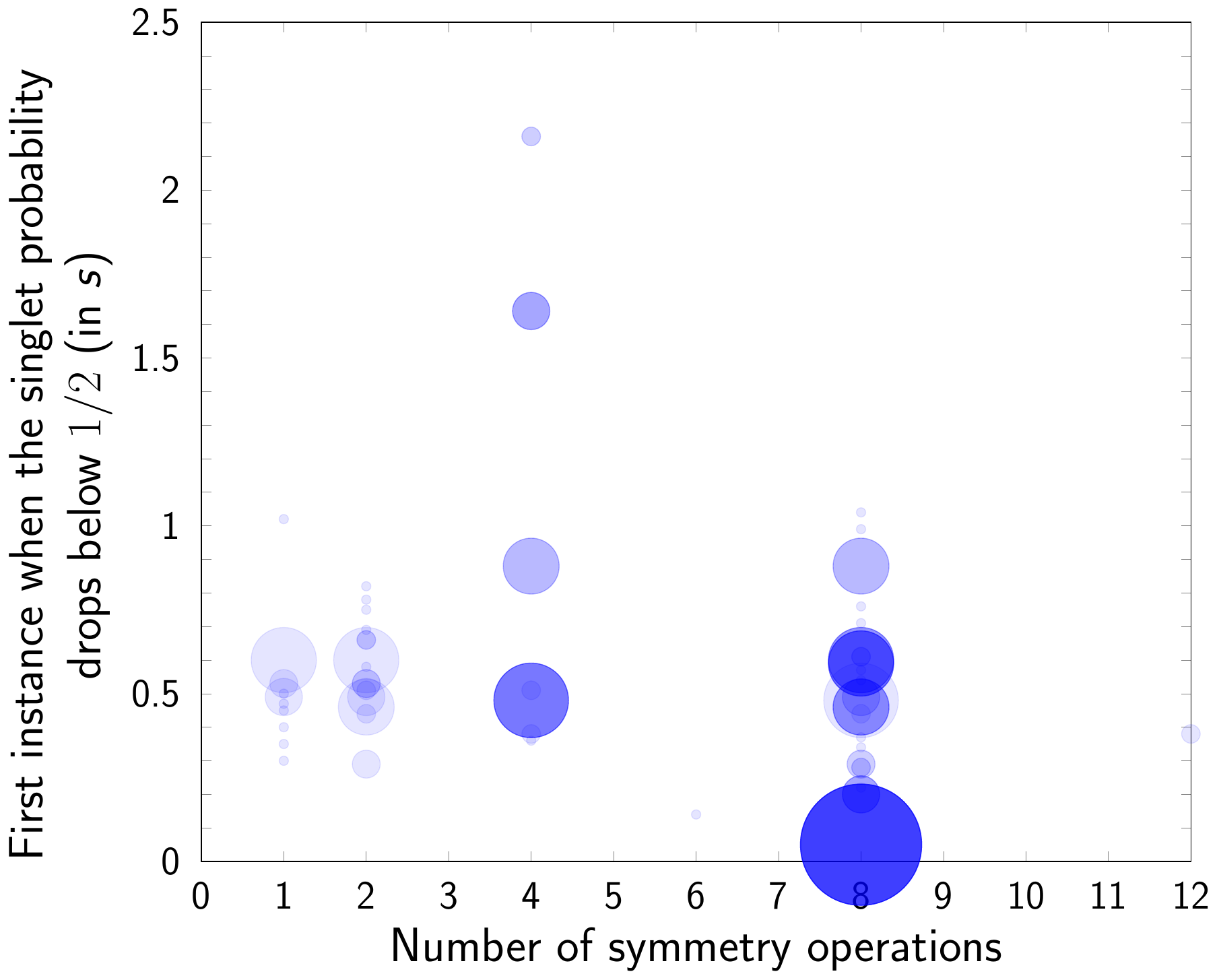}
}}
\hspace{1cm}
\subfloat[Last instances of the singlet probability being above $1/2$ versus the number of symmetry operations in the molecular point group.]{\scalebox{0.4}{
\includegraphics[width=1.1\textwidth]{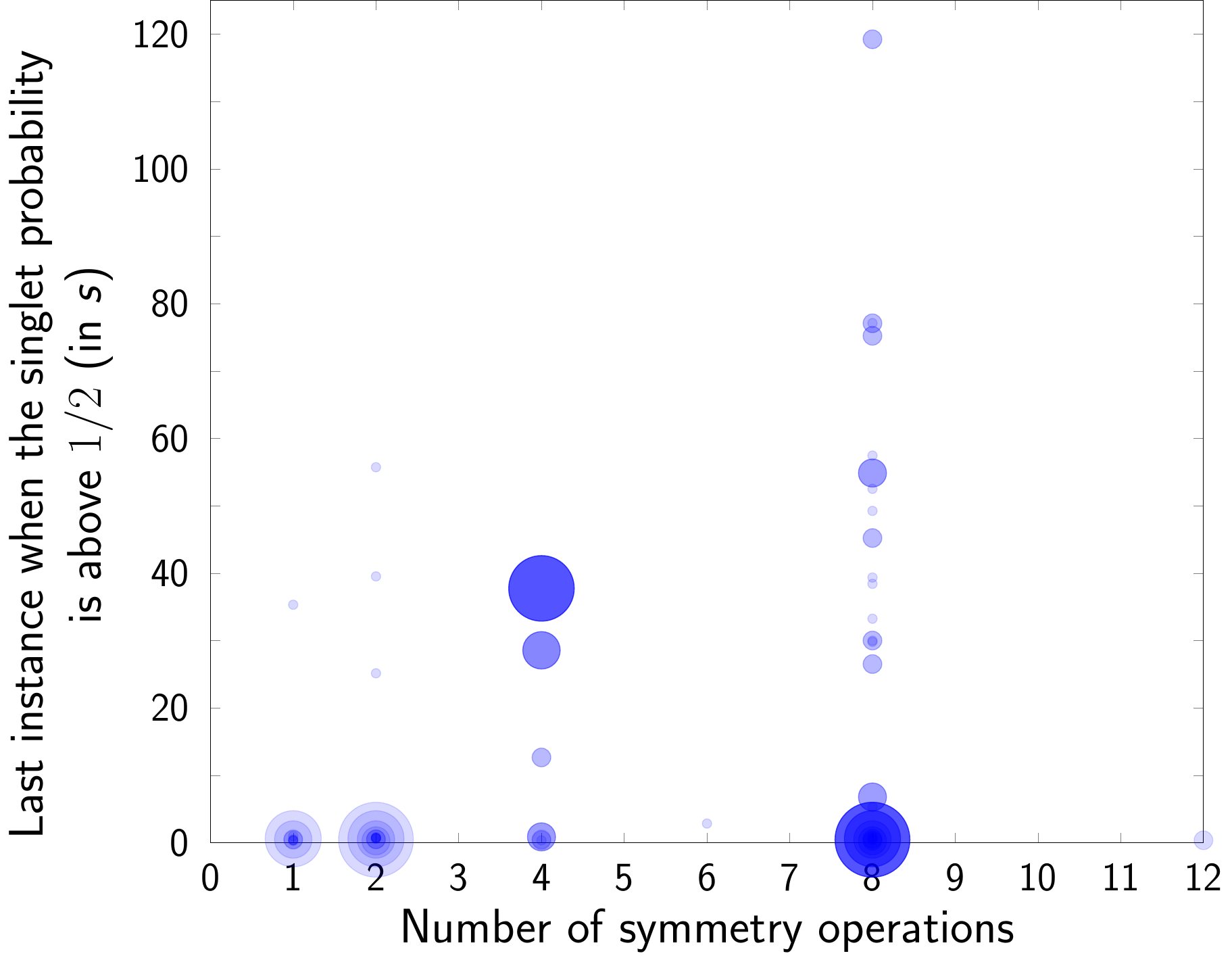}
}}
\caption{The symmetry of the molecule does not seem to play a major role in maintaining entanglement in the system, and Posner molecules generally have short entanglement lifetimes ($< 1$ s) irrespective of their point-group symmetry. The marker size in each plot is proportional to the number of data points at the location, whereas darker areas represent overlapping of markers.}
\label{fig:sym_ops}
\end{figure}

The above analysis suggests that the Posner molecule might not maintain long-lived singlet states for more than a second. However, Fig.~\ref{fig:max_sing} suggests that in a hypothetical high symmetry configuration, the molecule may sustain the nuclear singlet state for a longer duration, albeit only at specific instances when the singlet probability refocuses. Thus, to better understand the effect of symmetry on the longevity of the nuclear spin singlet state in the Posner molecule, we plotted the first and last instances when the singlet probability crosses the $1/2$ threshold against the point group symmetry of the molecule or, more specifically, the number of symmetry operations associated with the point group. This has been depicted in Fig.~\ref{fig:sym_ops}.

\begin{figure}[h!]
\centering
\includegraphics[width=0.5\textwidth]{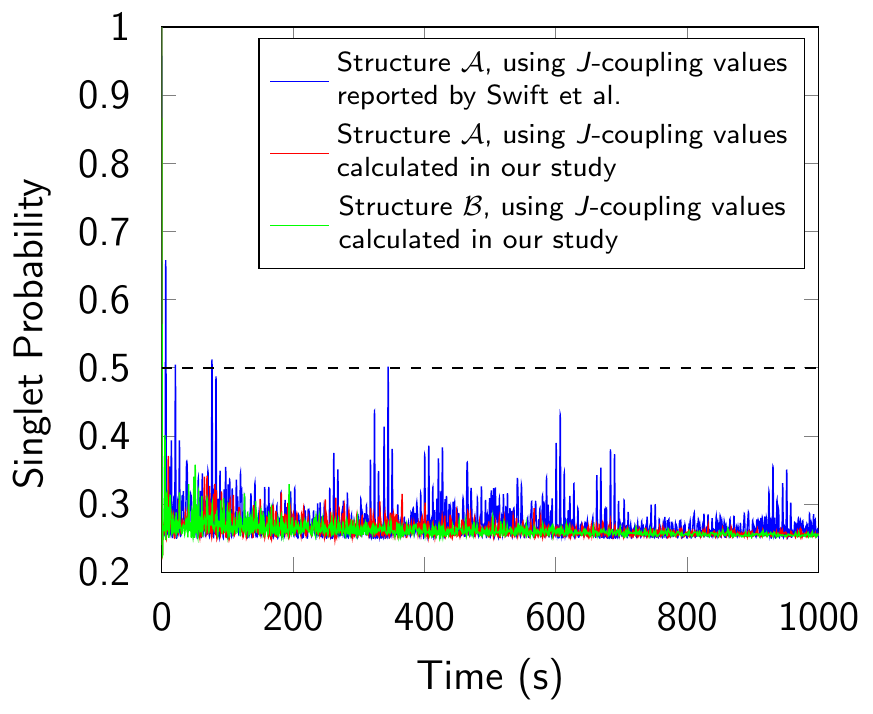}
\caption{A comparison of the singlet probabilities for two \ce{S6} structures --- $\mathcal{A}$, taken from Ref.\ \citenum{player2018posner}, and $\mathcal{B}$, taken from Ref.\ \citenum{swift2018posner} --- with coupling constants as reported in Ref. \citenum{swift2018posner} and as calculated in our study.}
\label{fig:comp}
\end{figure}

Contrary to the expectation that the entanglement lifetime would be longer for structures that have higher symmetry, we find that the singlet state in the Posner molecule is, on average, short-lived (up to a few seconds at best) irrespective of the symmetry. We would also like to mention in passing that in addition to the $102$ structures described above arising out of our AIMD simulations and subsequent analyses, we considered the two specific Posner molecule configurations explored in Ref.\ \citenum{swift2018posner} and Ref.\ \citenum{player2018posner}. These configurations were not found to be energetically or structurally relevant in our dynamical simulations \citep{agarwal2021dynamical}. However, for these structures too, we found that the singlet probability falls below the $1/2$ threshold within a second, and without any refocusing. On the other hand, if we make use of the \textit{J}-coupling constants as suggested in Ref.\ \citenum{swift2018posner} for one of these structures, our calculation broadly reproduces the results from Ref.\ \citenum{player2018posner}, albeit with minor differences due to slightly different values of the chemical shielding tensor and the rotational correlation time. A comparison of the singlet probabilities in the above cases has been depicted in Fig.~\ref{fig:comp}. Note that this comparison has only been done for two specific molecular configurations of the Posner molecule, and the results should not be generalized to other structures. Our observation above is consistent with the fact that, on average, symmetric molecules are expected to have a better entanglement yield. However, as we show in the SI, the values of the \textit{J}-coupling constants matter in the sense that it is possible for a molecular configuration to have a wide range of entanglement yields, depending on the calculated values of those constants. Therefore, it is important to be able to calculate these constants accurately.

Overall, we observe that the singlet state between a pair of \ce{^{31}P} nuclear spins in separate Posner molecules generally loses entanglement within a second. Only certain hypothetical or energetically unfavorable configurations of the molecule are found to preserve the state for more than a second. This is significantly shorter than previously suggested \cite{swift2018posner, player2018posner, fisher2015quantum}. Commensurate with these findings, the concurrence values are also observed to drop rapidly within the same period of time. Notably, these findings do not preclude the molecule from being relevant for biological processes taking place at time scales for which the nuclear spin coherences are indeed maintained. However, due to the small diffusion constant of the molecule, it is unable to traverse neuronally relevant length-scales within its short entanglement lifetime, and so it might not be suitable for entanglement mediated neural signal processing, as previously suggested \cite{fisher2015quantum}. Specifically, we estimated the translational diffusion properties of the molecule by analyzing its trajectory in a hydrated environment (details in the SI) and found the diffusion constant to be $\sim 1.01\times10^{-9}$ $\text{m}^2$/\;s. This implies that it would take the Posner molecule approximately $14$ hours to traverse the length of a neuronal cell axon ($\sim 1$ cm), much longer than the aforementioned entanglement lifetimes. We also note that the above estimated speeds are markedly slower than the average speed of nerve impulses \cite{nuwer2017monitoring}. 

\begin{figure}[h!]
    \centering
    \begin{subfigure}[b]{0.45\textwidth}
      \centering
      \includegraphics[width=\textwidth]{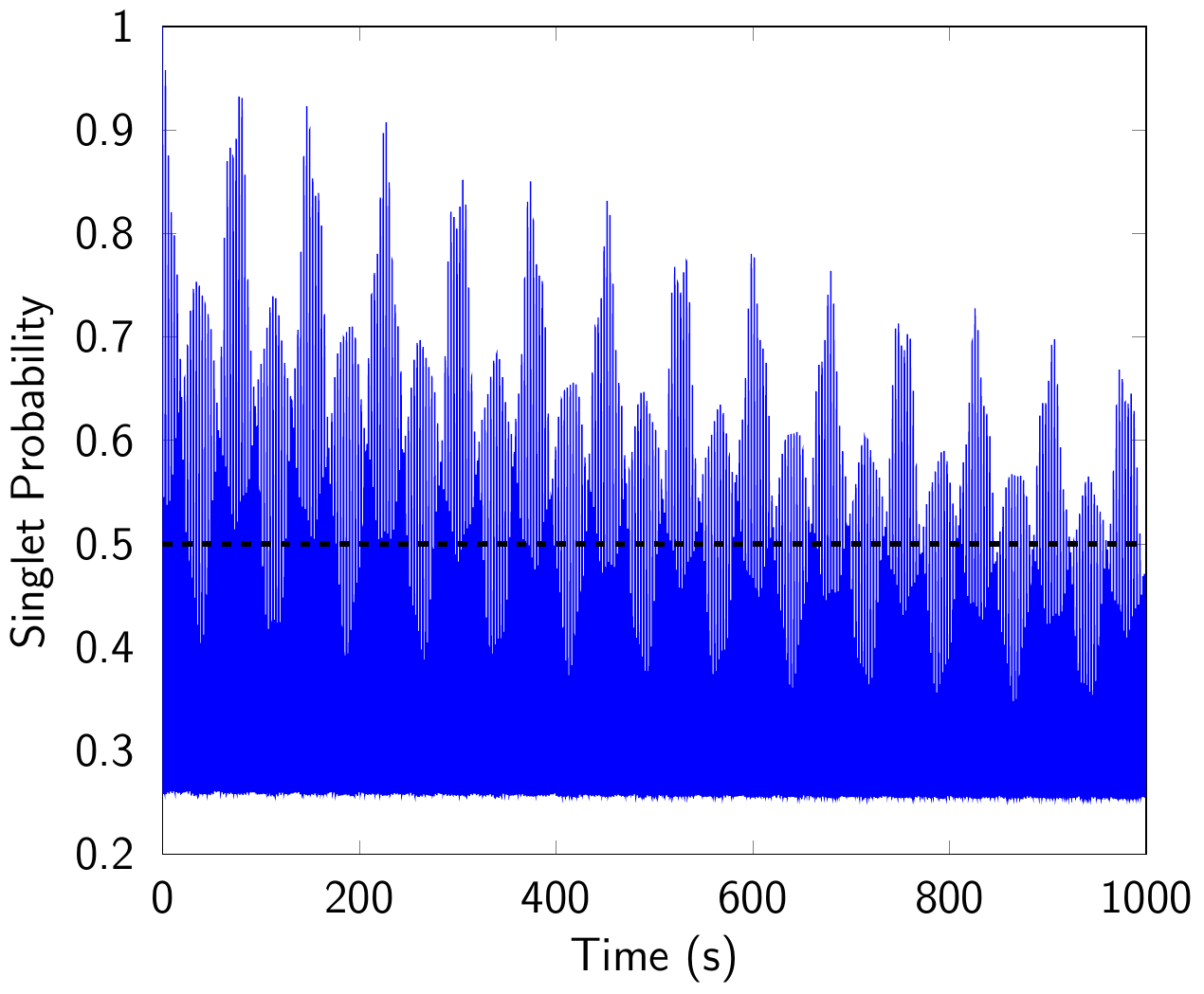}
      \captionsetup{font={small}}
      \caption{Singlet probability for a dimer structure with a \ce{C2} symmetry indicates a very long-lived coherence.}
      \label{fig:Td_sp}
    \end{subfigure}
    \hspace{1cm}
    \begin{subfigure}[b]{0.45\textwidth}
      \centering
      \includegraphics[width=\textwidth]{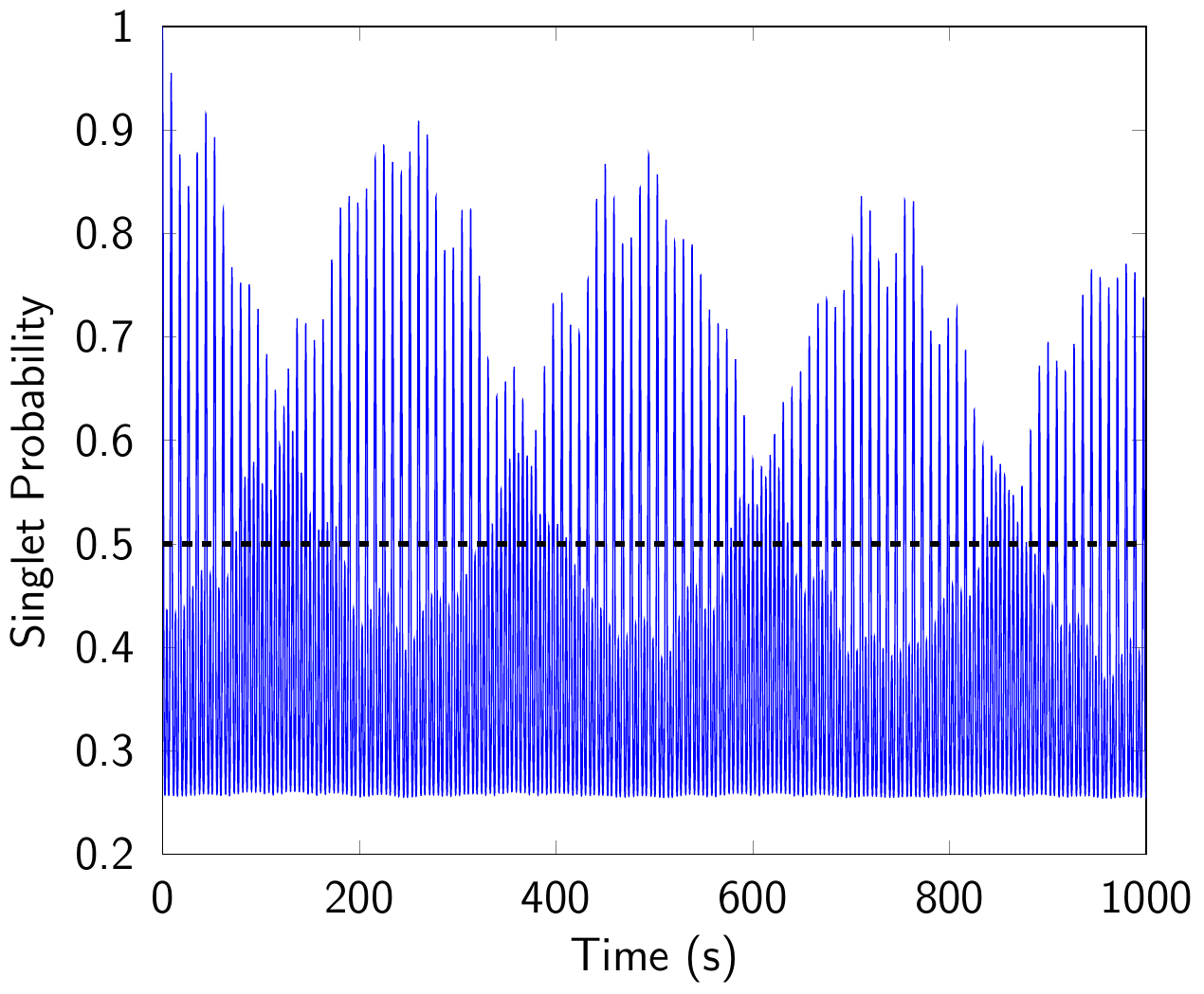}
      \captionsetup{font={small}}
      \caption{Singlet probability for a dimer structure with a \ce{C_s} symmetry also indicated a very long-lived coherence.}
      \label{fig:S4_sp}
    \end{subfigure}
    \caption{Singlet probabilities for two stable dimer structures. Long-lived singlet states are observed irrespective of the symmetry of the dimer, unlike the Posner molecule, potentially making it more suitable for quantum information processing. Similar plots for all dimer structures obtained in our previous study \cite{agarwal2021dynamical} can be found in the SI.}
    \label{fig:dimer_sp_main}
\end{figure}

Surprisingly, exploratory calculations on the tricalcium biphosphate dimer, \ce{Ca6(PO4)4}, revealed that the singlet state in the dimer is exceedingly long-lived, irrespective of its structure. As reported in earlier studies \cite{kanzaki2001calcium, agarwal2021dynamical}, although the dimer also exists in multiple configurations, its energetically favorable structures are known more conclusively. We were able to identify six stable structures for the dimer \cite{agarwal2021dynamical}, although many more may exist. Again, as for the Posner molecules, we initialized a pair of \ce{^{31}P} nuclei in two identical, spatially separated dimers in a singlet state, and then evaluated the singlet probability over time. Remarkably, even in the presence of relaxation, the singlet state in tricalcium biphosphate dimers is extremely long-lived (of the order of $10^3$ seconds, and with significant refocusing) regardless of the symmetry of the molecule. This has been depicted in Fig.~\ref{fig:dimer_sp_main} for two dimer structures. The plots for the remaining structures, the values of the coupling constants for each structure, as well as a detailed analysis of the transfer of coherence from one \ce{^{31}P} nuclear spin pair to another (for one structure), can be found in the SI. Despite these fascinating observations, we remark that the tricalcium biphosphate dimer does not appear to have been observed experimentally.

In general, we expect a spin system with a lesser number of coupled nuclear spins, such as the tricalcium biphosphate dimer, to be more favorable for maintaining long-lived singlet states than a spin system with a larger number of coupled nuclear spins, such as the Posner molecule. Considering systems composed from dimers -- with no symmetry (16 distinct energy levels per cluster) and with \ce{S4} symmetry (13 energy levels) -- and from Posner molecules -- with no symmetry (64 energy levels) and with \ce{S6} symmetry (44 energy levels) -- the number of unique positive frequencies contributing to the coherent evolution of the singlet probability are $7380$, $3159$, $2035152$, and $448877$, respectively (details of the calculation have been summarized in the SI). While some of these might be close to degenerate or zero or not significantly contribute to the singlet probability, these numbers illustrate that destructive interference of coherent contributions will be vastly more likely in larger spin systems, regardless of cluster symmetry. This is in line with our results. We confirm the above argument by evaluating the entanglement yield, in the absence of spin relaxation, for randomly coupled systems with different number of \ce{^{31}P} atoms and obtain the same trend (see SI). These calculations corroborate the insight that the preservation or recurrence of entangled states is primarily related to the number of coupled nuclei, and only secondarily influenced by the symmetry of the molecule and the actual \textit{J}-coupling constants, which impact the evolution through exact and accidental degeneracies of energy levels. Regardless of these observations, we maintain that accurate calculations to determine the \textit{J}-coupling constants for every molecular configuration are critical to evaluate the longevity of the spin singlet state of molecules in comparison to other structures of the same family, i.e.\ with the same number of coupled nuclear spins, and to assess effects of spin relaxation.

In view of the dimers' ability to maintain coherences for exceptionally long times, it is interesting to compare the formation energies per monomer unit of the trimer (Posner molecule) and the dimer. For the dimer configurations studied, the formation energy per monomer unit ranged from $-88.700$ eV to $-89.2007$ eV. For the trimer, the value ranges from $-88.332$ eV to $-90.517$ eV. In comparison, thermal energy at room temperature is $0.026$ eV. Thus, the dimer appears to be roughly as stable as the Posner molecule and is expected to exist as an independent, stable entity without coalescing into Posner molecules. These observations suggests that the nuclear spin entanglement in tricalcium biphosphate dimers could be worth exploring both theoretically and experimentally.

Our comprehensive study on the spin dynamics of the singlet state in pairs of Posner molecules, performed without any assumptions on the structure or the coupling constants of the molecule, shows that the singlet state is short-lived ($< 1$ second) irrespective of the structural symmetry of the molecule. This also indicates that accurate calculation of the \textit{J}-coupling constants, instead, are crucial to the study of spin dynamics pertaining to the Posner molecule. Our results suggest that the Posner molecule might be unsuitable as a biological quantum information processor since entanglement between \ce{^{31}P} nuclei in pairs of Posner molecules is not expected to last for more than a second. It may be possible that coherence is transferred from one pair of \ce{^{31}P} nuclei to the other, but as shown in earlier studies \cite{player2018posner} and, to a greater extent, in our SI, this does not increase the singlet probability yield over time in any way. Moreover, recent findings on \ce{^{31}P}-\ce{^{31}P} singlet lifetimes in common organic phosphorus compounds of interest (\textit{e.g.}, adenosine diphosphate \cite{devience2021nmr}, nicotinamide adenine dinucleotide \cite{devience2021nmr}, tetrabenzyle pyrophosphate \cite{korenchan202131}, pyrophosphate \cite{korenchan2022limits}), show that the singlet lifetime ranges from less than half a second in large diphosphates and pyrophosphates to few tens of seconds in small, highly symmetric pyrophosphates. Additionally, the latter study suggests that singlet lifetimes may be reduced further in a more realistic biological environment, and identifies spin-rotation relaxation as the dominant relaxation pathway \cite{korenchan2022limits}. While our model neglects spin-rotation relaxation (due to the larger size of the Posner molecule), it will be interesting to see how our results differ after its inclusion. The singlet lifetimes reported in the above studies are of the same order as that of the Posner molecule in the current study and suggest that, without conclusive evidence of the presence of isolated Posner molecules \textit{in vivo}, the longevity of the singlet state reported in this study is comparable to that found in other phosphorus compounds \textit{in vitro}. Additionally, while the current study assumed a weak net external magnetic field, it would be interesting to observe the singlet probability in the presence of stronger magnetic fields, and to see whether that has any effect on increasing the singlet longevity. Finally, we observe long-lived ($\sim$ 10$^2$--10$^3$ seconds) singlet states for the tricalcium biphosphate dimer, regardless of the structural symmetry of the molecule. This fascinating result is explained on the basis of the fact that any system with a smaller number of coupled nuclear spins is expected to better maintain entanglement within the singlet state, than a system with a larger number of coupled nuclear spins. This leads us to the conclusion that while the symmetry of the molecule and its coupling constants are critical for evaluating the longevity of the singlet state given a molecule, the behavior of the coherent oscillations is largely dictated by the system size. The fact that the dimers appear to be as energetically stable as the Posner molecule might suggest that the dimer could be a better candidate for a naturally occurring quantum information processor than the Posner molecule. However, confirmation of its presence \textit{in vivo} is necessary.

For the calculation of the scalar coupling constants, we used the pcJ-n basis set \cite{jensen2006basis} built specifically for the calculation of NMR scalar coupling constants. The pcseg-n basis set \cite{jensen2015segmented} was used for the calcium atoms. We used ORCA \cite{neese2020orca} for calculating the coupling constants, using the B3LYP exchange-correlation functional. The QuTiP library \cite{johansson2012qutip} was used to extract the Pauli matrices needed for the calculation of the singlet probabilities and concurrences. LAMMPS \cite{thompson2022lammps} was used for all molecular dynamics simulations, along with the force-fields developed by Demichelis et al.\ \cite{demichelis2018simulation}. Further details about the above simulations can be found in the SI.
\begin{acknowledgement}
The authors would like to thank Raffaella Demichelis for sharing the force fields that were used for the LAMMPS simulations. The authors would also like to thank Matthew Fisher and Michael Swift for insightful discussions and for providing an \ce{S6} - symmetric structure of the Posner molecule. Lastly, the authors would like to thank UCLA’s Institute for Digital Research and Education (IDRE) for making available the computing resources used in this work. D.\ R.\ K.\ would like to thank the Office of Naval Research for financial support (ONR award number N62909-21-1-2018).
\end{acknowledgement}


\newpage

\begin{center}
    {\LARGE \textbf{Supporting Information}}
\end{center}

\tableofcontents

\newpage

\section{Methods}

The calculation of the coupling constants between the \ce{^{31}P} nuclei is a critical part of the study, and was done using the ORCA program \cite{neese2020orca}. More specifically, we used the B3LYP exchange-correlation functional, along with basis sets specifically built for the calculation of NMR coupling constants. For the \ce{P} and \ce{O} atoms, we used the pcJ-2 basis set \cite{jensen2006basis} (the largest such basis set that we could afford within computational constraint limitations), whereas the pcseg-2 basis set \cite{jensen2015segmented} was used for the \ce{Ca} atoms, because the pcJ-n basis sets are not supported for atoms beyond \ce{Ar}. Both these basis sets are polarization consistent. The chemical shielding tensors needed for including the chemical shift anisotropy effect were calculated using Gaussian \cite{g16} using the cc-pV5Z basis set. Following the calculation of the scalar coupling constants and the chemical shielding tensors, we simulated the spin dynamics of the \ce{^{31}P} nuclear pairs, and calculated the singlet probabilities and the concurrences. The matrix representations of the various operators needed for this calculation were constructed using the QuTiP \cite{johansson2012qutip} library. A singlet state consisting of  corresponding \ce{^{31}P} nuclei in separate but identical Posner molecules was evolved through time, much like in Ref.\ \citenum{player2018posner}. For the molecular dynamics (MD) simulations, we used the LAMMPS software \cite{thompson2022lammps}. Following the work of Demichelis et al.\ \cite{demichelis2018simulation}, we placed the Posner molecule in a box of length $56$ \AA\;, with $\sim 4000$ water molecules. The force fields introduced in Ref.\ \citenum{demichelis2018simulation} were used. The canonical NVT ensemble was utilized at a temperature of $300$ K, with a timestep of $0.1$ fs. The atoms were initially given a Gaussian-distributed velocity corresponding to a temperature of $300$ K, and the initial linear momentum of the Posner molecule was set to zero. Following the MD simulation that proceeded for a total of $340790000$ timesteps (equivalent to $34.07$ ns), the trajectory was analyzed to calculate the rotational correlation constant, $\tau_c$, and the translational diffusion constant of the molecule. The Visual Molecular Dynamics (VMD) software \cite{HUMP96} was used for calculating the point group symmetries of relevant molecular structures.

\newpage

\section{Intramolecular dipole-dipole interactions between \texorpdfstring{\ce{^{31}P}{ n}} nuclear spins}

The dominant relaxation mechanism considered here is the intramolecular dipole-dipole interactions. As highlighted in the main text, the intermolecular dipole-dipole interactions, dipolar coupling with the solvent, and spin-rotation relaxation have all been considered to be negligible and not included in our calculations. The chemical shielding anisotropy (CSA) contribution, although negligible in the geomagnetic field, has been included. The intramolecular dipole-dipole interactions for the $15$ \ce{^{31}P} pairs is given by:
\begin{align}
    \hat{H}_{iDD}(t)=-\sum_{j<k}\sum_{k}\frac{\mu_0\gamma_j\gamma_k\hbar}{4\pi|\textbf{r}_{jk}|^3}\left(\hat{I}_{j}\cdot\hat{I}_{k} - \frac{3}{|\textbf{r}_{jk}|^2}\left(\hat{I}_{j}\cdot\textbf{r}_{jk}\right)\left(\hat{I}_{k}\cdot\textbf{r}_{jk}\right)\right)
\end{align}
where $\theta$ is the angle made by the vector $\textbf{r}_{jk}$ joining the two nuclear spins with the \textit{z}-axis. Here, $\gamma_k$ is the gyromagnetic ratio of the \ce{^{31}P} nucleus, $\hat{I}_{k}$ is the spin angular momentum operator for nucleus $k$, and $\hat{I}_{k,z}$ is the $z$-component of the spin angular momentum operator.

\newpage

\section{Calculation of concurrence}

In addition to the evolution of the singlet state between pairs of \ce{^{31}P} nuclear spins, we also study the evolution of the two-qubit concurrence for a given pair. This acts as a measure of entanglement, and is calculated as \cite{wootters1998entanglement,player2018posner}:
\begin{align}
    \mathcal{C}_{0_A,0_B}=\text{max}\left(0,\lambda_1-\lambda_2-\lambda_3-\lambda_4\right)
\end{align}
where $\lambda_i$ are the eigenvalues of the matrix $\hat{\rho}(t)\left(\sigma_y\otimes\sigma_y\right)\hat{\rho}(t)\left(\sigma_y\otimes\sigma_y\right)$ in decreasing order, with $\sigma_y$ being the second Pauli matrix. $\hat{\rho}(t)$ is already know after having solved the Liouville-von Neumann equation (see main text).

\newpage

\section[Singlet probability for a pair of calcium phosphate monomers]{Singlet probability for a pair of calcium phosphate monomers}

Given the structure of the calcium phosphate monomer, \ce{Ca3(PO4)2}, \cite{agarwal2021dynamical, treboux2000existence, kanzaki2001calcium} we can calculate the singlet probability for a pair of \ce{^{31}P} nuclei in two spatially-separated calcium phosphate monomers (with the remaining uncorrelated nuclear spins serving as background). Unsurprisingly, the singlet probability is very well-behaved, i.e., it decays extremely slowly and has well-defined oscillations at a single frequency. However, this is to be expected since the monomer only has one unique scalar coupling constant, which suggests that the singlet probability --- which is dominated by $J$-coupling constants --- will recur at a single frequency. This is unlike the Posner molecule wherein the singlet probability recurs at multiple frequencies as determined by the various $J$-coupling constants, leading to destructive interference and a significantly faster relaxation of the singlet state \cite{player2018posner}.

\begin{figure}[h!]
      \centering
      \includegraphics[width=0.45\textwidth]{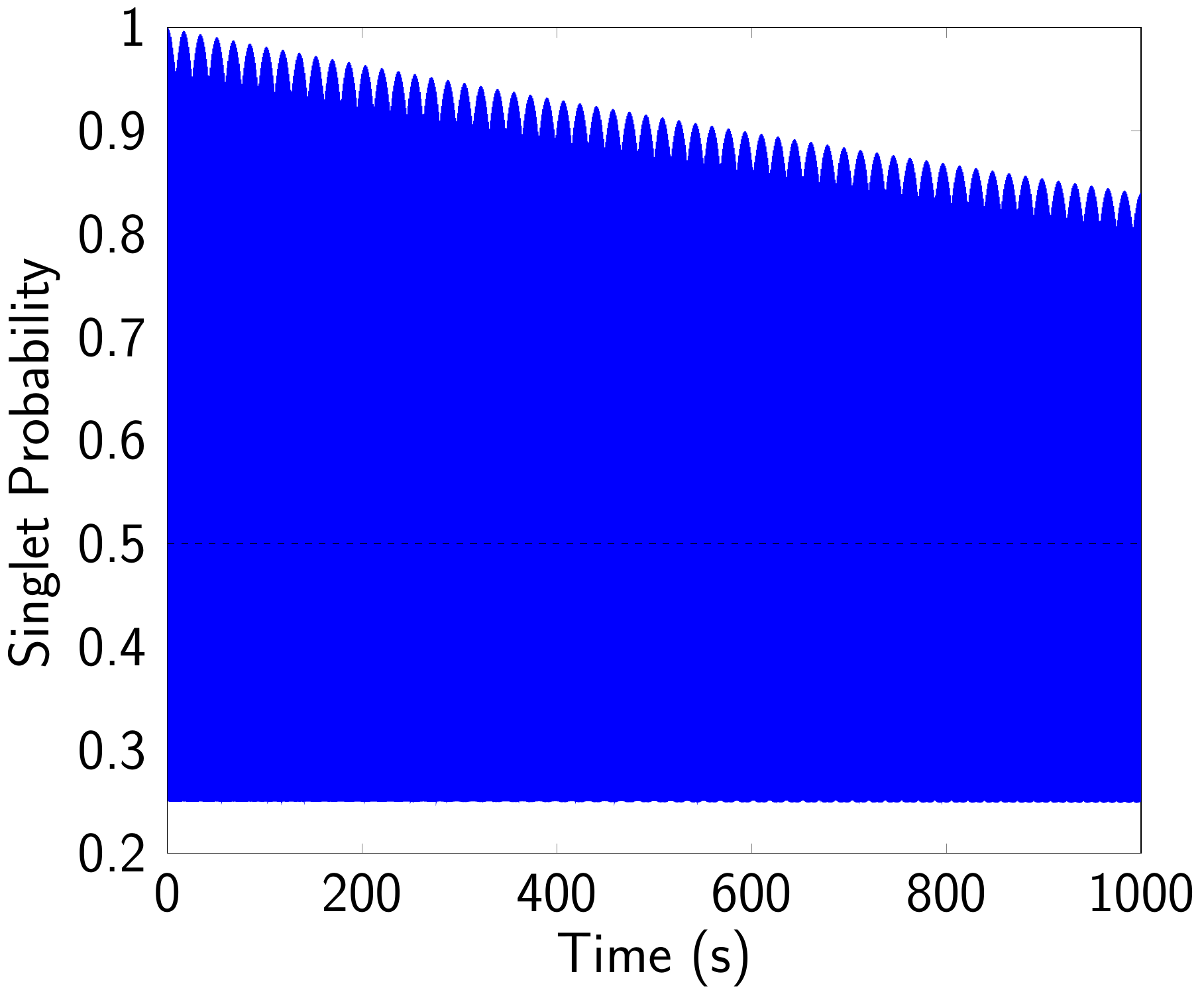}
      \caption{Singlet probability for the calcium phosphate monomer}
      \label{fig:Td_sp_1}
\end{figure}

However, these findings related to the entanglement behavior of the monomers is not expected to be of practical relevance due to energetic considerations. For the monomer, the formation energy was calculated to be $-84.244$ eV, whereas the formation energy of the Posner molecule per monomer unit ranged from $-88.332$ eV to $-90.517$ eV. Clearly, the Posner molecule is much more energetically stable, suggesting that two monomers would very likely coalesce to form a lumped, aggregate molecule, which may or may not result in the formation of calcium phosphate dimers and trimers (Posner molecules).

\newpage

\section{Singlet probability for pairs of calcium phosphate dimers}

The singlet probability for a pair of \ce{^{31}P} nuclei in two identical, spatially separated dimers, \ce{Ca6(PO4)4}, initialized in the singlet state has been shown below. All the dimer structures considered here have been obtained by us in our previous study \cite{agarwal2021dynamical}. We see exceedingly long-lived singlet states even in the presence of relaxation, regardless of the symmetry of the molecule. The coupling constants used for these calculations have been shown in Table \ref{tab:dimer_Jconsts}.

\begin{figure}[h!]
    \centering
    \begin{subfigure}[b]{0.45\textwidth}
      \centering
      \includegraphics[width=\textwidth]{plots-figure6.pdf}
      \captionsetup{font={small}}
      \caption{Singlet probability for a dimer structure with a \ce{C2} symmetry}
      \label{fig:C2_ms_sp_SI}
    \end{subfigure}
    \hspace{1cm}
    \begin{subfigure}[b]{0.45\textwidth}
      \centering
      \includegraphics[width=\textwidth]{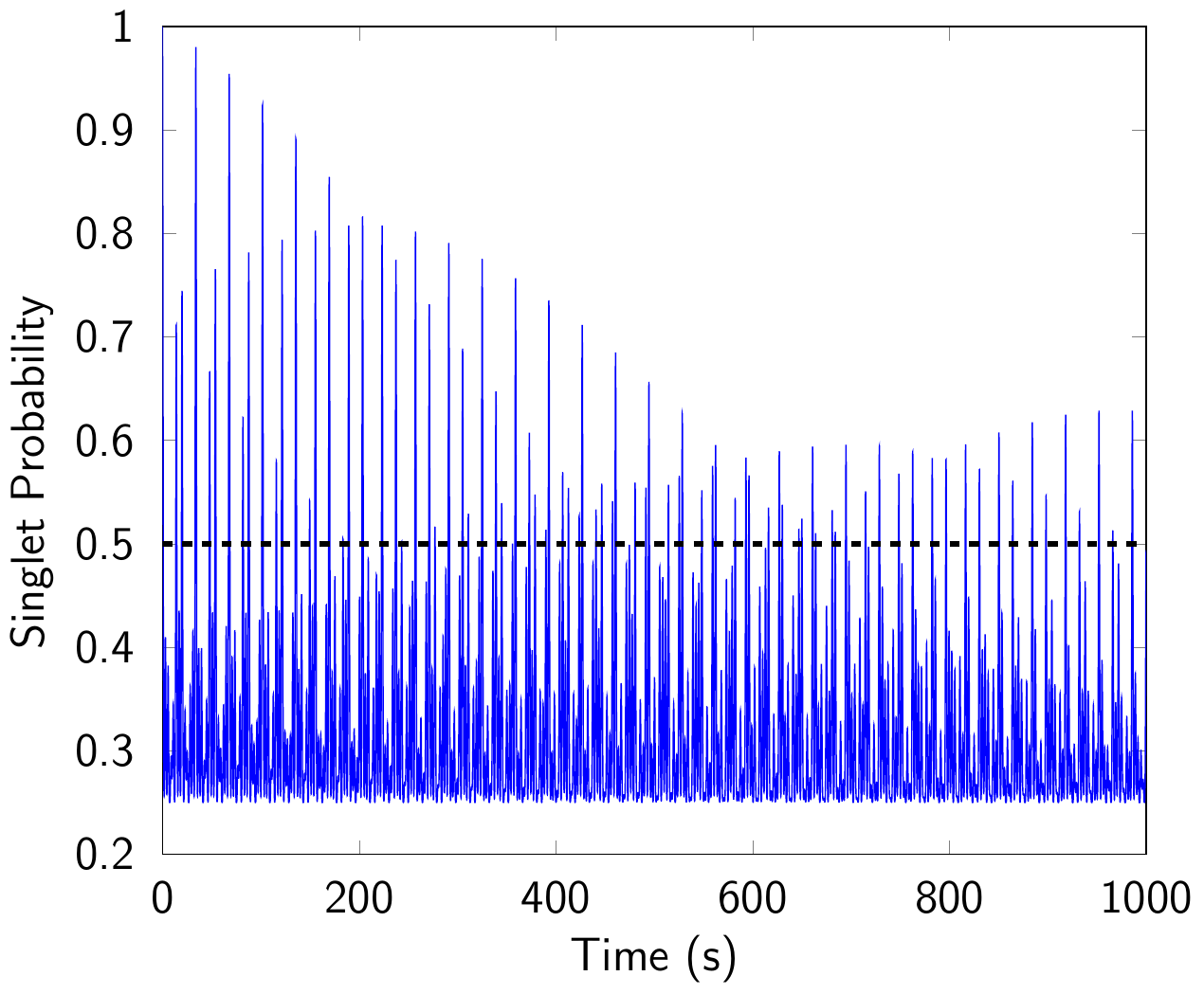}
      \captionsetup{font={small}}
      \caption{Singlet probability for a dimer structure with a \ce{S4} symmetry}
      \label{fig:S4_sp_SI}
    \end{subfigure}
    \\
    \vspace{1cm}
    \begin{subfigure}[b]{0.45\textwidth}
      \centering
      \includegraphics[width=\textwidth]{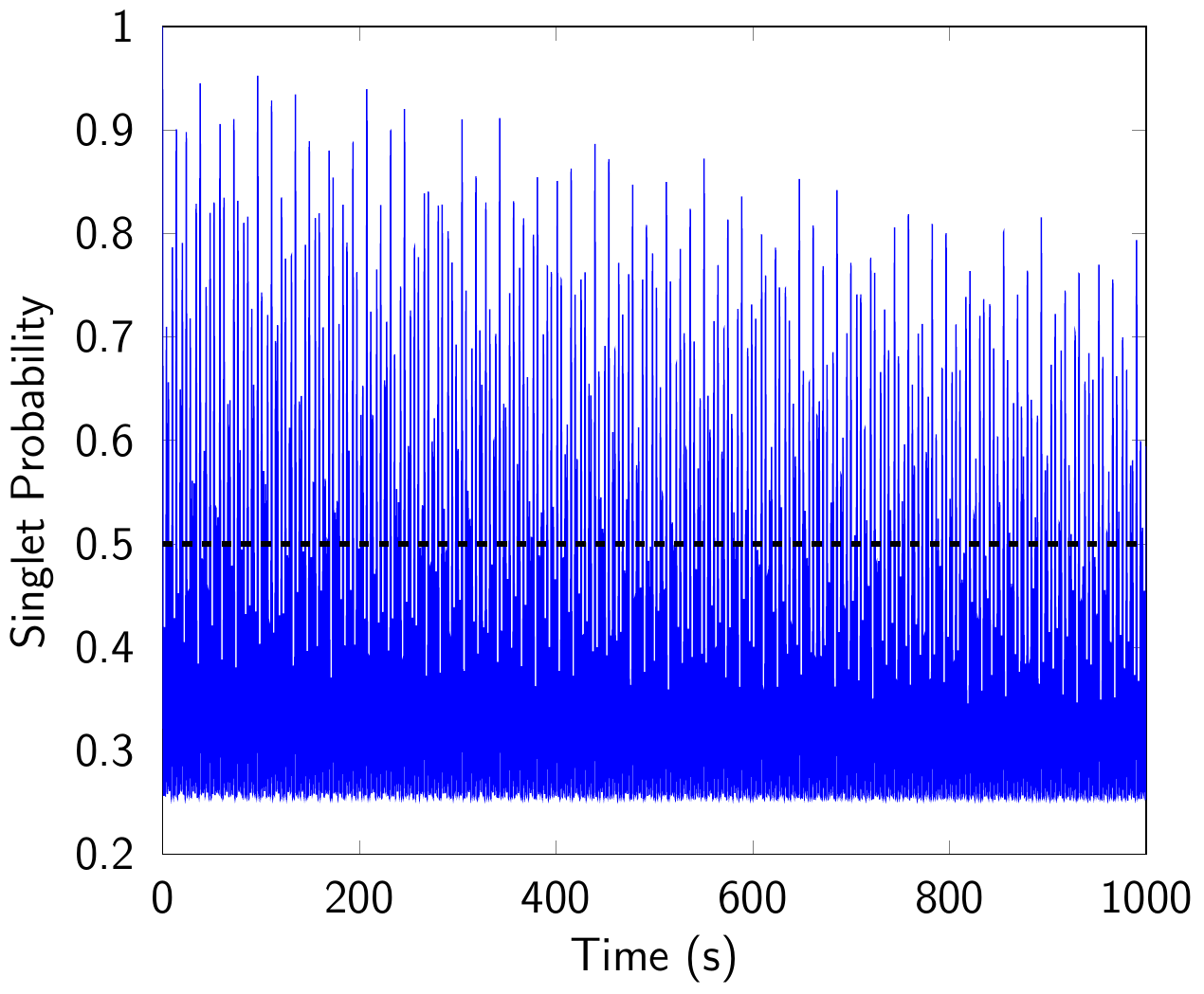}
      \captionsetup{font={small}}
      \caption{Singlet probability for a dimer structure with a \ce{C2} symmetry}
      \label{fig:C2_sp_SI}
    \end{subfigure}
    \hspace{1cm}
    \begin{subfigure}[b]{0.45\textwidth}
      \centering
      \includegraphics[width=\textwidth]{plots-figure9.pdf}
      \captionsetup{font={small}}
      \caption{Singlet probability for a dimer structure with a \ce{C_s} symmetry}
      \label{fig:Cs_sp_SI}
    \end{subfigure}
\end{figure}

\newpage

\begin{figure}[h!]\ContinuedFloat
    \begin{subfigure}[b]{0.45\textwidth}
      \centering
      \includegraphics[width=\textwidth]{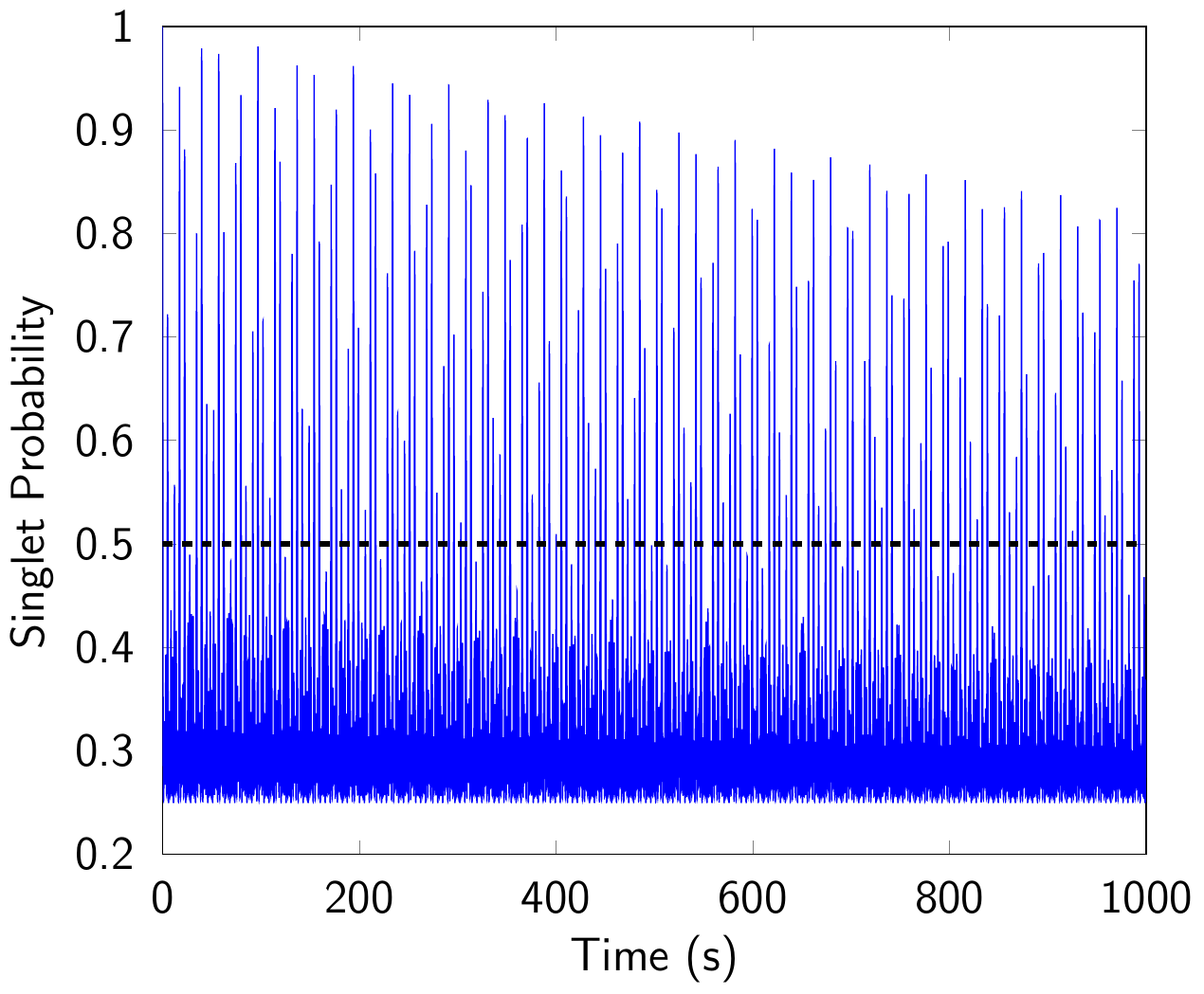}
      \captionsetup{font={small}}
      \caption{Singlet probability for a dimer structure with a \ce{C_{2v}} symmetry}
      \label{fig:C2v_sp_SI}
    \end{subfigure}
    \hspace{1cm}
    \begin{subfigure}[b]{0.45\textwidth}
      \centering
      \includegraphics[width=\textwidth]{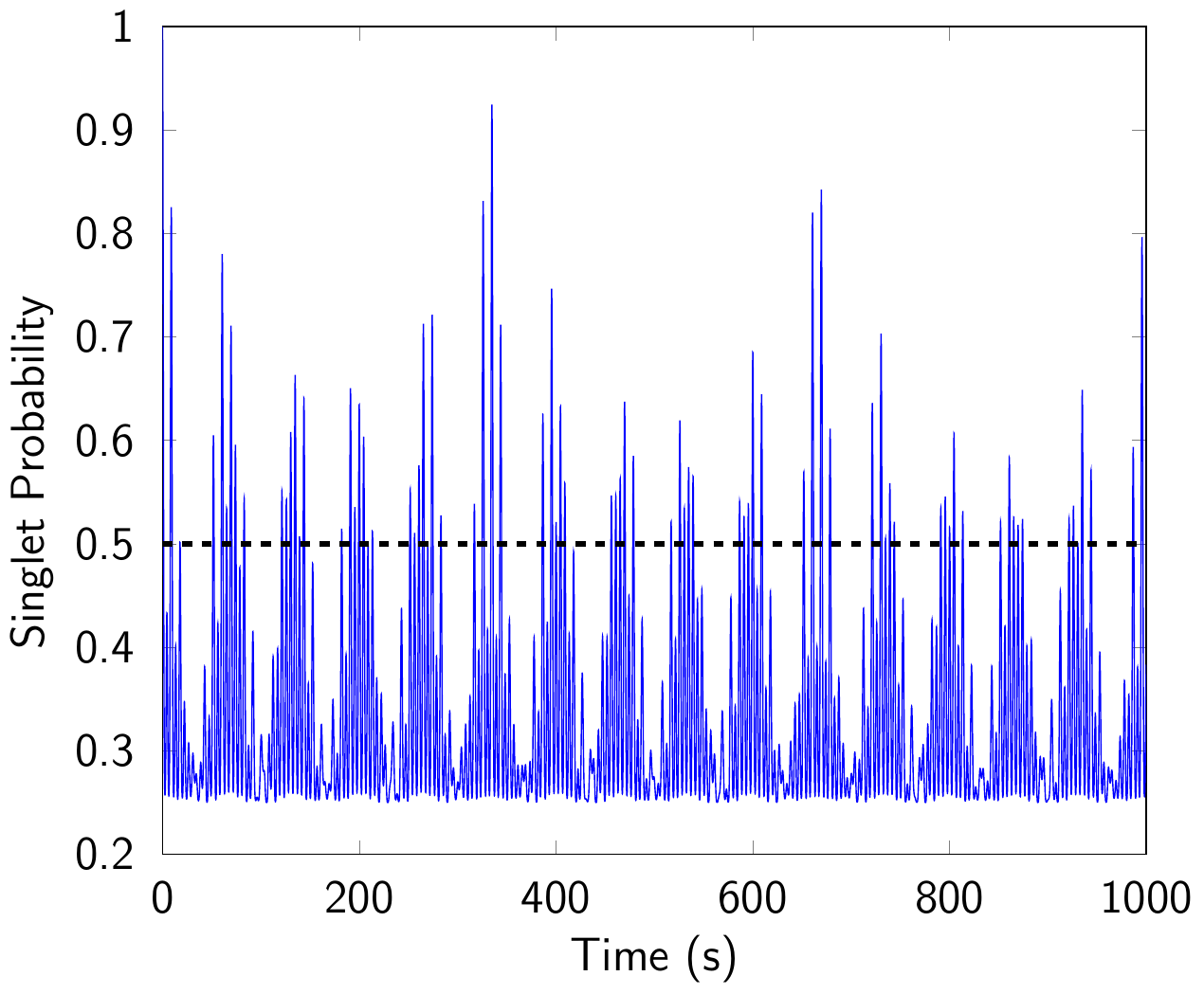}
      \captionsetup{font={small}}
      \caption{Singlet probability for a dimer structure with a \ce{T_d} symmetry}
      \label{fig:Td_sp_SI}
    \end{subfigure}
    \caption{Singlet probabilities for the six stable dimer structures explored in our previous study \cite{agarwal2021dynamical}. Long-lived singlet states are observed irrespective of the symmetry of the dimer, unlike the Posner molecule, potentially making it more suitable for quantum information processing. For clarity, the plots (a) through (f) have been arranged in decreasing order of the stability of the dimer structure}
    \label{fig:dimer_sp}
\end{figure}

\begin{figure}[h!]
    \small
    \begin{tabularx}{\textwidth}{|@{}C|*{6}{C|}@{}}
      \hline
      \textbf{Coupling constants (in Hz)} & \footnotesize\textbf{Structure corresponding to Fig. \ref{fig:C2_ms_sp_SI}} & \footnotesize\textbf{Structure corresponding to Fig. \ref{fig:S4_sp_SI}} & \footnotesize\textbf{Structure corresponding to Fig. \ref{fig:C2_sp_SI}} & \footnotesize\textbf{Structure corresponding to Fig. \ref{fig:Cs_sp_SI}} & \footnotesize\textbf{Structure corresponding to Fig. \ref{fig:C2v_sp_SI}} & \footnotesize\textbf{Structure corresponding to Fig. \ref{fig:Td_sp_SI}} \\
     \hline
     $\text{J}_{ab}$ & -0.394 & -0.193 & -0.200 & 0.142 & -0.163 & 0.147 \\ 
     $\text{J}_{ac}$ & -0.415 & -0.139 & -0.209 & 0.147 & -0.161 & 0.154 \\ 
     $\text{J}_{ad}$ & -0.163 & -0.136 & -0.267 & 0.152 & -0.374 & 0.135 \\ 
     $\text{J}_{bc}$ & -0.046 & -0.143 & -0.404 & 0.140 & 0.042 & 0.144 \\ 
     $\text{J}_{bd}$ & -0.082 & -0.142 & -0.218 & 0.124 & -0.162 & 0.150 \\ 
     $\text{J}_{cd}$ & 0.221 & -0.189 & -0.065 & 0.148 & -0.161 & 0.142 \\ 
     \hline
    \end{tabularx}
    \caption{The \textit{J}-coupling constants for each of the $6$ dimer structures considered in this study. All values are in Hz. The coordinates of the \ce{^{31}P} atoms corresponding to subscripts $a,b,c,d$ are shown in Table \ref{tab:dimer_cords}.}
    \label{tab:dimer_Jconsts}
\end{figure}

\begin{figure}[!ht]
    \small
    \begin{tabularx}{\textwidth}{|@{}C|*{4}{C|}@{}}
      \hline
      \textbf{Figure corresponding to the structure} & \textbf{Coordinates (in \r{A}) of \ce{^{31}P} atom $a$} & \textbf{Coordinates (in \r{A}) of \ce{^{31}P} atom $b$} & \textbf{Coordinates (in \r{A}) of \ce{^{31}P} atom $c$} & \textbf{Coordinates (in \r{A}) of \ce{^{31}P} atom $d$} \\
     \hline
     Fig. \ref{fig:C2_ms_sp_SI} & \footnotesize{[-2.82, 0.51, 0.06]} & \footnotesize{[0.47, 1.25, -3.39]} & \footnotesize{[0.44, 0.54, 3.21]} & \footnotesize{[1.98, -2.20, -0.01]} \\ 
     Fig. \ref{fig:S4_sp_SI} & \footnotesize{[1.34, -1.75, 2.03]} & \footnotesize{[-1.31, 1.69, 2.11]} & \footnotesize{[-1.74, -1.29, -2.09]} & \footnotesize{[1.71, 1.36, -2.08]} \\ 
     Fig. \ref{fig:C2_sp_SI} & \footnotesize{[-0.97, -0.24, 2.92]} & \footnotesize{[0.99, -2.86, -1.96]} & \footnotesize{[-2.69, 1.36, -1.58]} & \footnotesize{[2.77, 1.87, 0.30]} \\ 
     Fig. \ref{fig:Cs_sp_SI} & \footnotesize{[2.95, -1.15, 1.05]} & \footnotesize{[-2.48, -1.75, 1.46]} & \footnotesize{[-0.28, 3.24, 0.81]} & \footnotesize{[-0.20, -0.34, -3.35]} \\ 
     Fig. \ref{fig:C2v_sp_SI} & \footnotesize{[-3.42, -0.27, 0.06]} & \footnotesize{[0.88, 1.60, -2.80]} & \footnotesize{[0.99, 1.67, 2.73]} & \footnotesize{[1.51, -3.08, 0.01]} \\ 
     Fig. \ref{fig:Td_sp_SI} & \footnotesize{[-2.57, 0.28, 2.14]} & \footnotesize{[0.25, 2.80, -1.83]} & \footnotesize{[2.84, -0.47, 1.73]} & \footnotesize{[-0.52, -2.62, -2.04]} \\ 
     \hline
    \end{tabularx}
    \caption{The coordinates of the \ce{^{31}P} atoms corresponding to each figure in Fig.\ \ref{fig:dimer_sp}}
    \label{tab:dimer_cords}
\end{figure}

\begin{figure}
\setlength\tabcolsep{2pt}
\begin{tabularx}{\textwidth}{|@{}c*{10}{|C|}@{}}
		\hline
		& $00$ & $01$ & $02$ & $03$ & $11$ & $12$ & $13$ & $22$ & $23$ & $33$\\
		\hline 
		$00$ & \vspace{0.2cm}
		\includegraphics[ width=\linewidth, height=\linewidth, keepaspectratio]{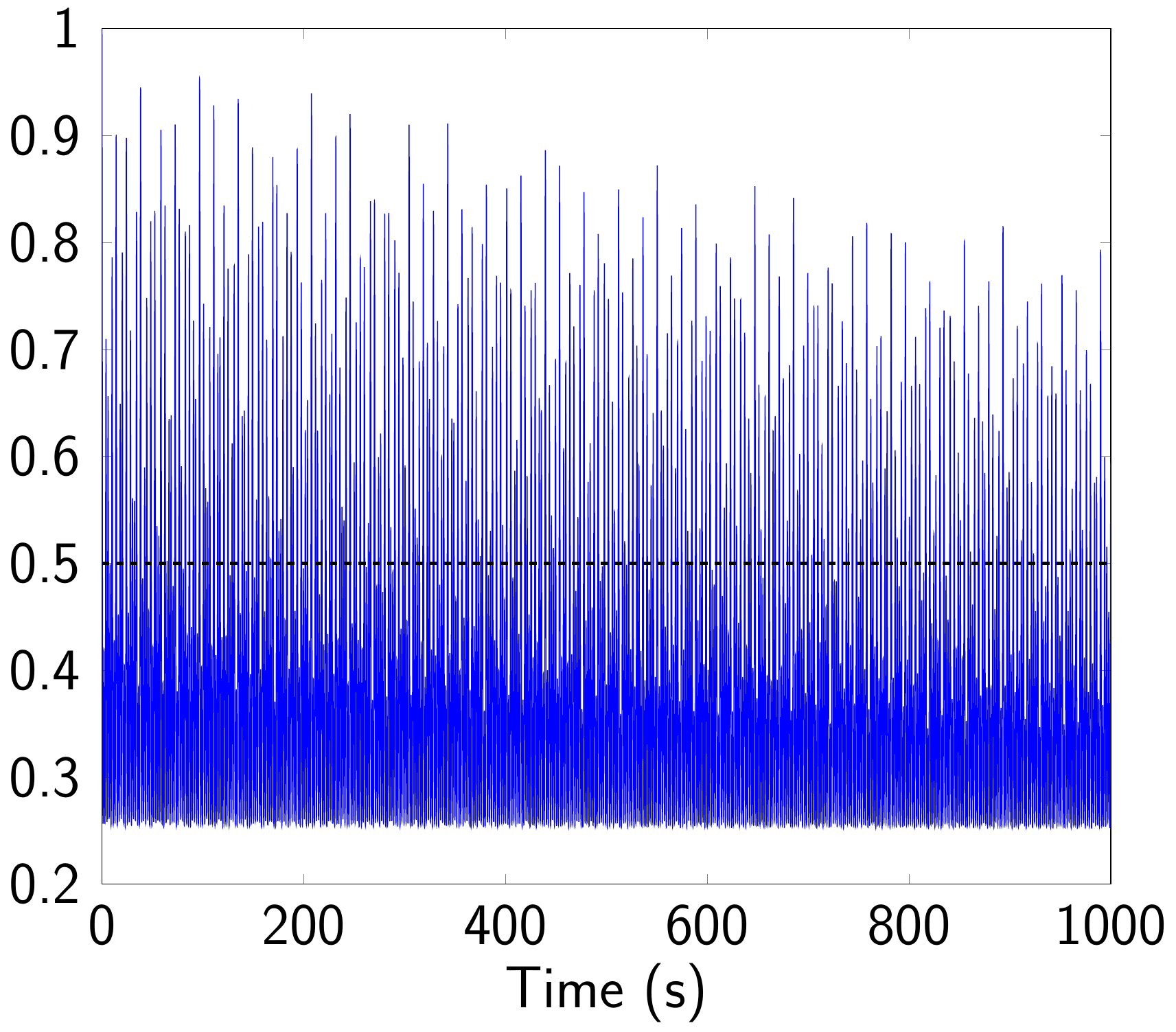} & \vspace{0.2cm}
		\includegraphics[ width=\linewidth, height=\linewidth, keepaspectratio]{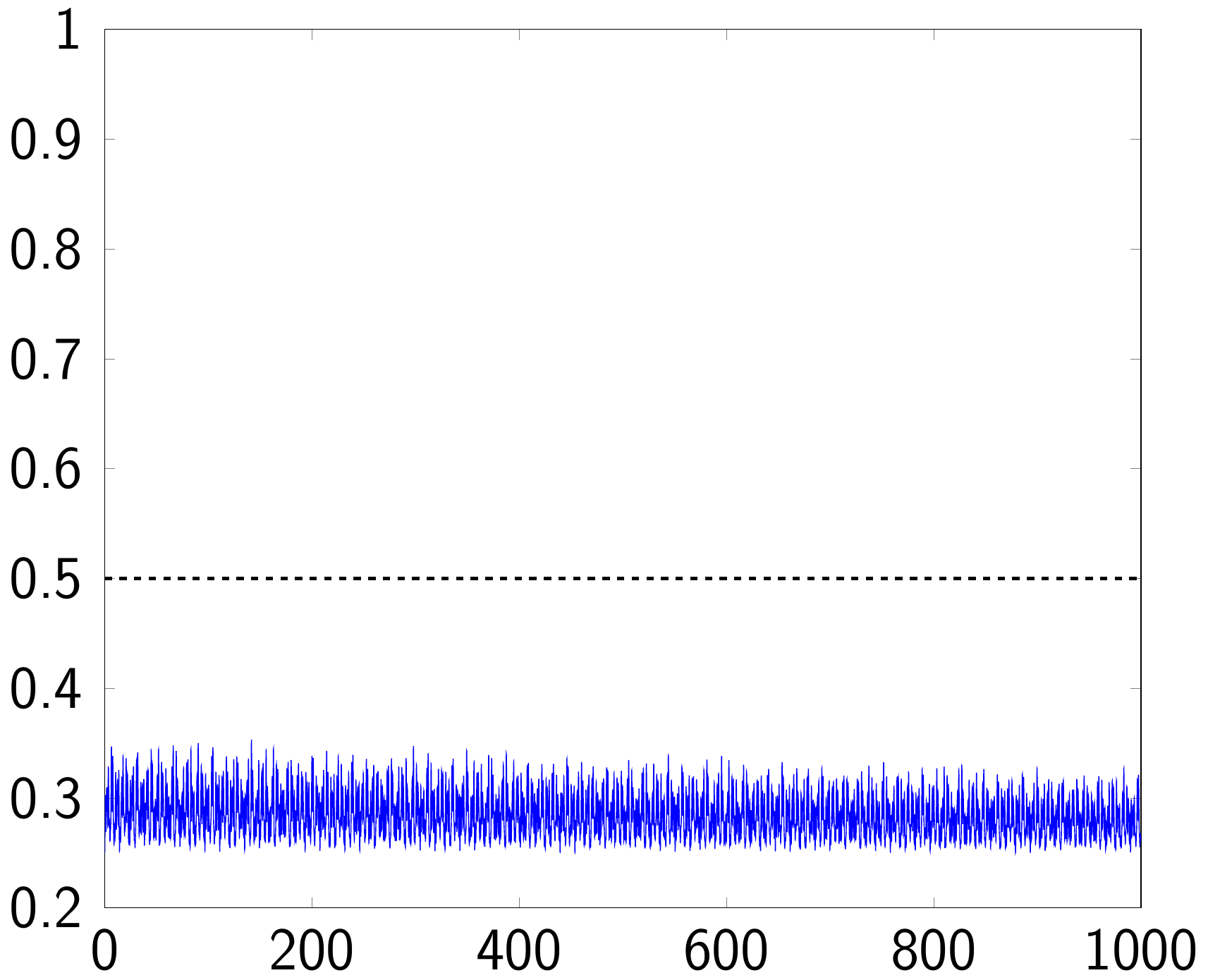} & \vspace{0.2cm}
		\includegraphics[ width=\linewidth, height=\linewidth, keepaspectratio]{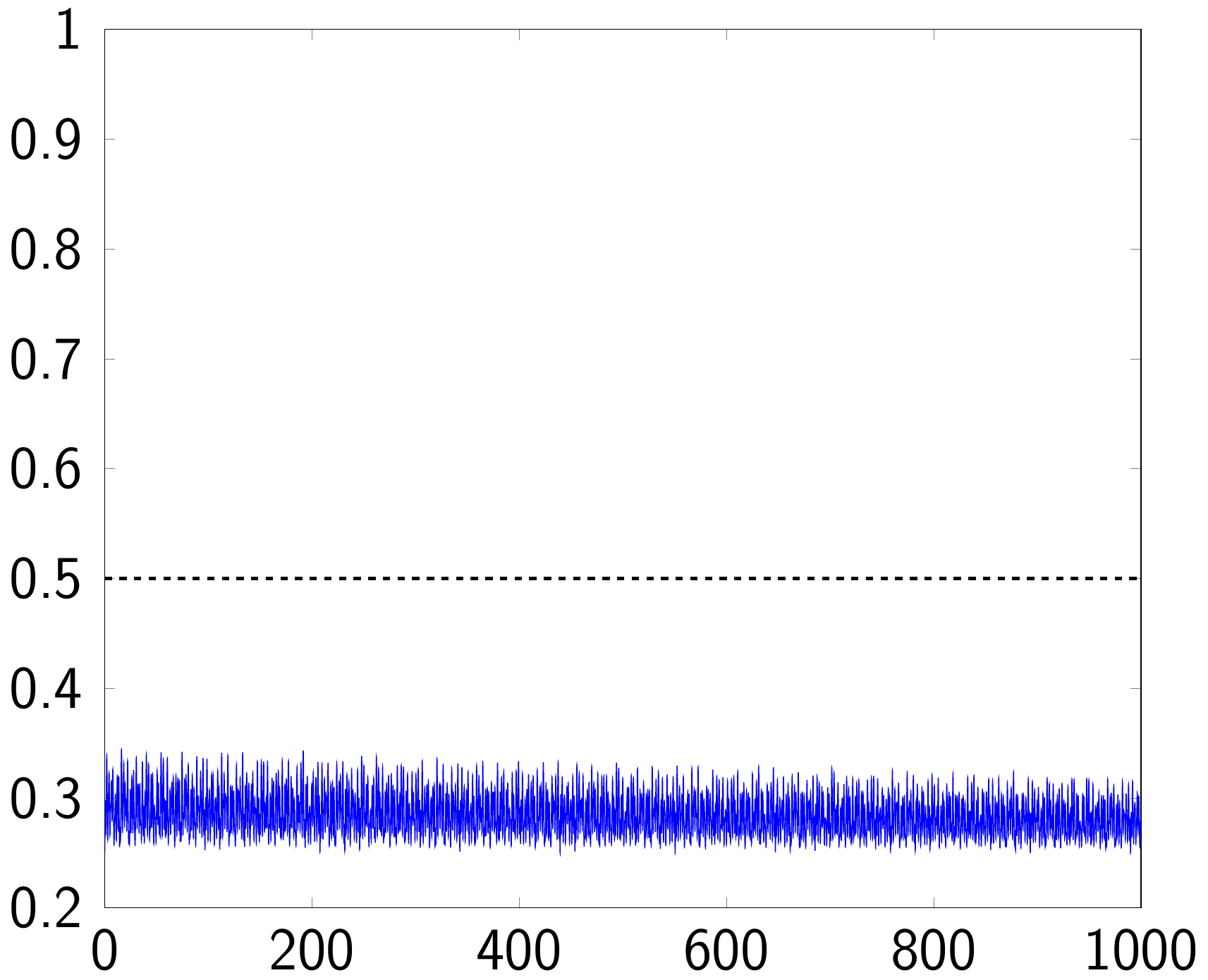} & \vspace{0.2cm}
		\includegraphics[ width=\linewidth, height=\linewidth, keepaspectratio]{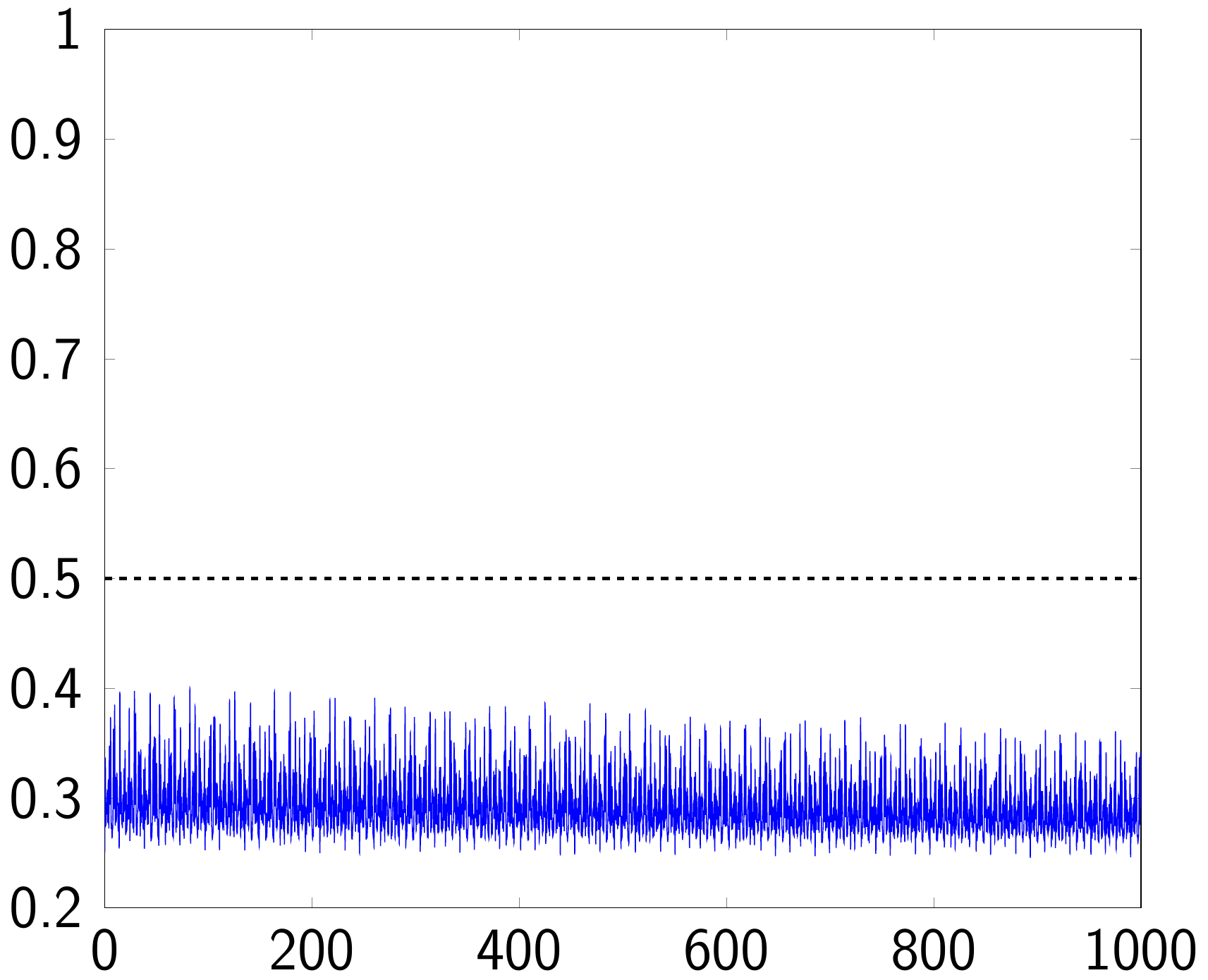} & \vspace{0.2cm}
		\includegraphics[ width=\linewidth, height=\linewidth, keepaspectratio]{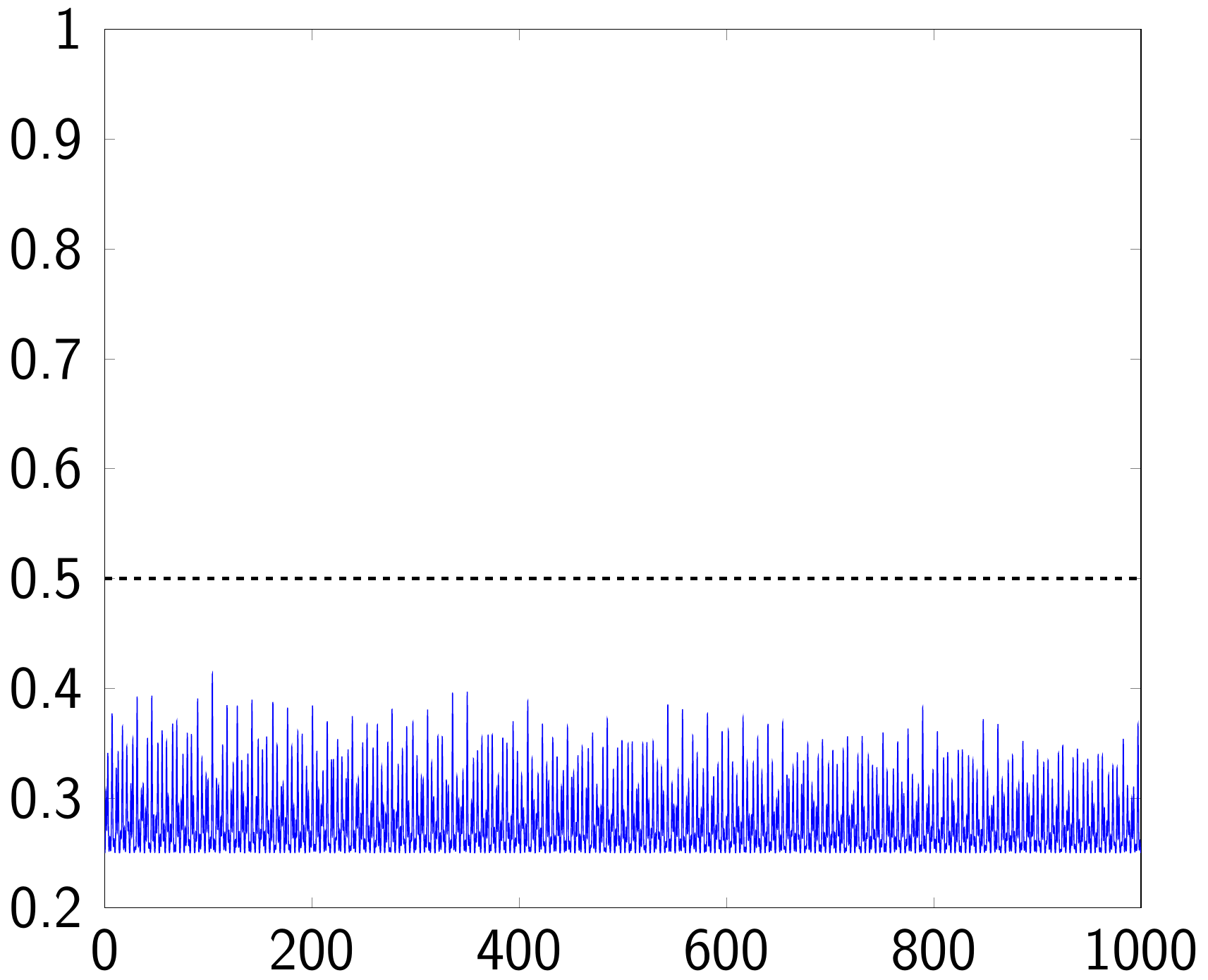} & \vspace{0.2cm}
		\includegraphics[ width=\linewidth, height=\linewidth, keepaspectratio]{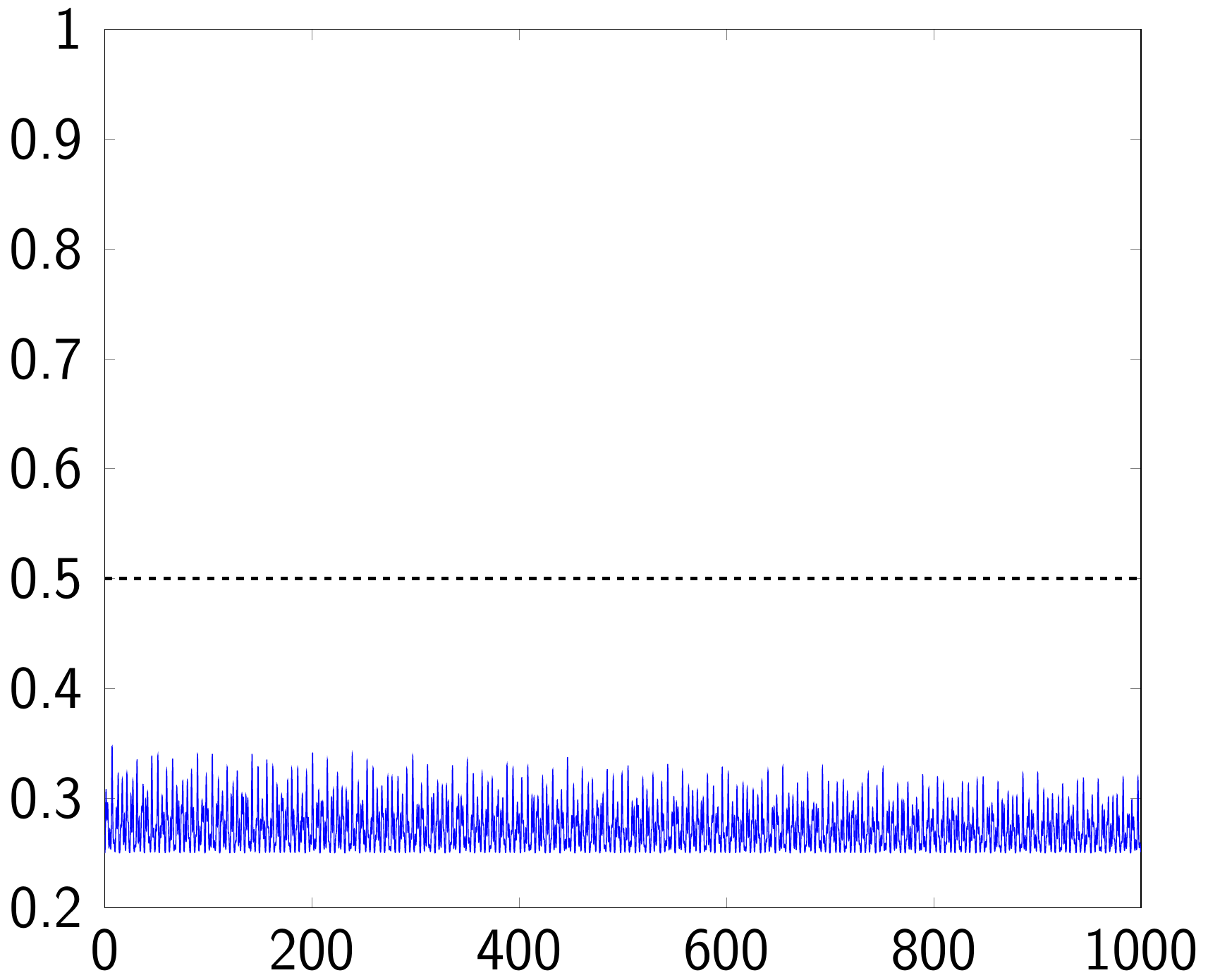} & \vspace{0.2cm}
		\includegraphics[ width=\linewidth, height=\linewidth, keepaspectratio]{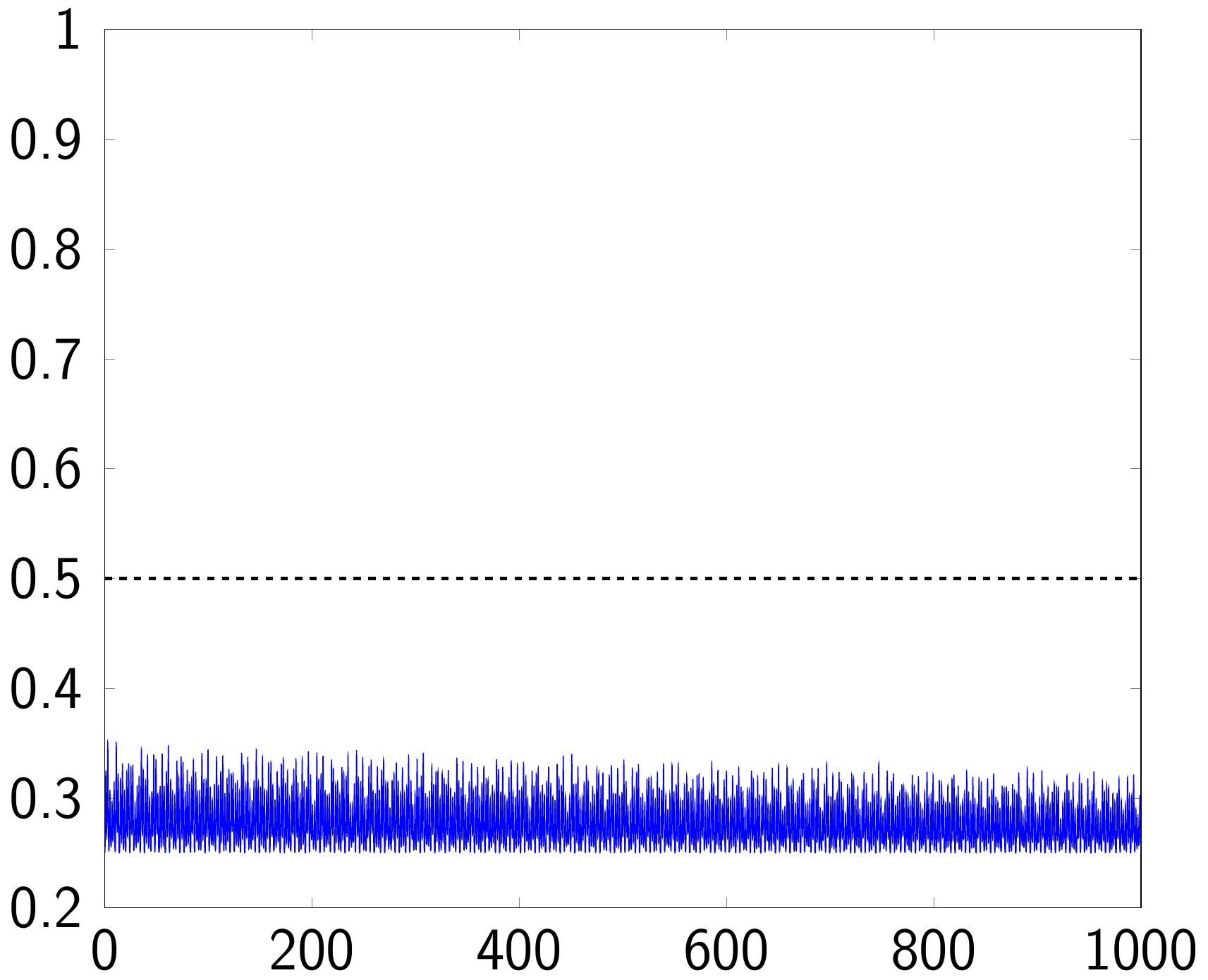} & \vspace{0.2cm}
		\includegraphics[ width=\linewidth, height=\linewidth, keepaspectratio]{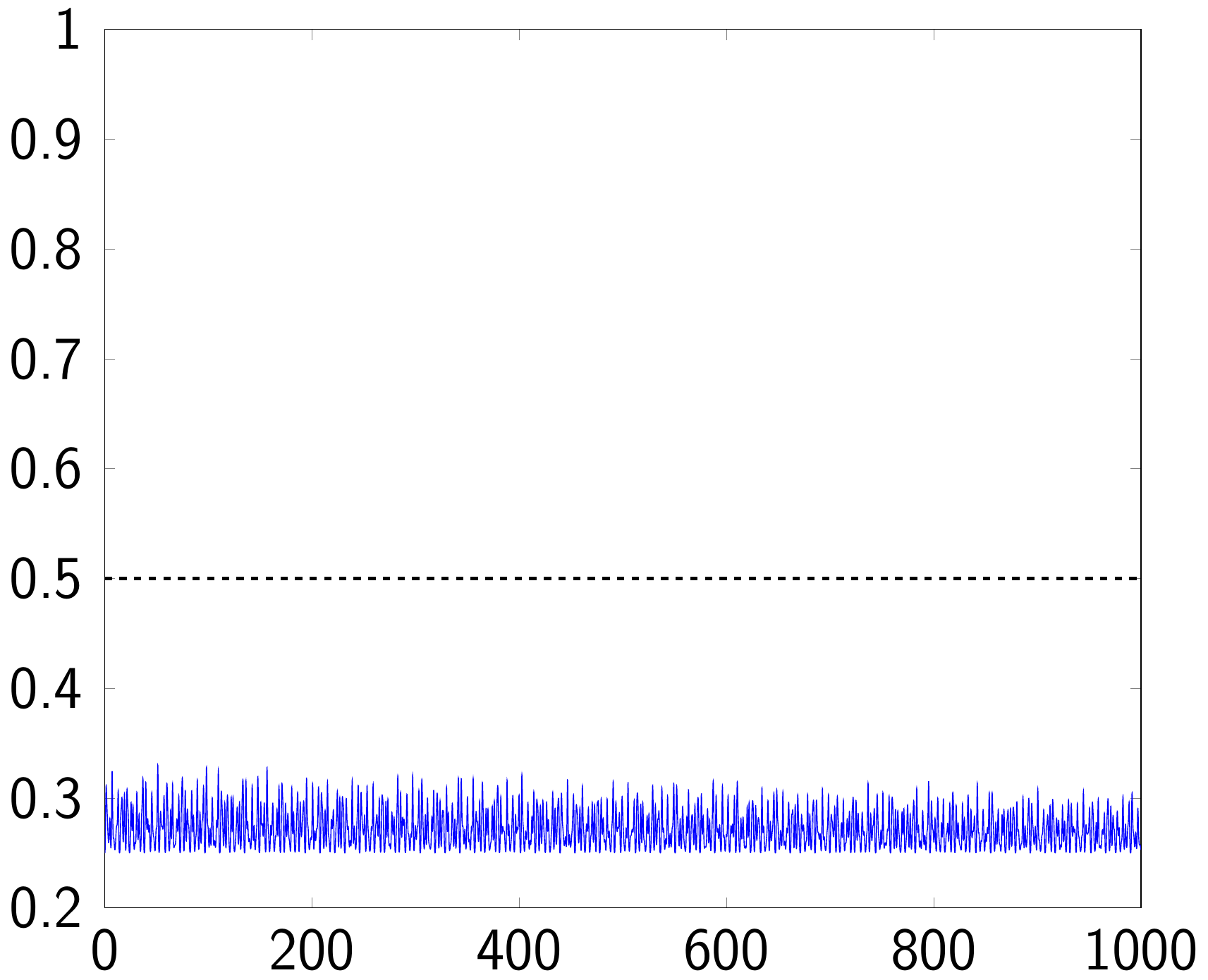} & \vspace{0.2cm}
		\includegraphics[ width=\linewidth, height=\linewidth, keepaspectratio]{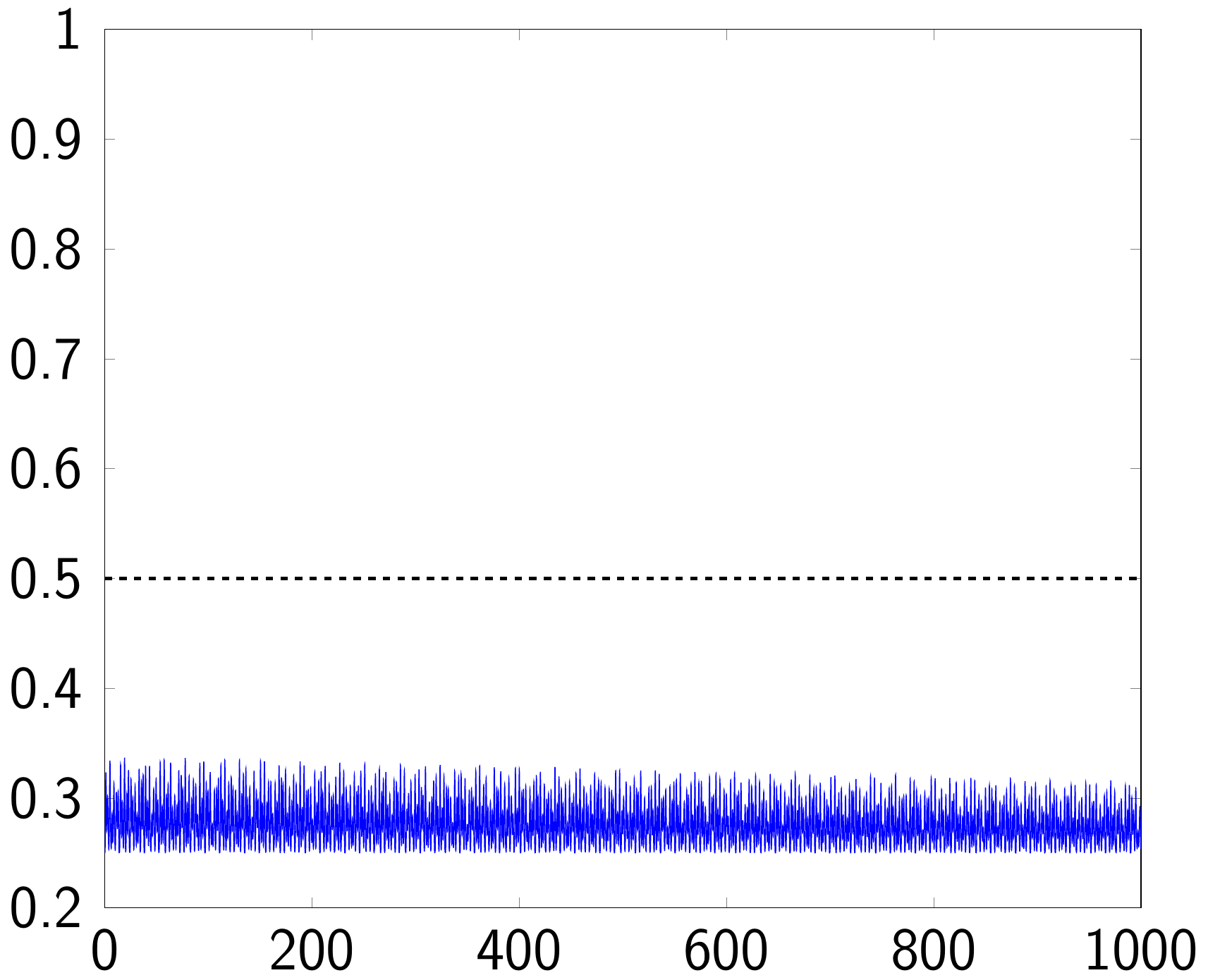} & \vspace{0.2cm}
		\includegraphics[ width=\linewidth, height=\linewidth, keepaspectratio]{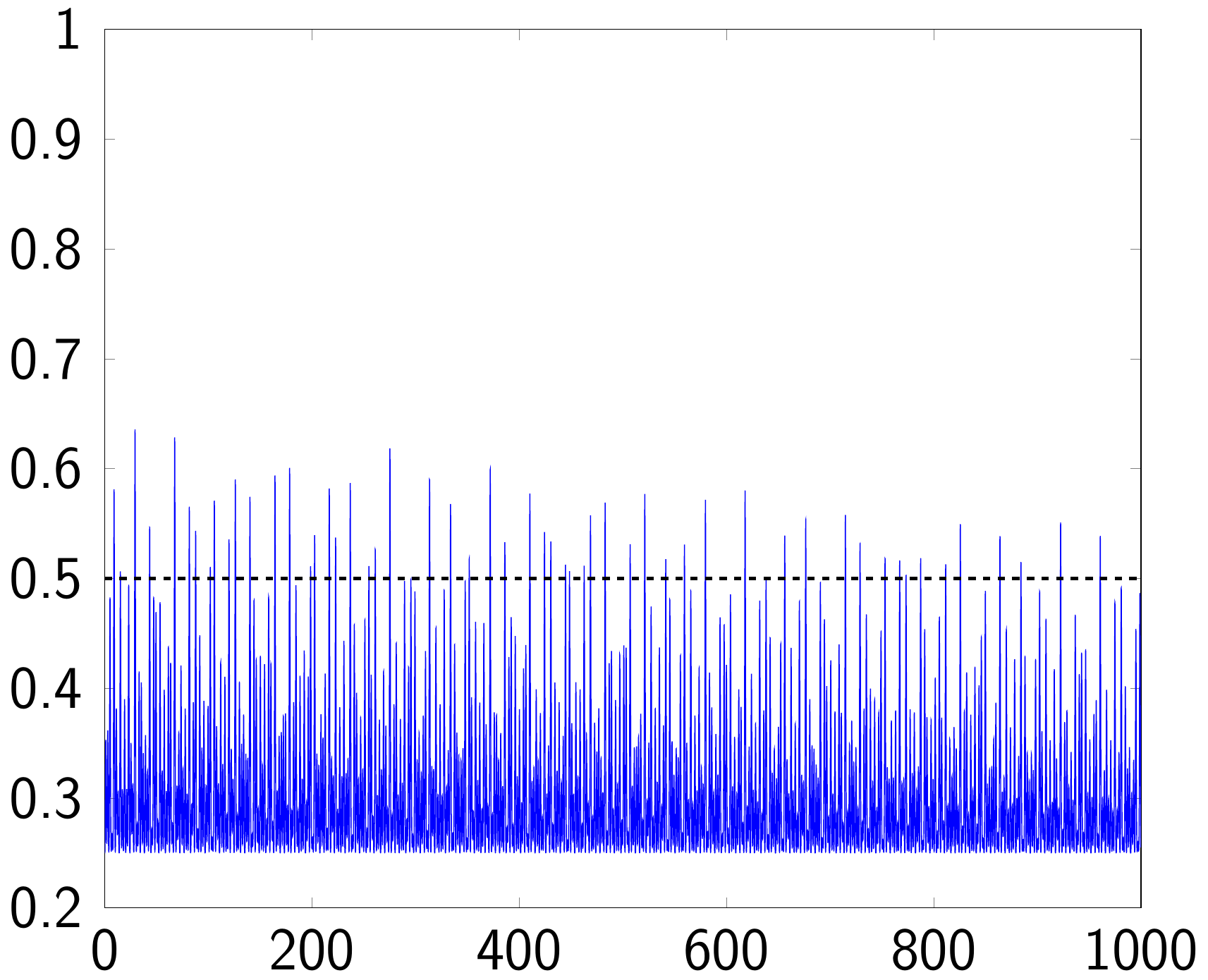} \\
		\hline
		$01$ & \vspace{0.2cm}
		\includegraphics[ width=\linewidth, height=\linewidth, keepaspectratio]{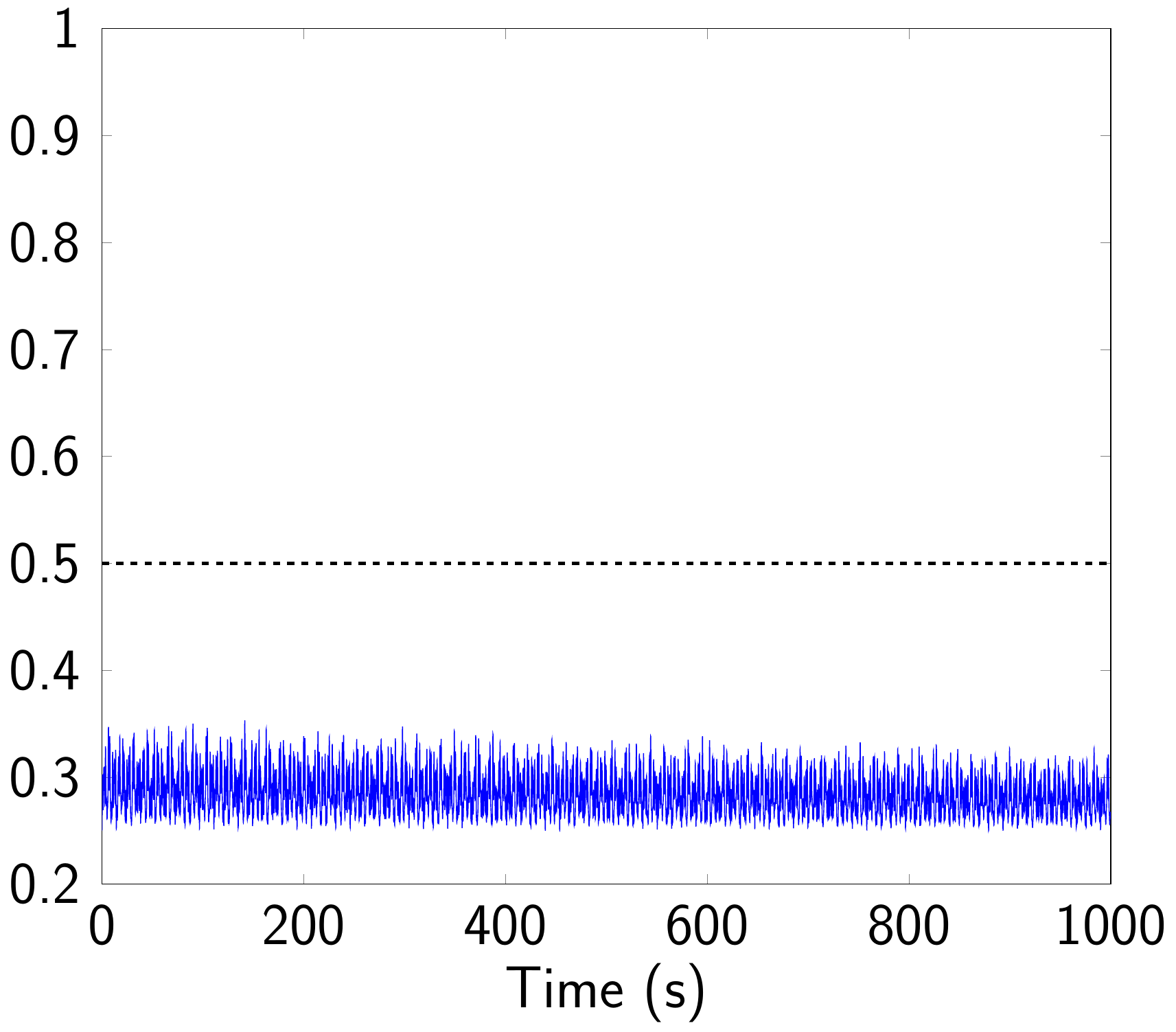} & \vspace{0.2cm}
		\includegraphics[ width=\linewidth, height=\linewidth, keepaspectratio]{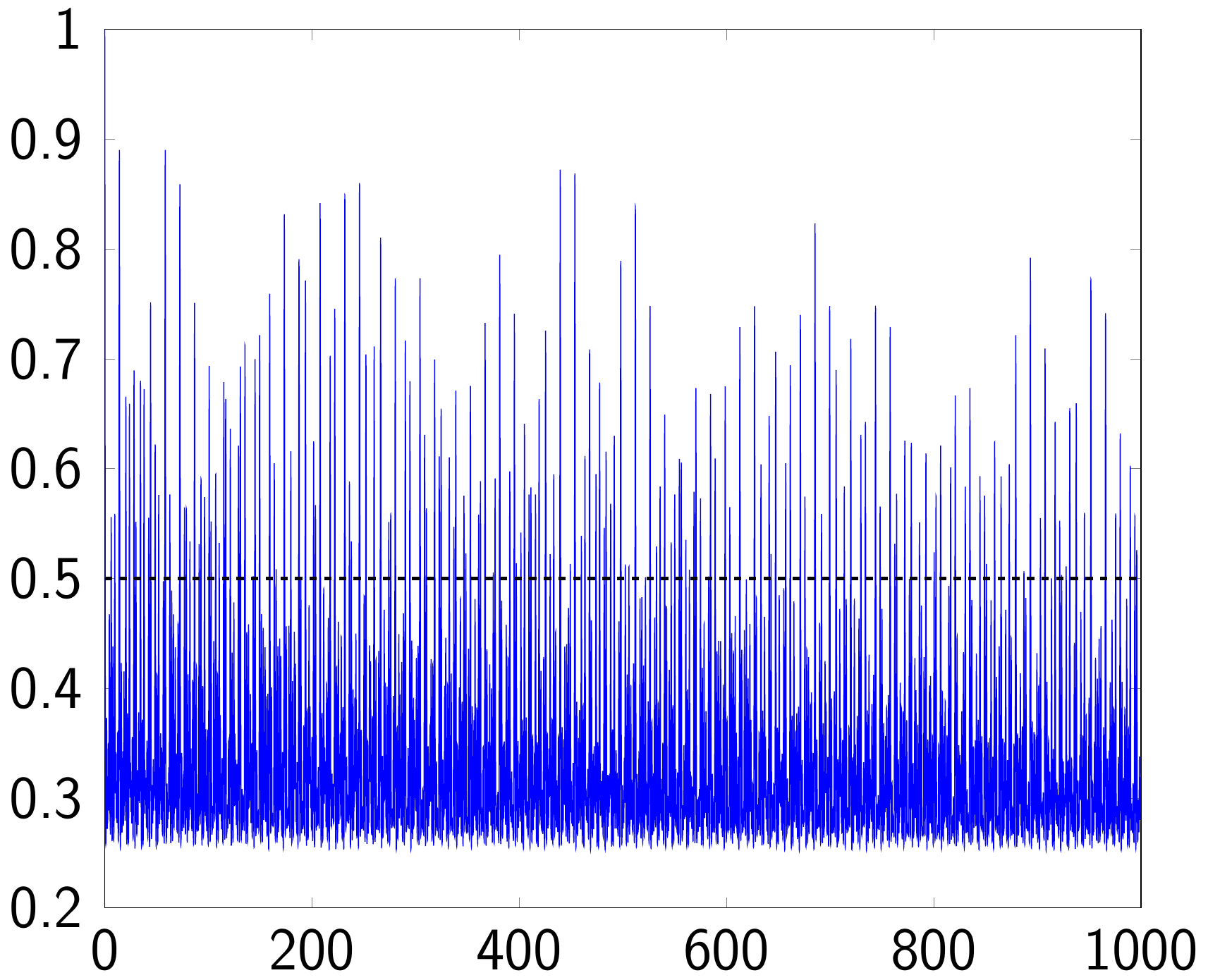} & \vspace{0.2cm}
		\includegraphics[ width=\linewidth, height=\linewidth, keepaspectratio]{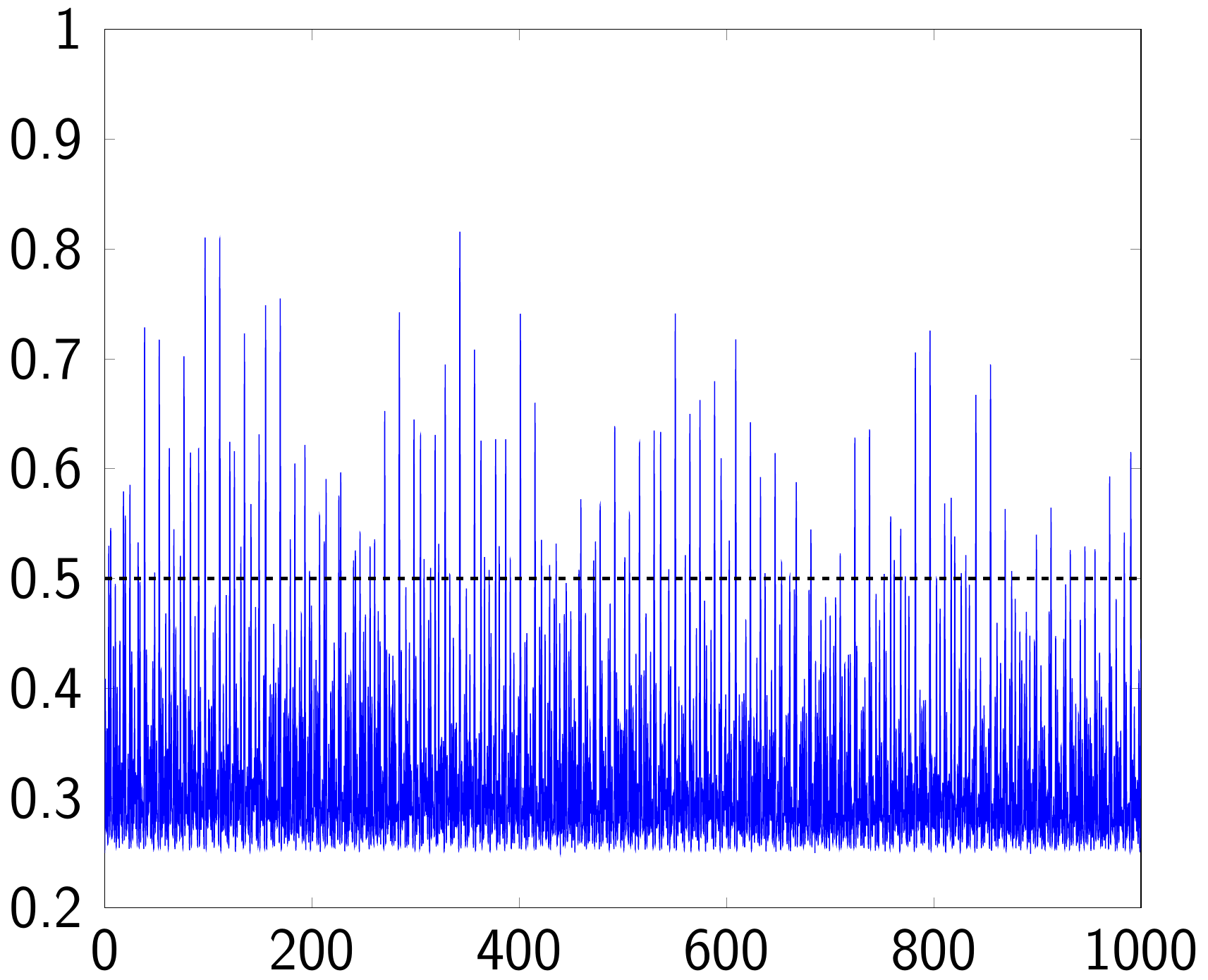} & \vspace{0.2cm}
		\includegraphics[ width=\linewidth, height=\linewidth, keepaspectratio]{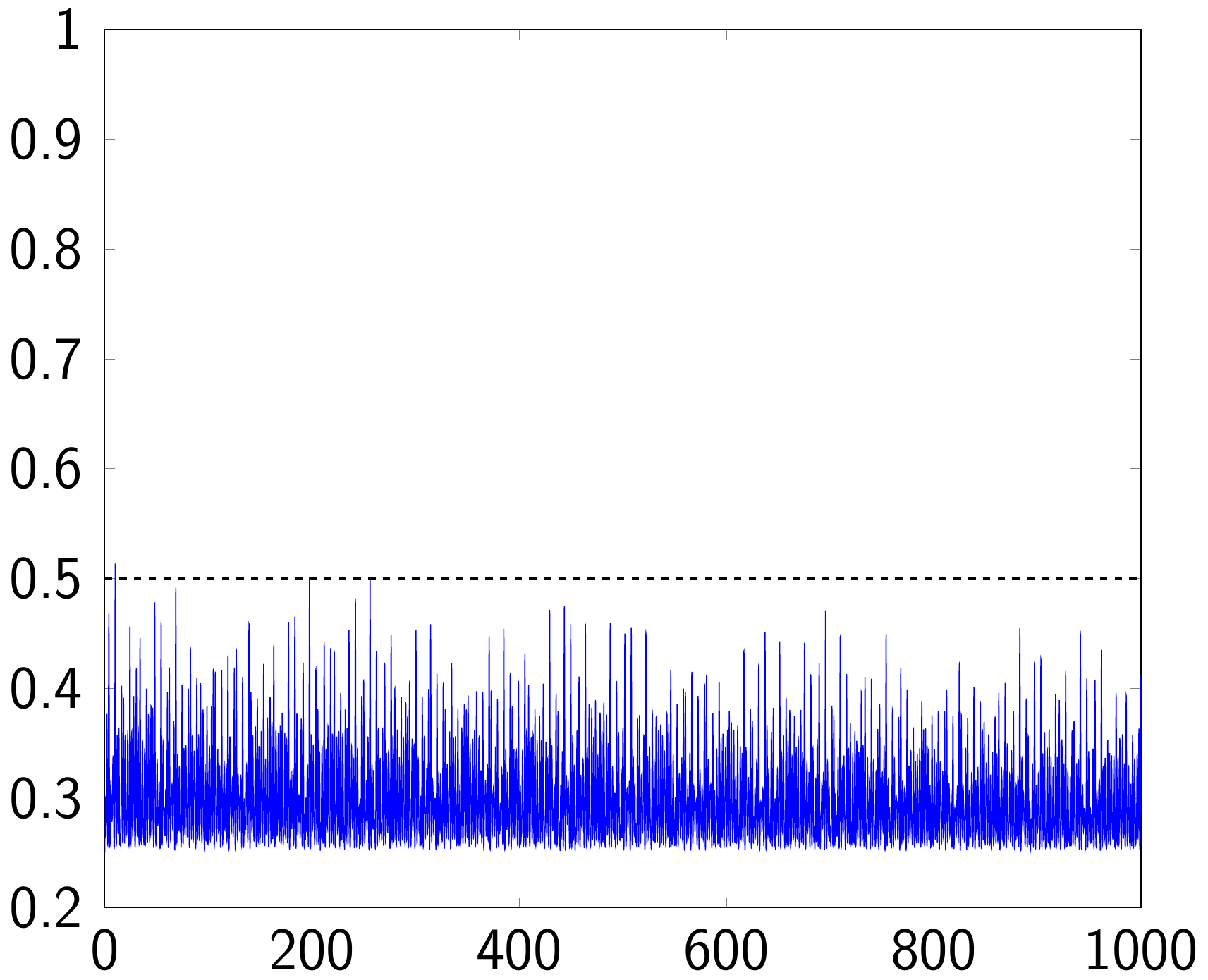} & \vspace{0.2cm}
		\includegraphics[ width=\linewidth, height=\linewidth, keepaspectratio]{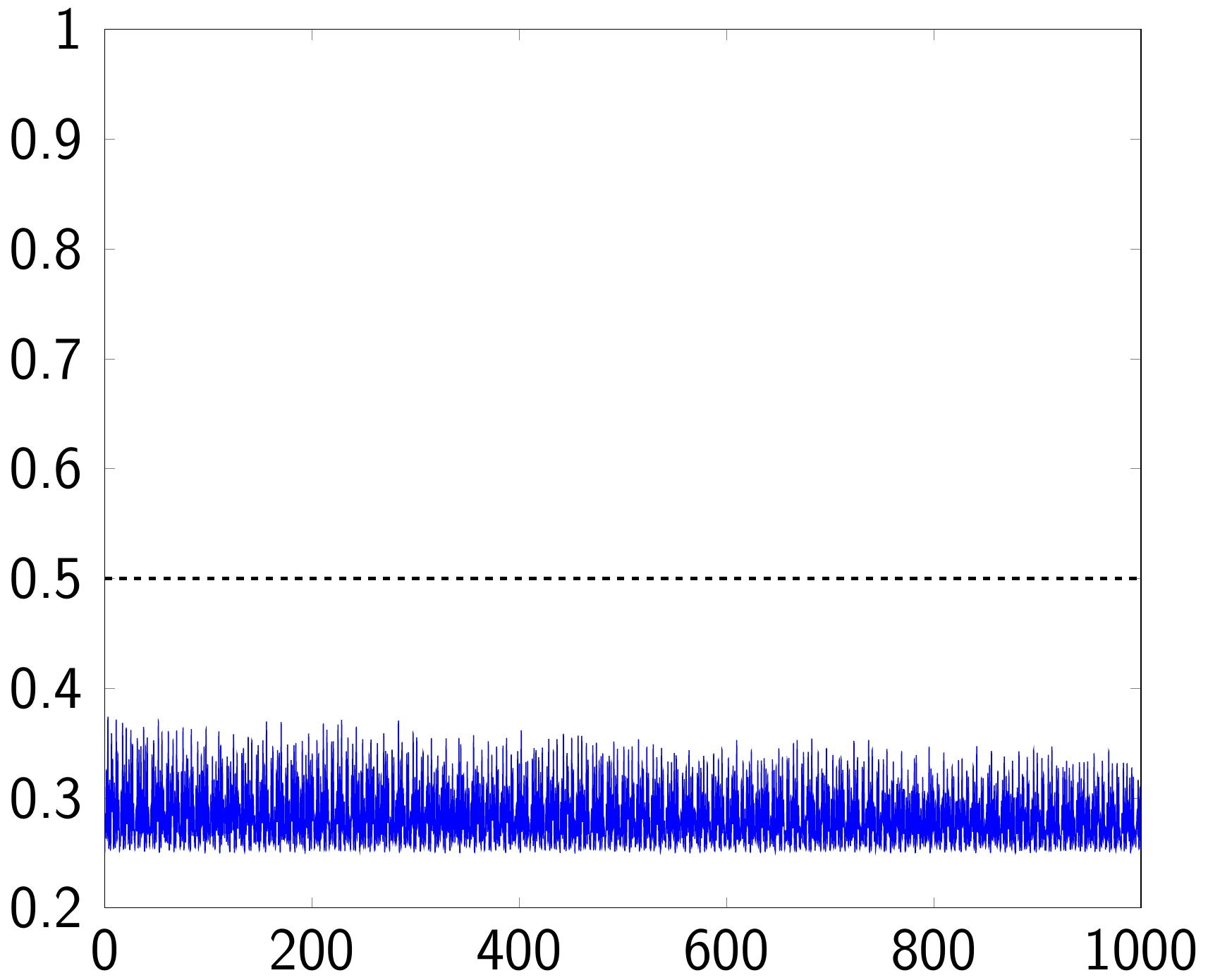} & \vspace{0.2cm}
		\includegraphics[ width=\linewidth, height=\linewidth, keepaspectratio]{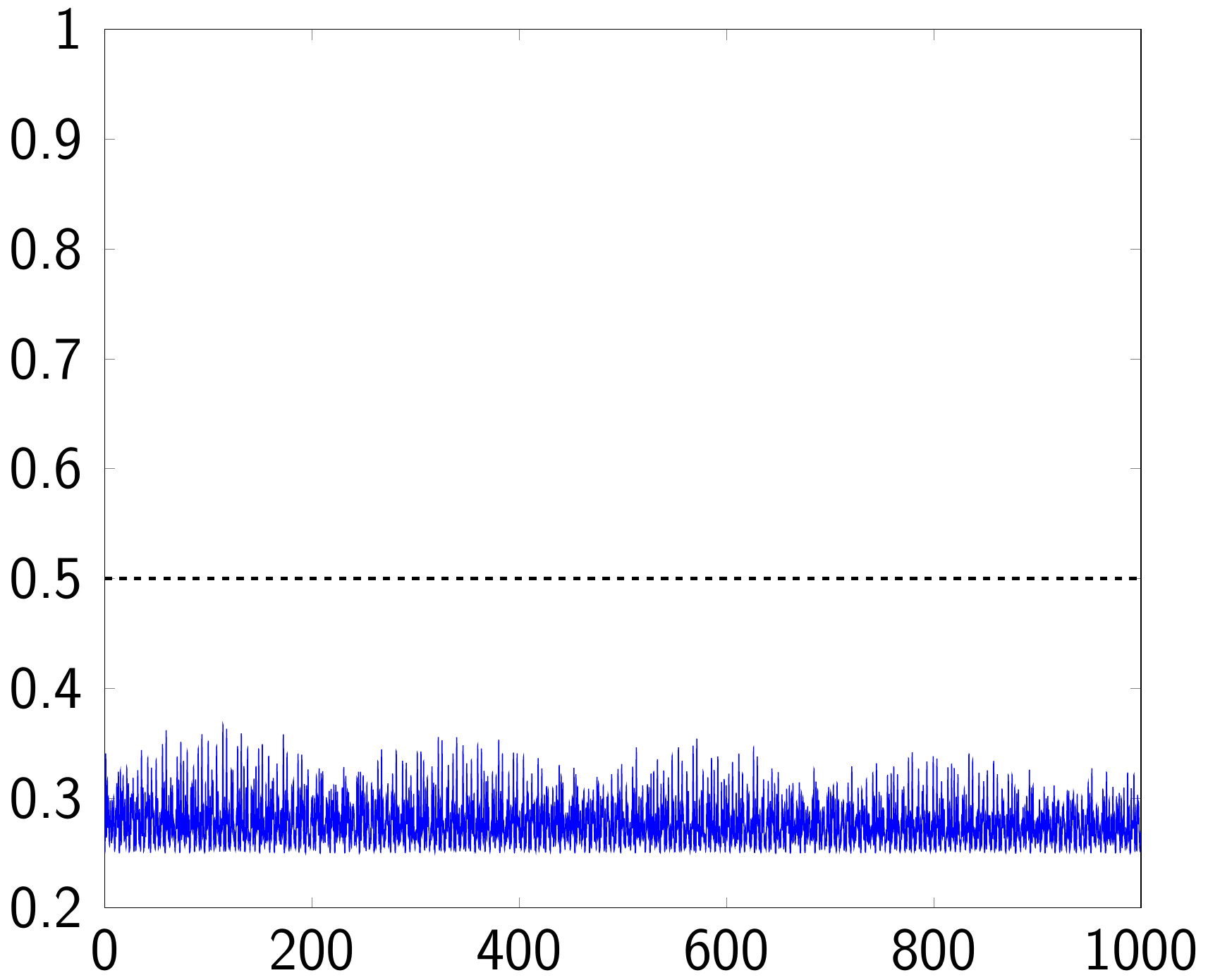} & \vspace{0.2cm}
		\includegraphics[ width=\linewidth, height=\linewidth, keepaspectratio]{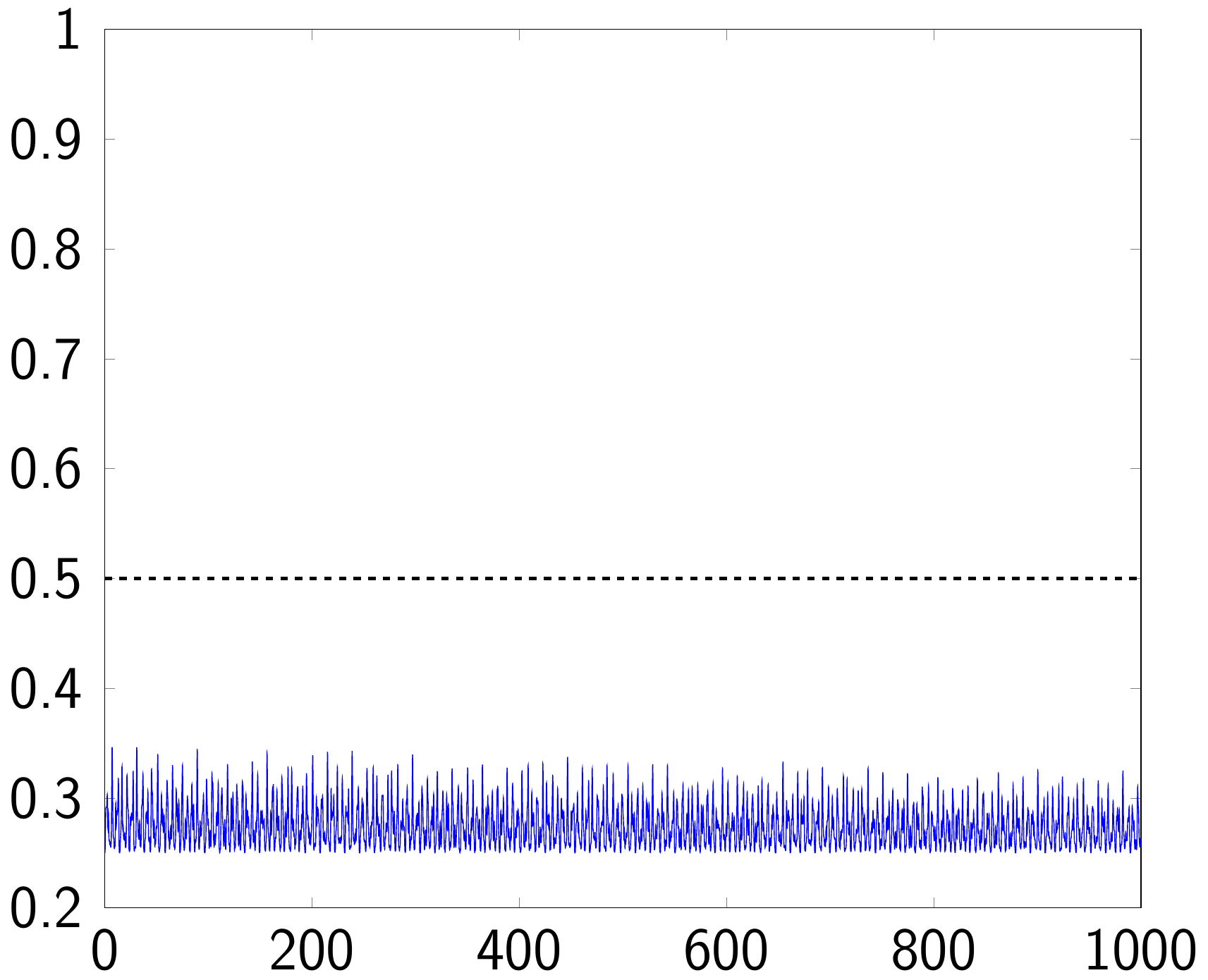} & \vspace{0.2cm}
		\includegraphics[ width=\linewidth, height=\linewidth, keepaspectratio]{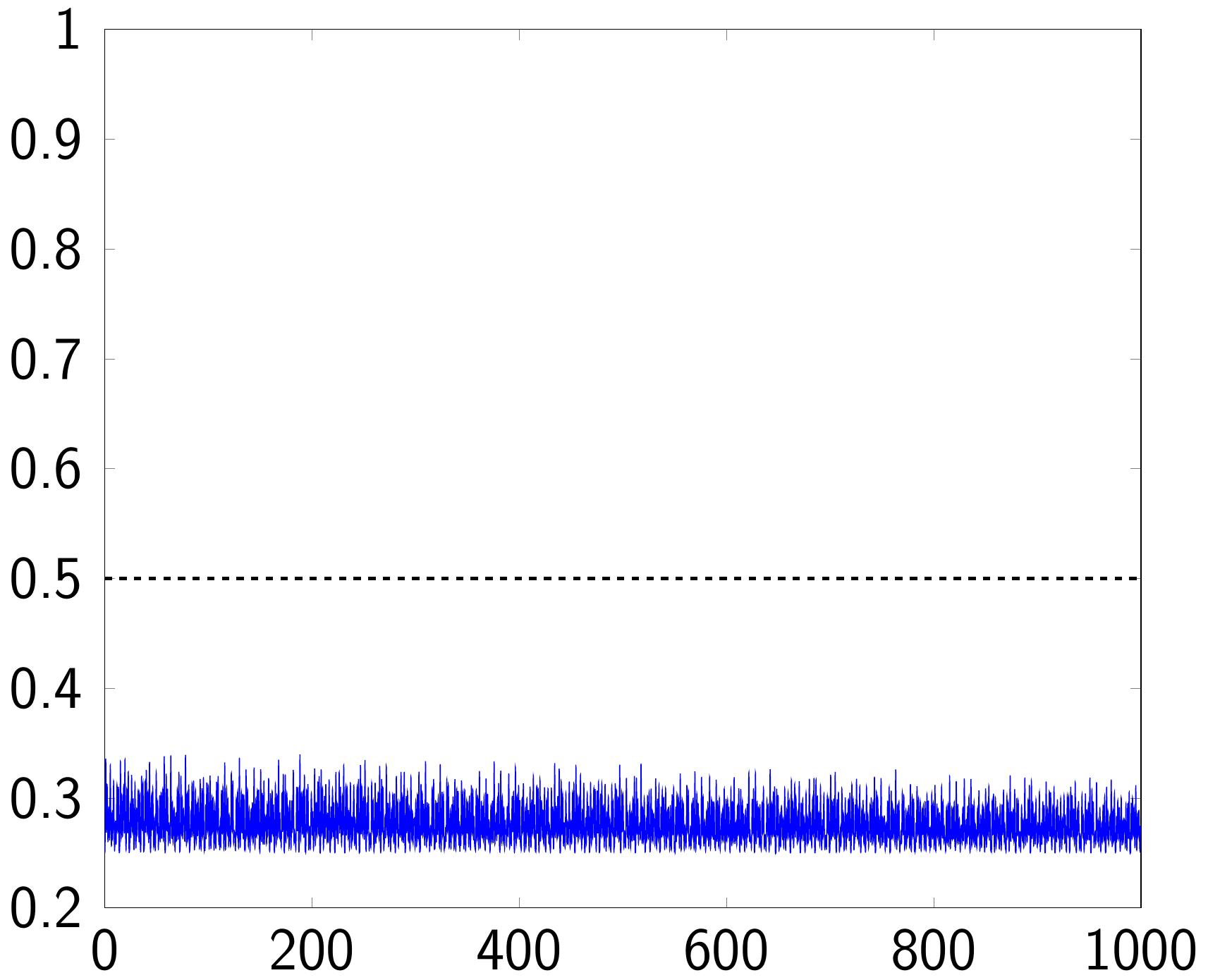} & \vspace{0.2cm}
		\includegraphics[ width=\linewidth, height=\linewidth, keepaspectratio]{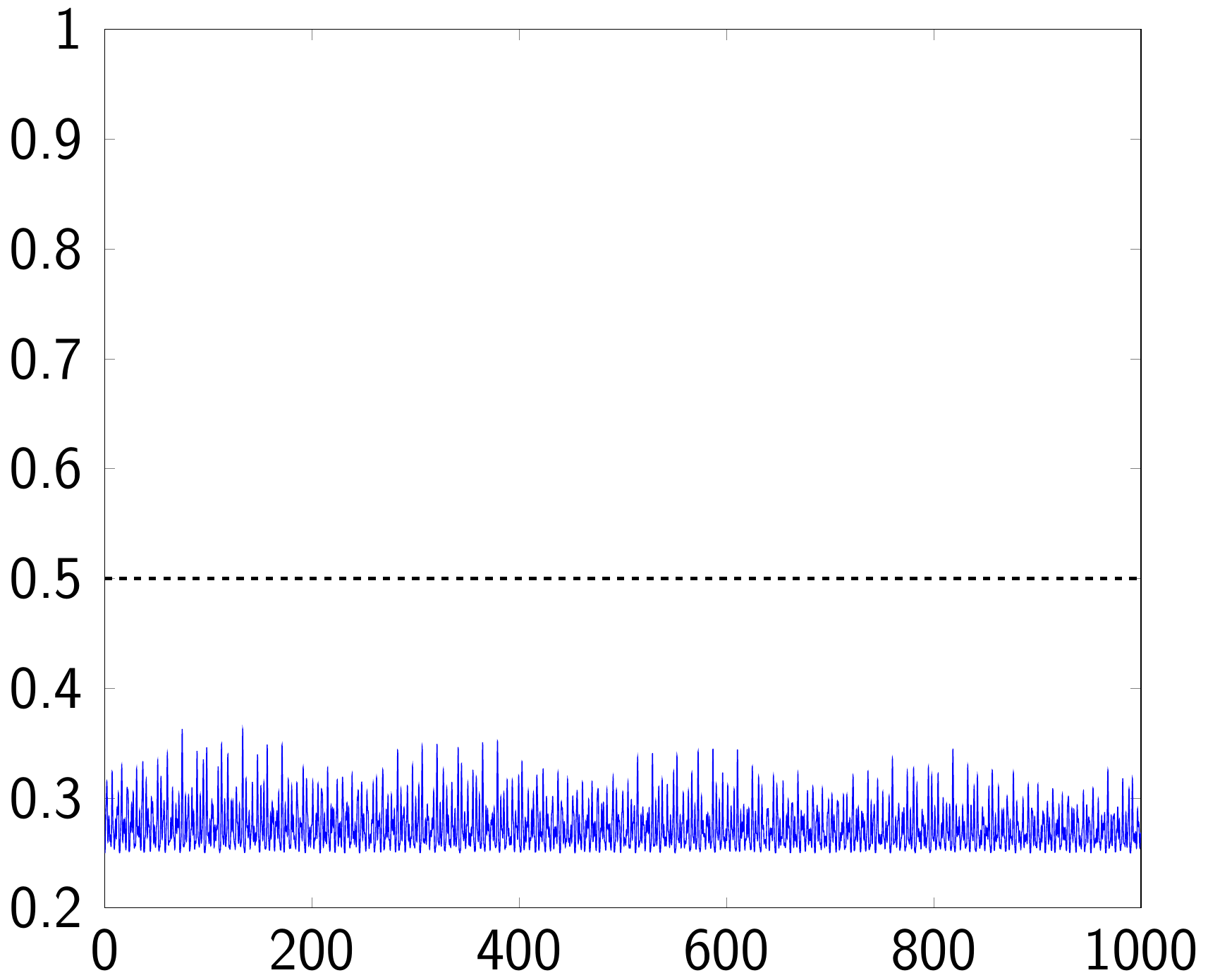} & \vspace{0.2cm}
		\includegraphics[ width=\linewidth, height=\linewidth, keepaspectratio]{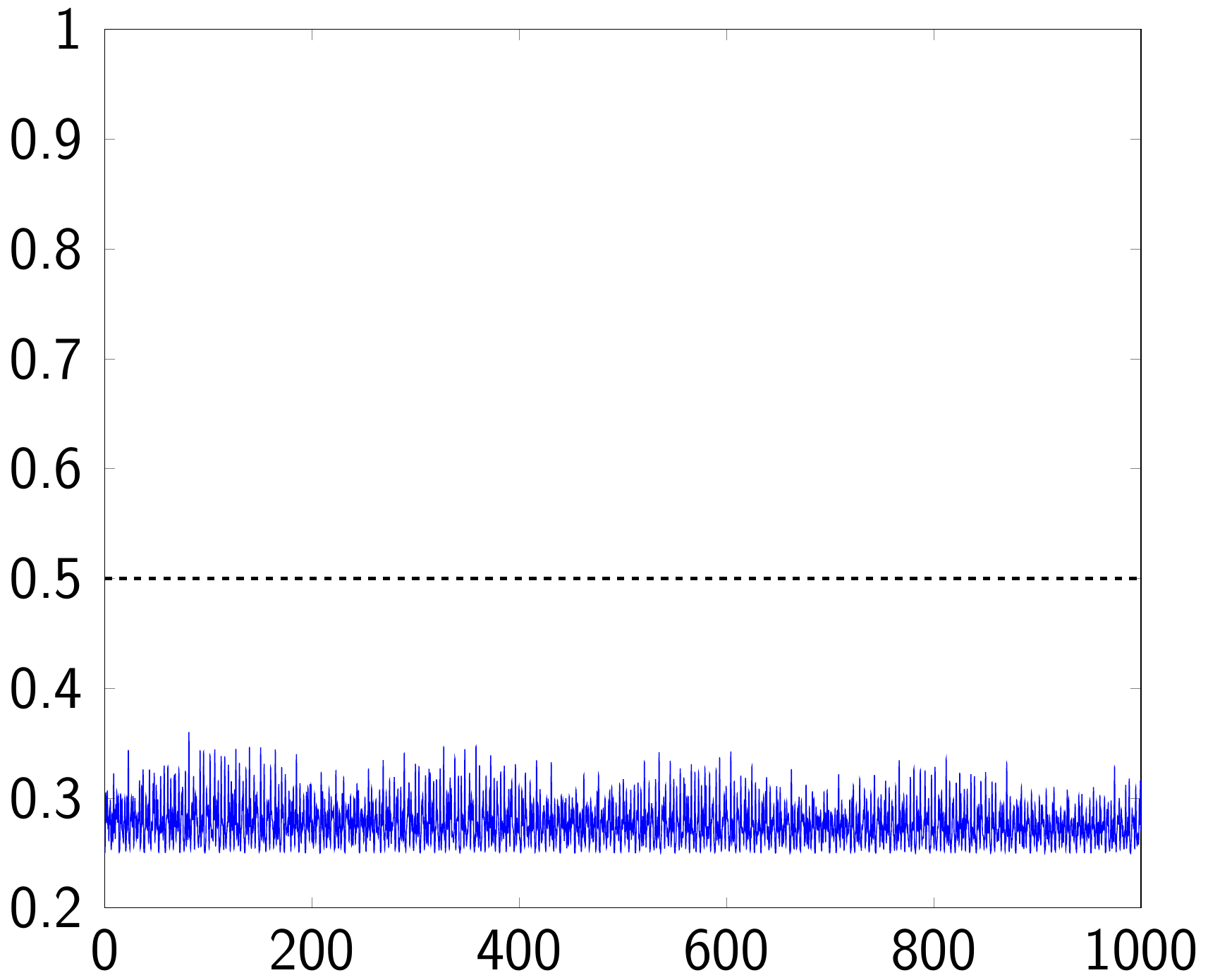} \\
		\hline
		$02$ & \vspace{0.2cm}
		\includegraphics[ width=\linewidth, height=\linewidth, keepaspectratio]{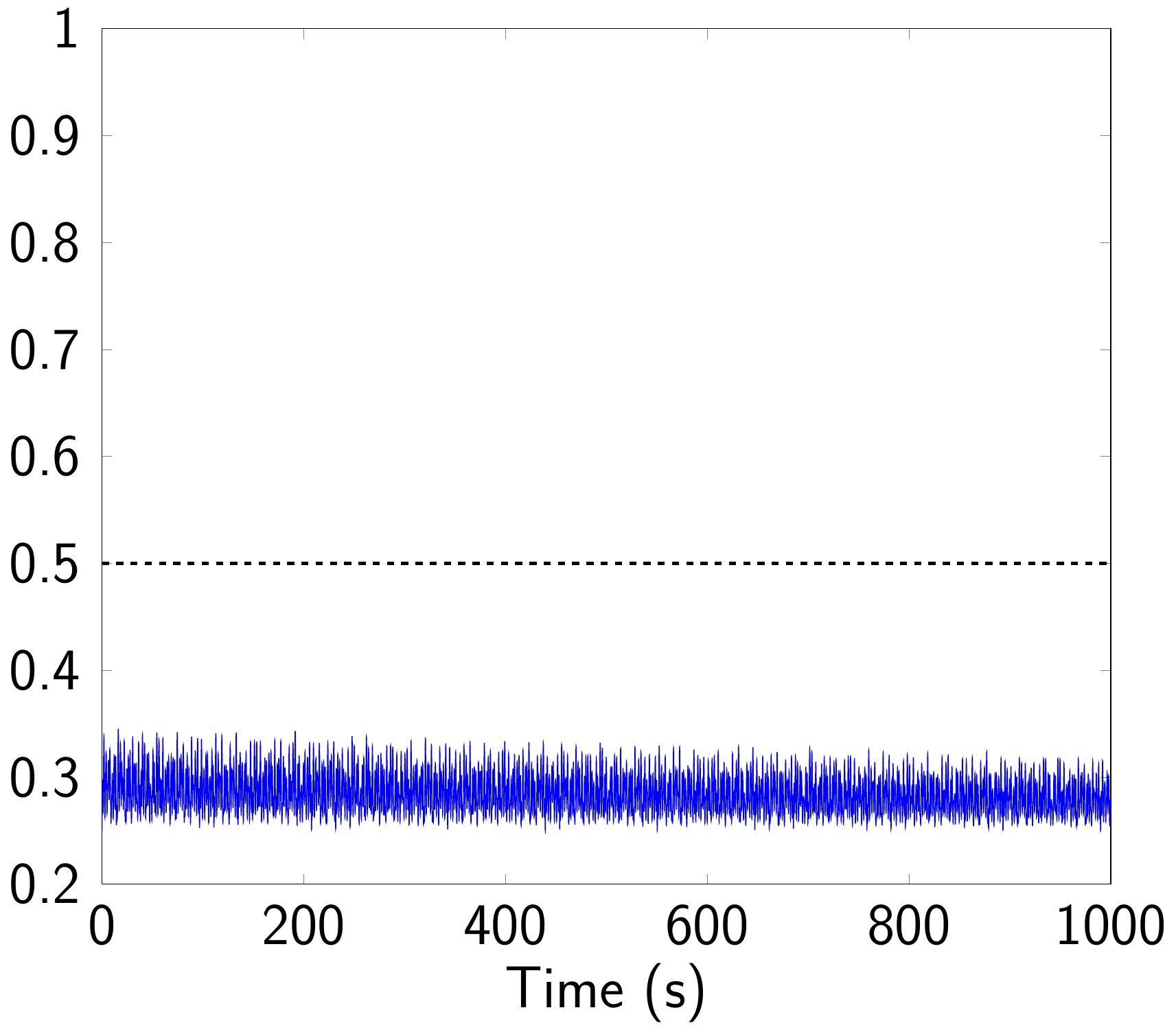} & \vspace{0.2cm}
		\includegraphics[ width=\linewidth, height=\linewidth, keepaspectratio]{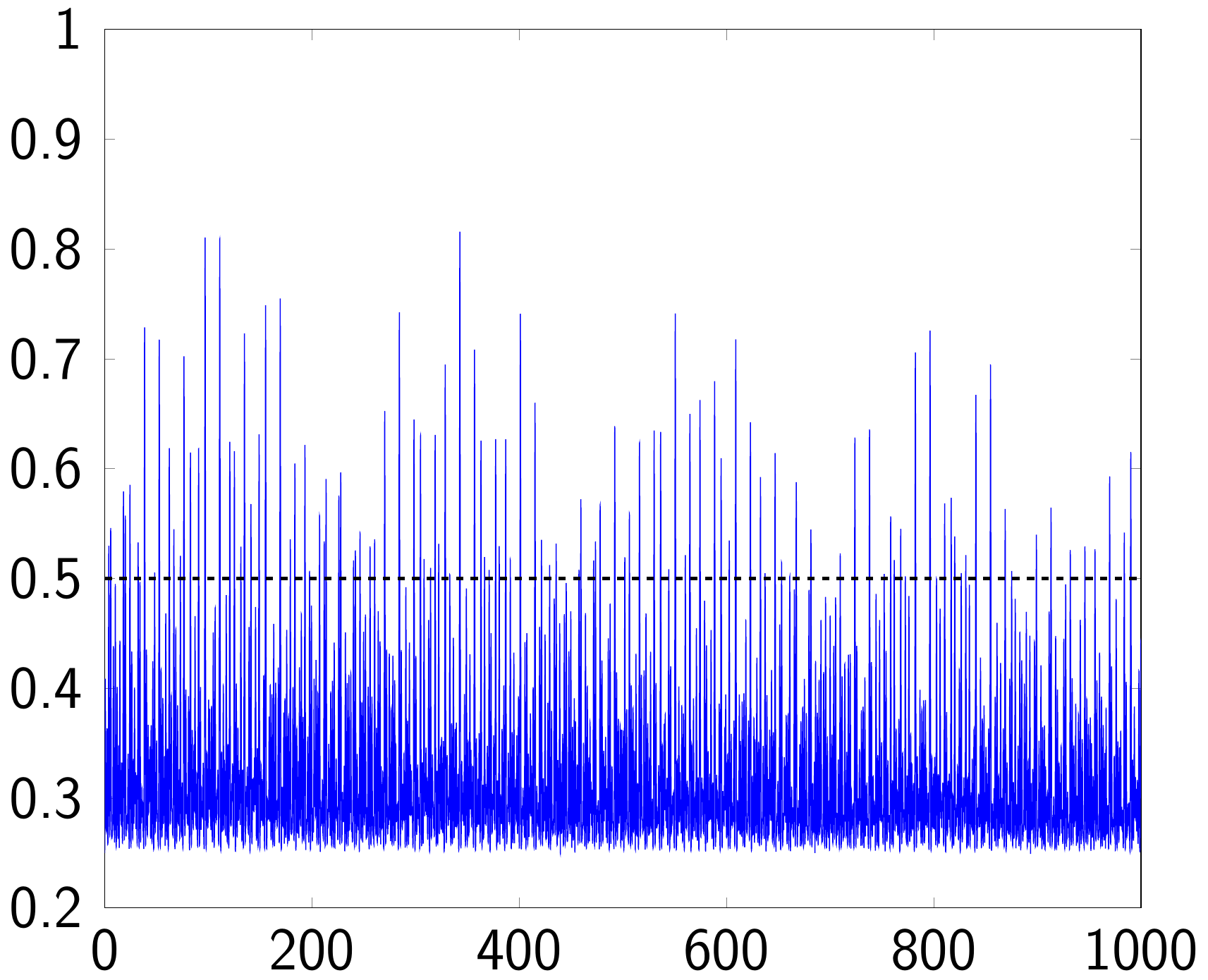} & \vspace{0.2cm}
		\includegraphics[ width=\linewidth, height=\linewidth, keepaspectratio]{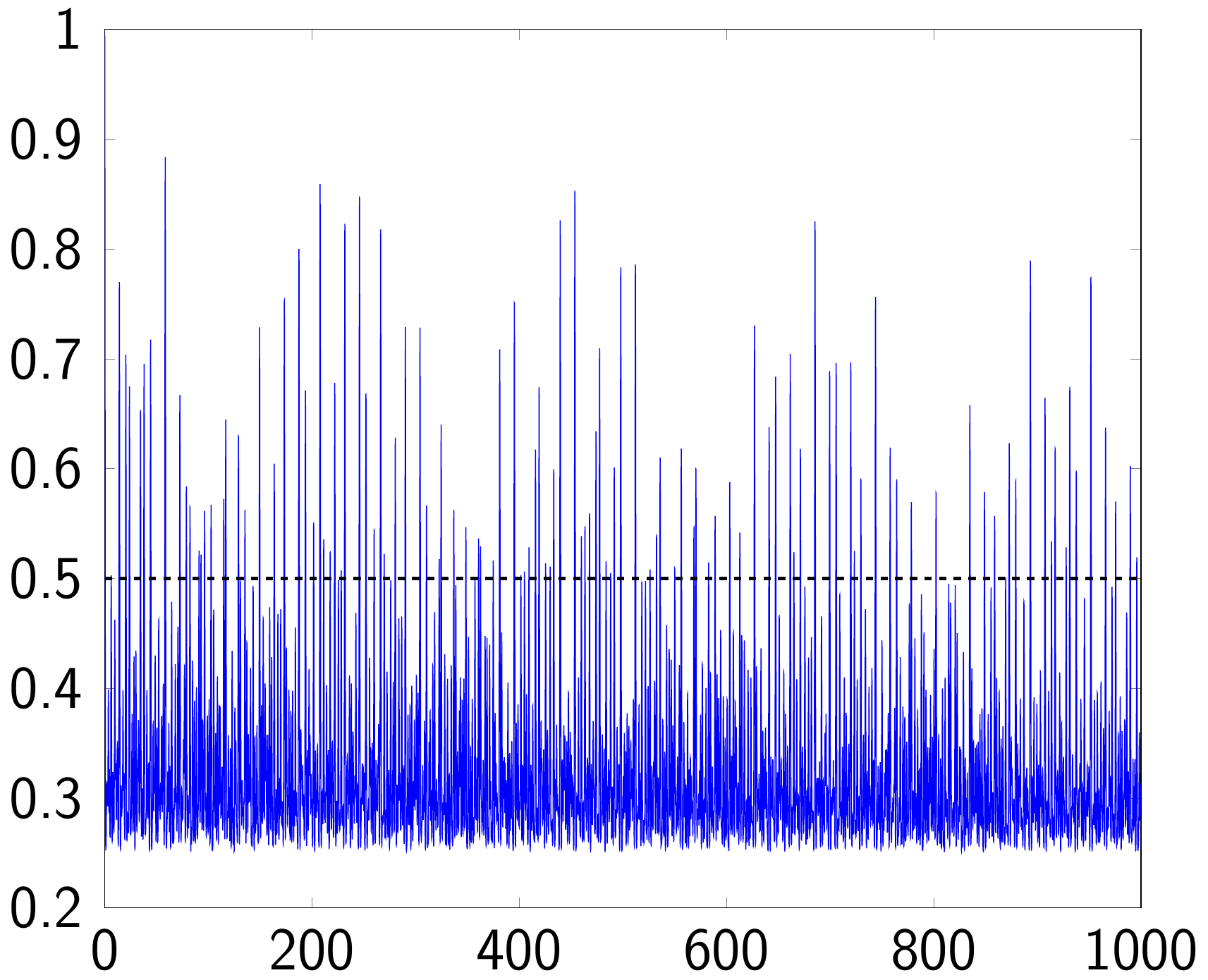} & \vspace{0.2cm}
		\includegraphics[ width=\linewidth, height=\linewidth, keepaspectratio]{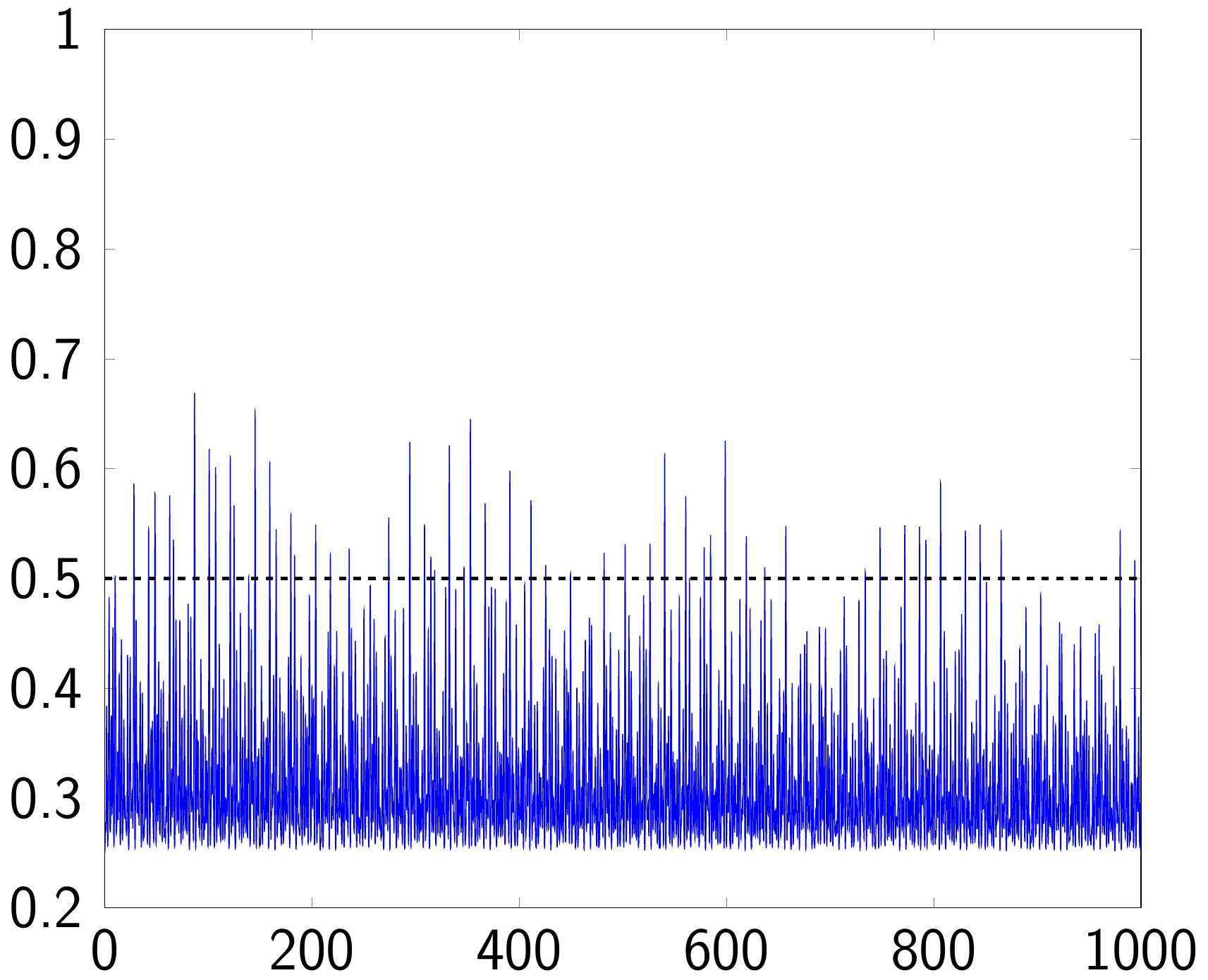} & \vspace{0.2cm}
		\includegraphics[ width=\linewidth, height=\linewidth, keepaspectratio]{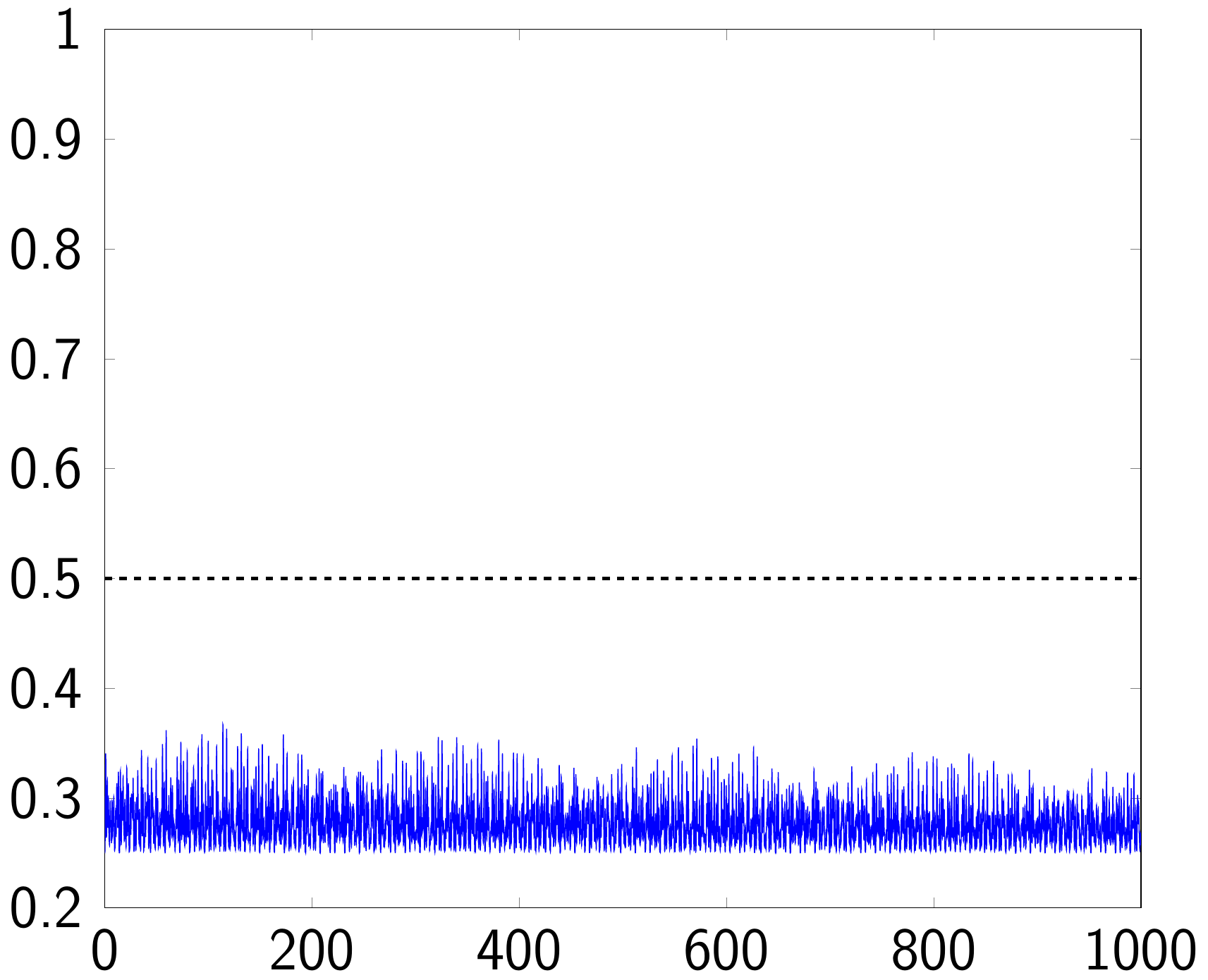} & \vspace{0.2cm}
		\includegraphics[ width=\linewidth, height=\linewidth, keepaspectratio]{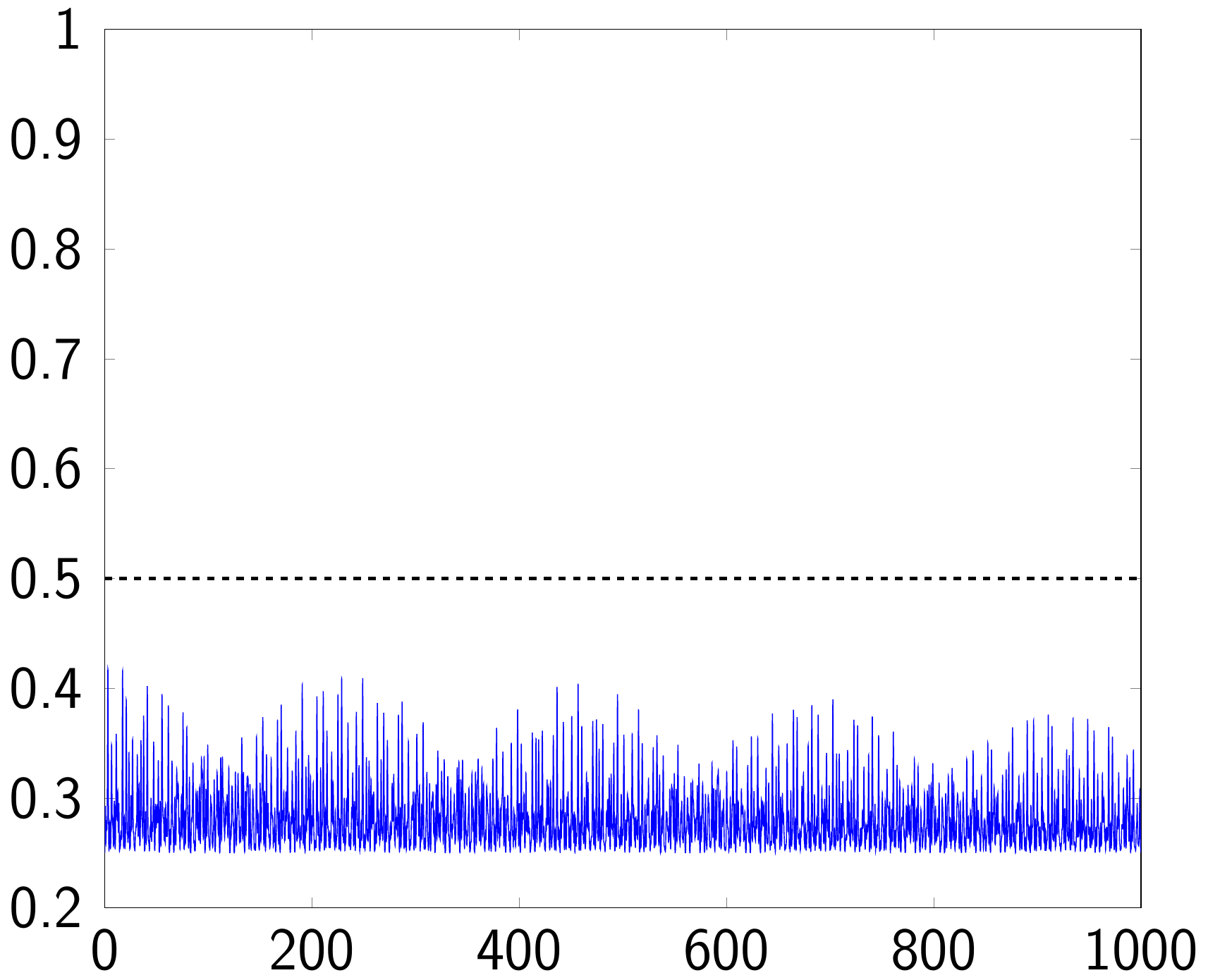} & \vspace{0.2cm}
		\includegraphics[ width=\linewidth, height=\linewidth, keepaspectratio]{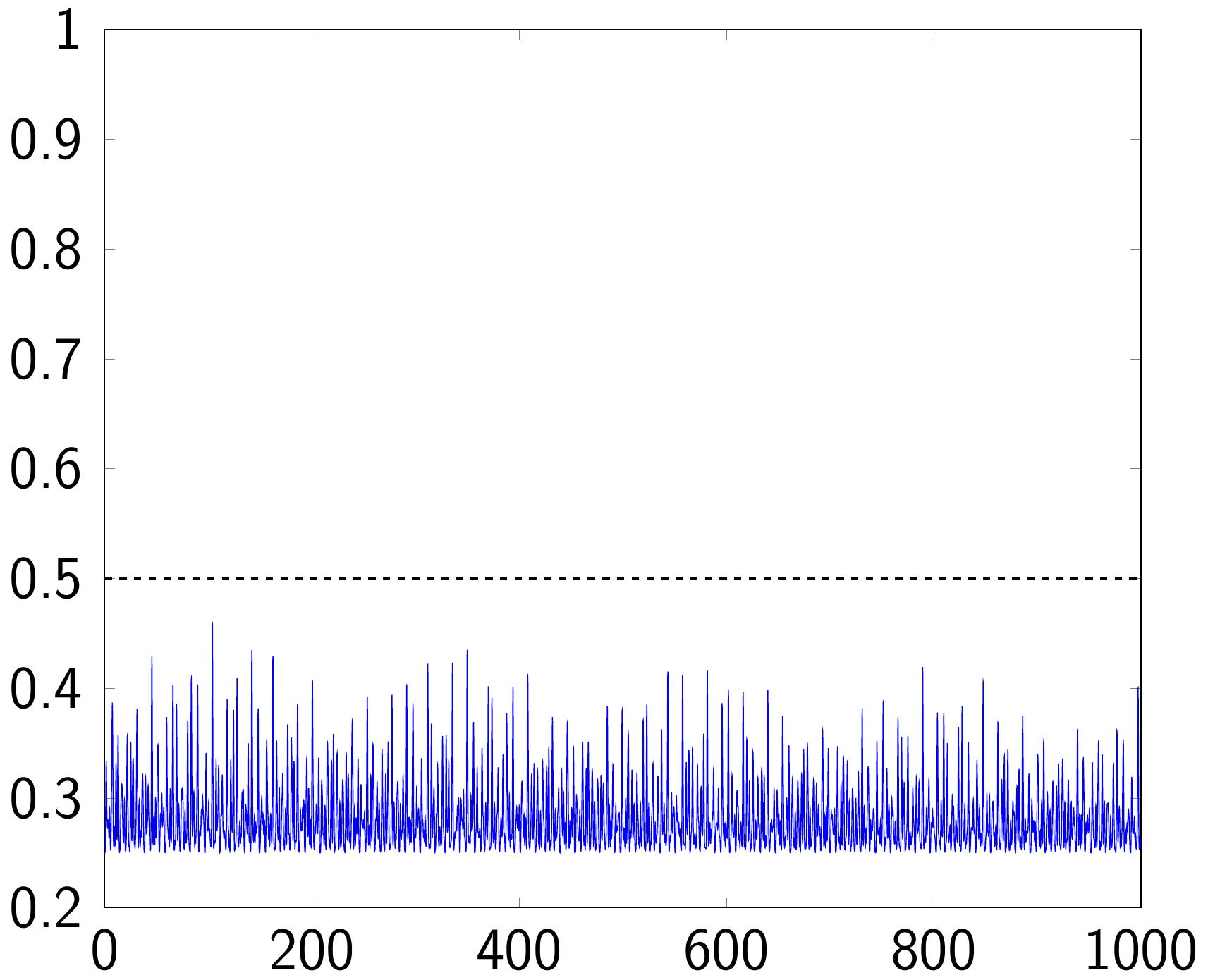} & \vspace{0.2cm}
		\includegraphics[ width=\linewidth, height=\linewidth, keepaspectratio]{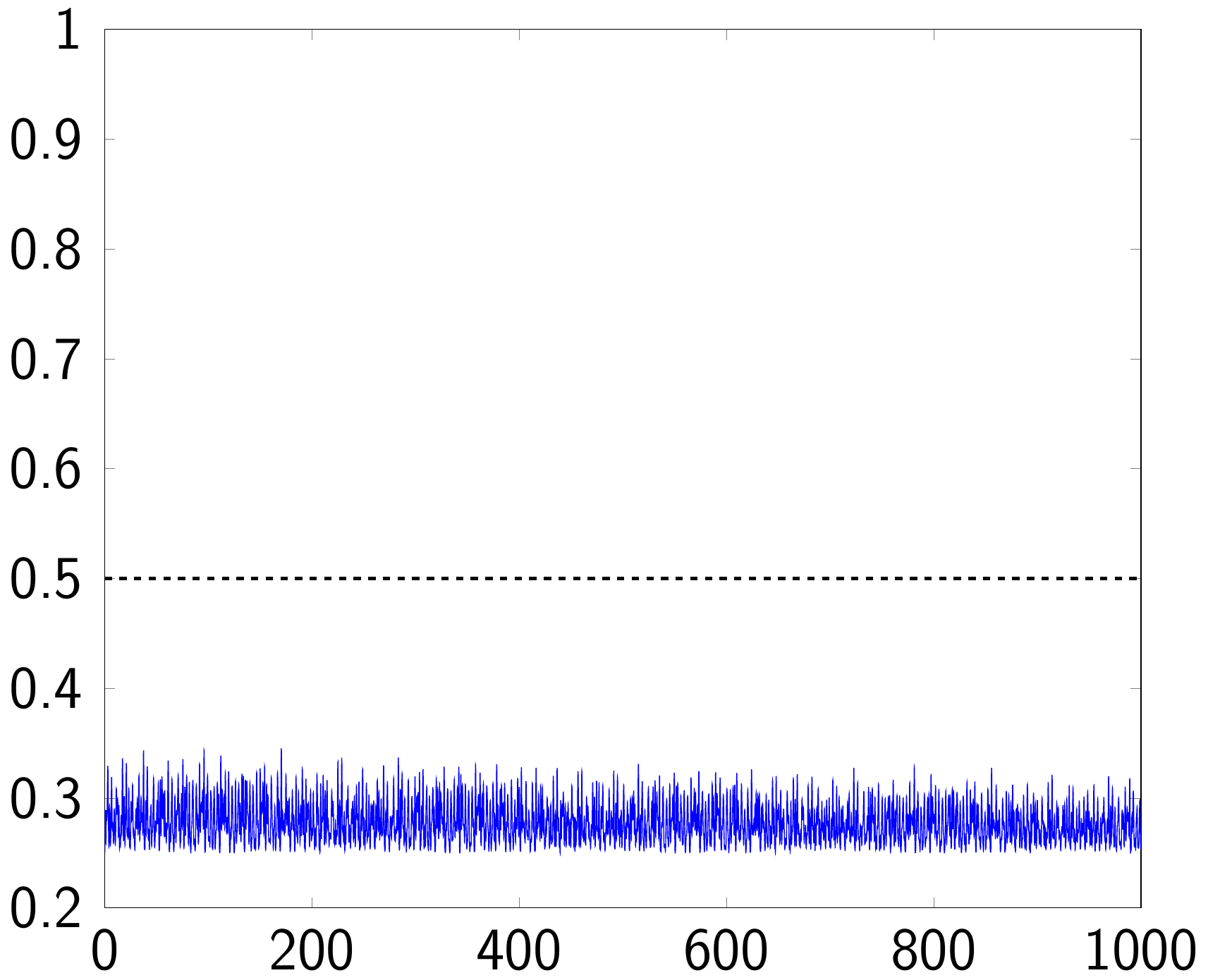} & \vspace{0.2cm}
		\includegraphics[ width=\linewidth, height=\linewidth, keepaspectratio]{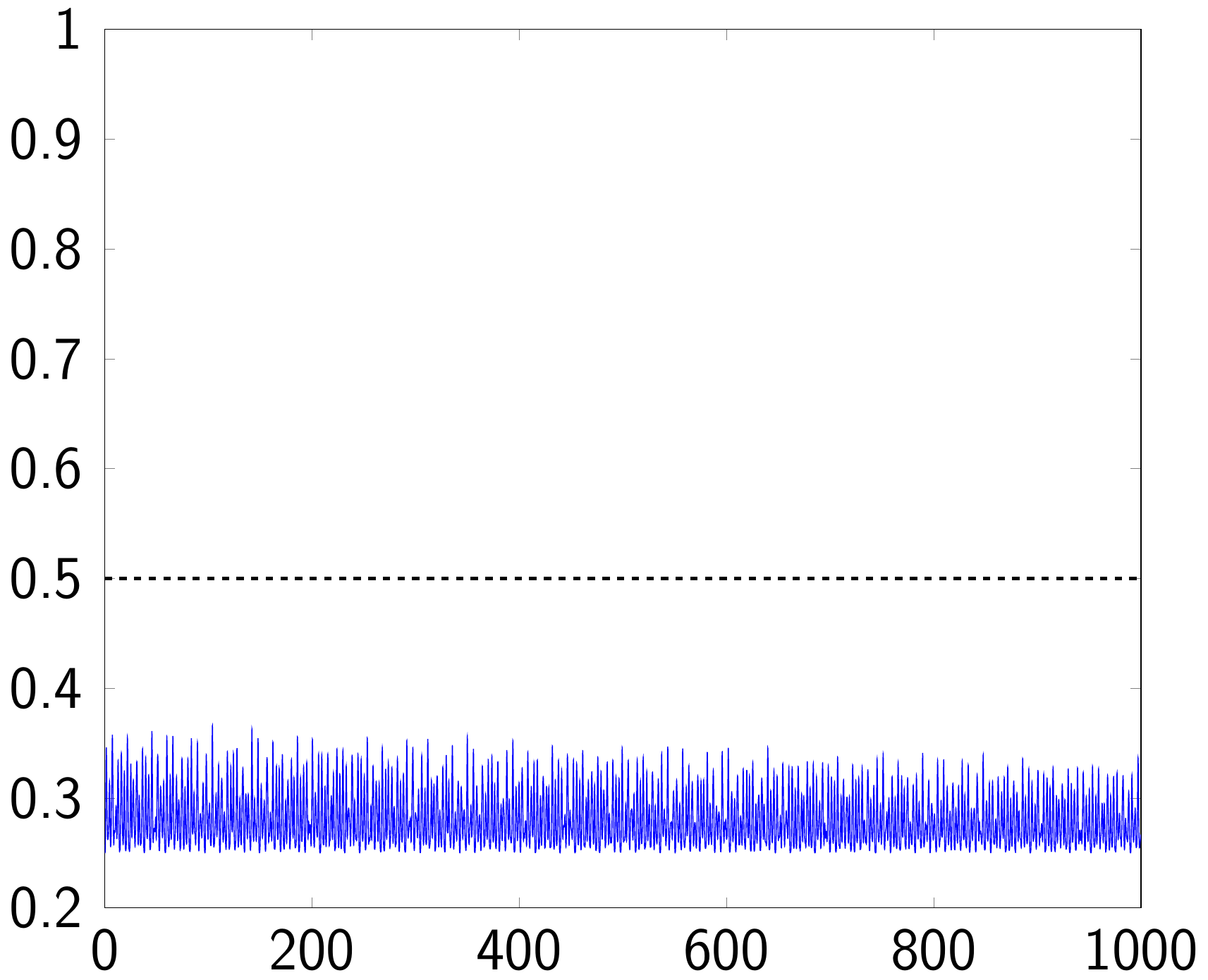} & \vspace{0.2cm}
		\includegraphics[ width=\linewidth, height=\linewidth, keepaspectratio]{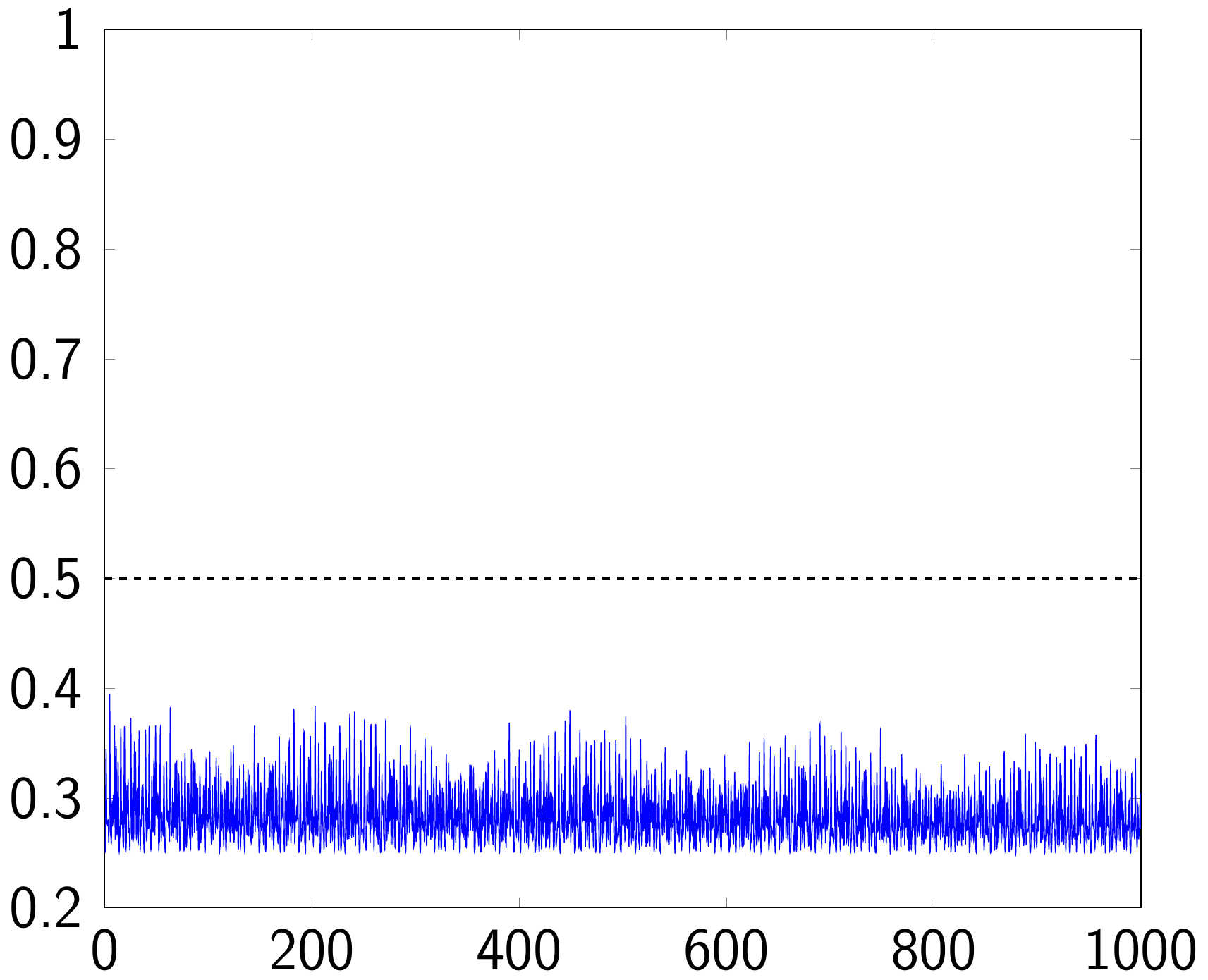} \\
		\hline
		$03$ & \vspace{0.2cm}
		\includegraphics[ width=\linewidth, height=\linewidth, keepaspectratio]{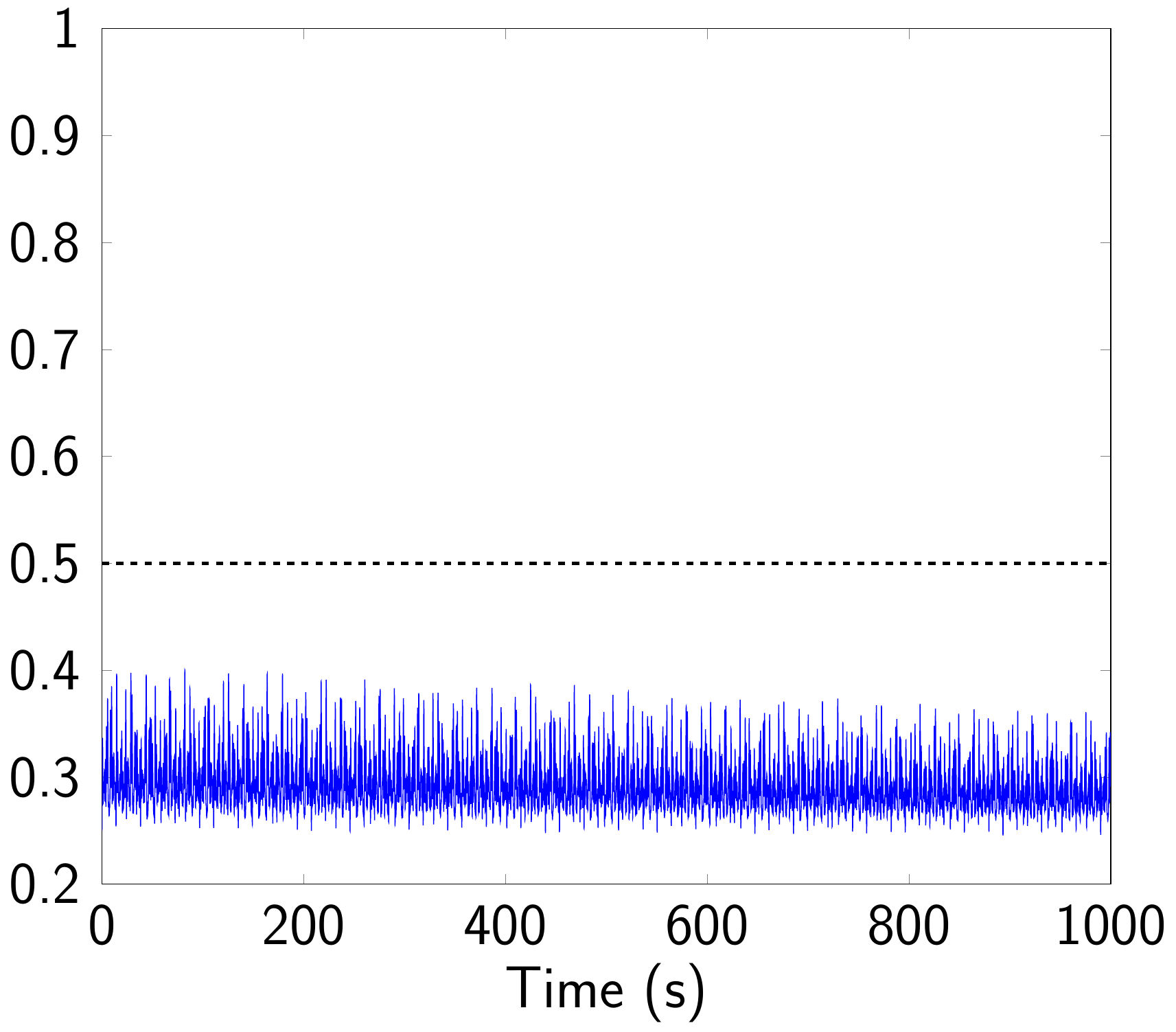} & \vspace{0.2cm}
		\includegraphics[ width=\linewidth, height=\linewidth, keepaspectratio]{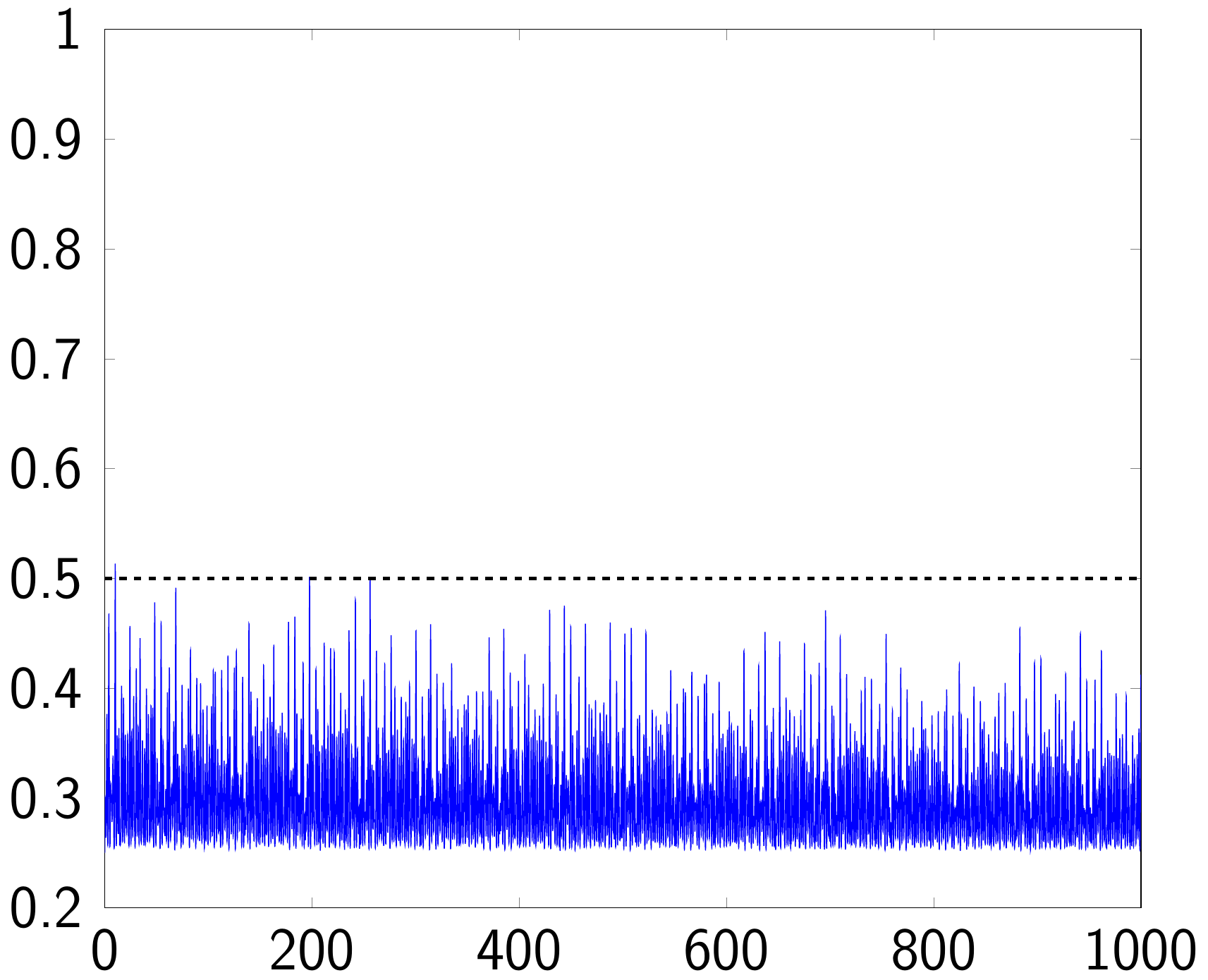} & \vspace{0.2cm}
		\includegraphics[ width=\linewidth, height=\linewidth, keepaspectratio]{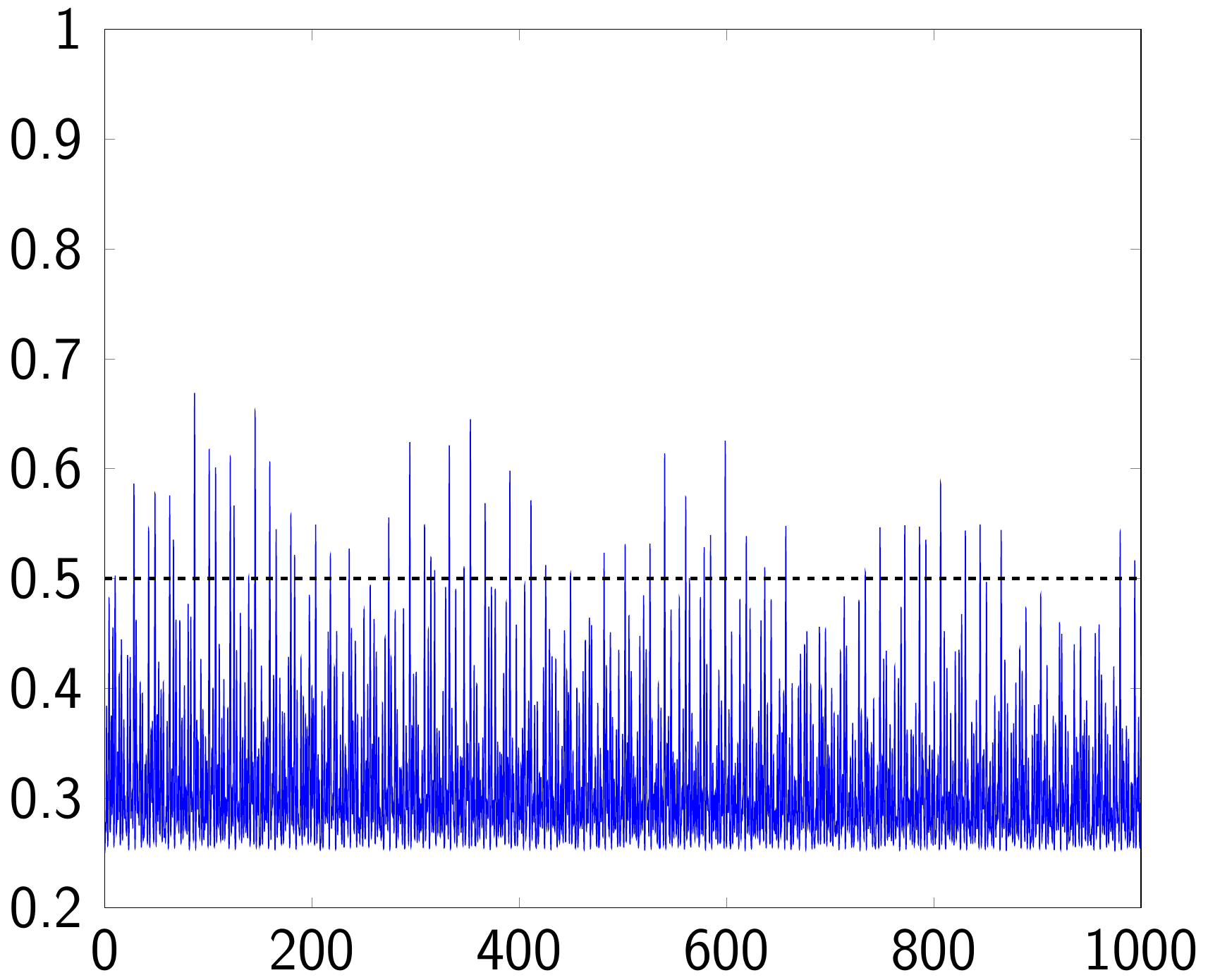} & \vspace{0.2cm}
		\includegraphics[ width=\linewidth, height=\linewidth, keepaspectratio]{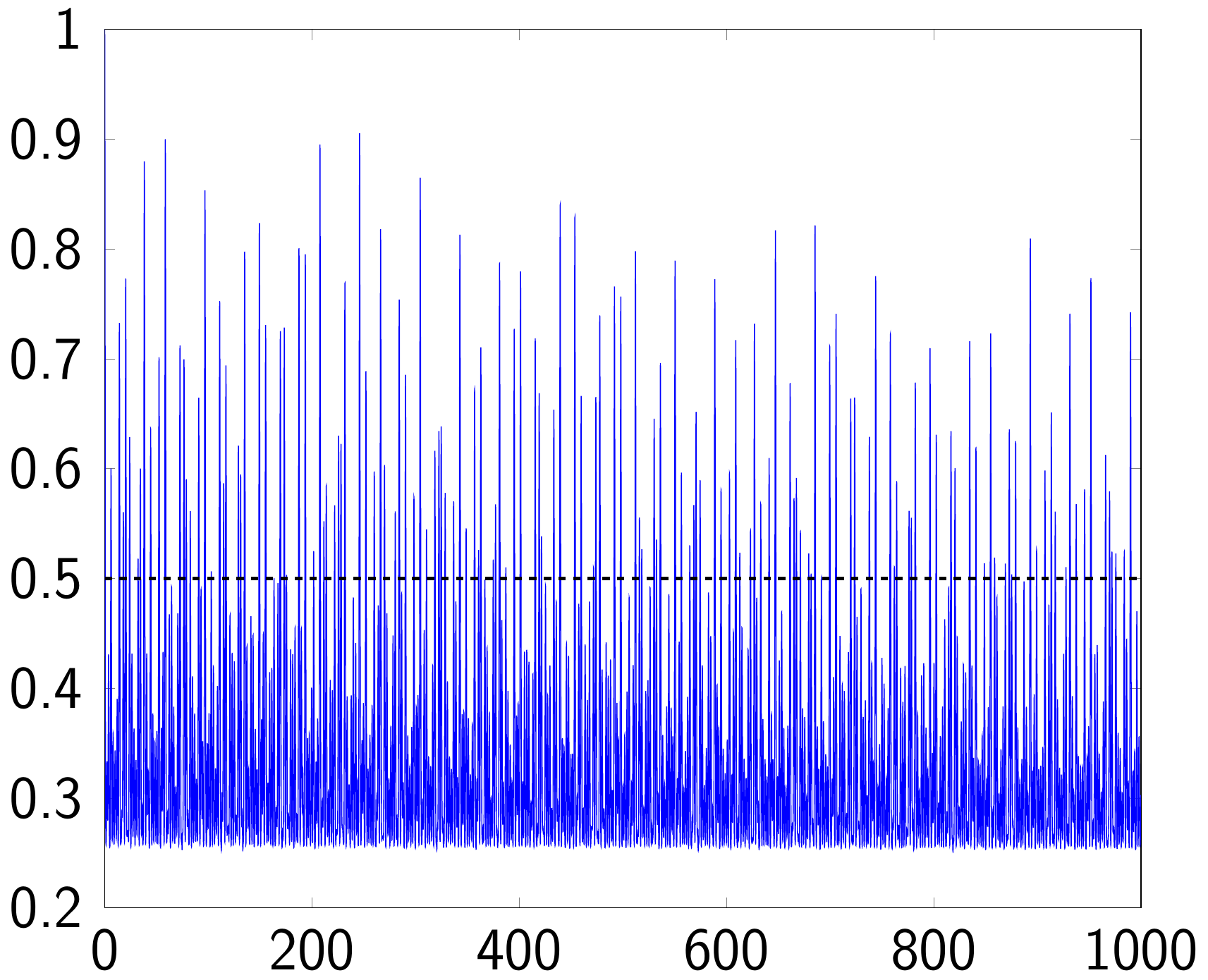} & \vspace{0.2cm}
		\includegraphics[ width=\linewidth, height=\linewidth, keepaspectratio]{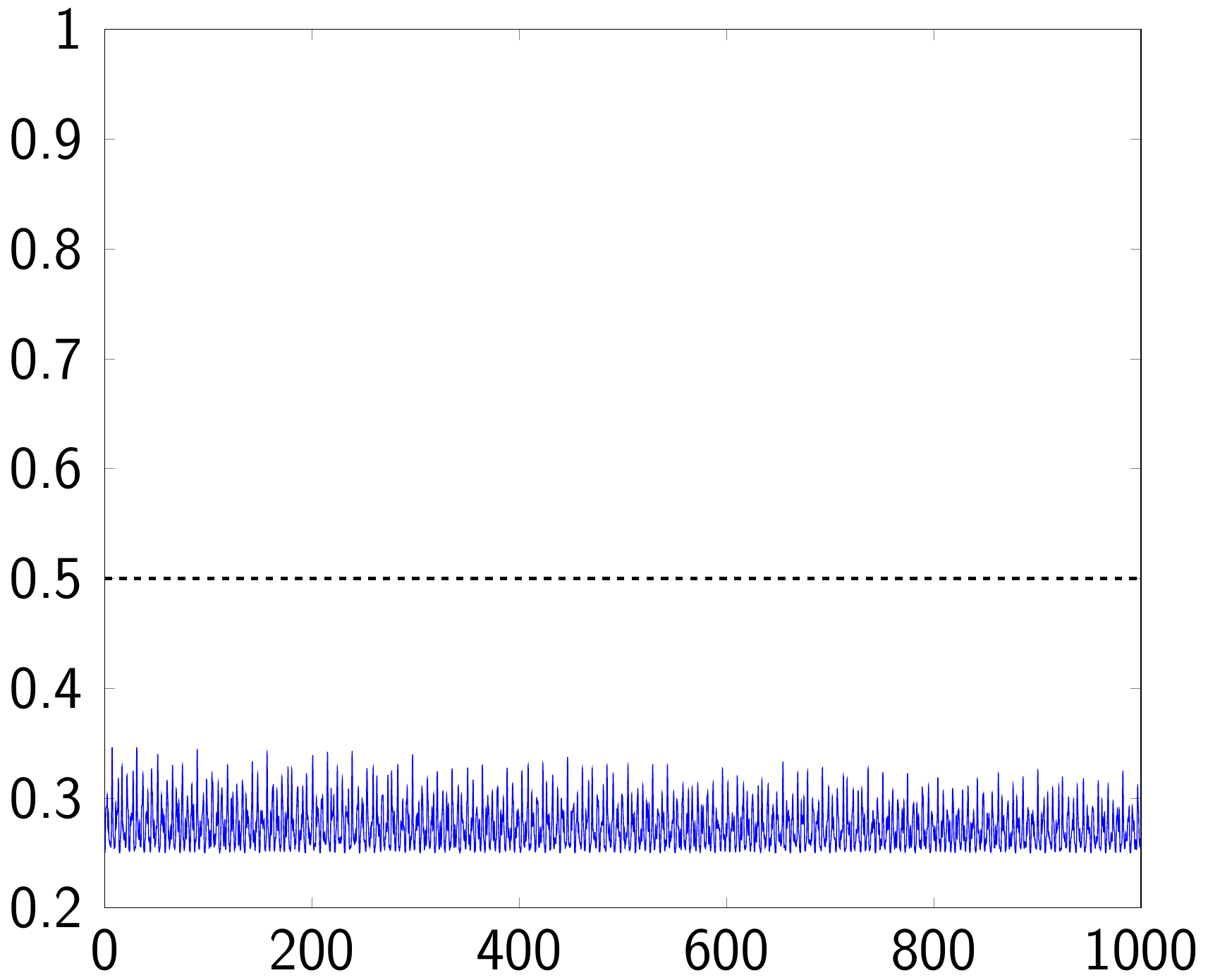} & \vspace{0.2cm}
		\includegraphics[ width=\linewidth, height=\linewidth, keepaspectratio]{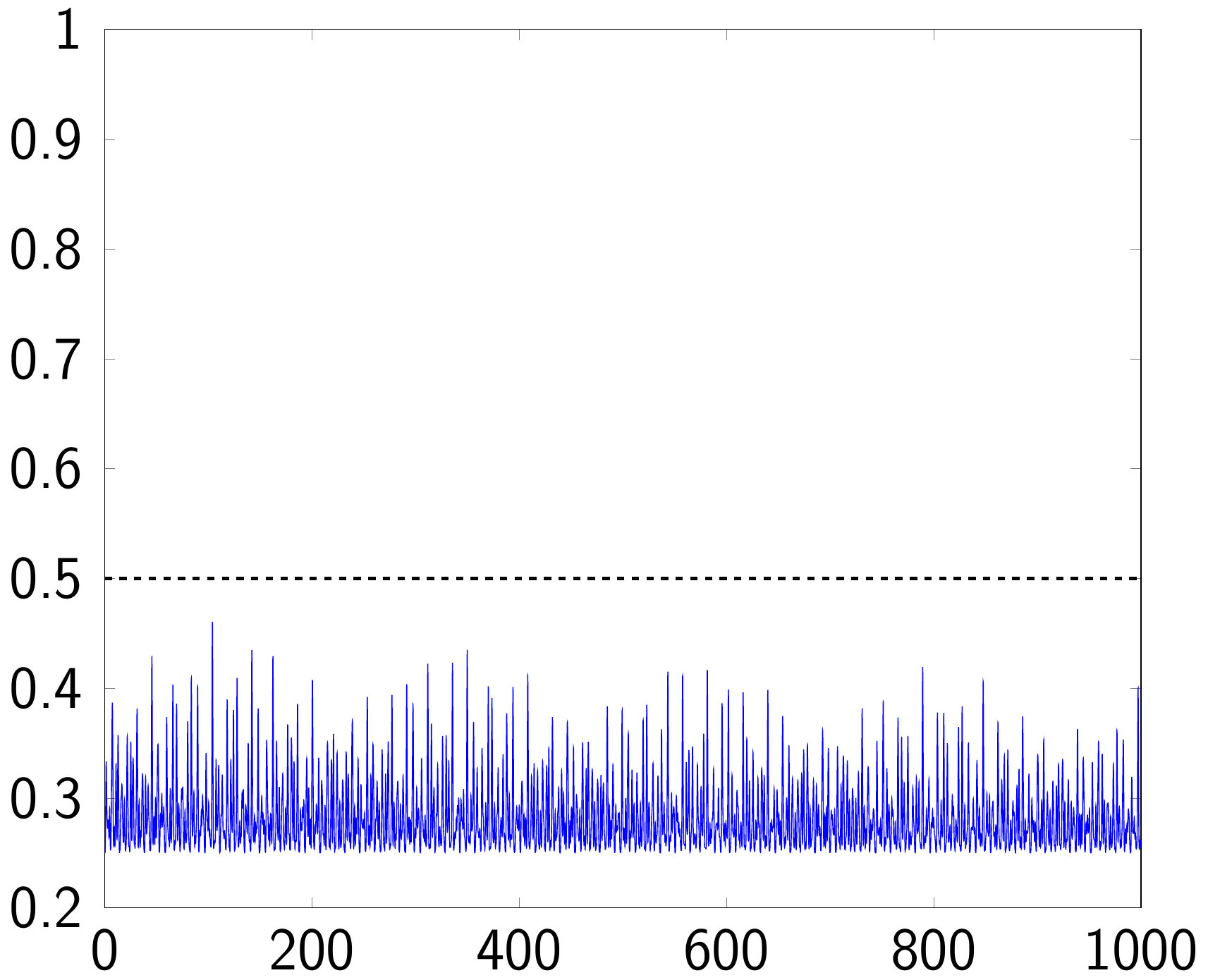} & \vspace{0.2cm}
		\includegraphics[ width=\linewidth, height=\linewidth, keepaspectratio]{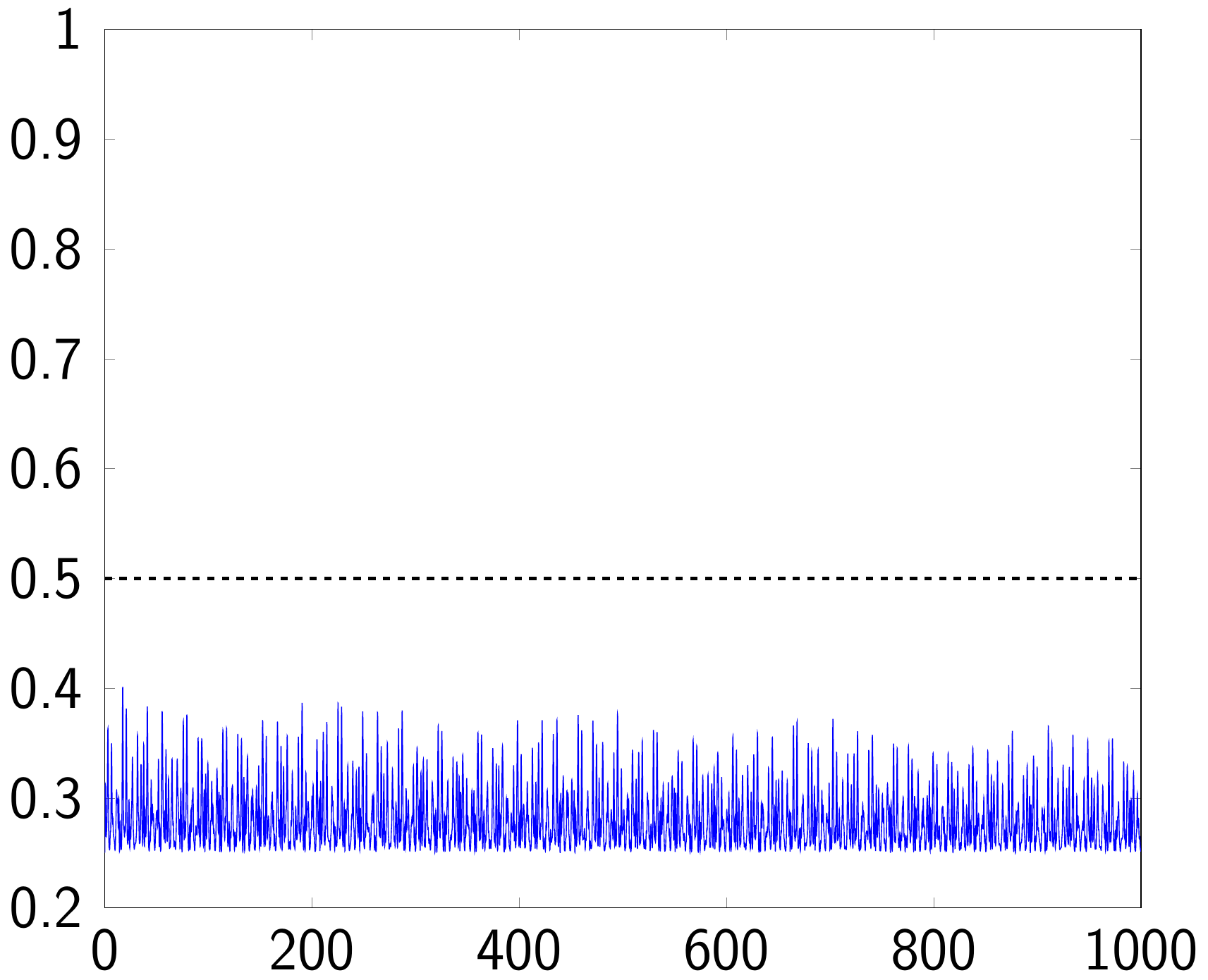} & \vspace{0.2cm}
		\includegraphics[ width=\linewidth, height=\linewidth, keepaspectratio]{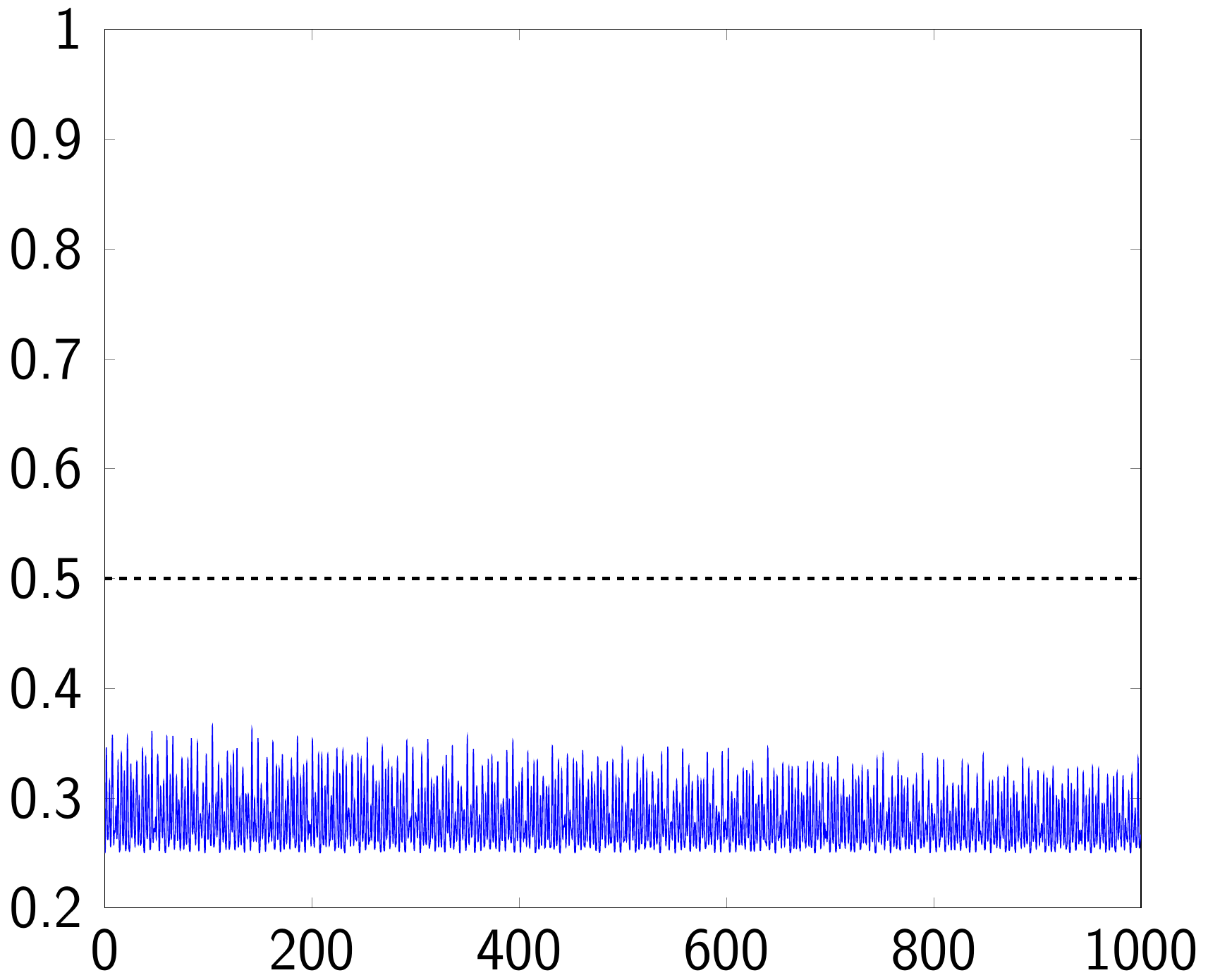} & \vspace{0.2cm}
		\includegraphics[ width=\linewidth, height=\linewidth, keepaspectratio]{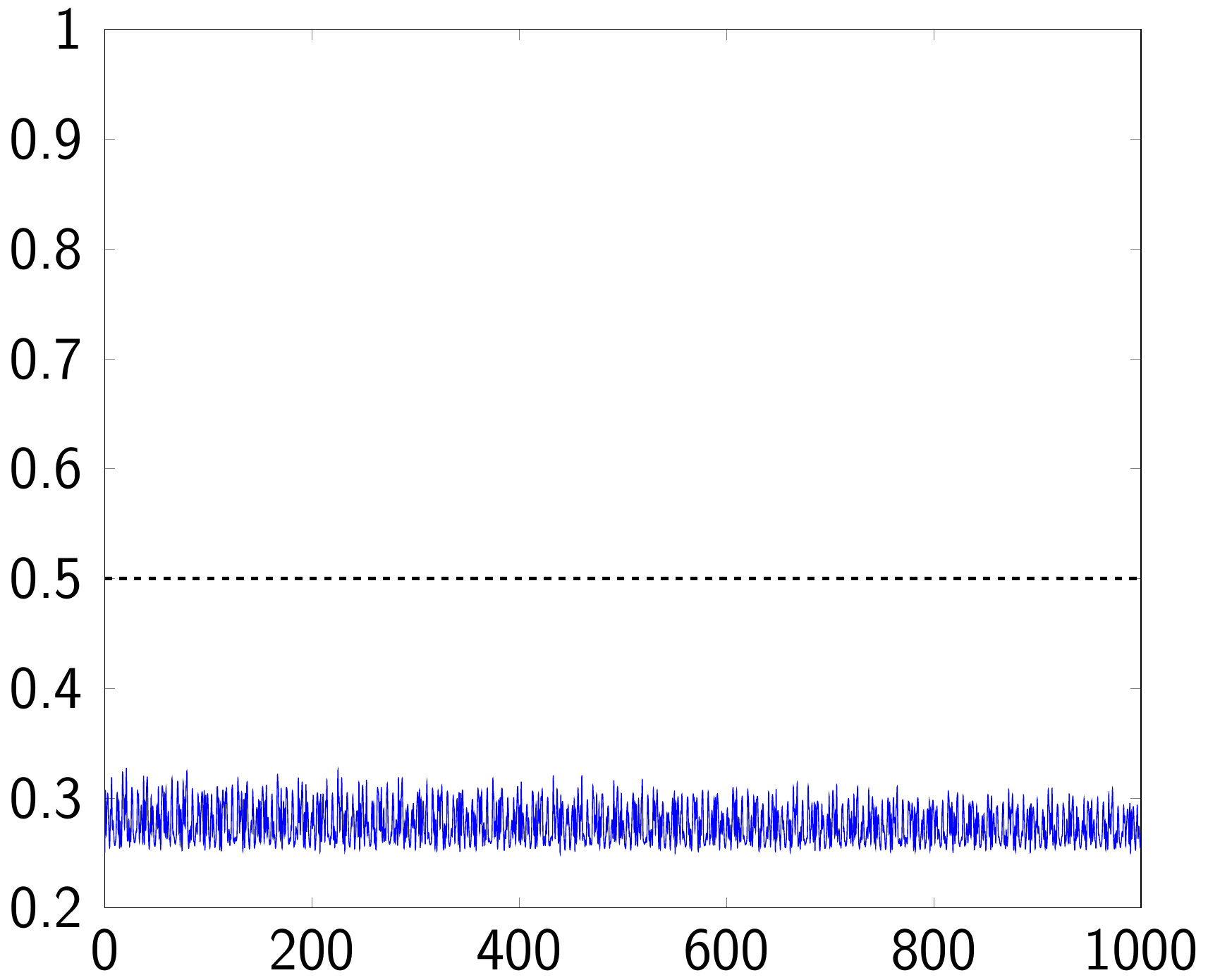} & \vspace{0.2cm}
		\includegraphics[ width=\linewidth, height=\linewidth, keepaspectratio]{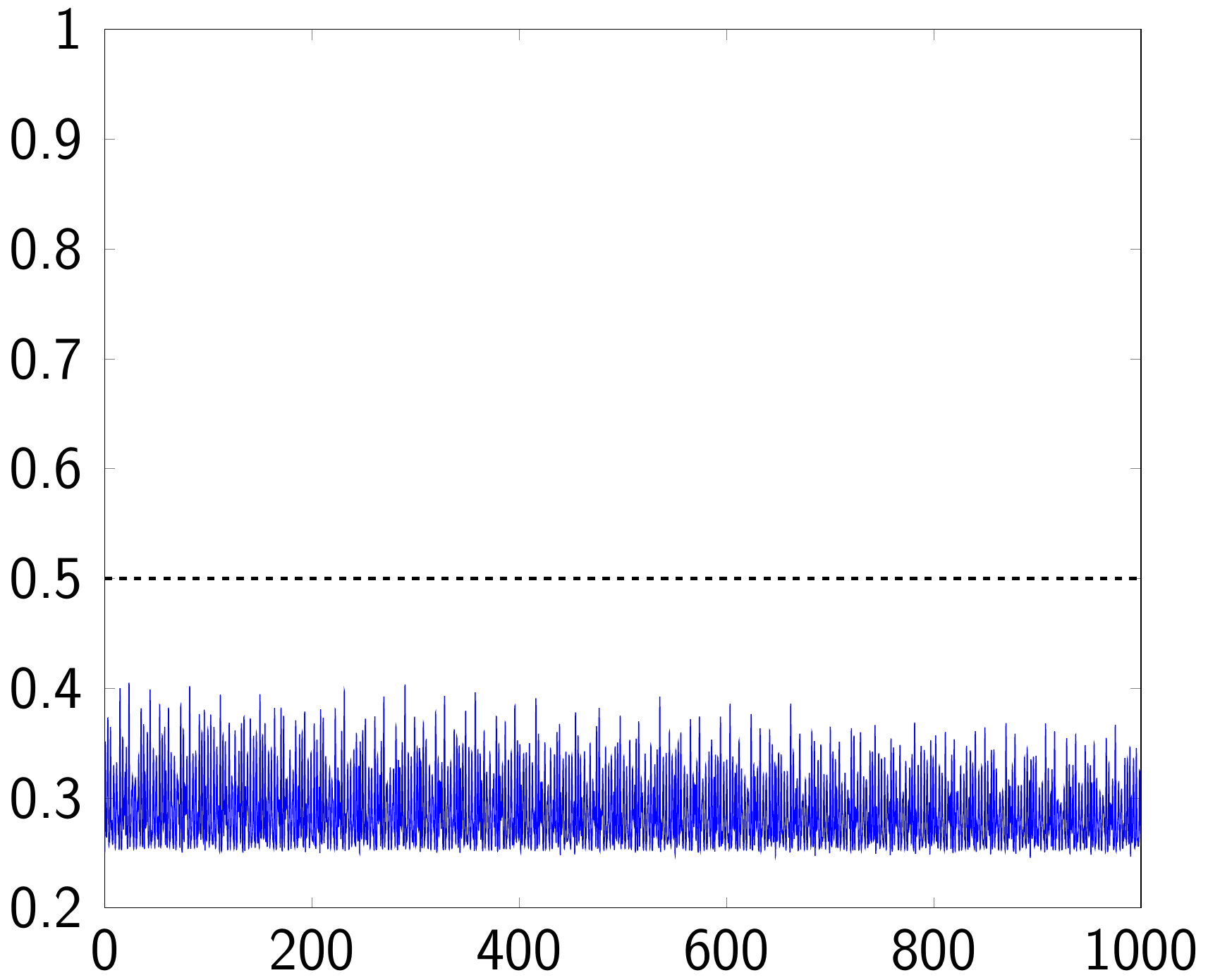} \\
		\hline
		$11$ & \vspace{0.2cm}
		\includegraphics[ width=\linewidth, height=\linewidth, keepaspectratio]{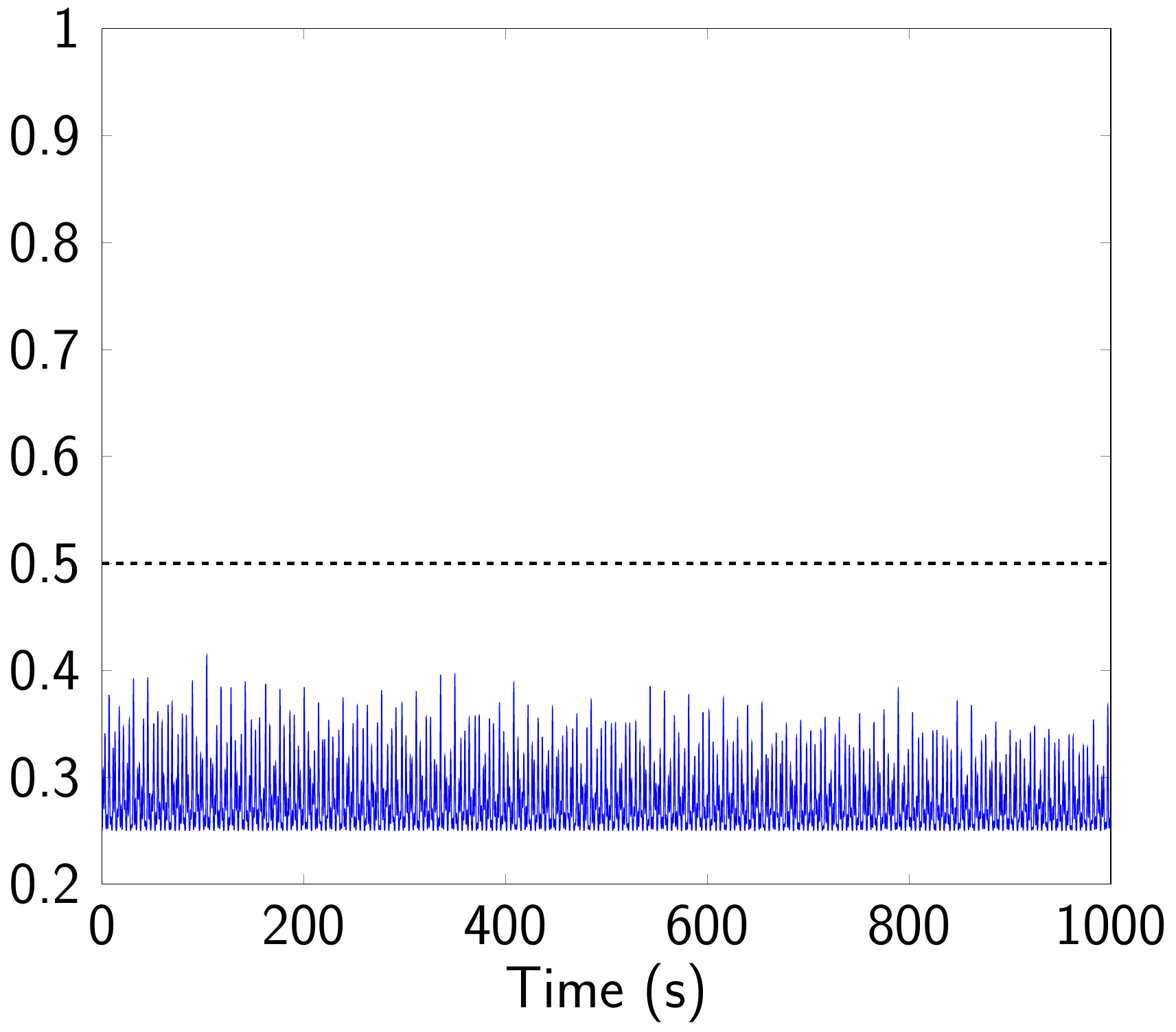} & \vspace{0.2cm}
		\includegraphics[ width=\linewidth, height=\linewidth, keepaspectratio]{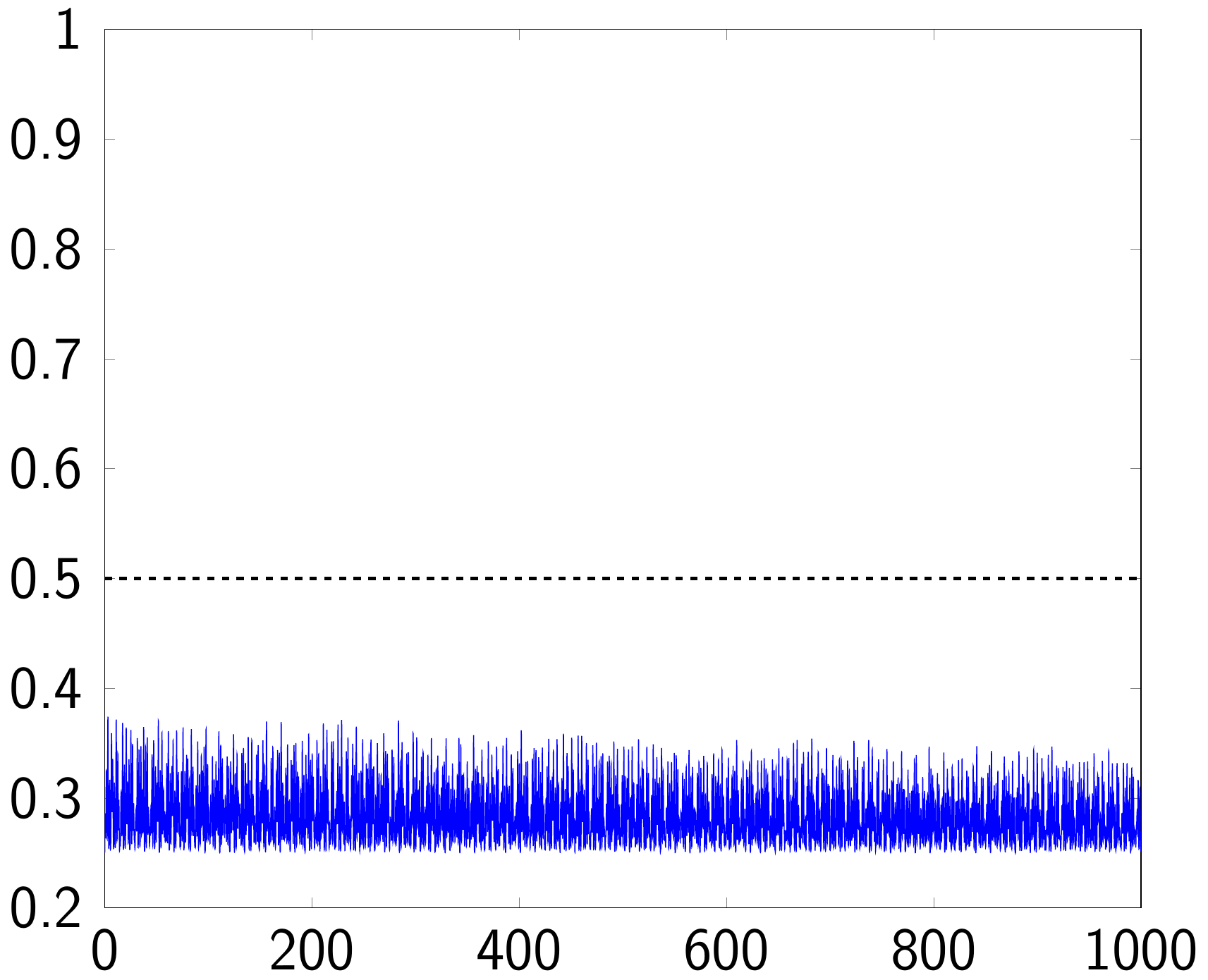} & \vspace{0.2cm}
		\includegraphics[ width=\linewidth, height=\linewidth, keepaspectratio]{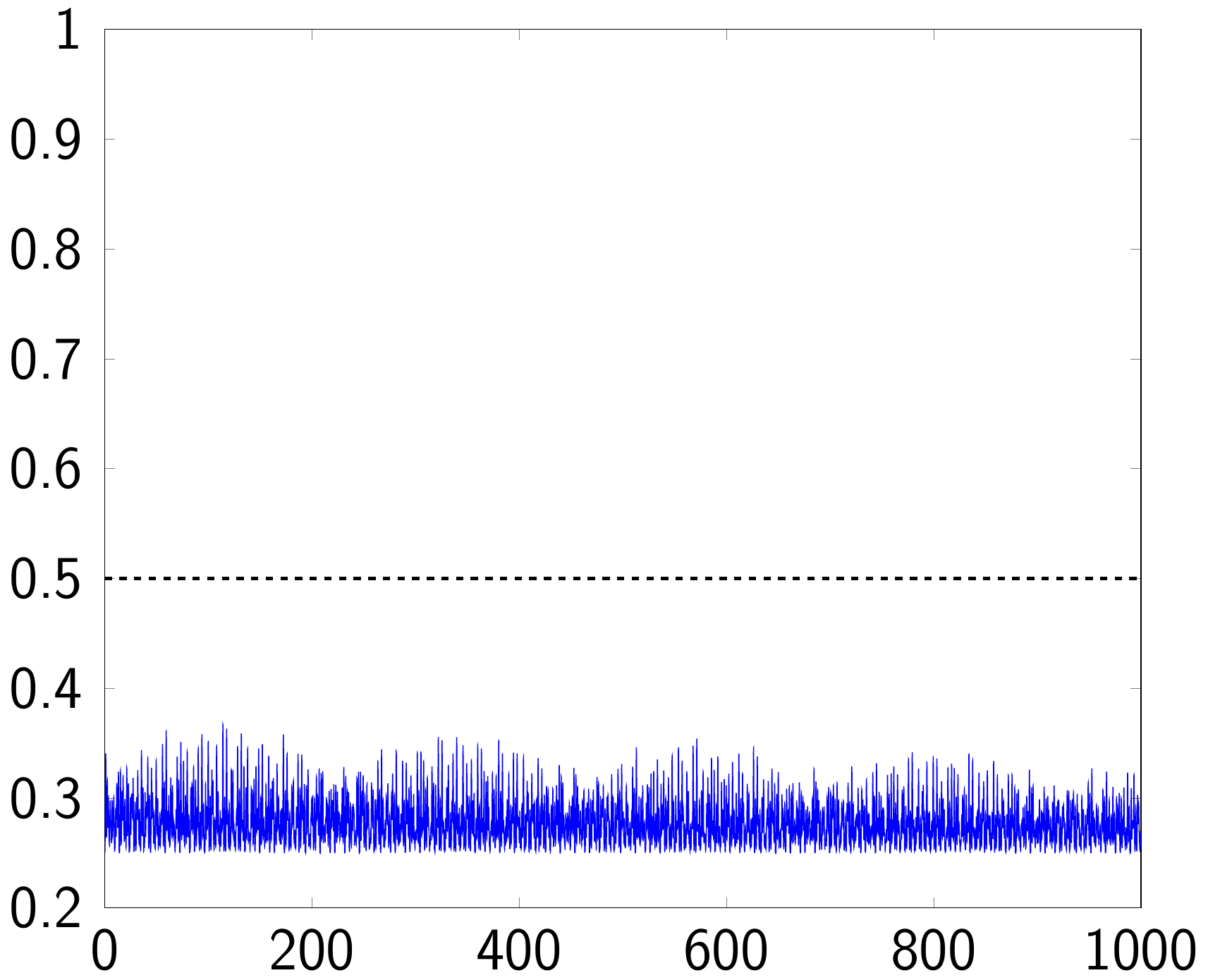} & \vspace{0.2cm}
		\includegraphics[ width=\linewidth, height=\linewidth, keepaspectratio]{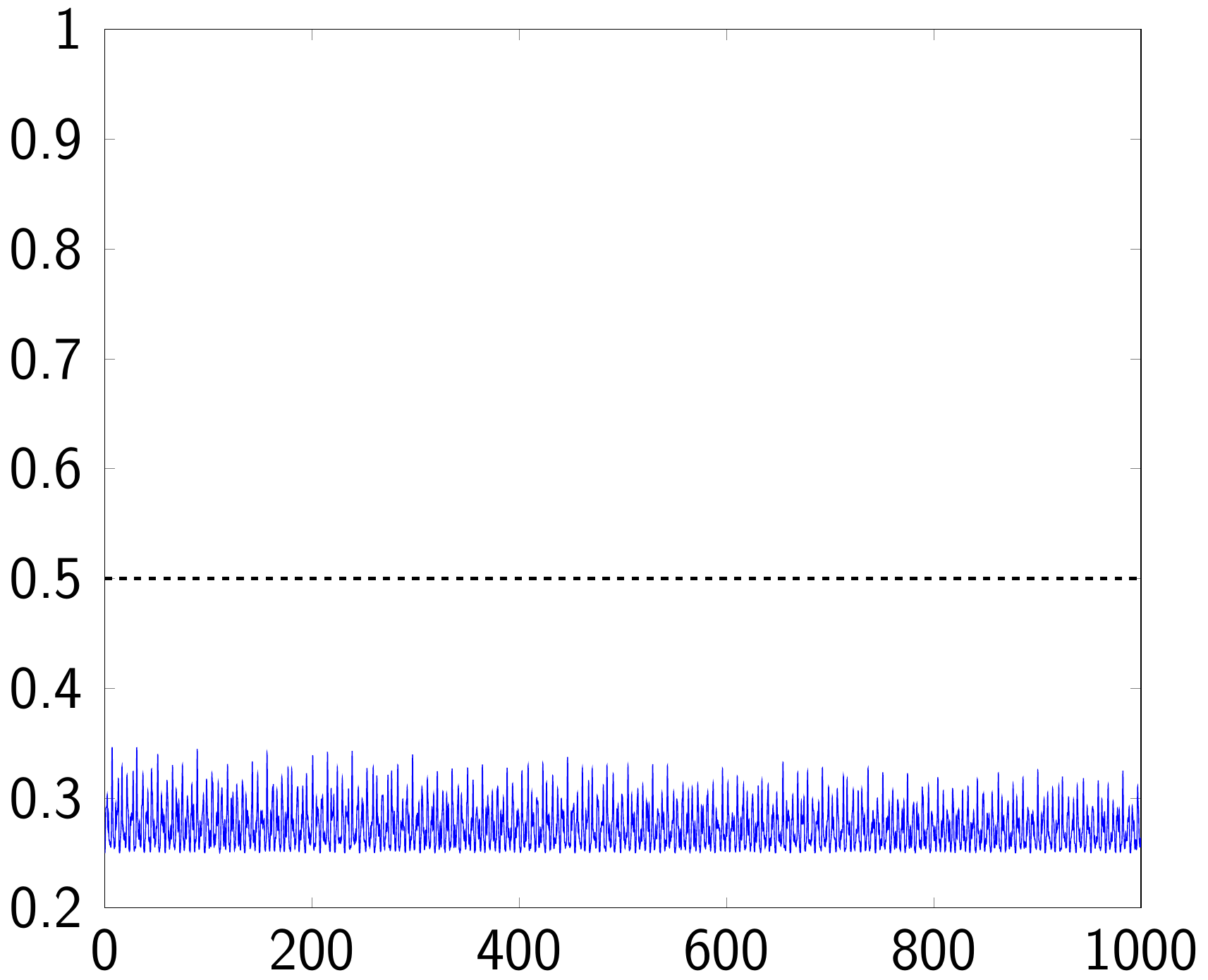} & \vspace{0.2cm}
		\includegraphics[ width=\linewidth, height=\linewidth, keepaspectratio]{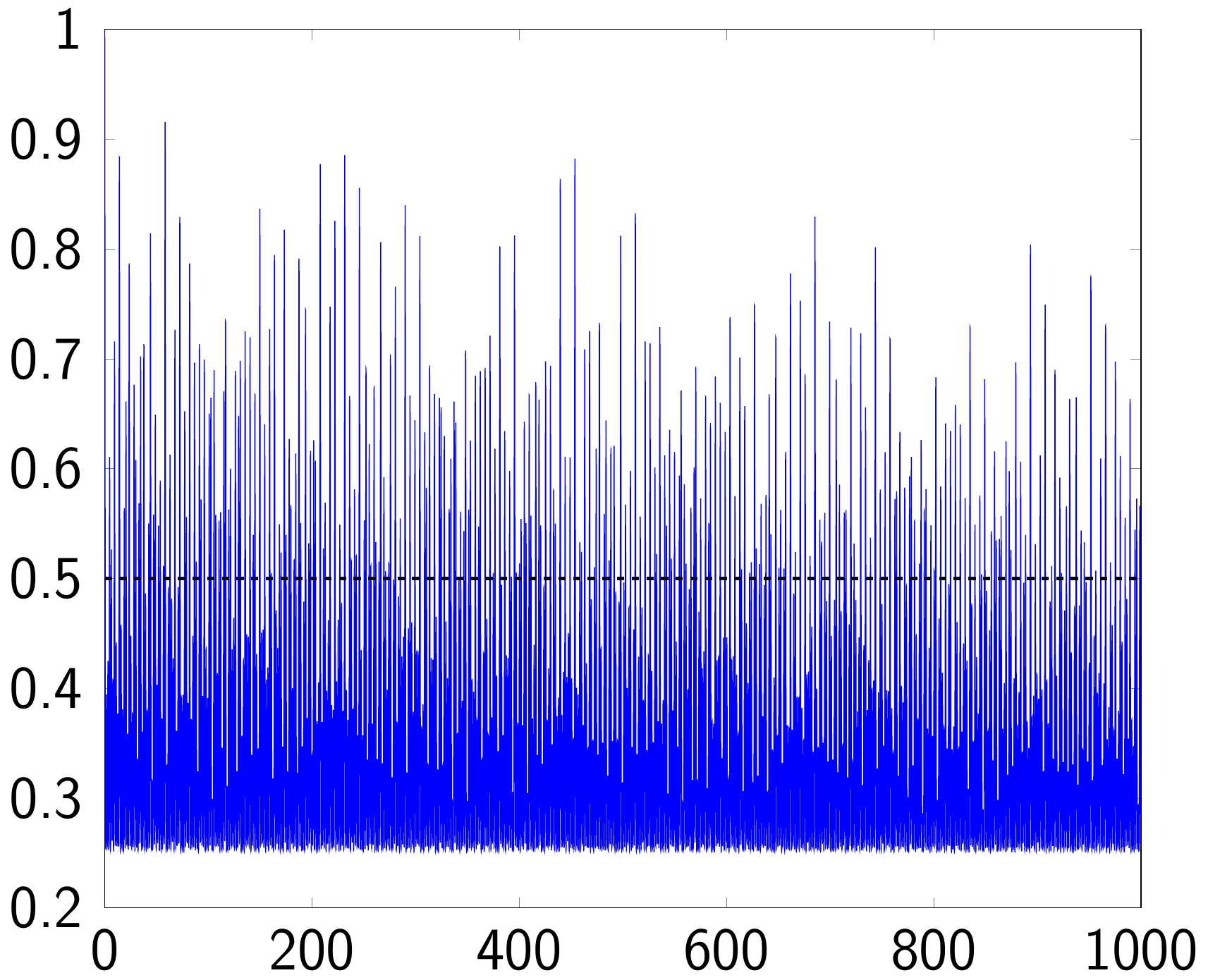} & \vspace{0.2cm}
		\includegraphics[ width=\linewidth, height=\linewidth, keepaspectratio]{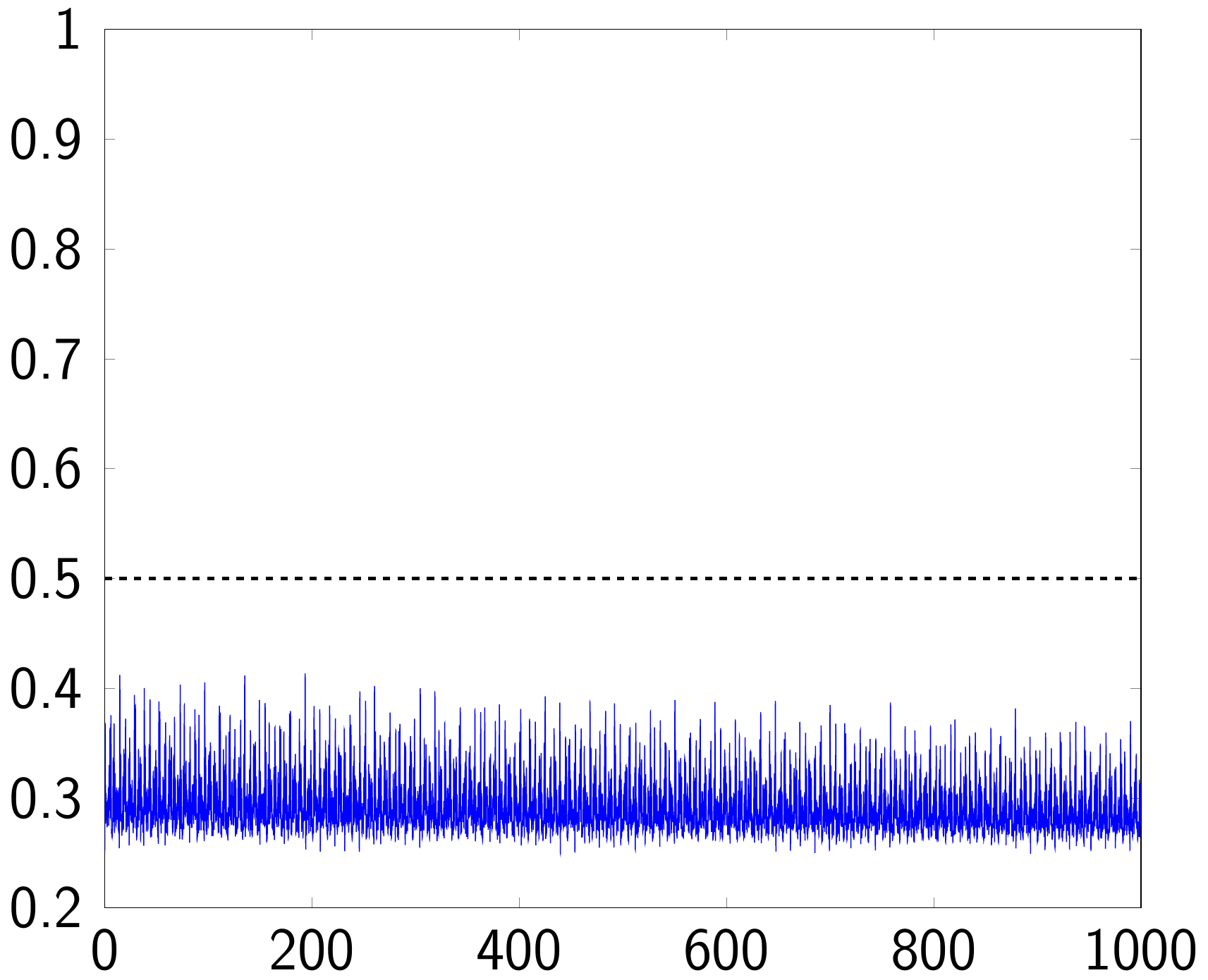} & \vspace{0.2cm}
		\includegraphics[ width=\linewidth, height=\linewidth, keepaspectratio]{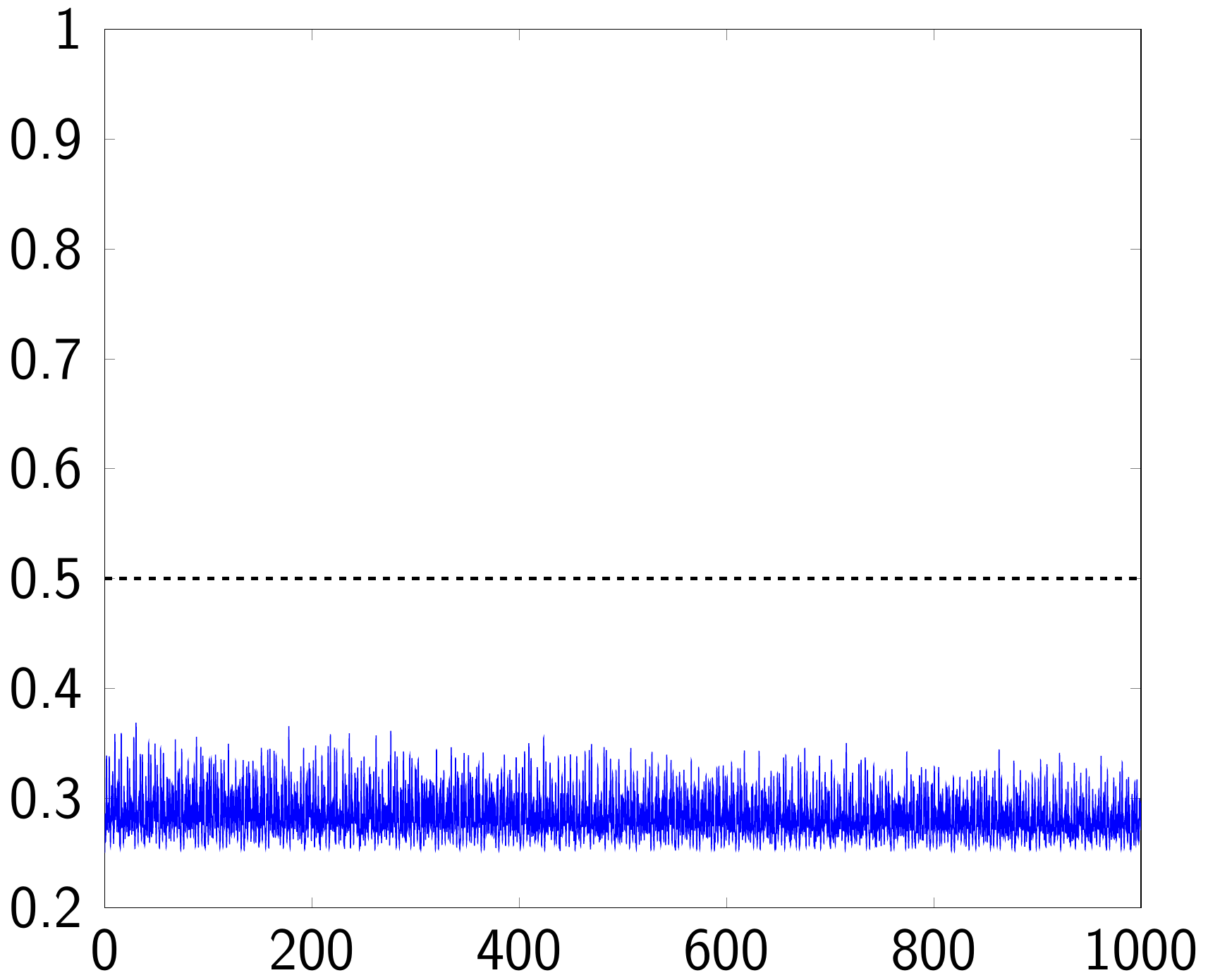} & \vspace{0.2cm}
		\includegraphics[ width=\linewidth, height=\linewidth, keepaspectratio]{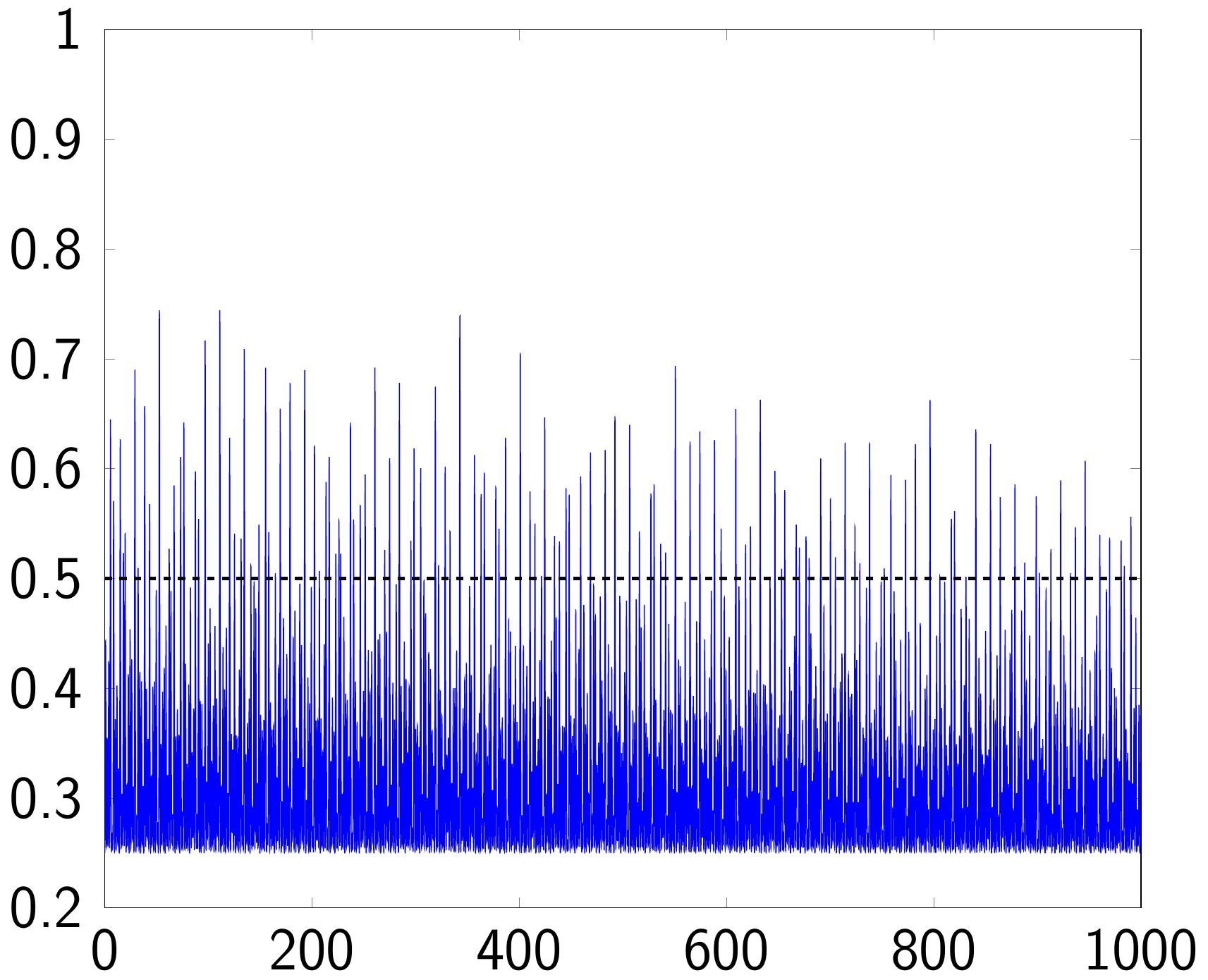} & \vspace{0.2cm}
		\includegraphics[ width=\linewidth, height=\linewidth, keepaspectratio]{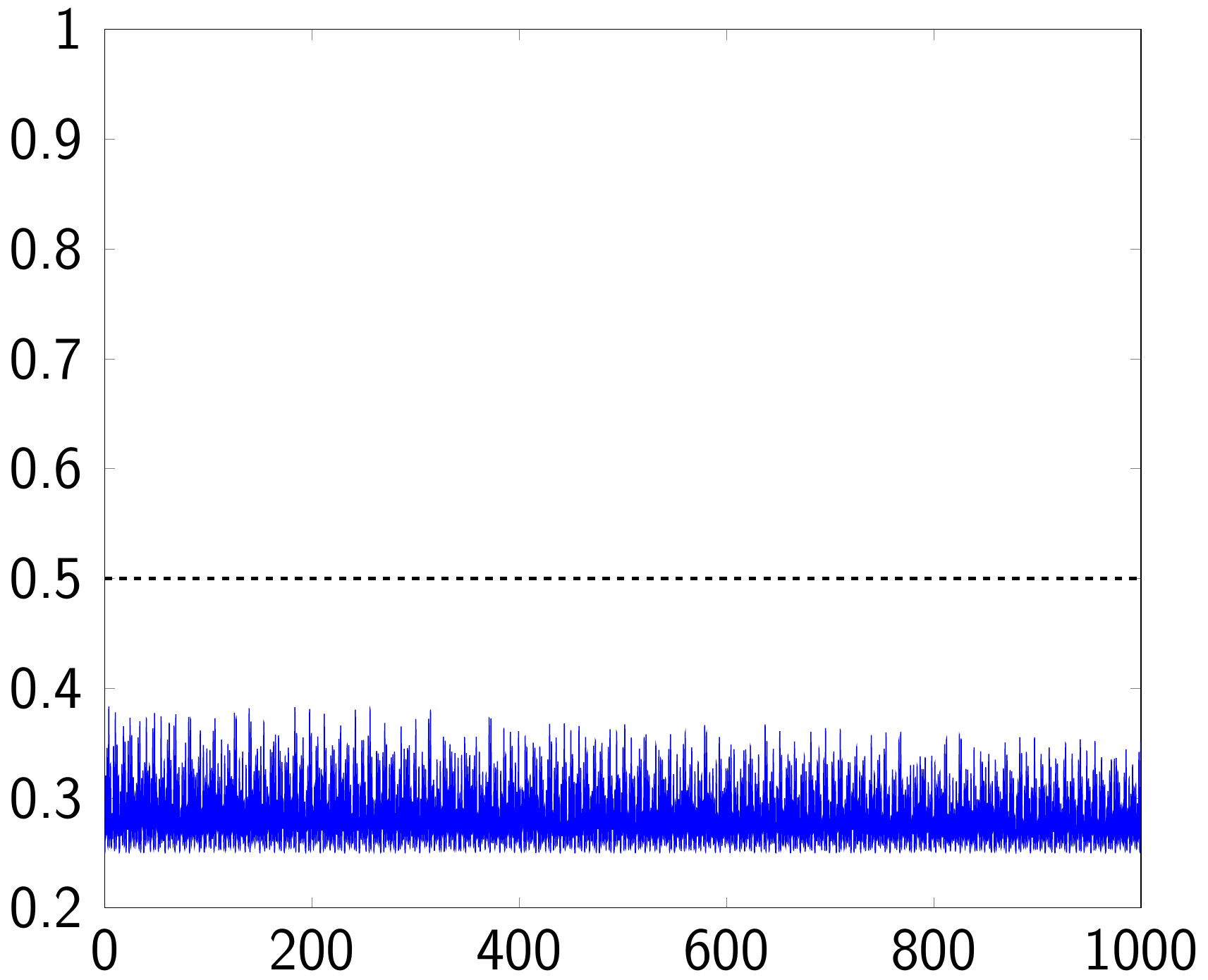} & \vspace{0.2cm}
		\includegraphics[ width=\linewidth, height=\linewidth, keepaspectratio]{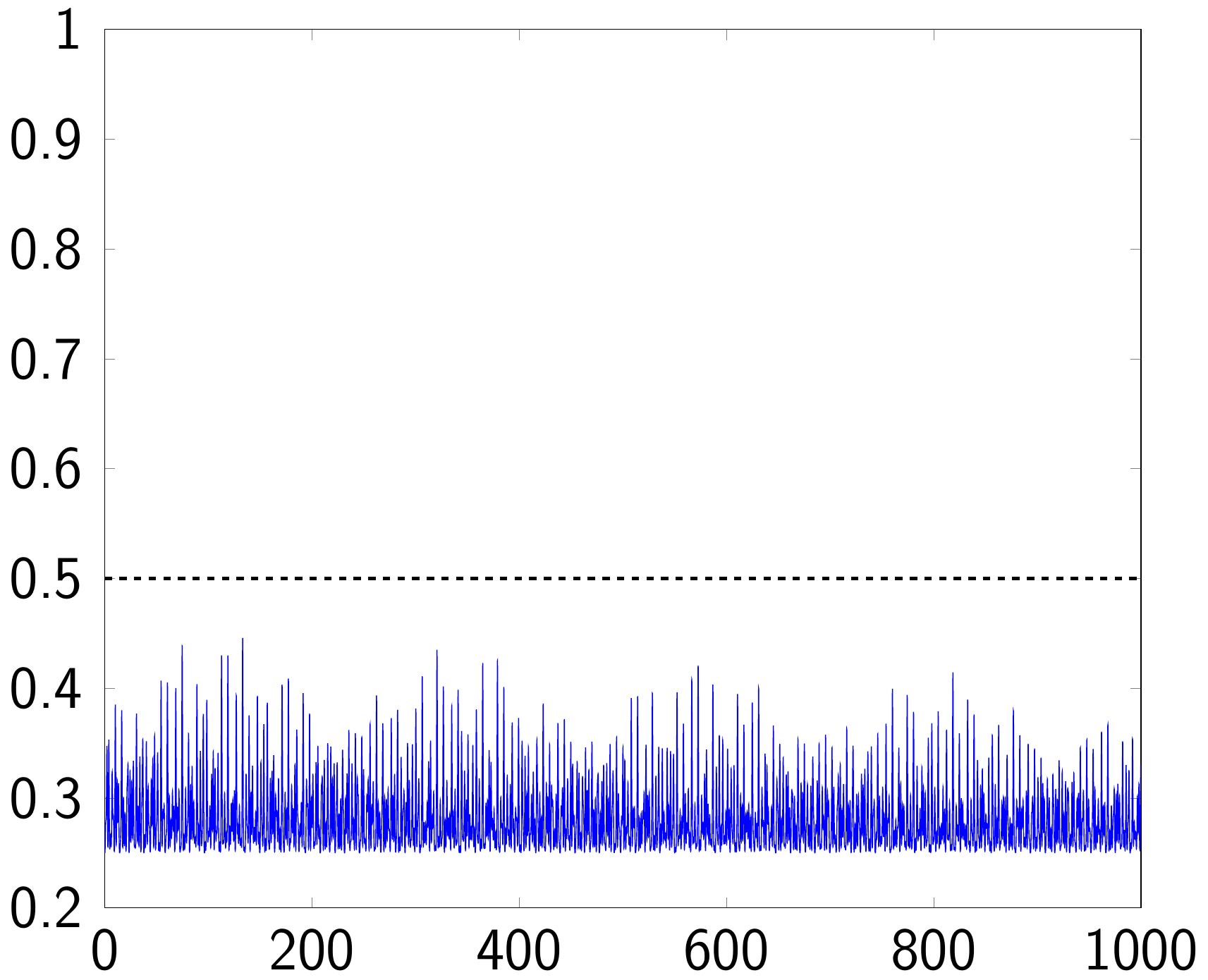} \\
		\hline
		$12$ & \vspace{0.2cm}
		\includegraphics[ width=\linewidth, height=\linewidth, keepaspectratio]{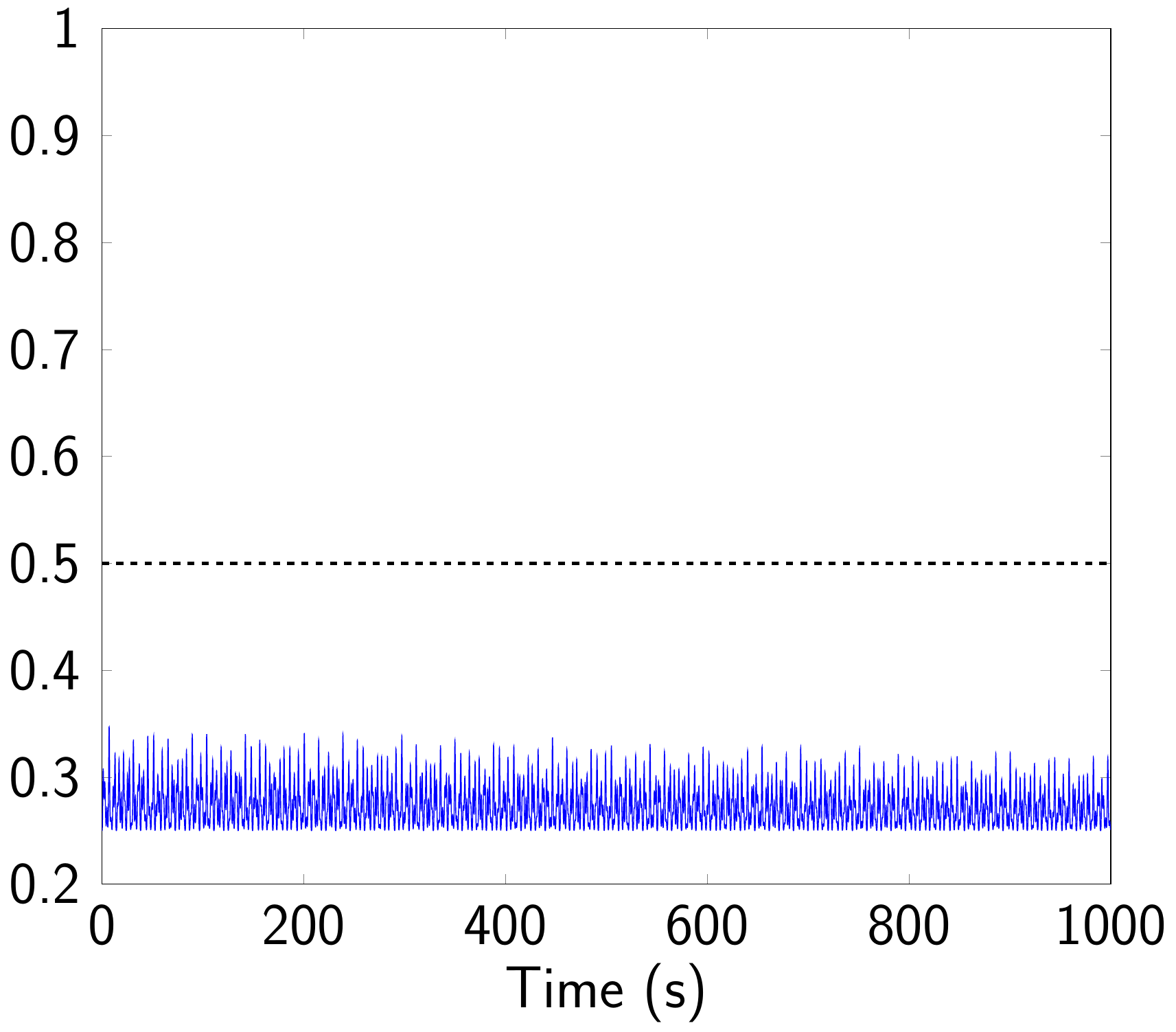} & \vspace{0.2cm}
		\includegraphics[ width=\linewidth, height=\linewidth, keepaspectratio]{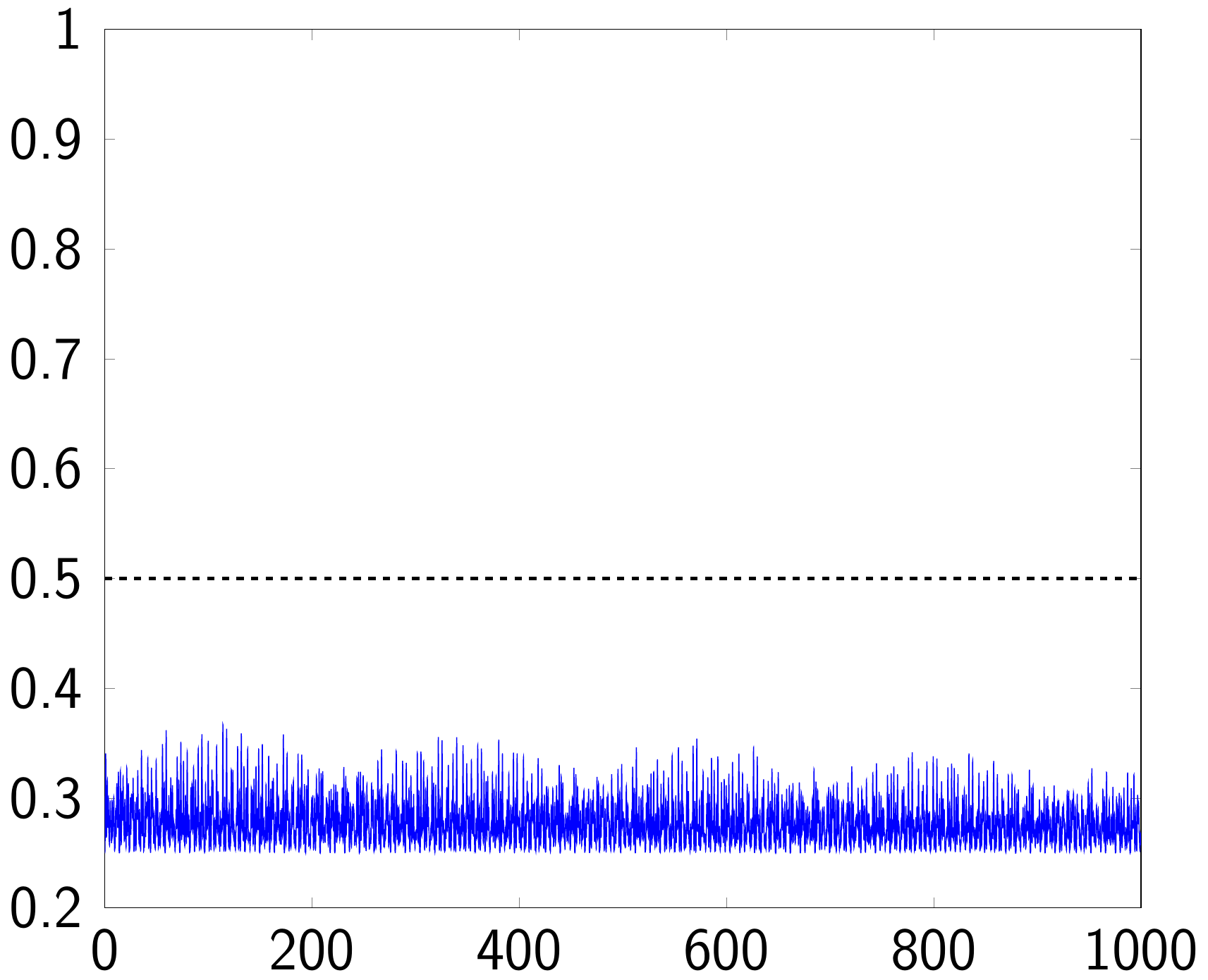} & \vspace{0.2cm}
		\includegraphics[ width=\linewidth, height=\linewidth, keepaspectratio]{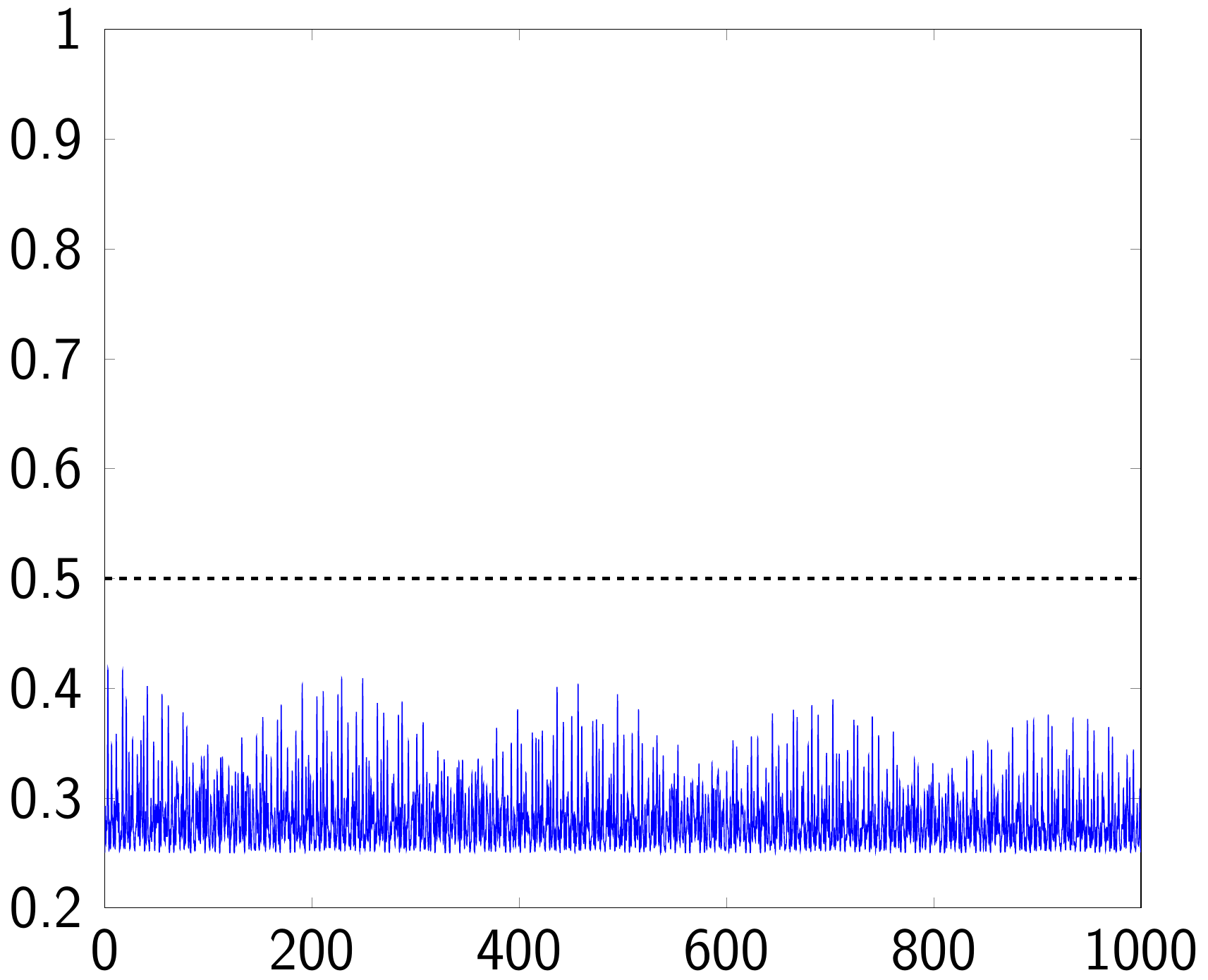} & \vspace{0.2cm}
		\includegraphics[ width=\linewidth, height=\linewidth, keepaspectratio]{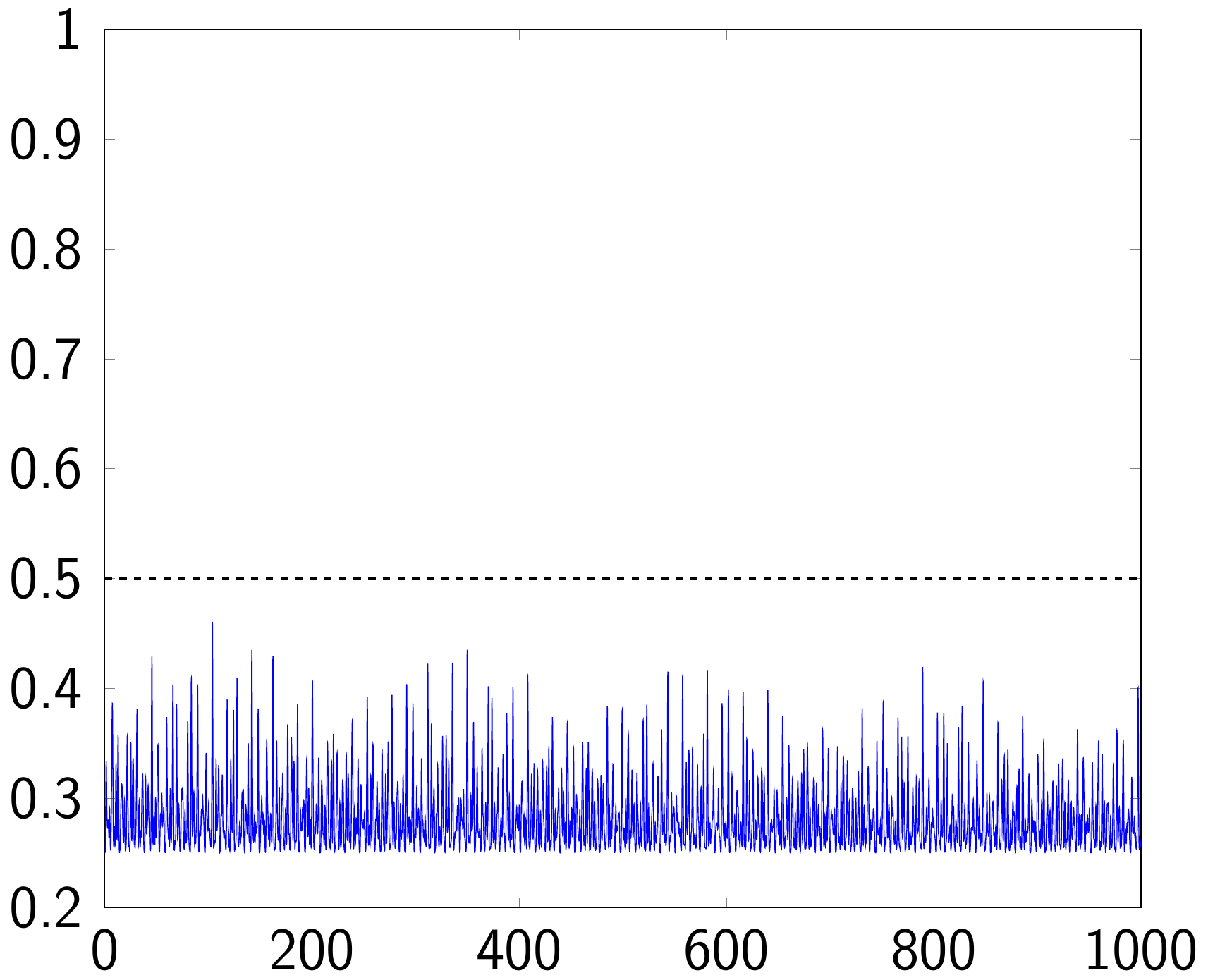} & \vspace{0.2cm}
		\includegraphics[ width=\linewidth, height=\linewidth, keepaspectratio]{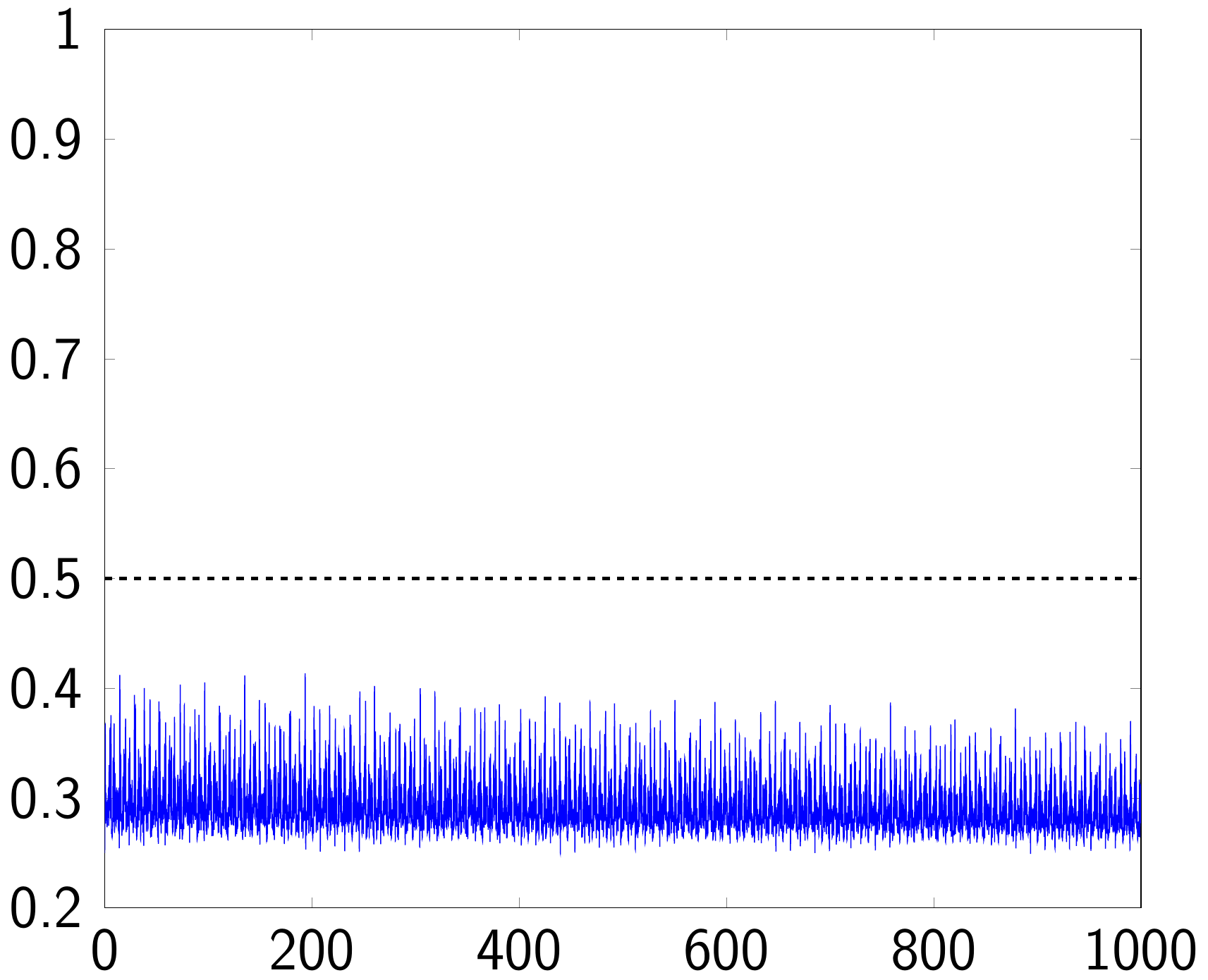} & \vspace{0.2cm}
		\includegraphics[ width=\linewidth, height=\linewidth, keepaspectratio]{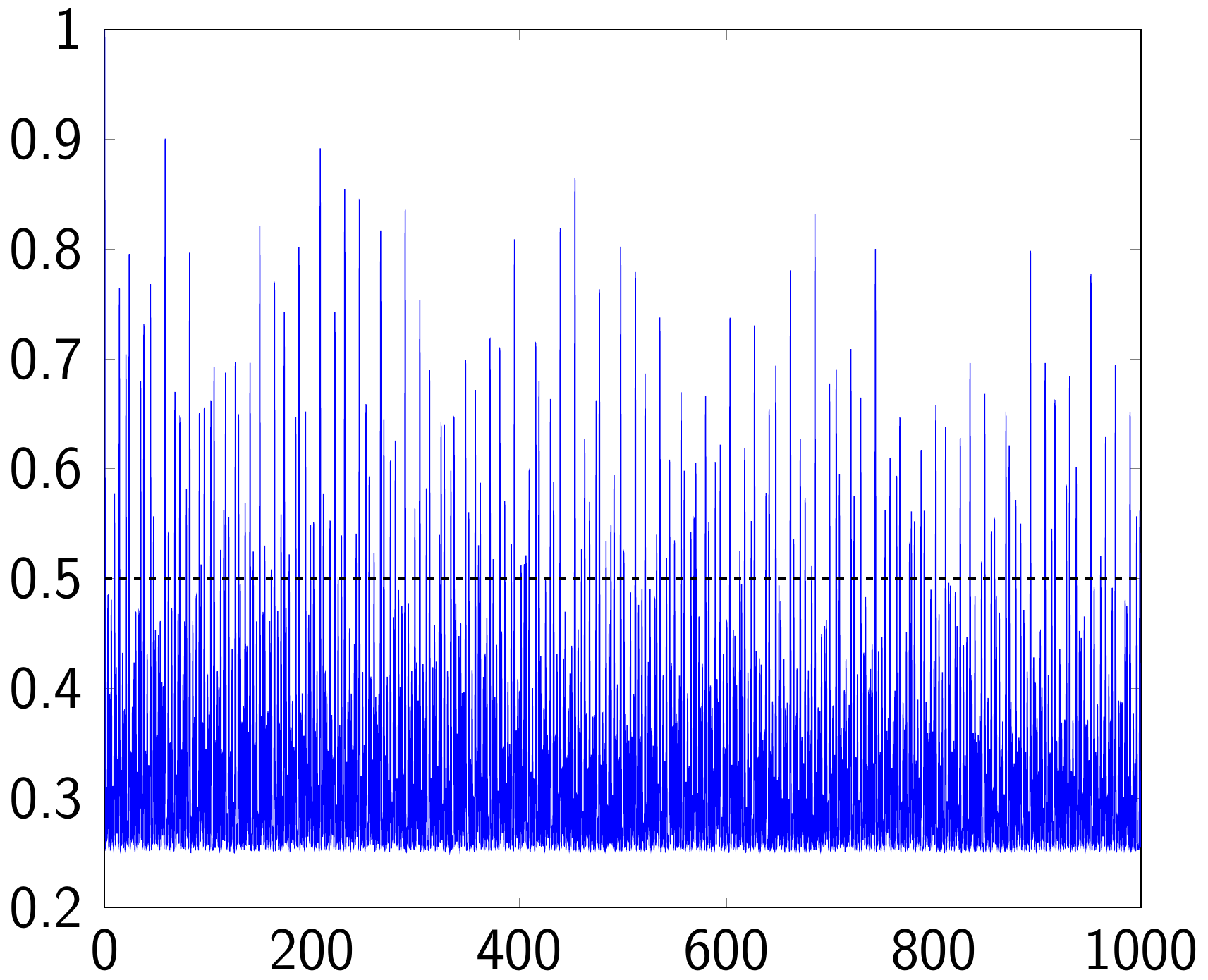} & \vspace{0.2cm}
		\includegraphics[ width=\linewidth, height=\linewidth, keepaspectratio]{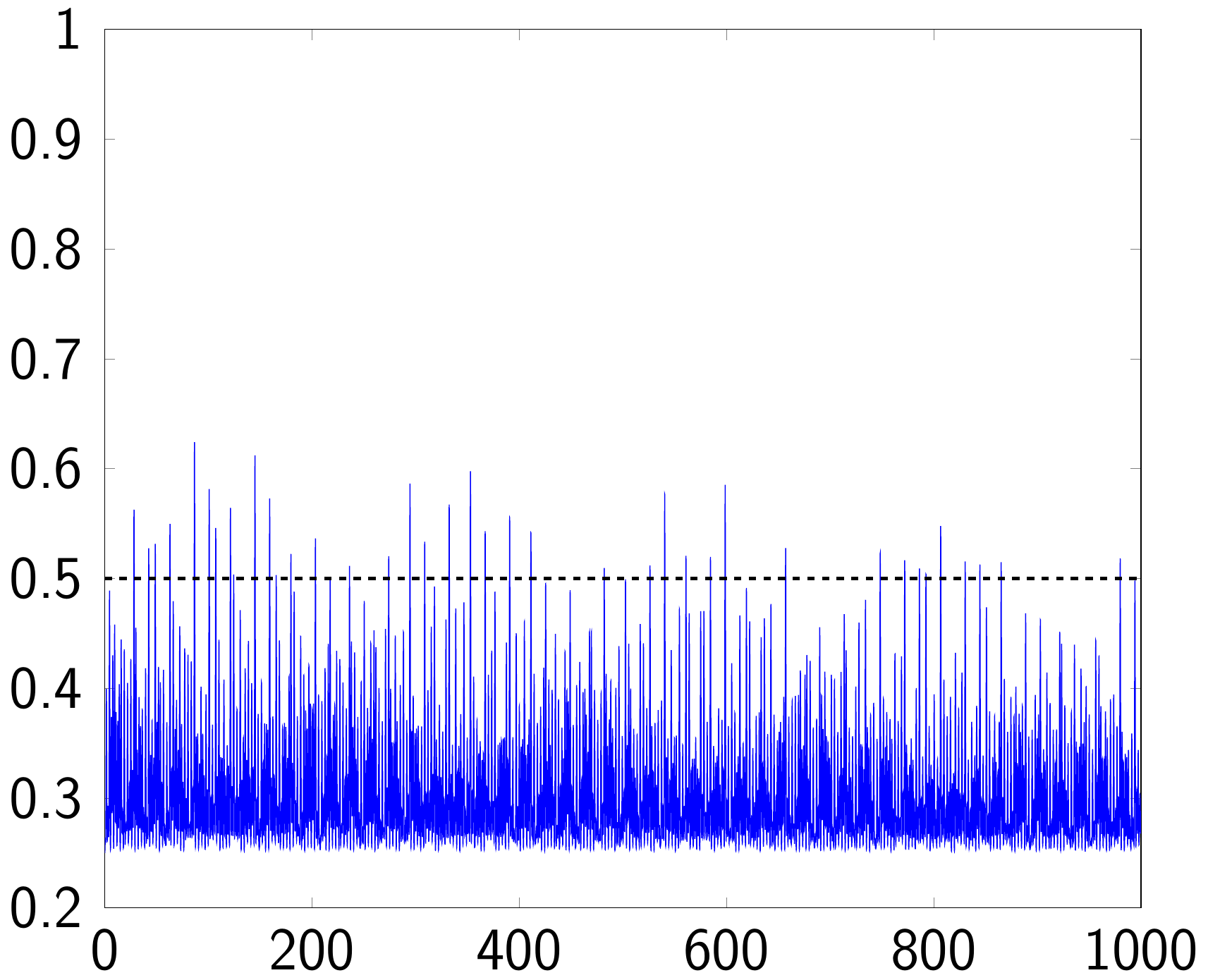} & \vspace{0.2cm}
		\includegraphics[ width=\linewidth, height=\linewidth, keepaspectratio]{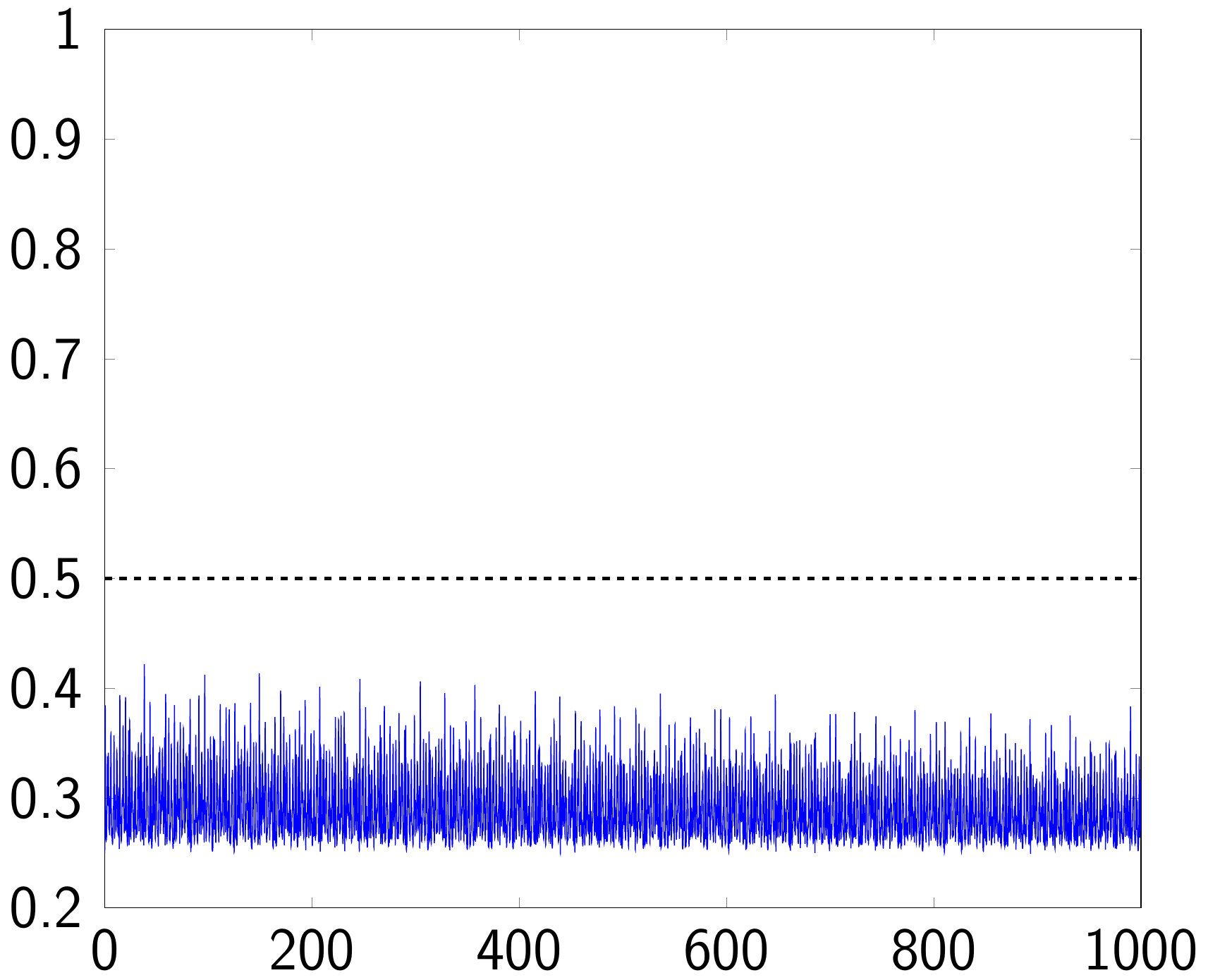} & \vspace{0.2cm}
		\includegraphics[ width=\linewidth, height=\linewidth, keepaspectratio]{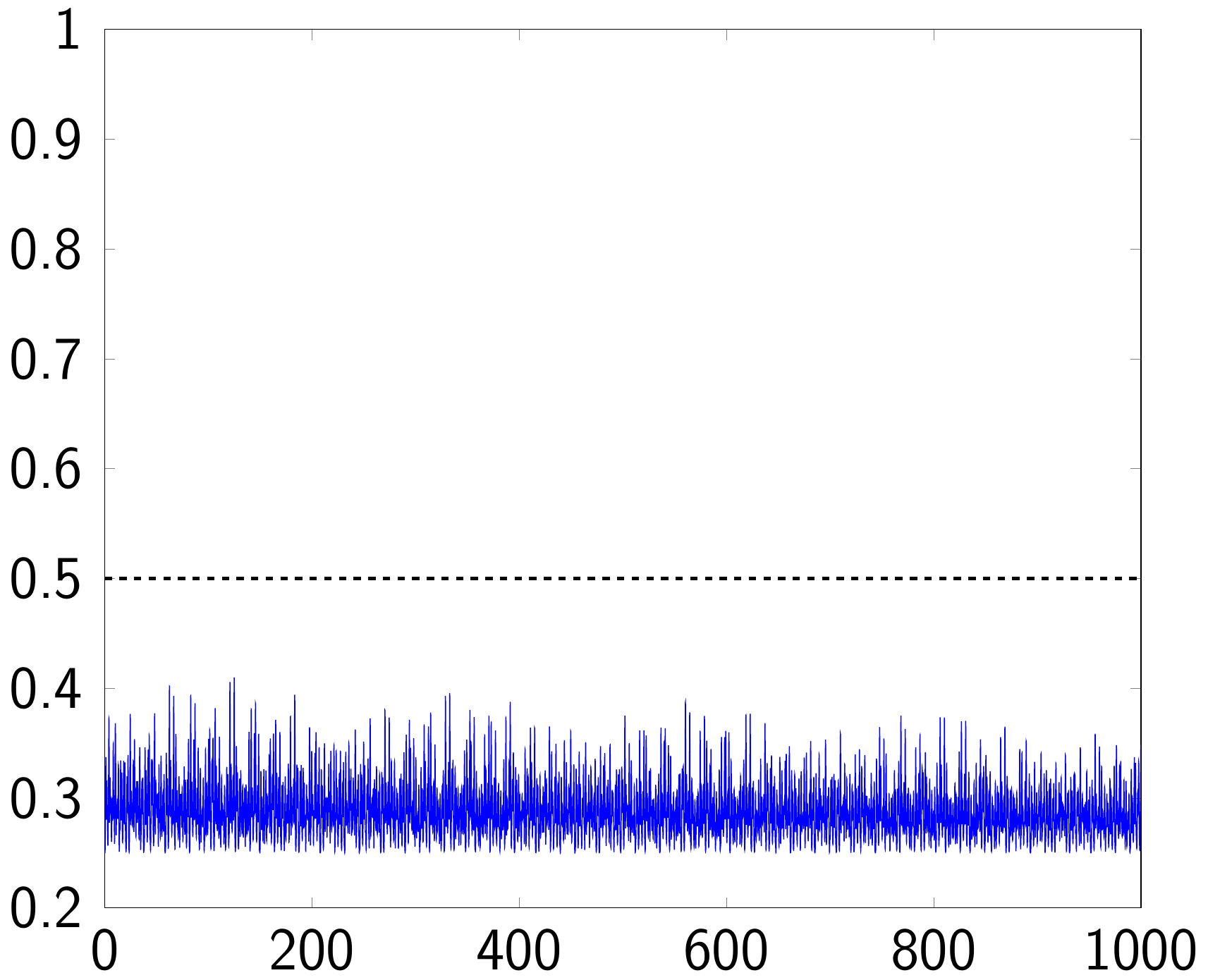} & \vspace{0.2cm}
		\includegraphics[ width=\linewidth, height=\linewidth, keepaspectratio]{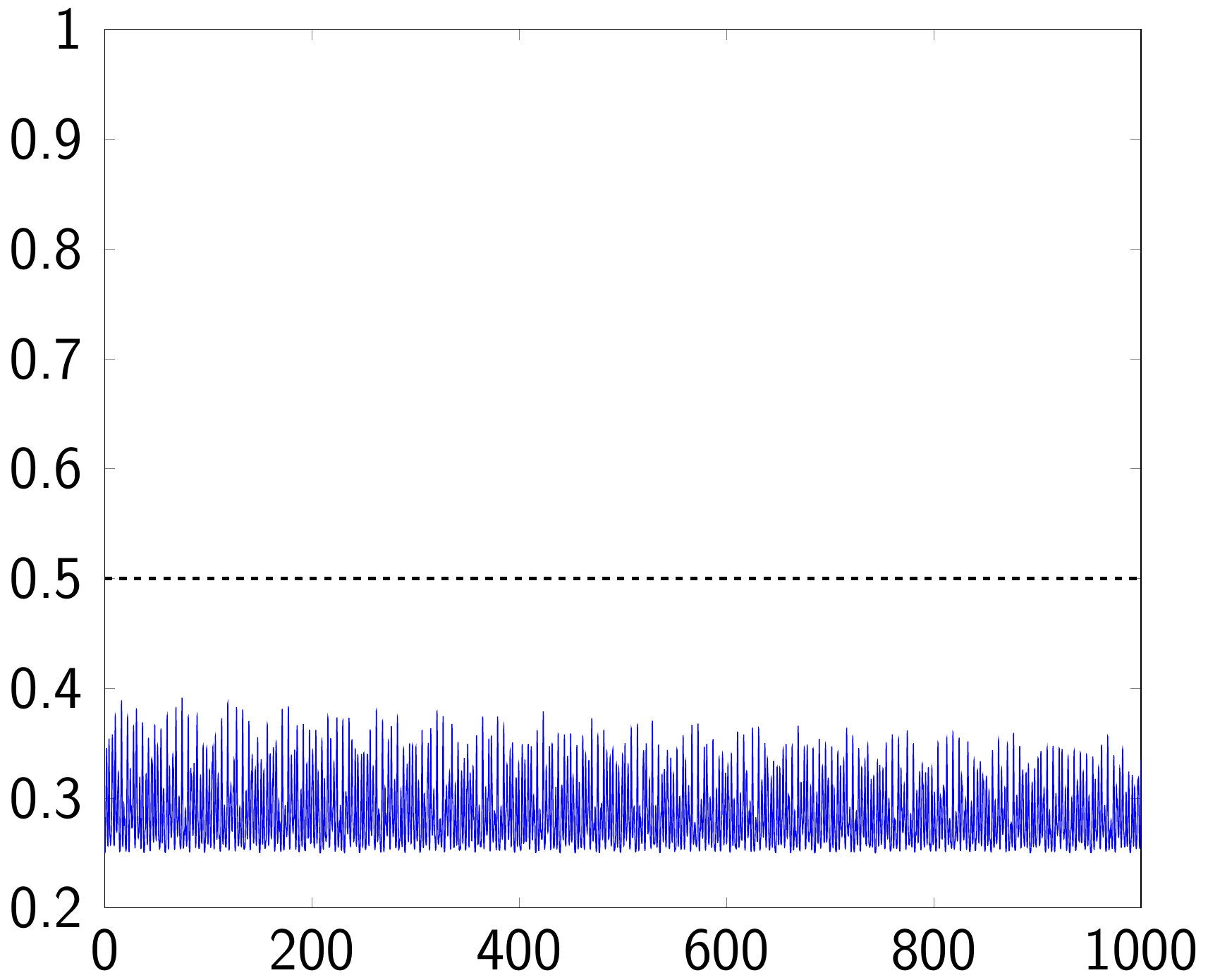} \\
		\hline
		$13$ & \vspace{0.2cm}
		\includegraphics[ width=\linewidth, height=\linewidth, keepaspectratio]{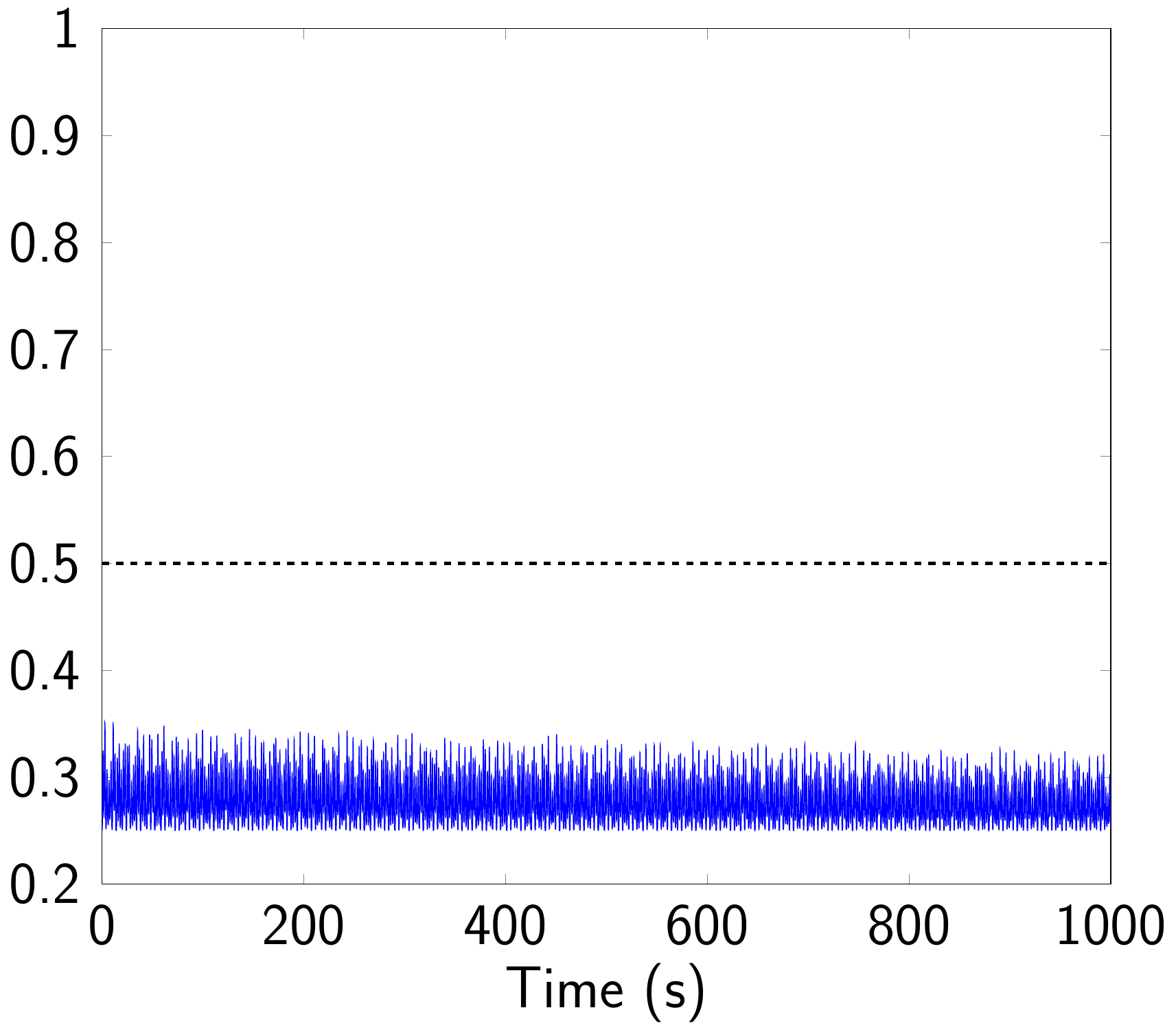} & \vspace{0.2cm}
		\includegraphics[ width=\linewidth, height=\linewidth, keepaspectratio]{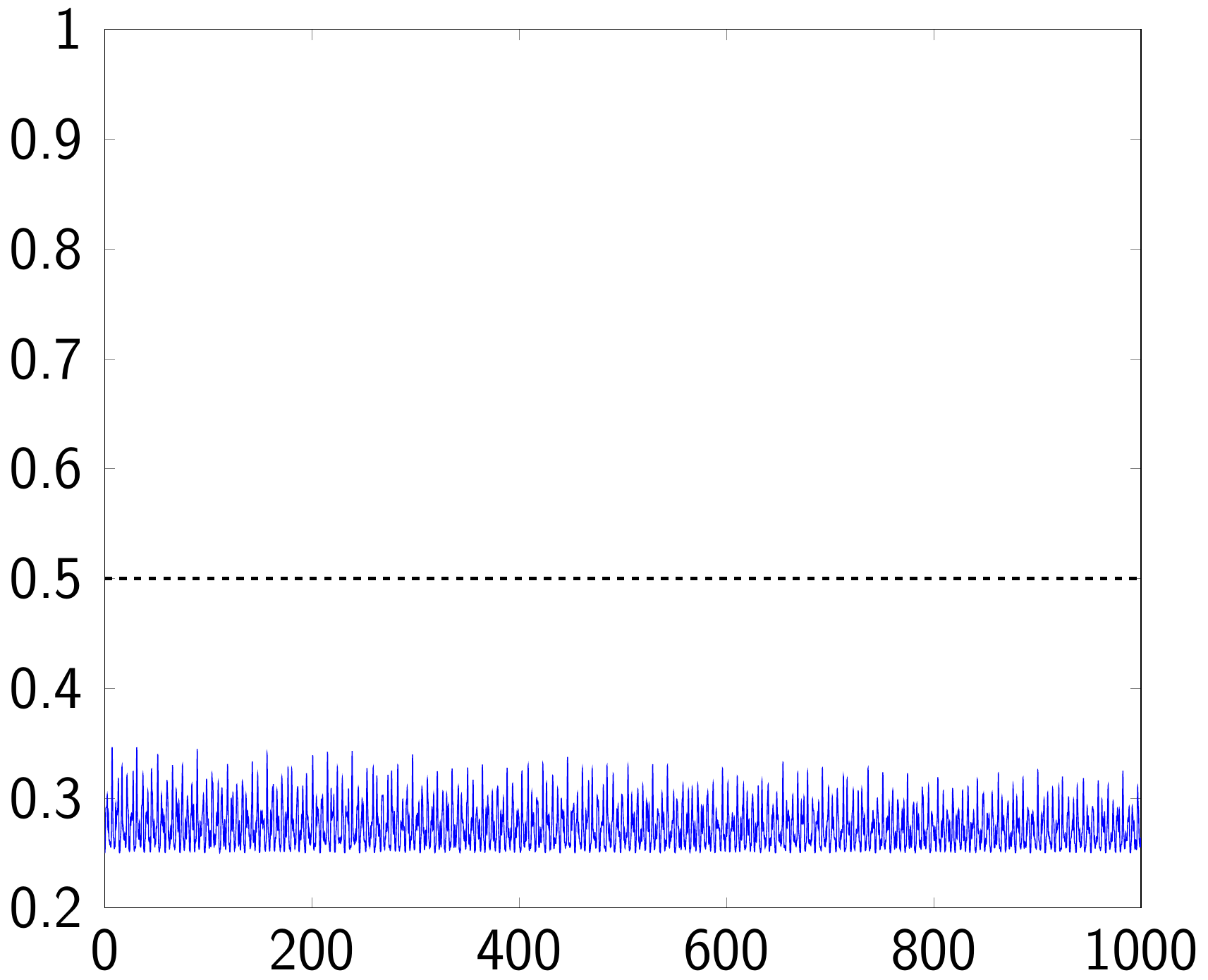} & \vspace{0.2cm}
		\includegraphics[ width=\linewidth, height=\linewidth, keepaspectratio]{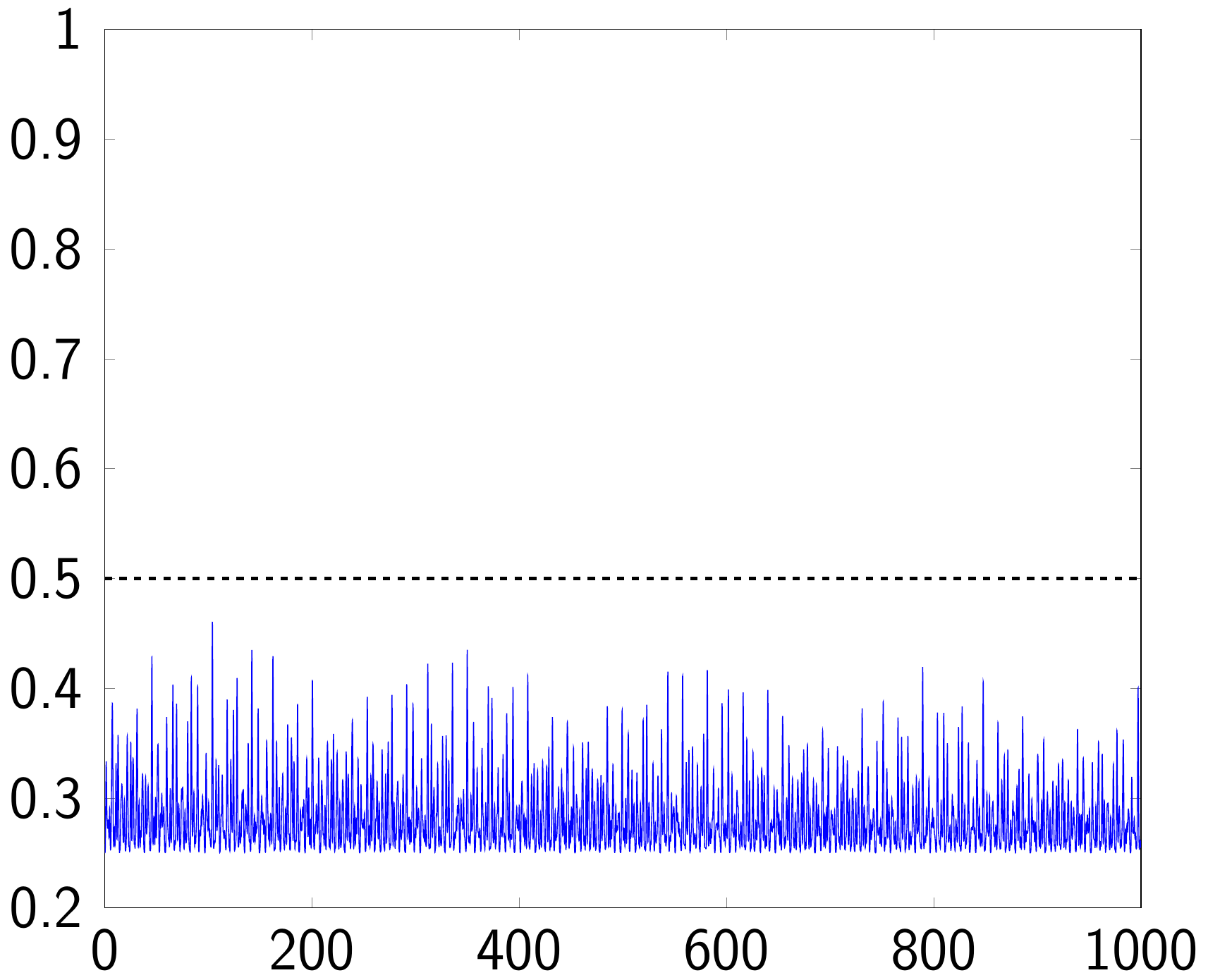} & \vspace{0.2cm}
		\includegraphics[ width=\linewidth, height=\linewidth, keepaspectratio]{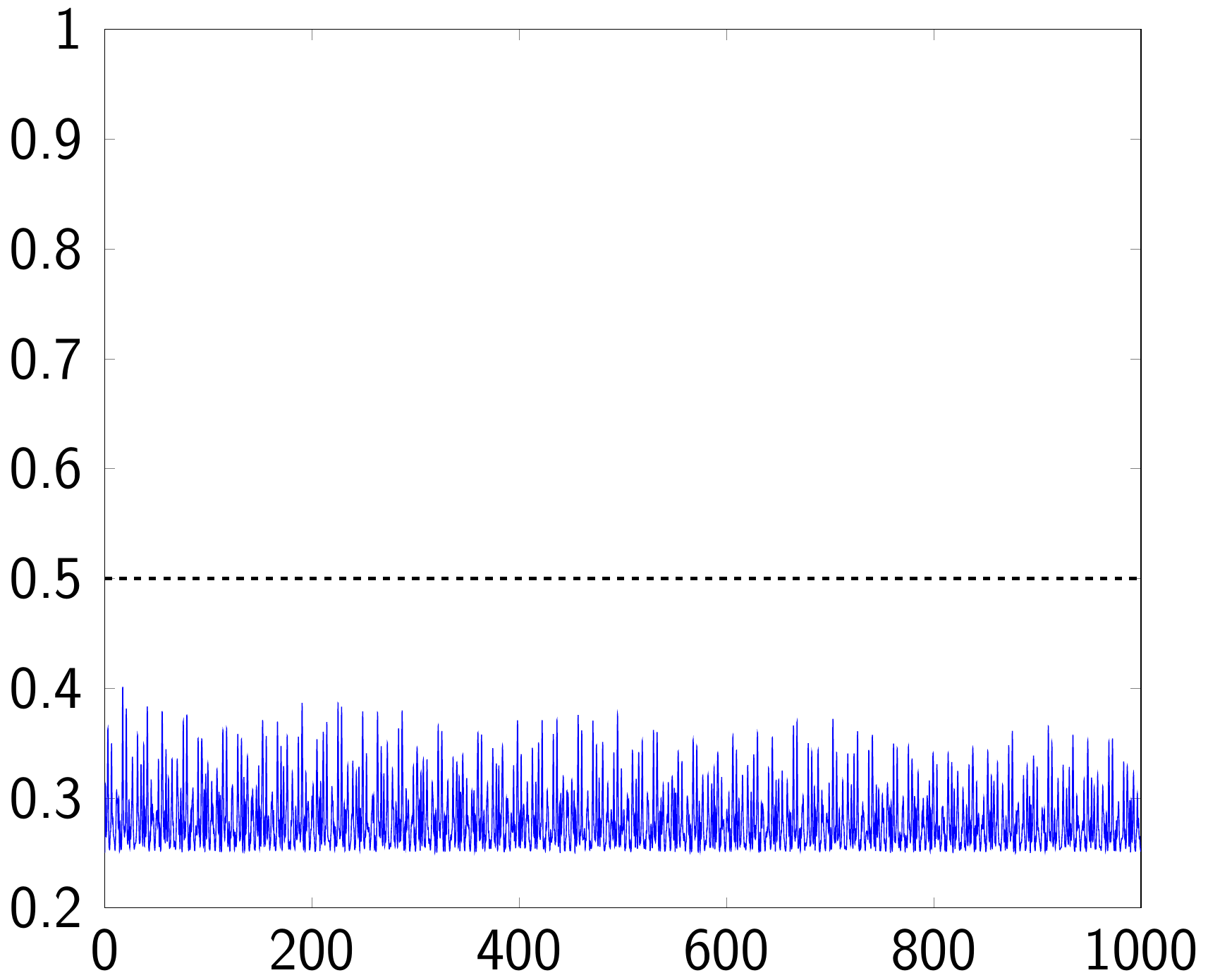} & \vspace{0.2cm}
		\includegraphics[ width=\linewidth, height=\linewidth, keepaspectratio]{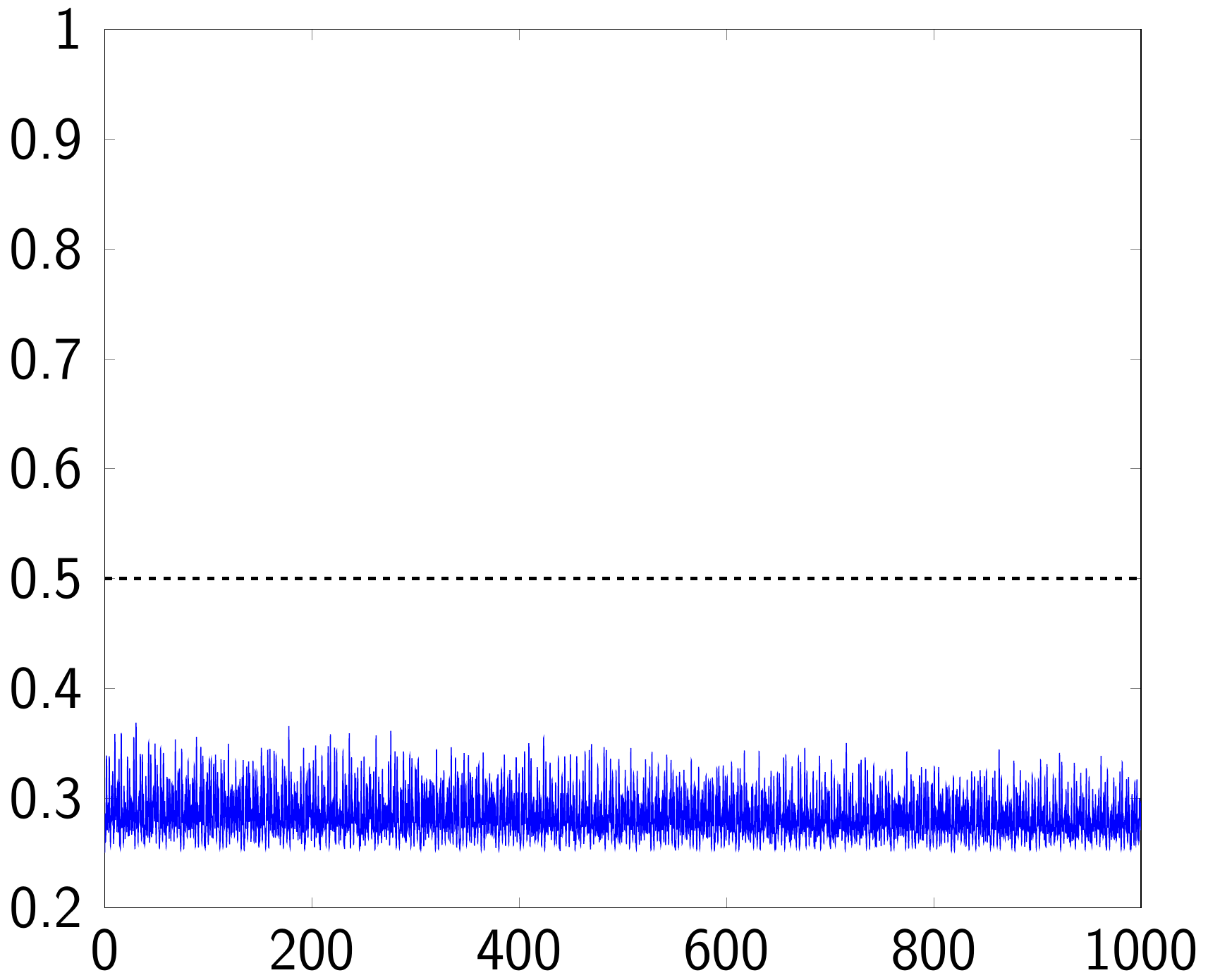} & \vspace{0.2cm}
		\includegraphics[ width=\linewidth, height=\linewidth, keepaspectratio]{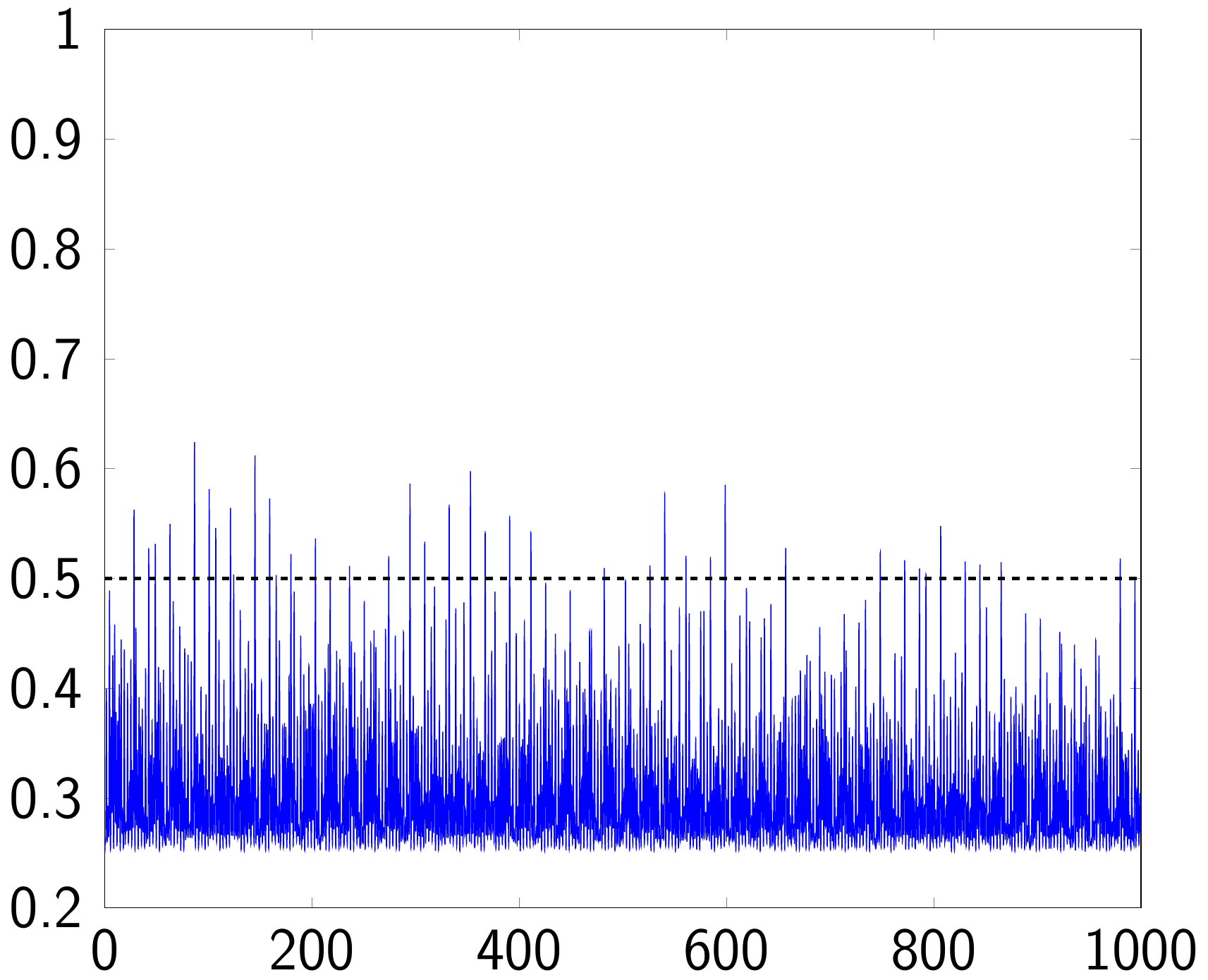} & \vspace{0.2cm}
		\includegraphics[ width=\linewidth, height=\linewidth, keepaspectratio]{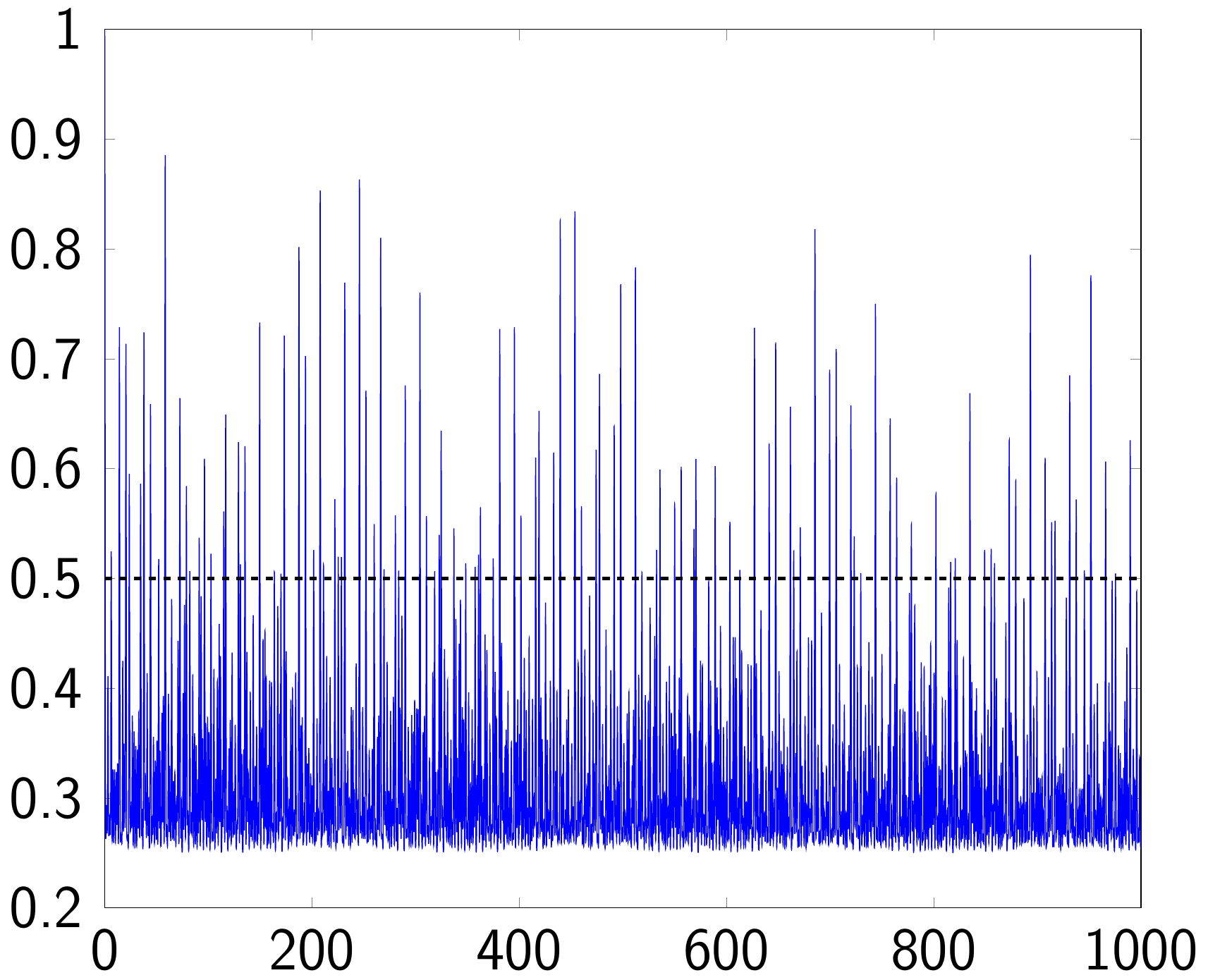} & \vspace{0.2cm}
		\includegraphics[ width=\linewidth, height=\linewidth, keepaspectratio]{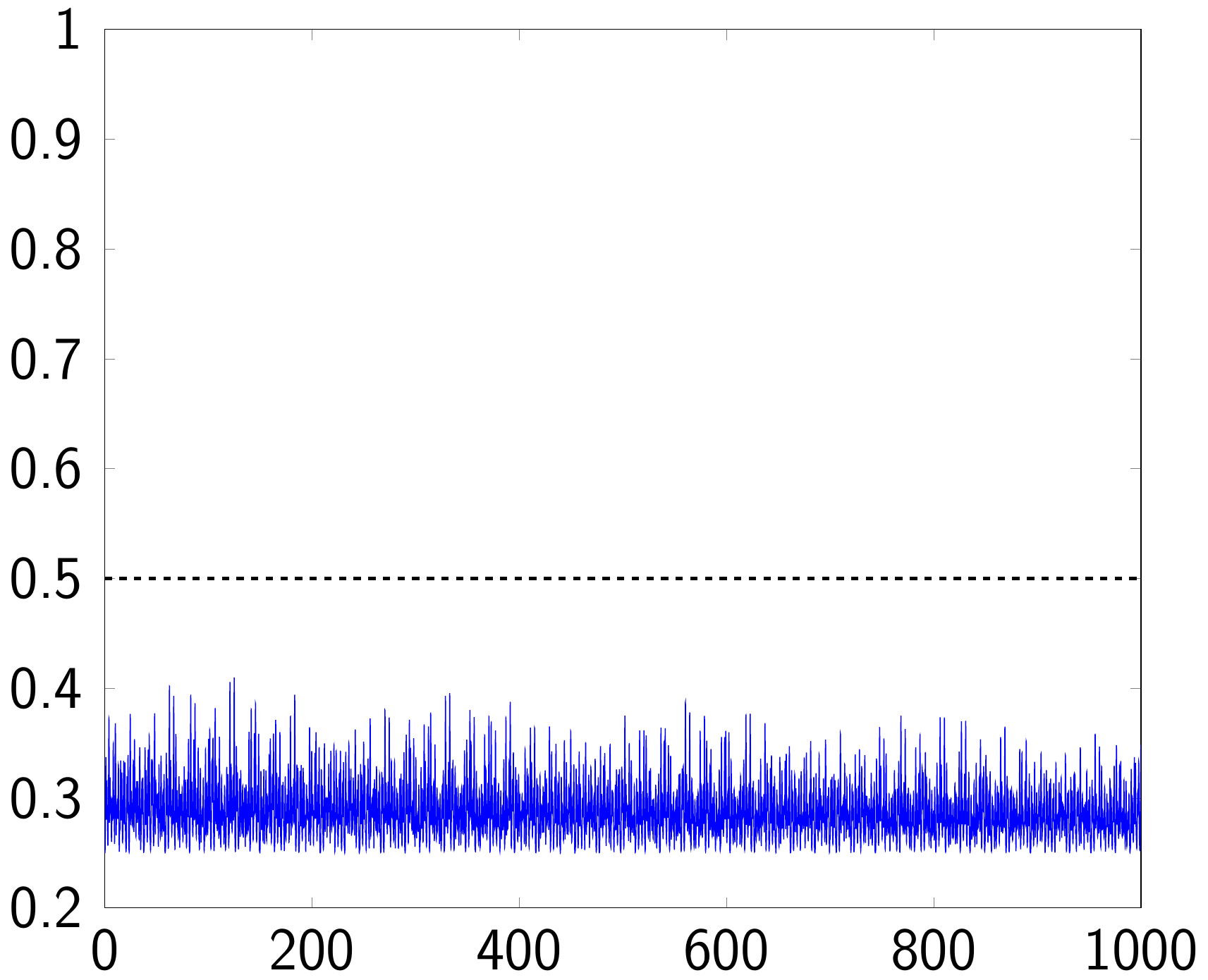} & \vspace{0.2cm}
		\includegraphics[ width=\linewidth, height=\linewidth, keepaspectratio]{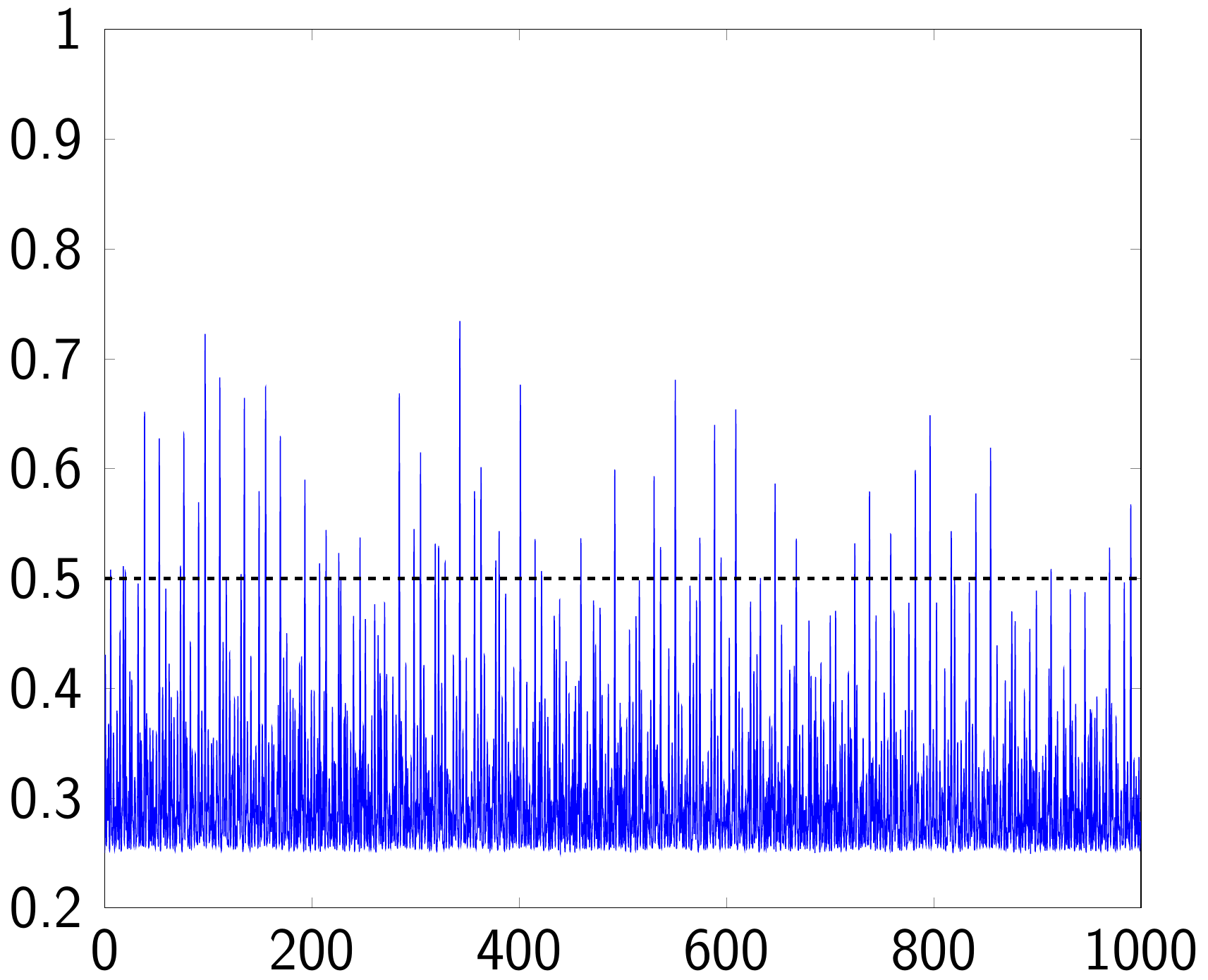} & \vspace{0.2cm}
		\includegraphics[ width=\linewidth, height=\linewidth, keepaspectratio]{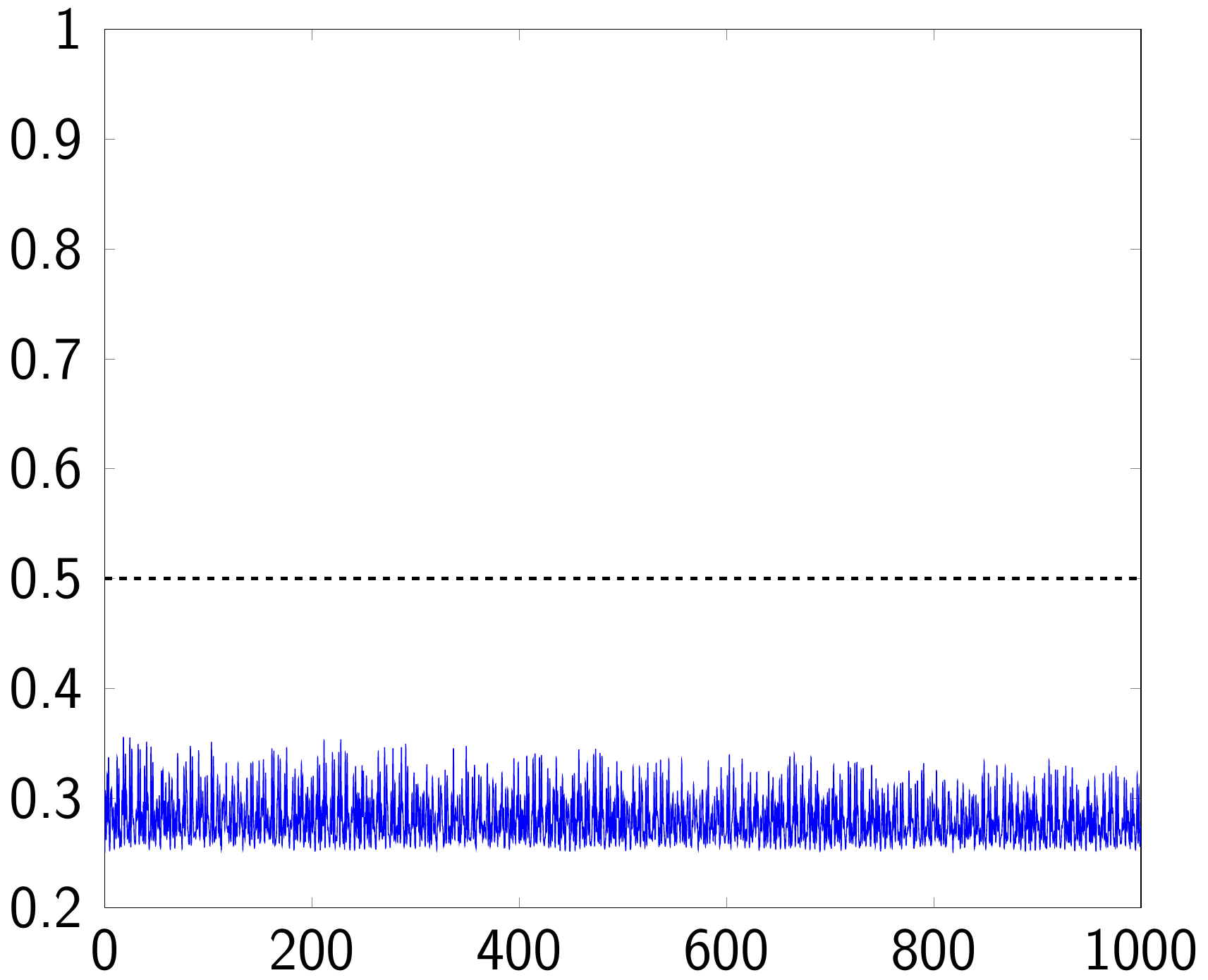} \\
		\hline
		$22$ & \vspace{0.2cm}
		\includegraphics[ width=\linewidth, height=\linewidth, keepaspectratio]{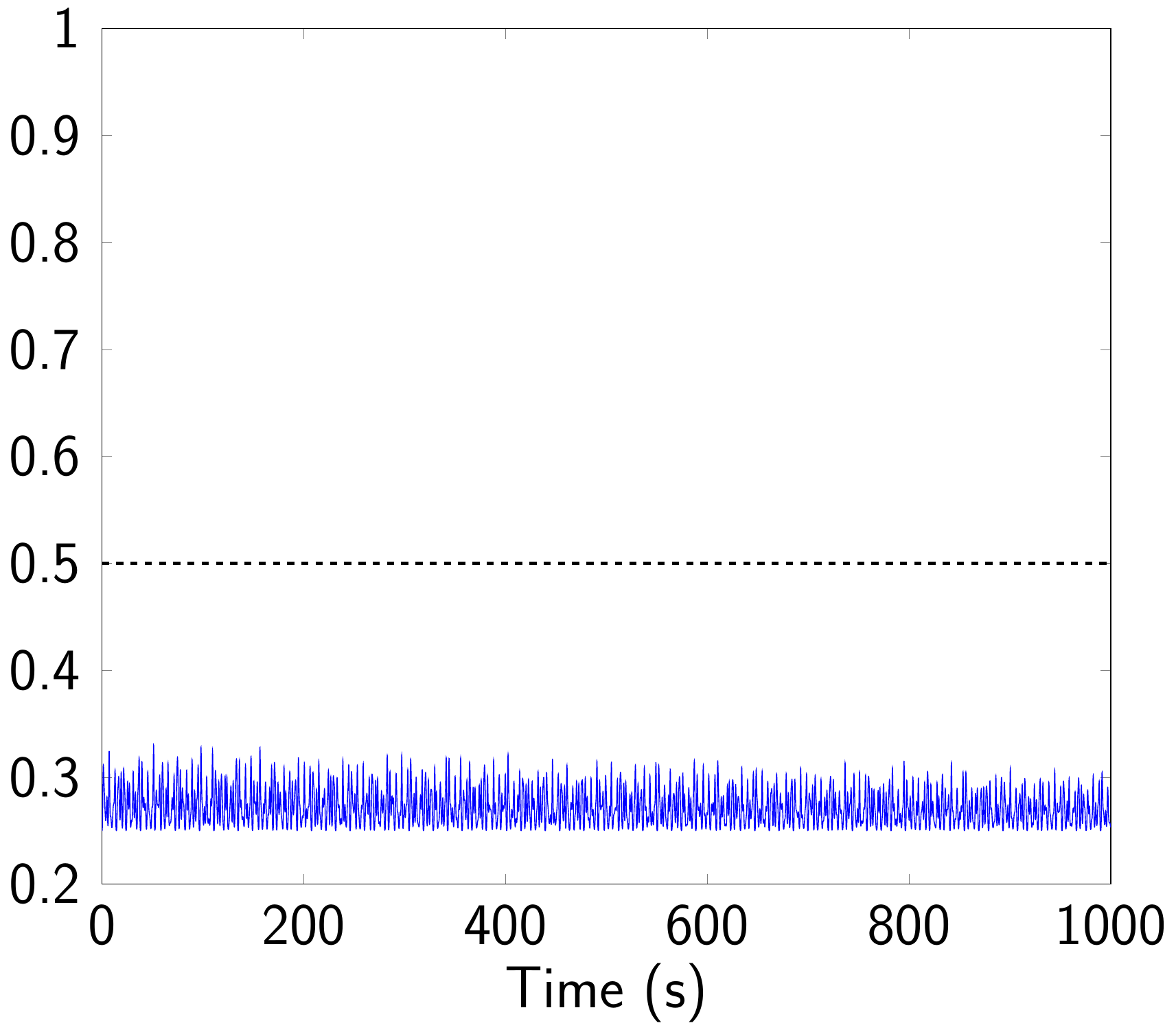} & \vspace{0.2cm}
		\includegraphics[ width=\linewidth, height=\linewidth, keepaspectratio]{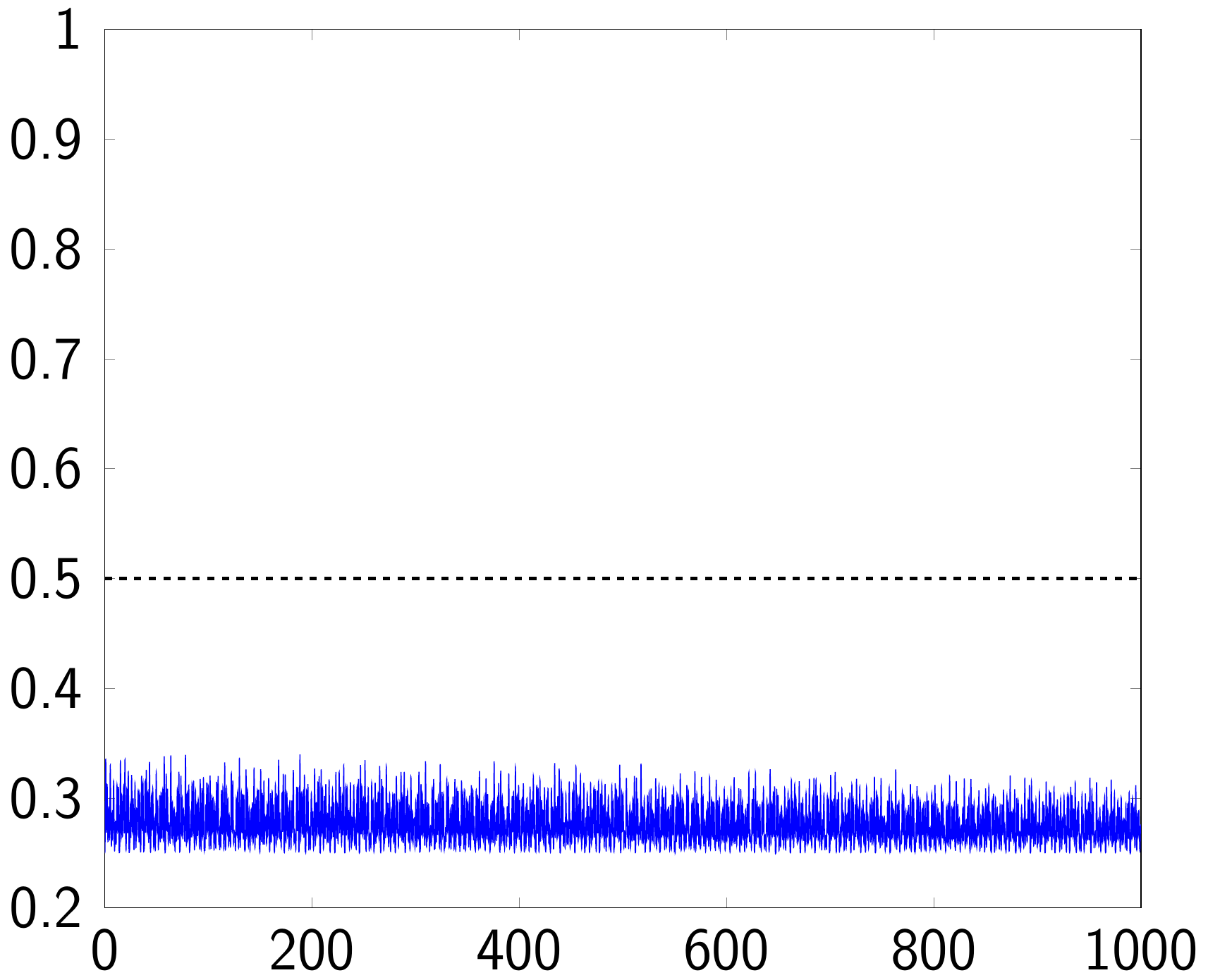} & \vspace{0.2cm}
		\includegraphics[ width=\linewidth, height=\linewidth, keepaspectratio]{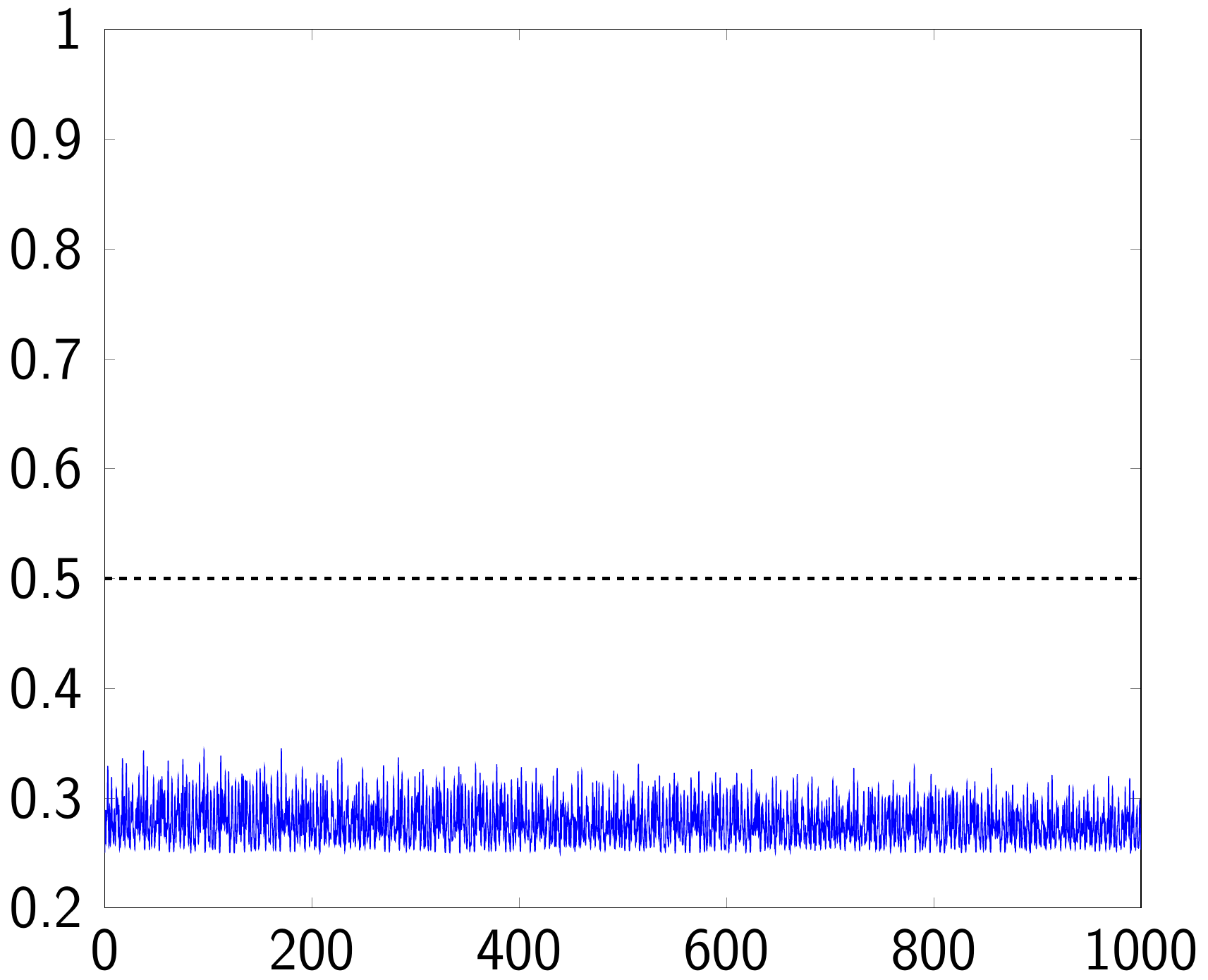} & \vspace{0.2cm}
		\includegraphics[ width=\linewidth, height=\linewidth, keepaspectratio]{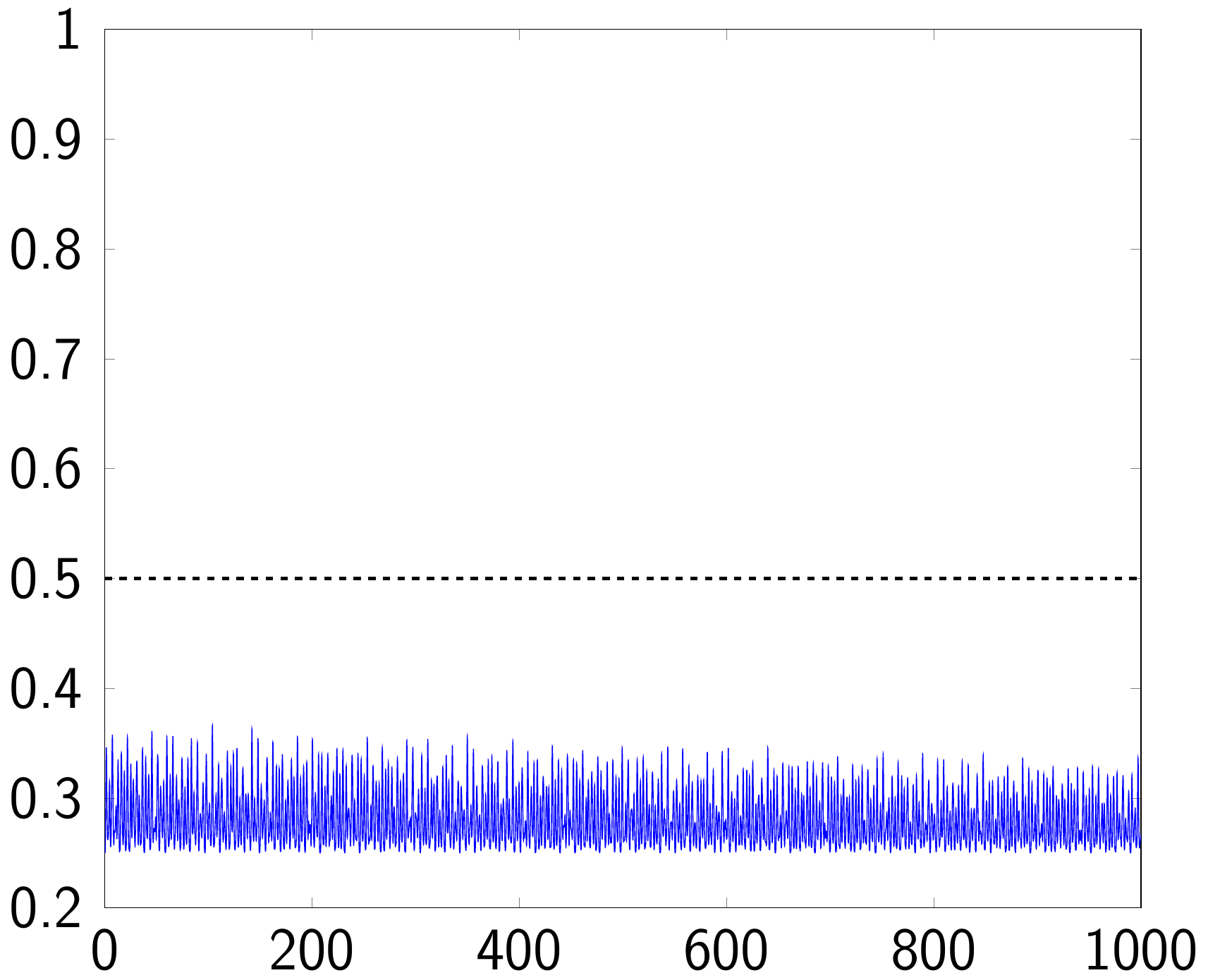} & \vspace{0.2cm}
		\includegraphics[ width=\linewidth, height=\linewidth, keepaspectratio]{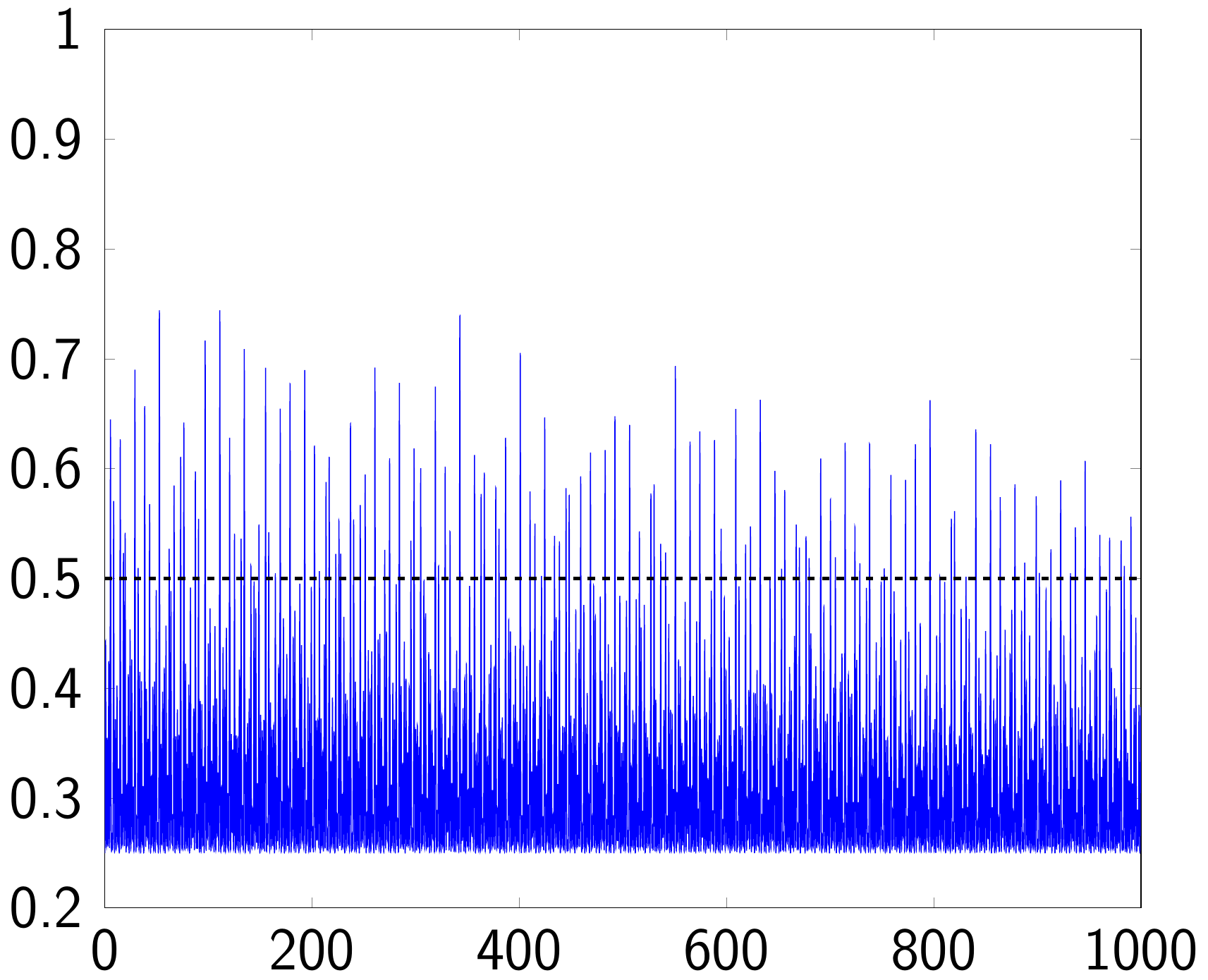} & \vspace{0.2cm}
		\includegraphics[ width=\linewidth, height=\linewidth, keepaspectratio]{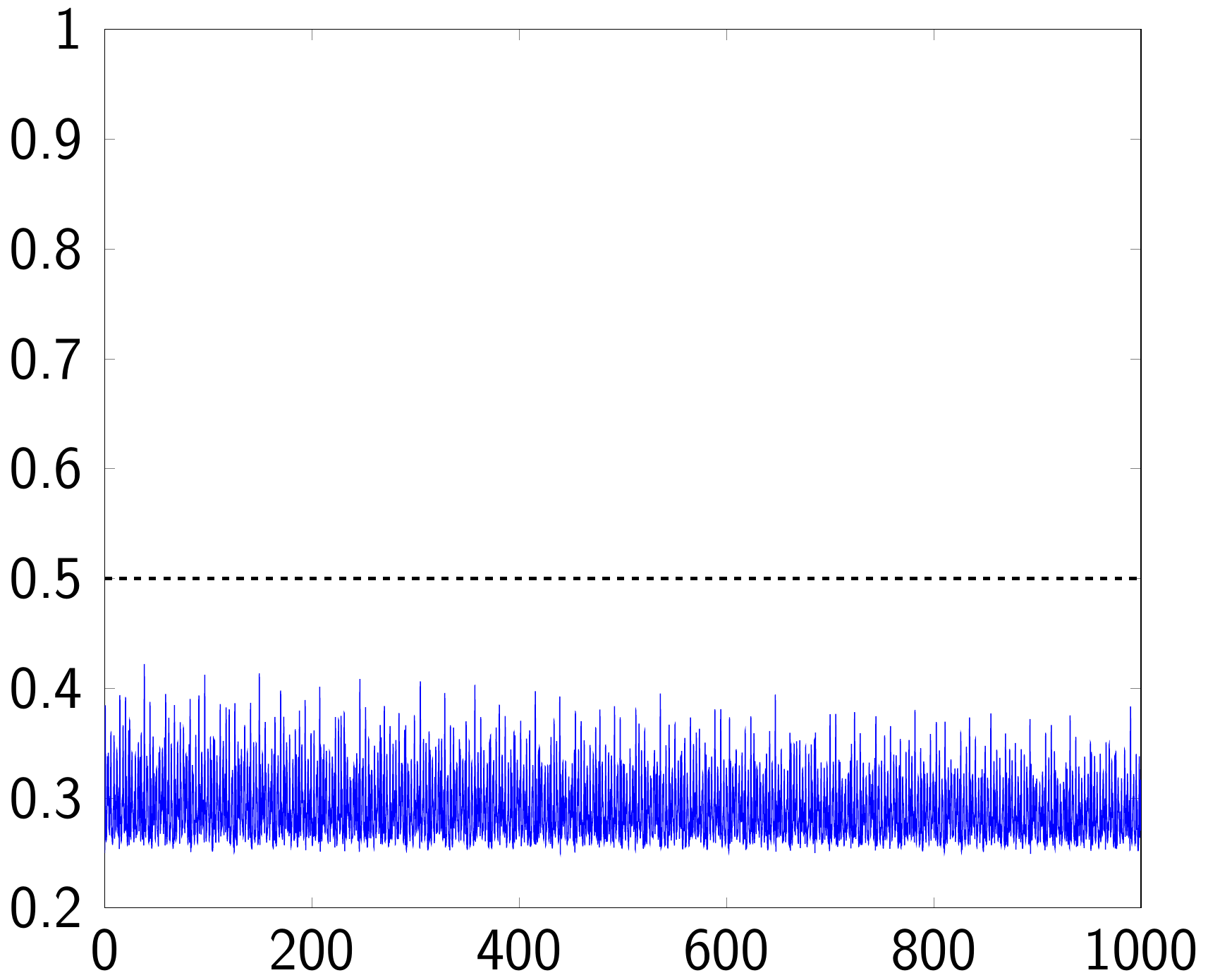} & \vspace{0.2cm}
		\includegraphics[ width=\linewidth, height=\linewidth, keepaspectratio]{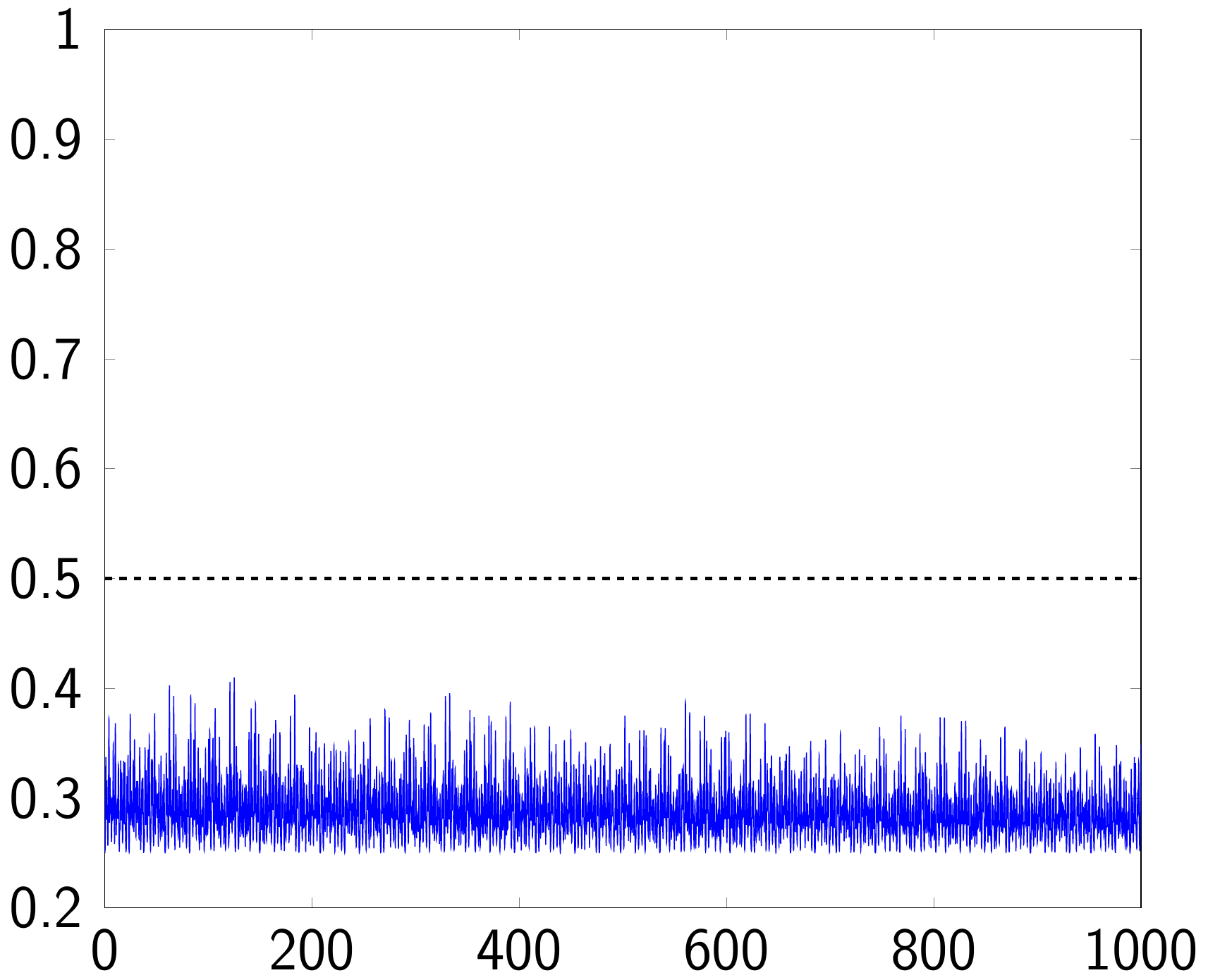} & \vspace{0.2cm}
		\includegraphics[ width=\linewidth, height=\linewidth, keepaspectratio]{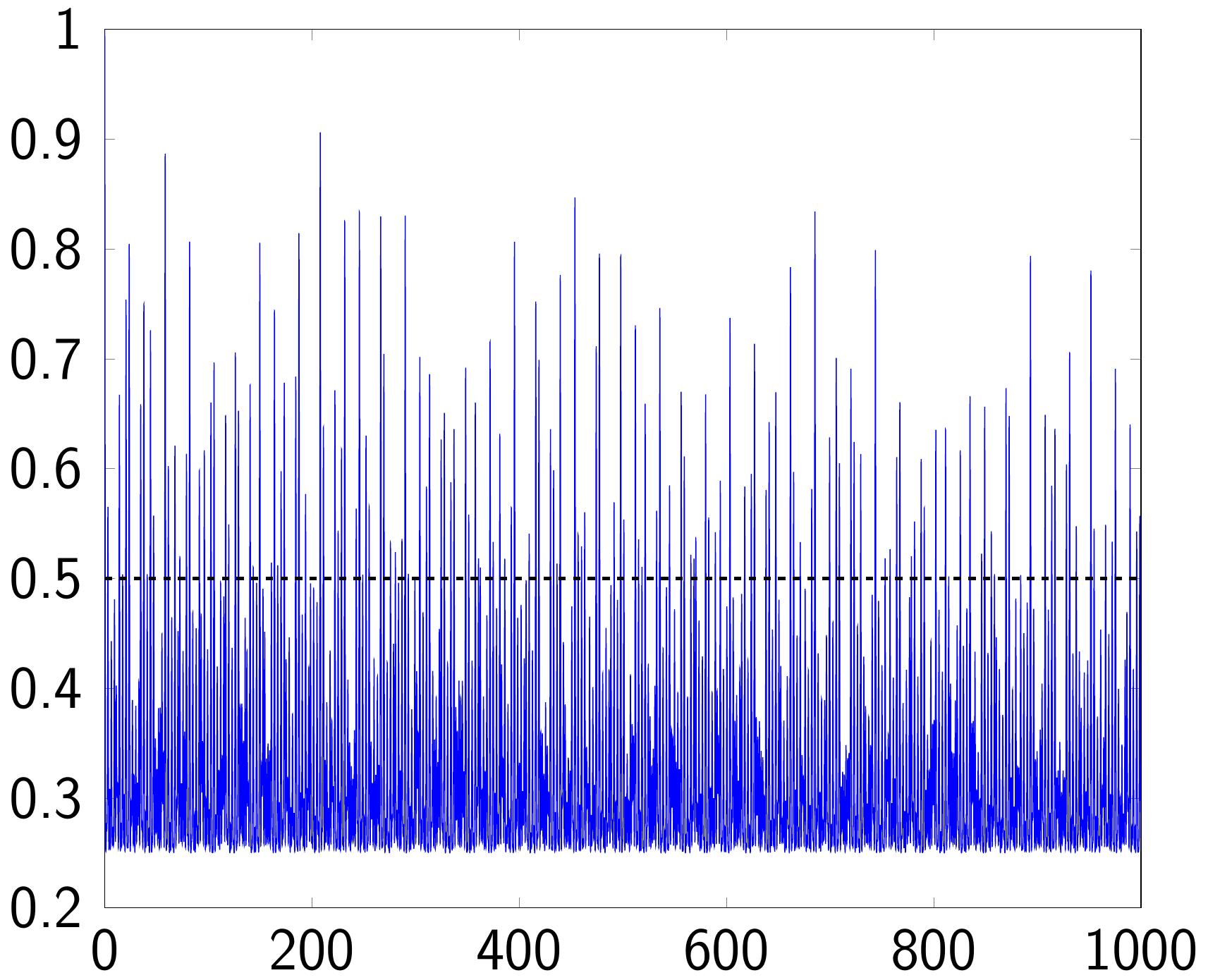} & \vspace{0.2cm}
		\includegraphics[ width=\linewidth, height=\linewidth, keepaspectratio]{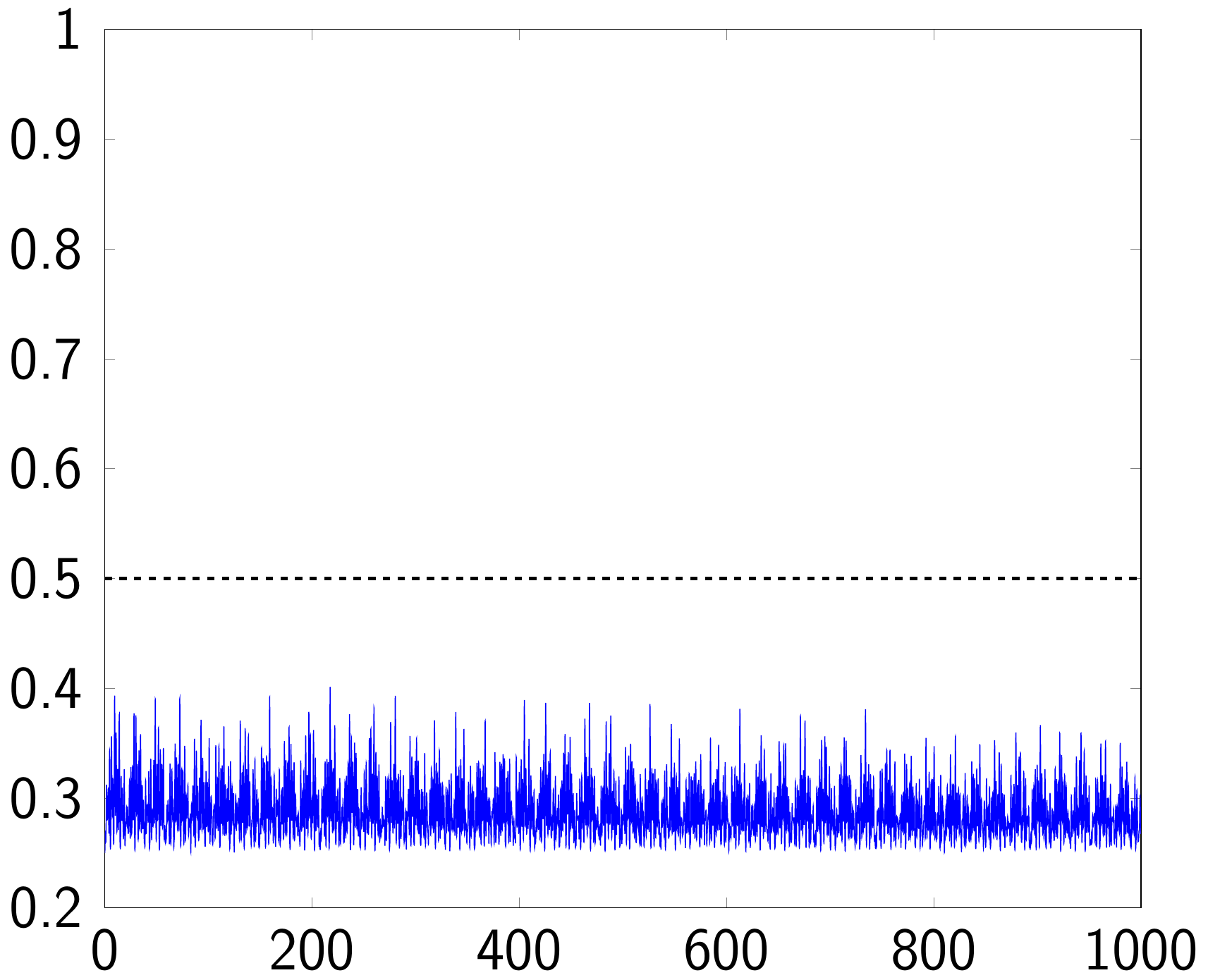} & \vspace{0.2cm}
		\includegraphics[ width=\linewidth, height=\linewidth, keepaspectratio]{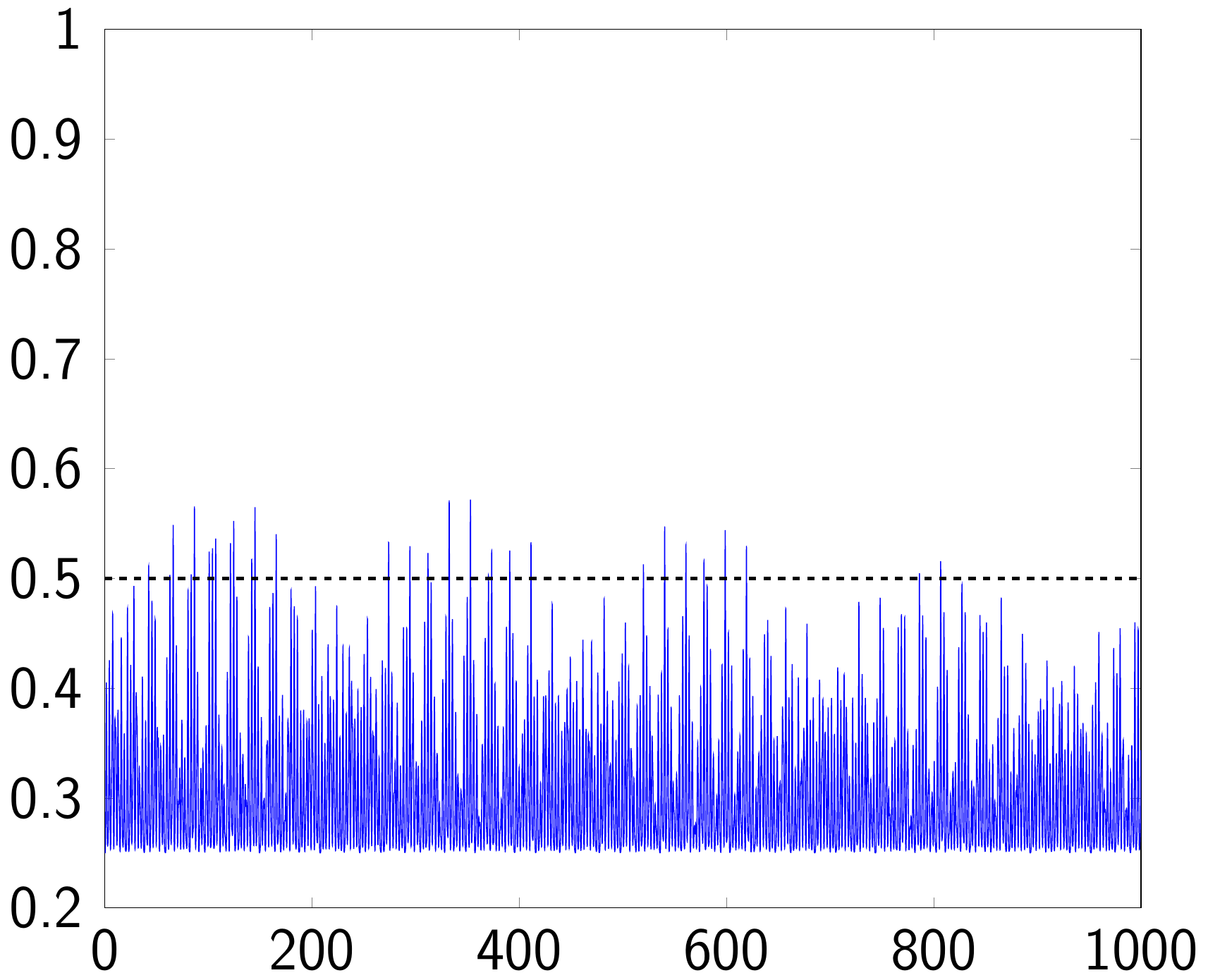} \\
		\hline
		$23$ & \vspace{0.2cm}
		\includegraphics[ width=\linewidth, height=\linewidth, keepaspectratio]{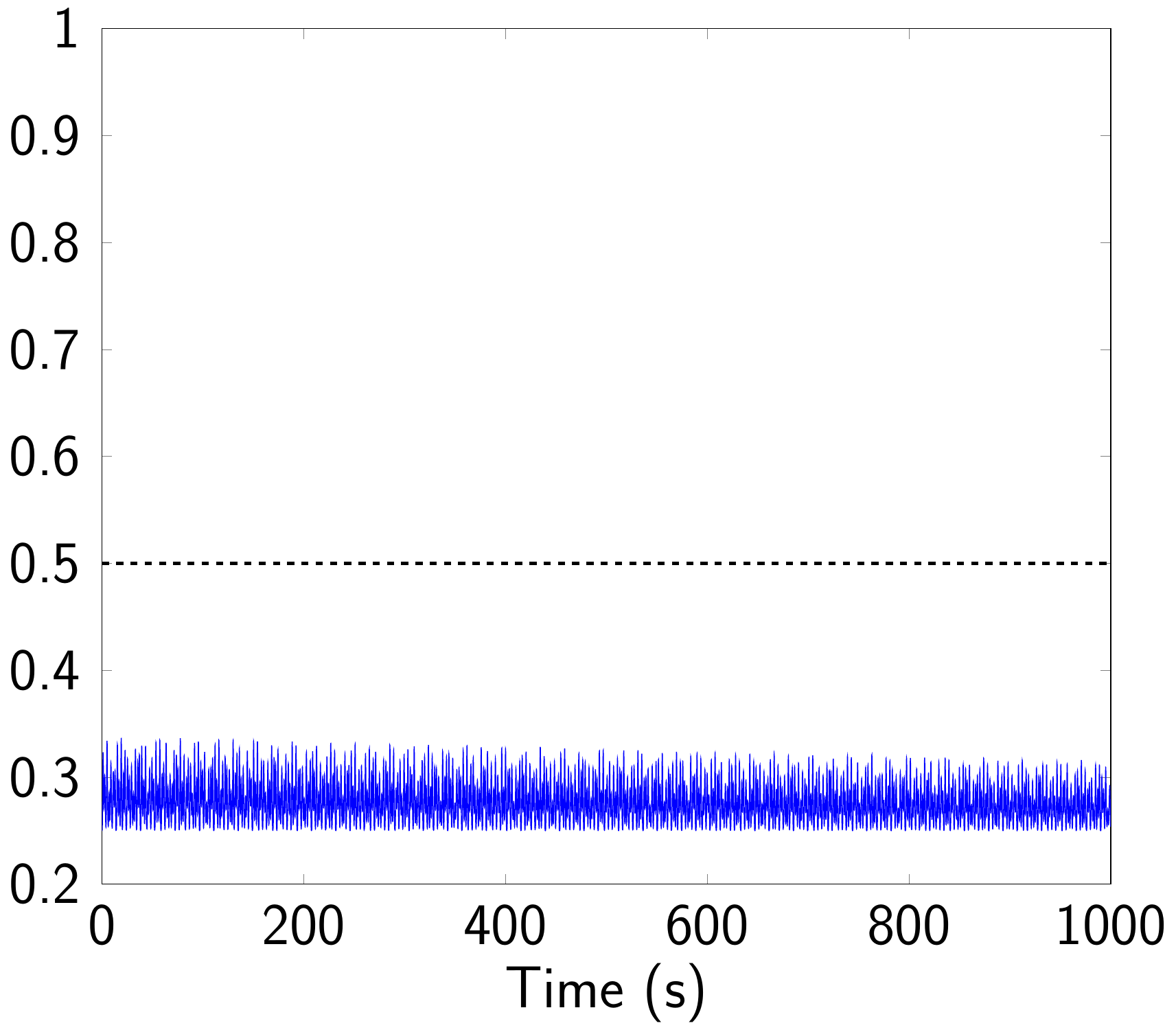} & \vspace{0.2cm}
		\includegraphics[ width=\linewidth, height=\linewidth, keepaspectratio]{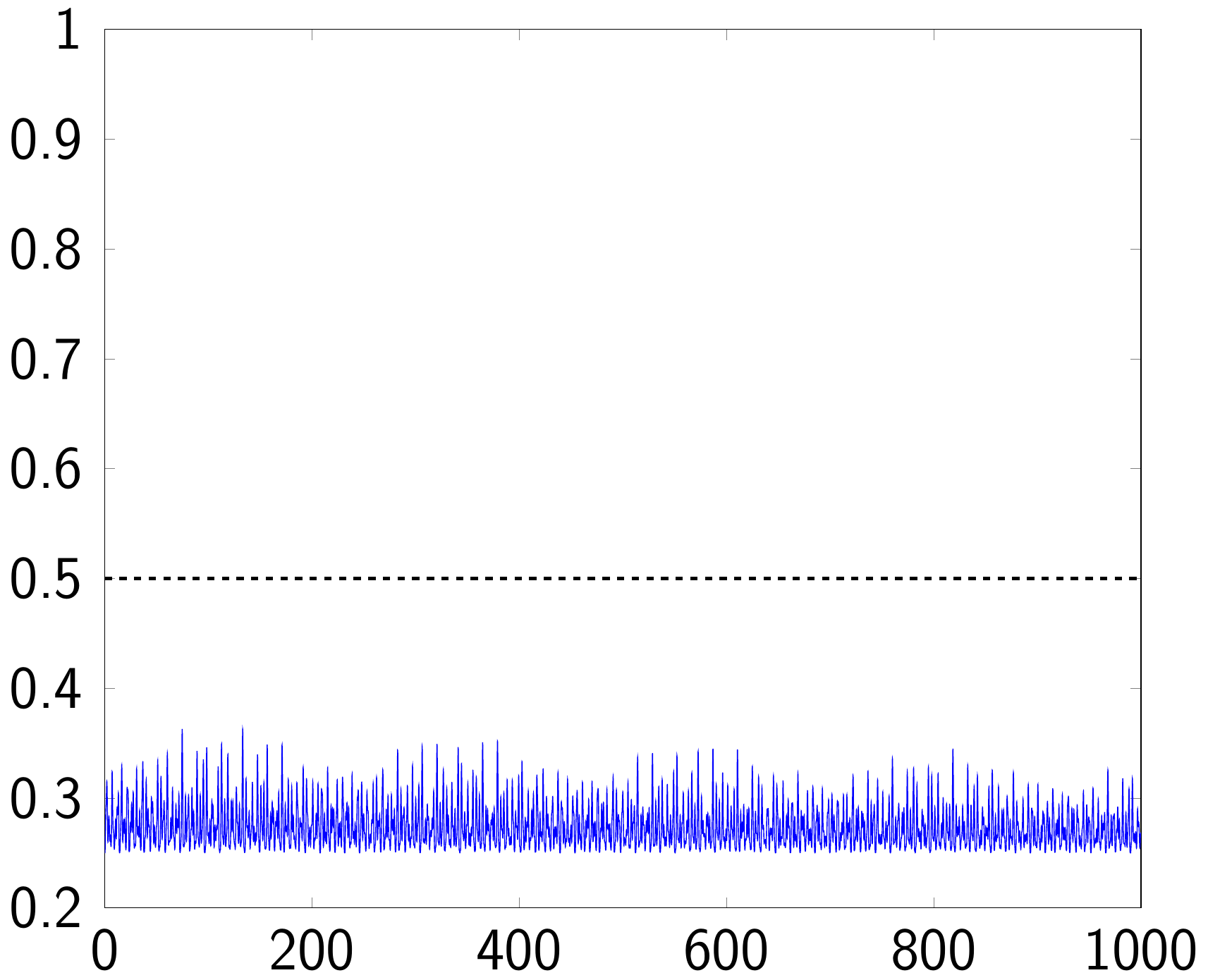} & \vspace{0.2cm}
		\includegraphics[ width=\linewidth, height=\linewidth, keepaspectratio]{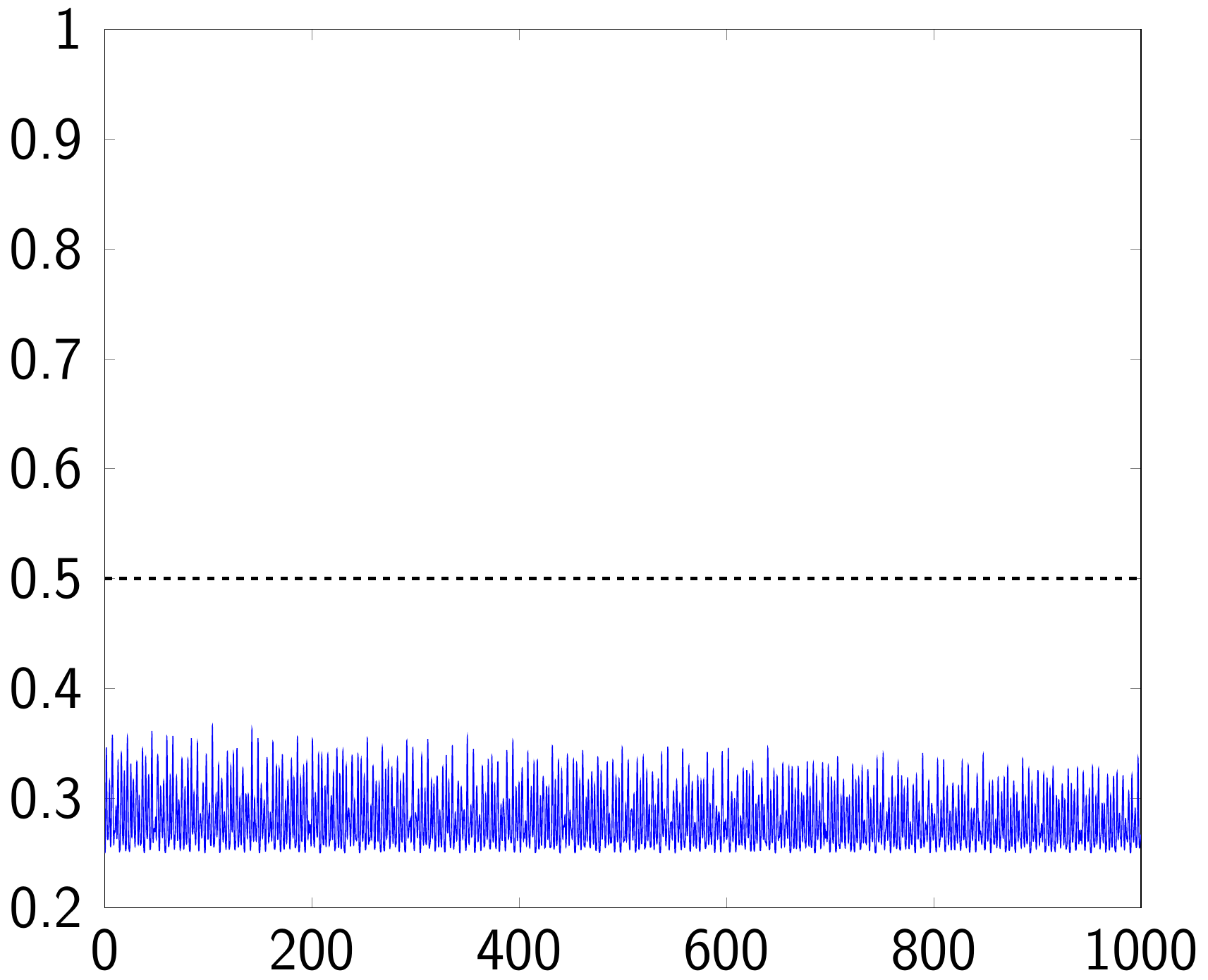} & \vspace{0.2cm}
		\includegraphics[ width=\linewidth, height=\linewidth, keepaspectratio]{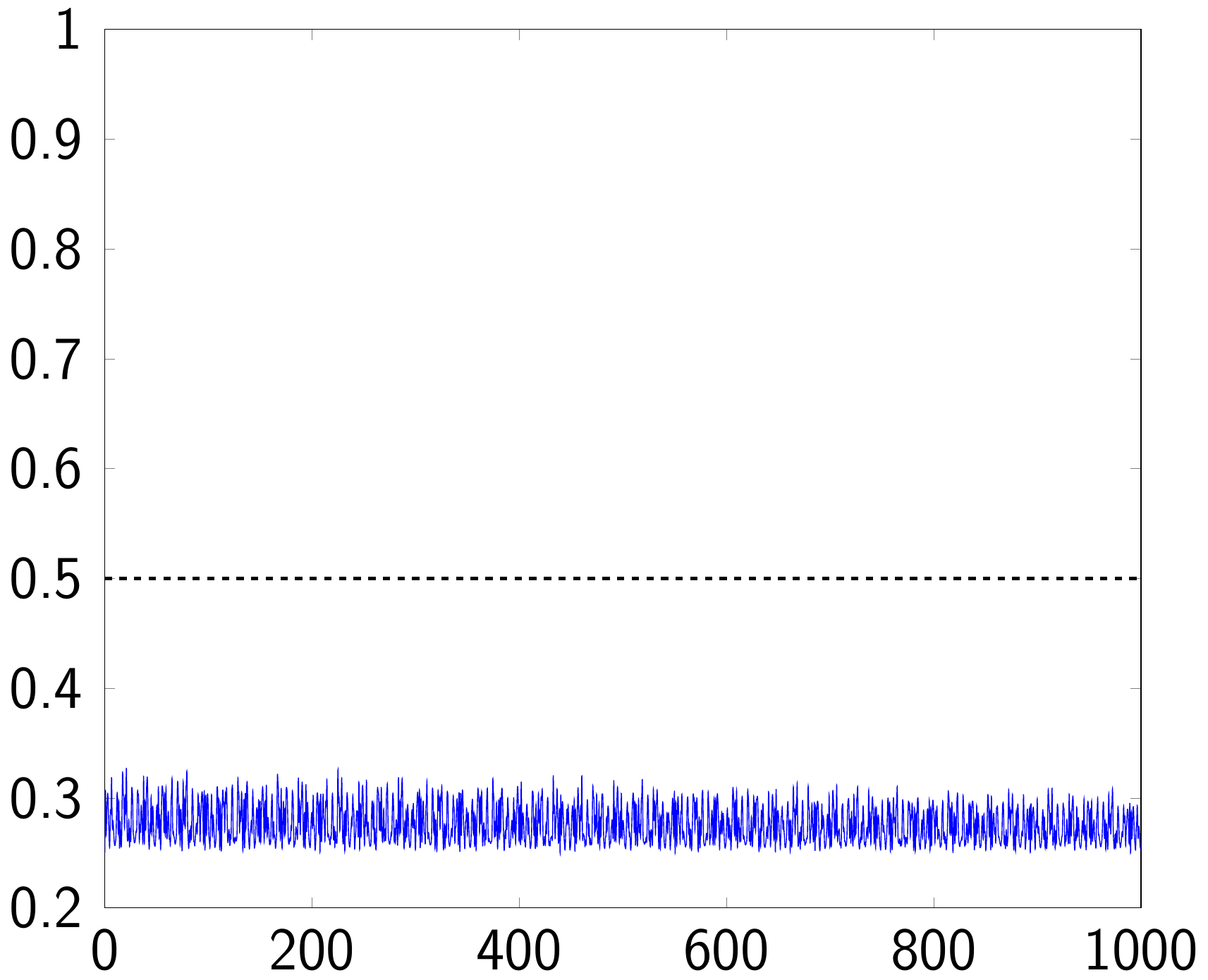} & \vspace{0.2cm}
		\includegraphics[ width=\linewidth, height=\linewidth, keepaspectratio]{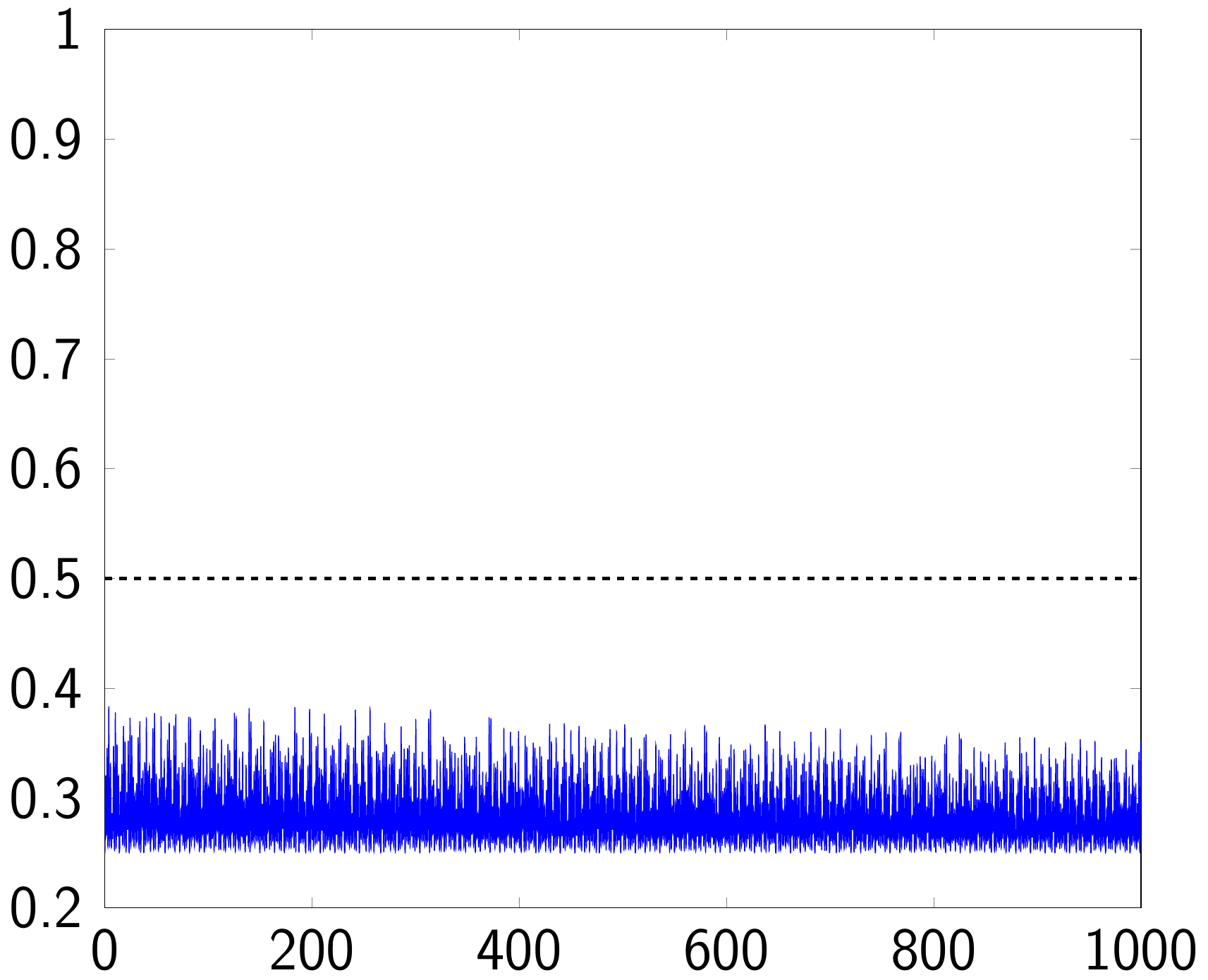} & \vspace{0.2cm}
		\includegraphics[ width=\linewidth, height=\linewidth, keepaspectratio]{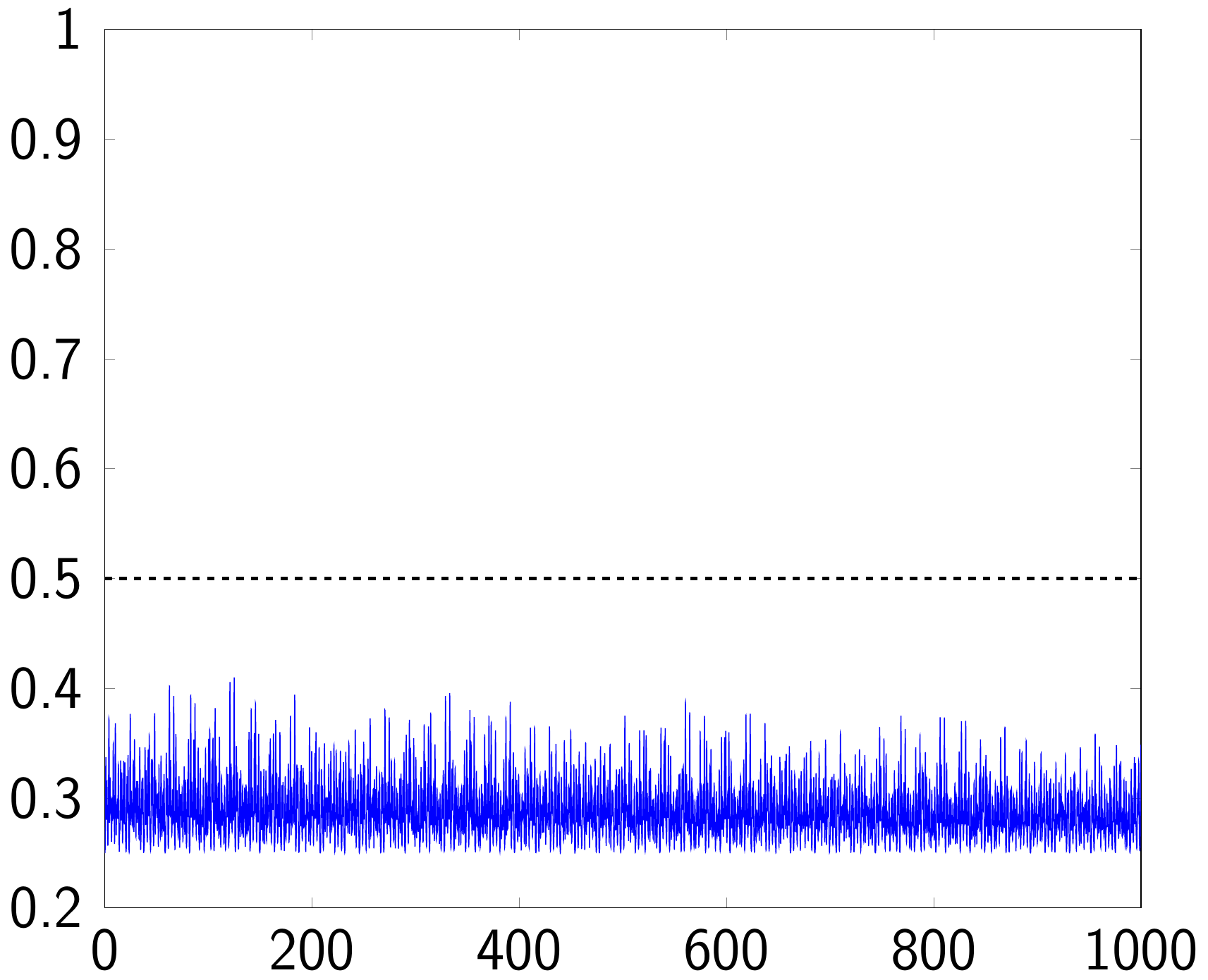} & \vspace{0.2cm}
		\includegraphics[ width=\linewidth, height=\linewidth, keepaspectratio]{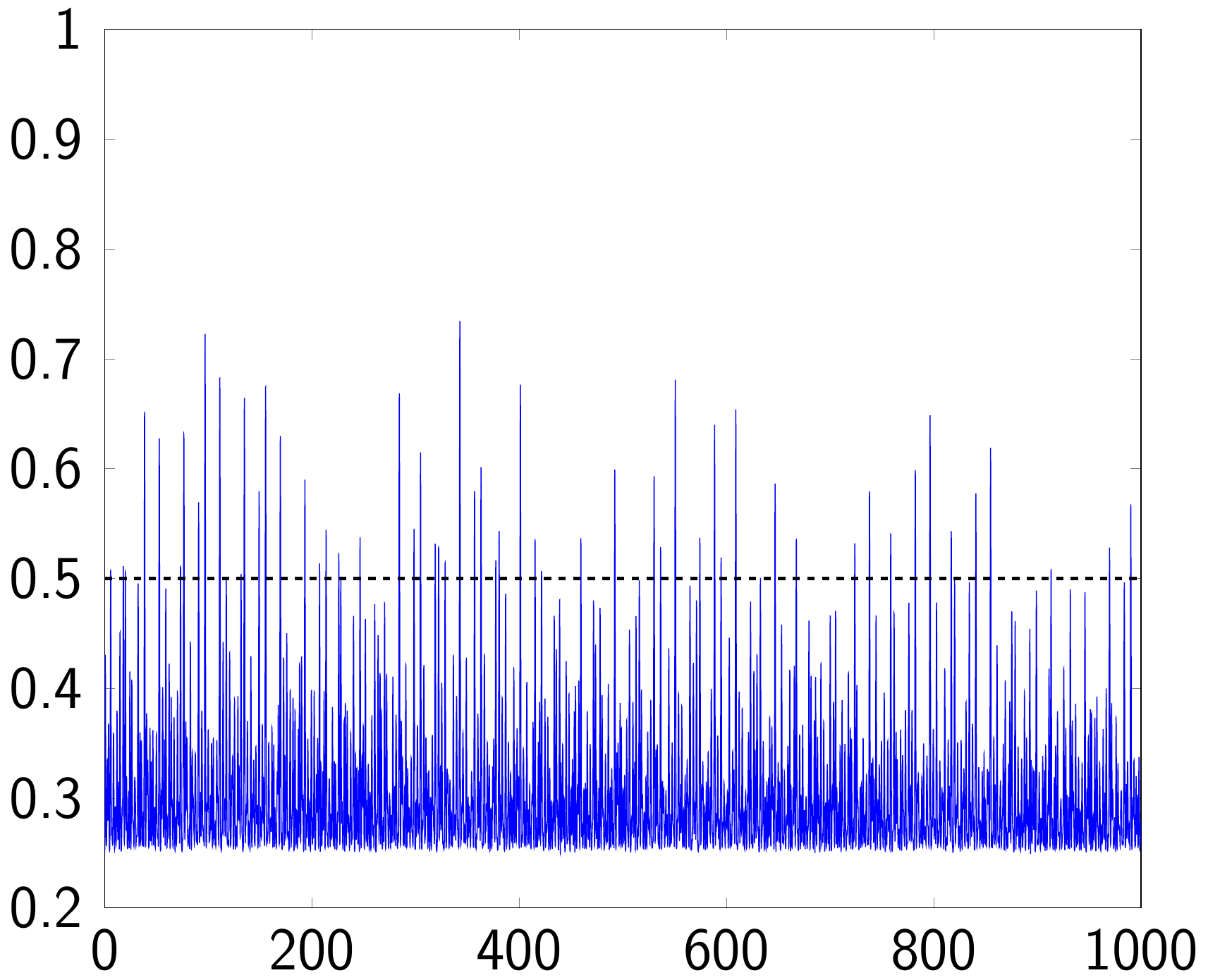} & \vspace{0.2cm}
		\includegraphics[ width=\linewidth, height=\linewidth, keepaspectratio]{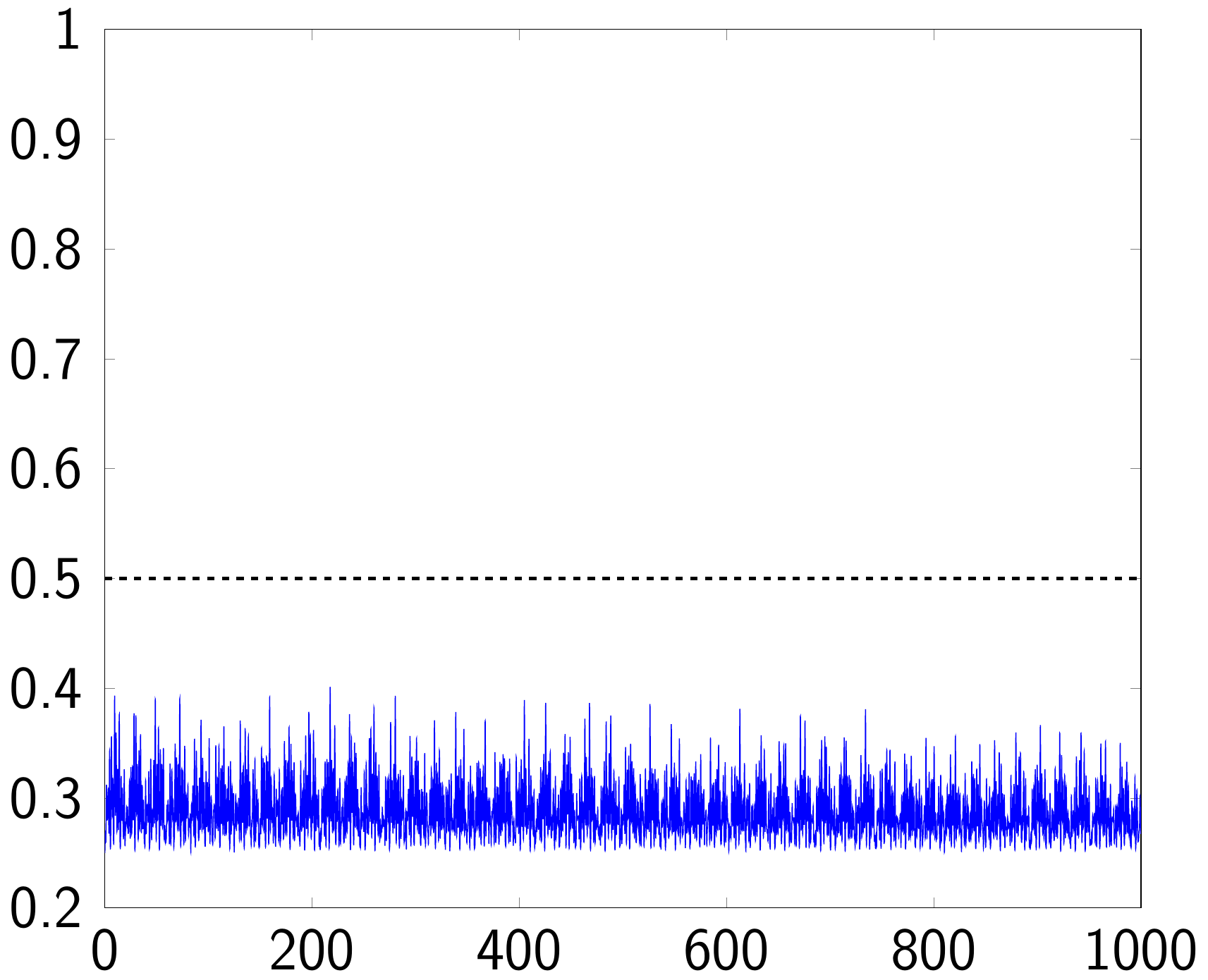} & \vspace{0.2cm}
		\includegraphics[ width=\linewidth, height=\linewidth, keepaspectratio]{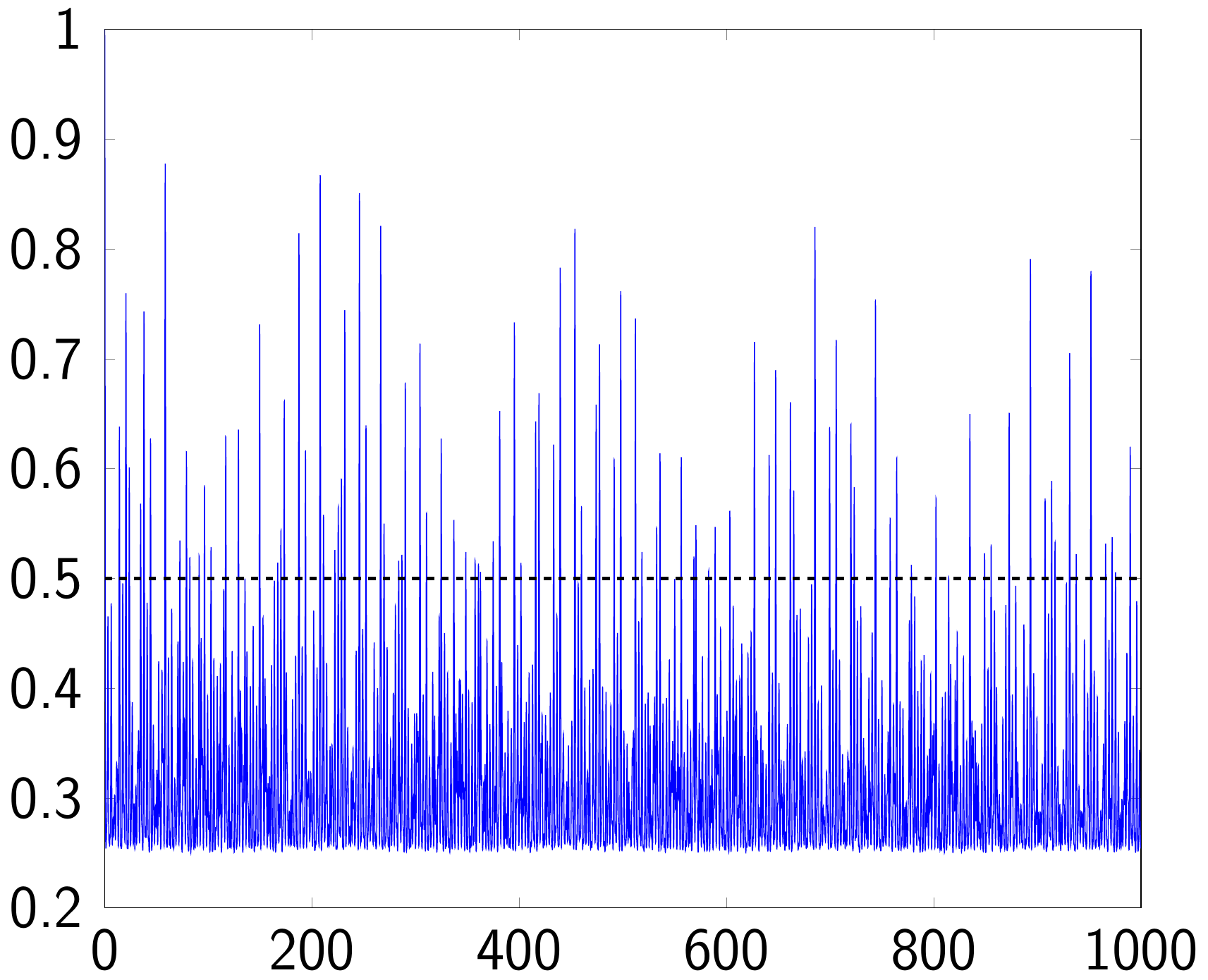} & \vspace{0.2cm}
		\includegraphics[ width=\linewidth, height=\linewidth, keepaspectratio]{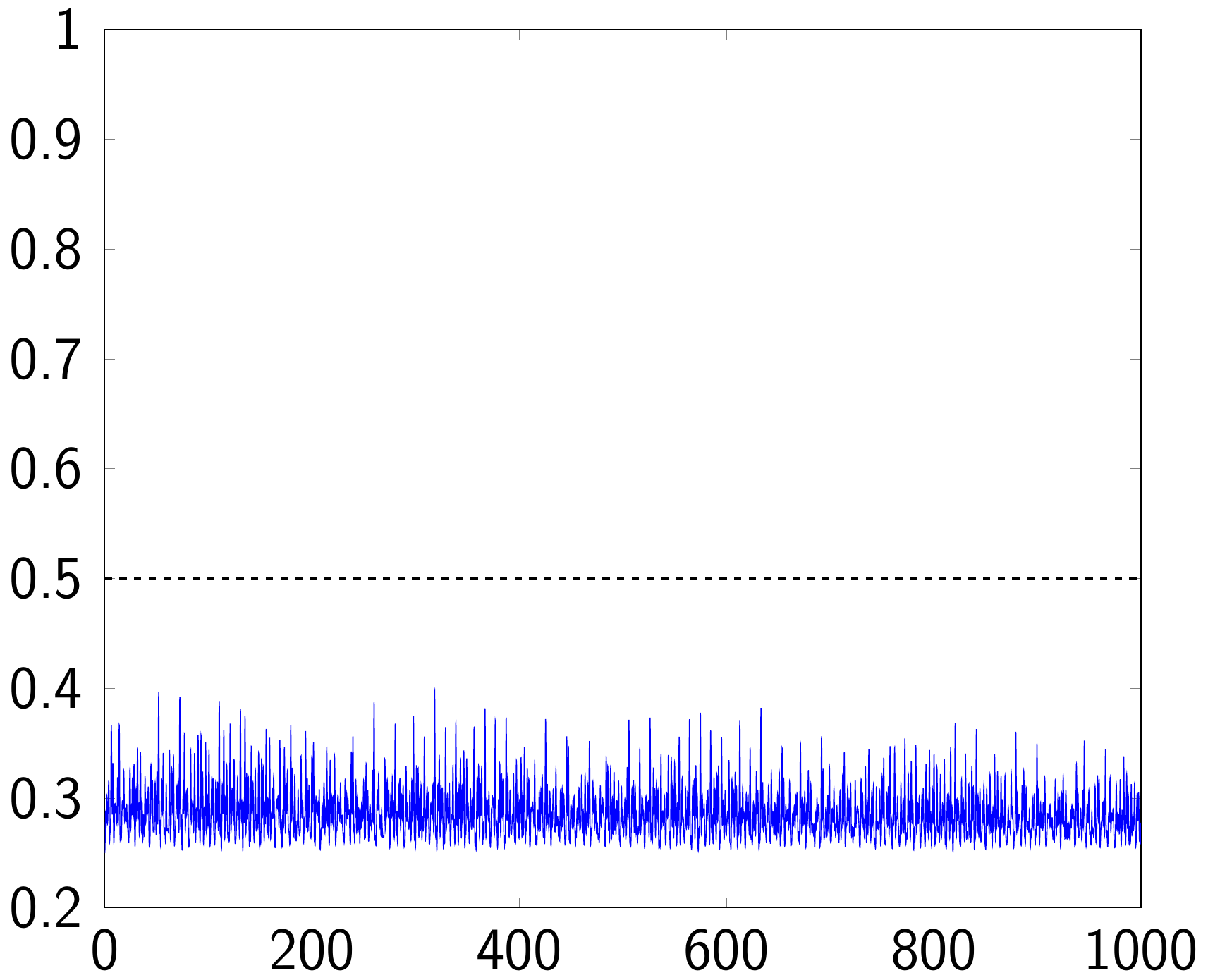} \\
		\hline
		$33$ & \vspace{0.2cm}
		\includegraphics[ width=\linewidth, height=\linewidth, keepaspectratio]{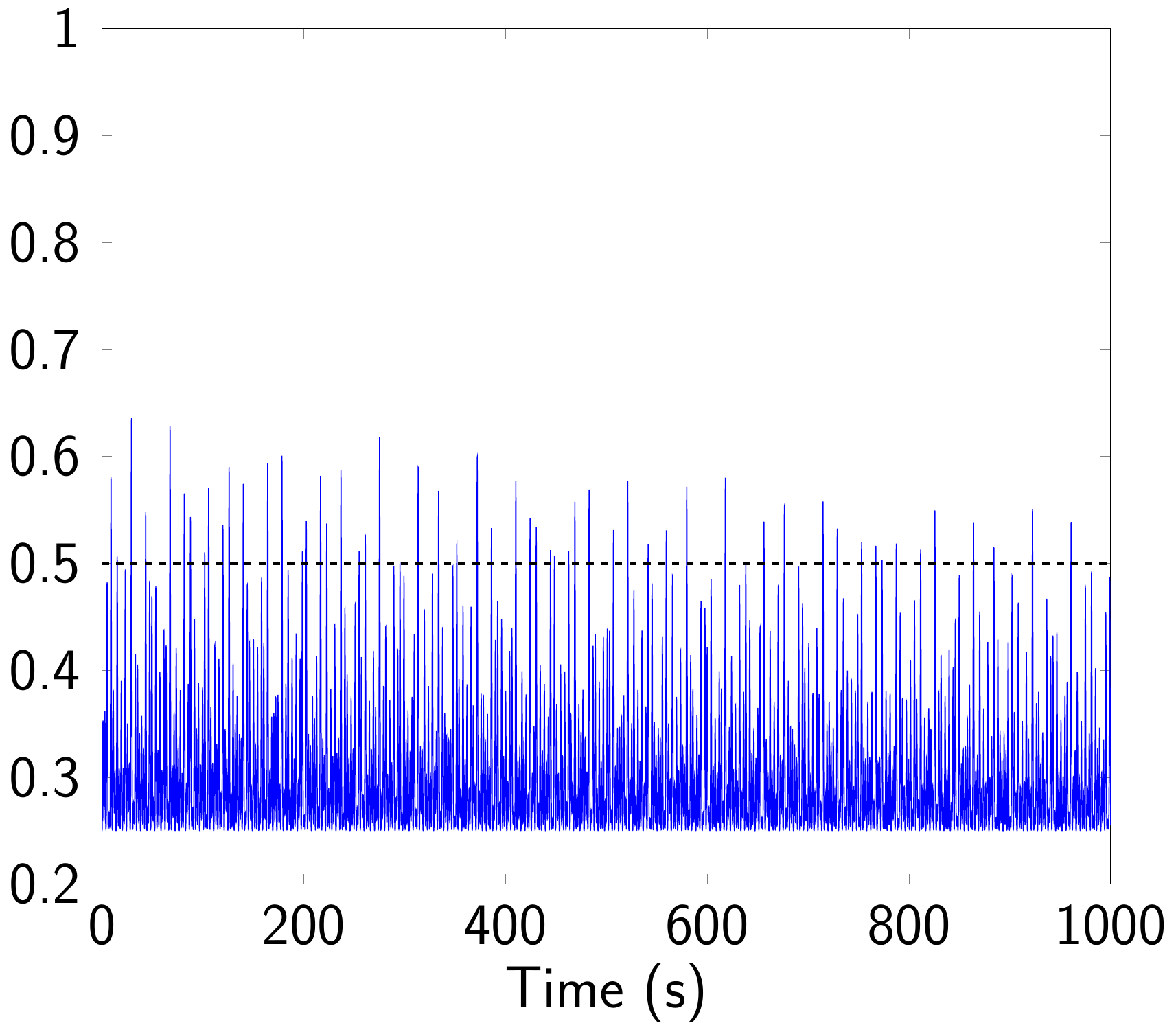} & \vspace{0.2cm}
		\includegraphics[ width=\linewidth, height=\linewidth, keepaspectratio]{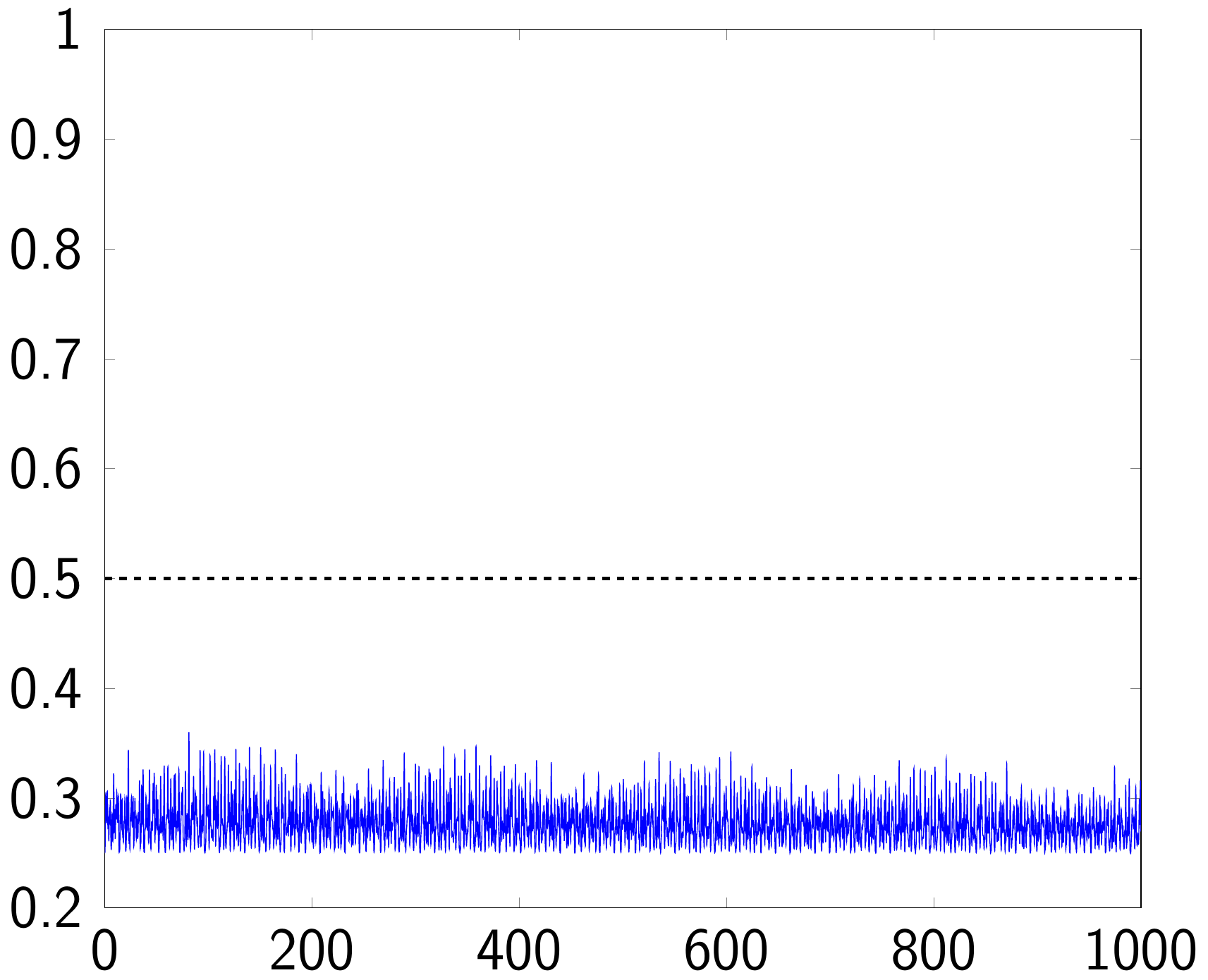} & \vspace{0.2cm}
		\includegraphics[ width=\linewidth, height=\linewidth, keepaspectratio]{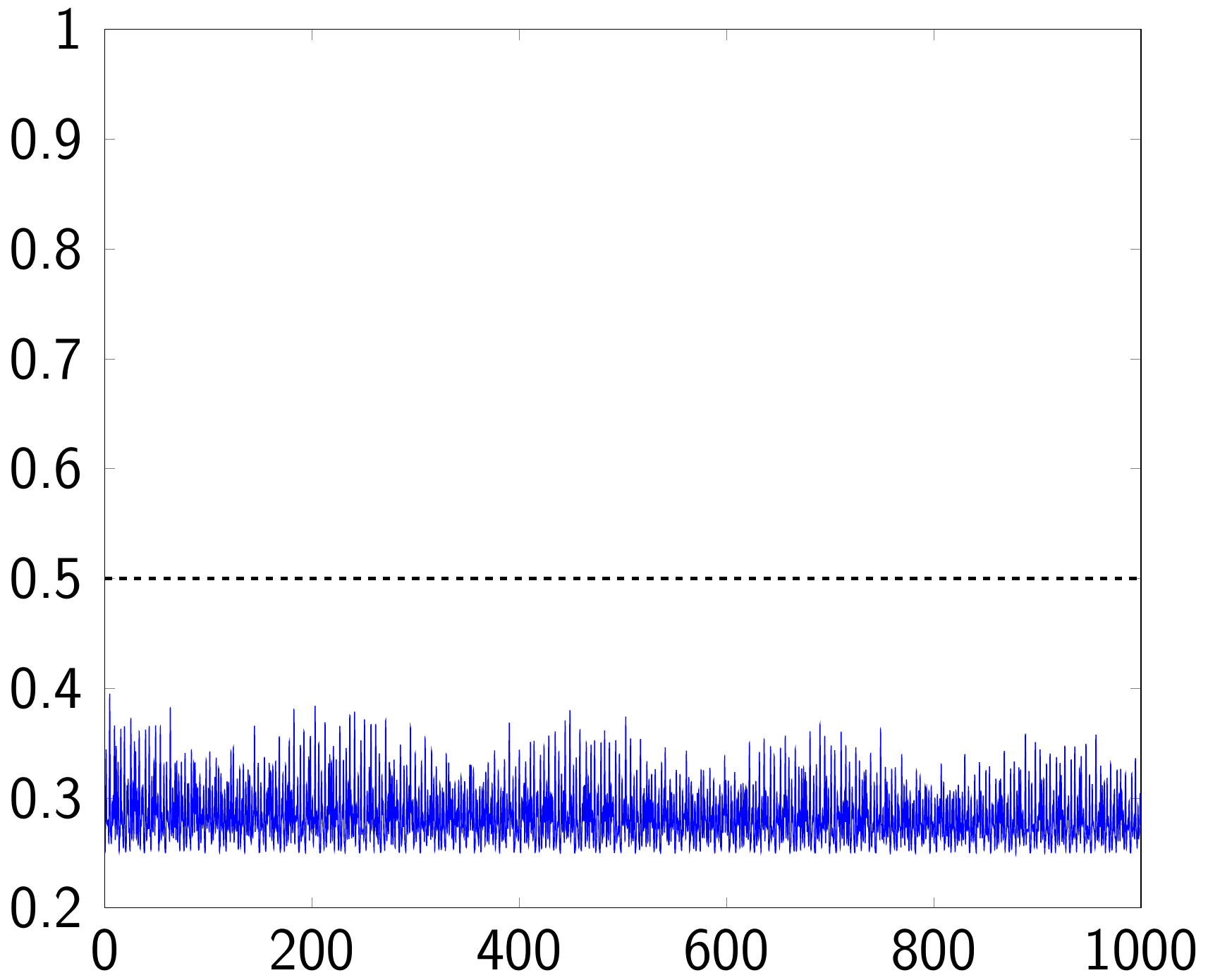} & \vspace{0.2cm}
		\includegraphics[ width=\linewidth, height=\linewidth, keepaspectratio]{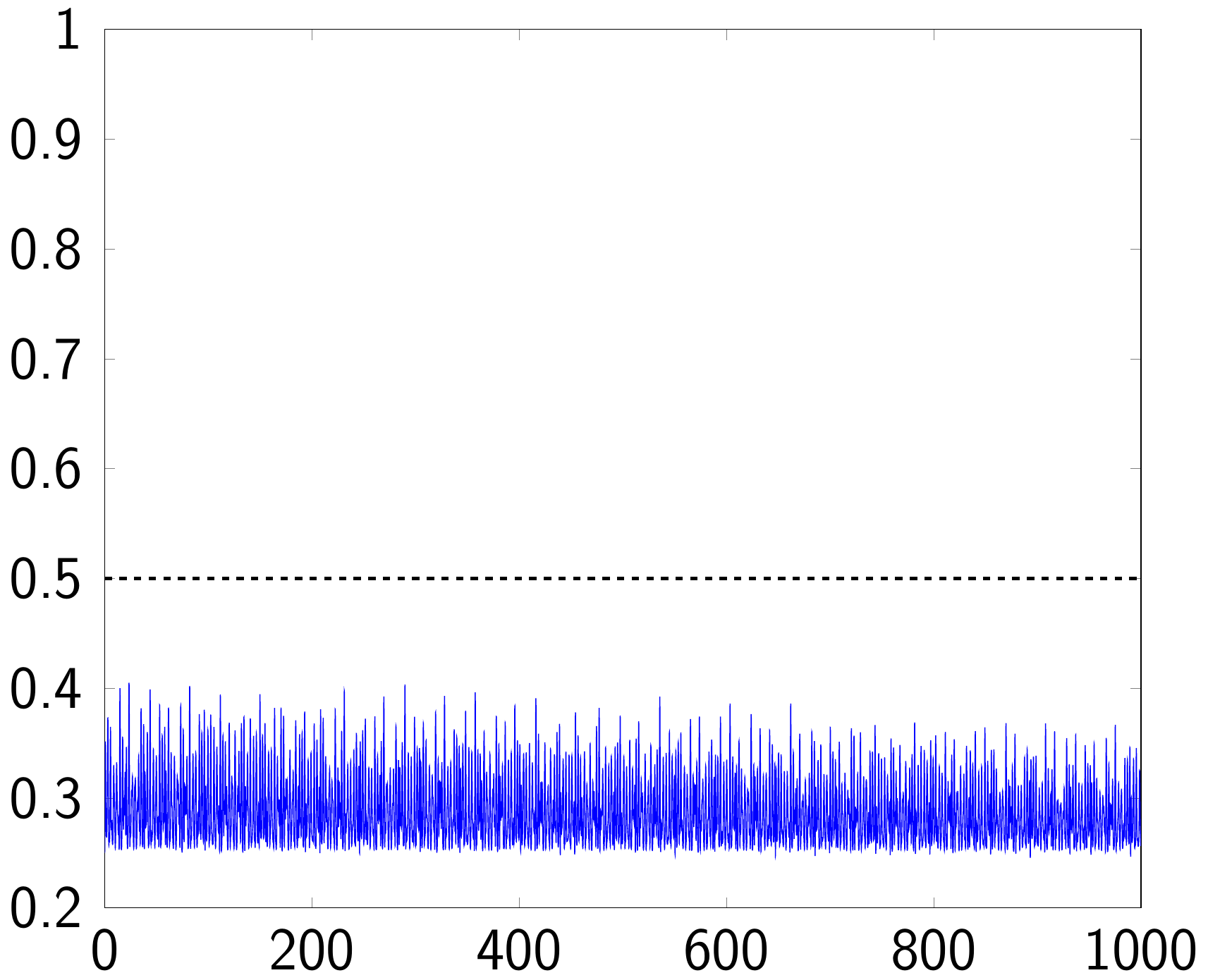} & \vspace{0.2cm}
		\includegraphics[ width=\linewidth, height=\linewidth, keepaspectratio]{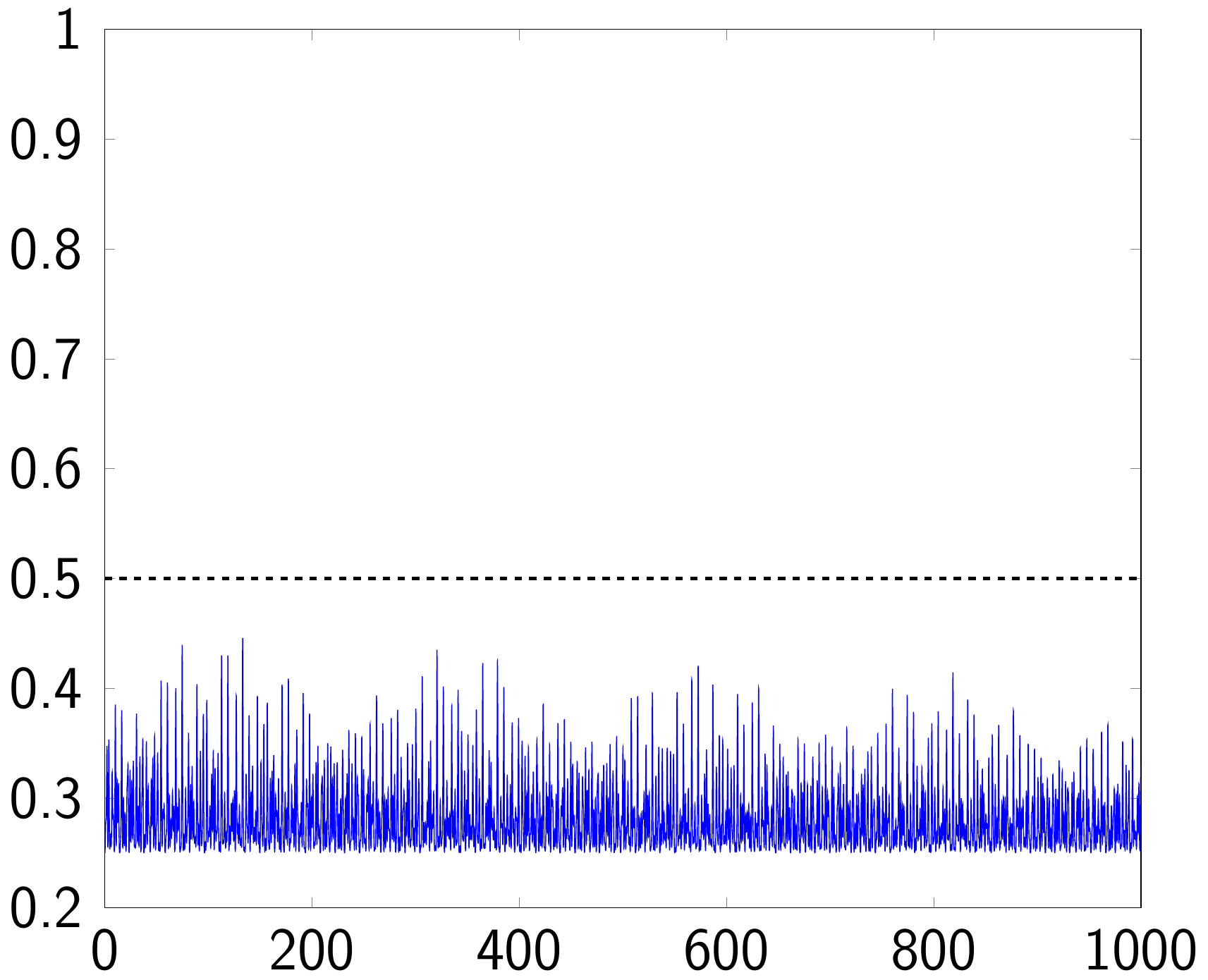} & \vspace{0.2cm}
		\includegraphics[ width=\linewidth, height=\linewidth, keepaspectratio]{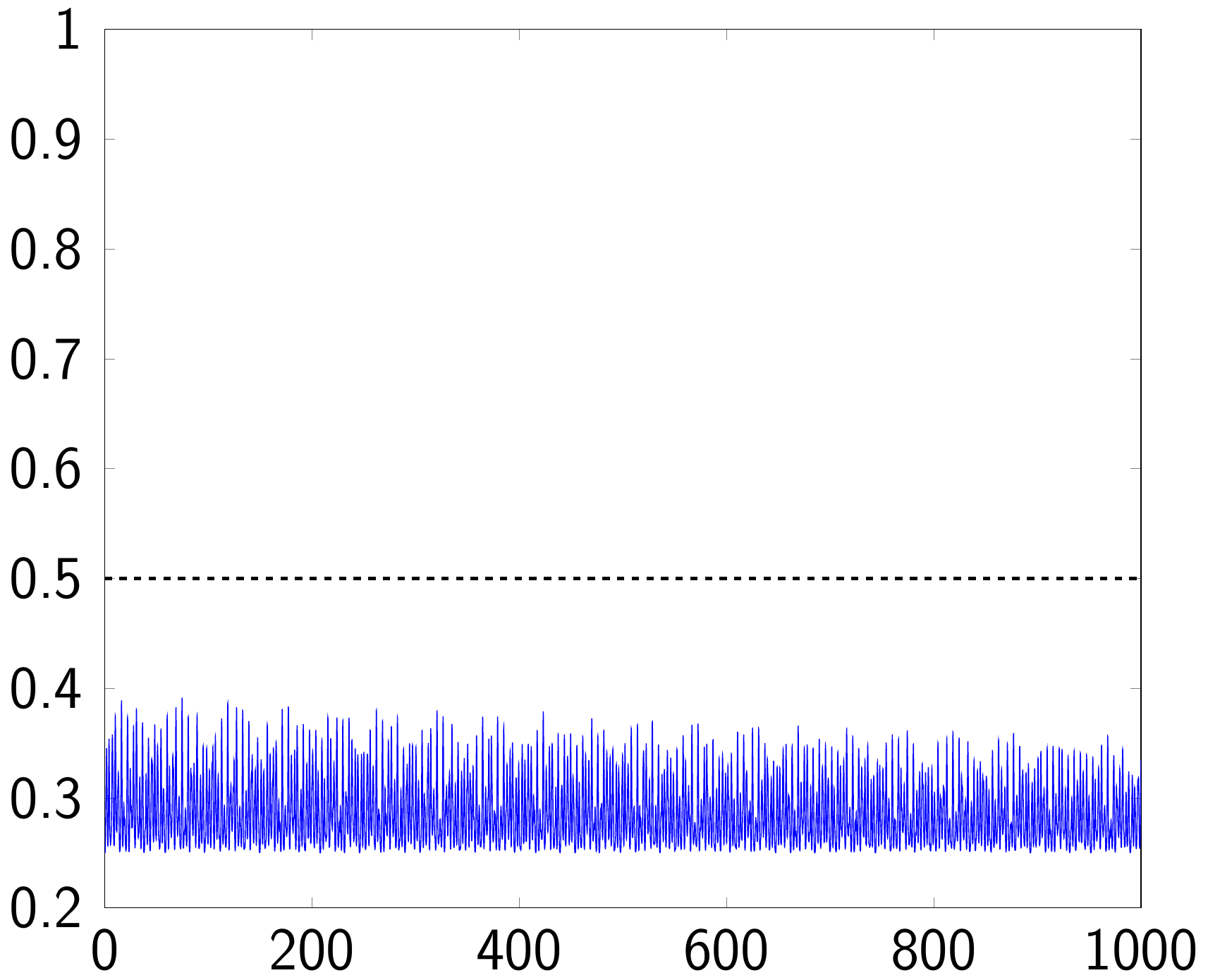} & \vspace{0.2cm}
		\includegraphics[ width=\linewidth, height=\linewidth, keepaspectratio]{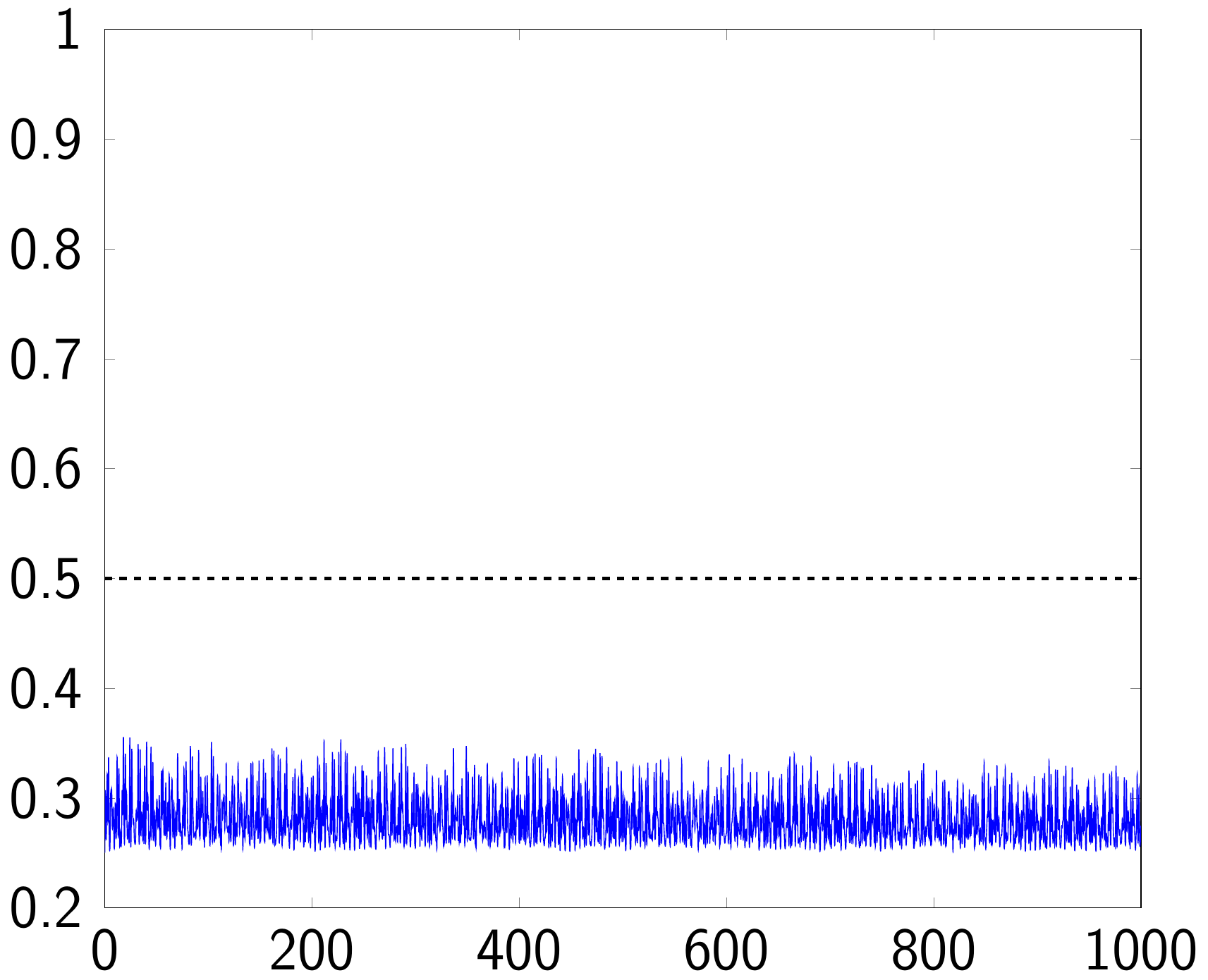} & \vspace{0.2cm}
		\includegraphics[ width=\linewidth, height=\linewidth, keepaspectratio]{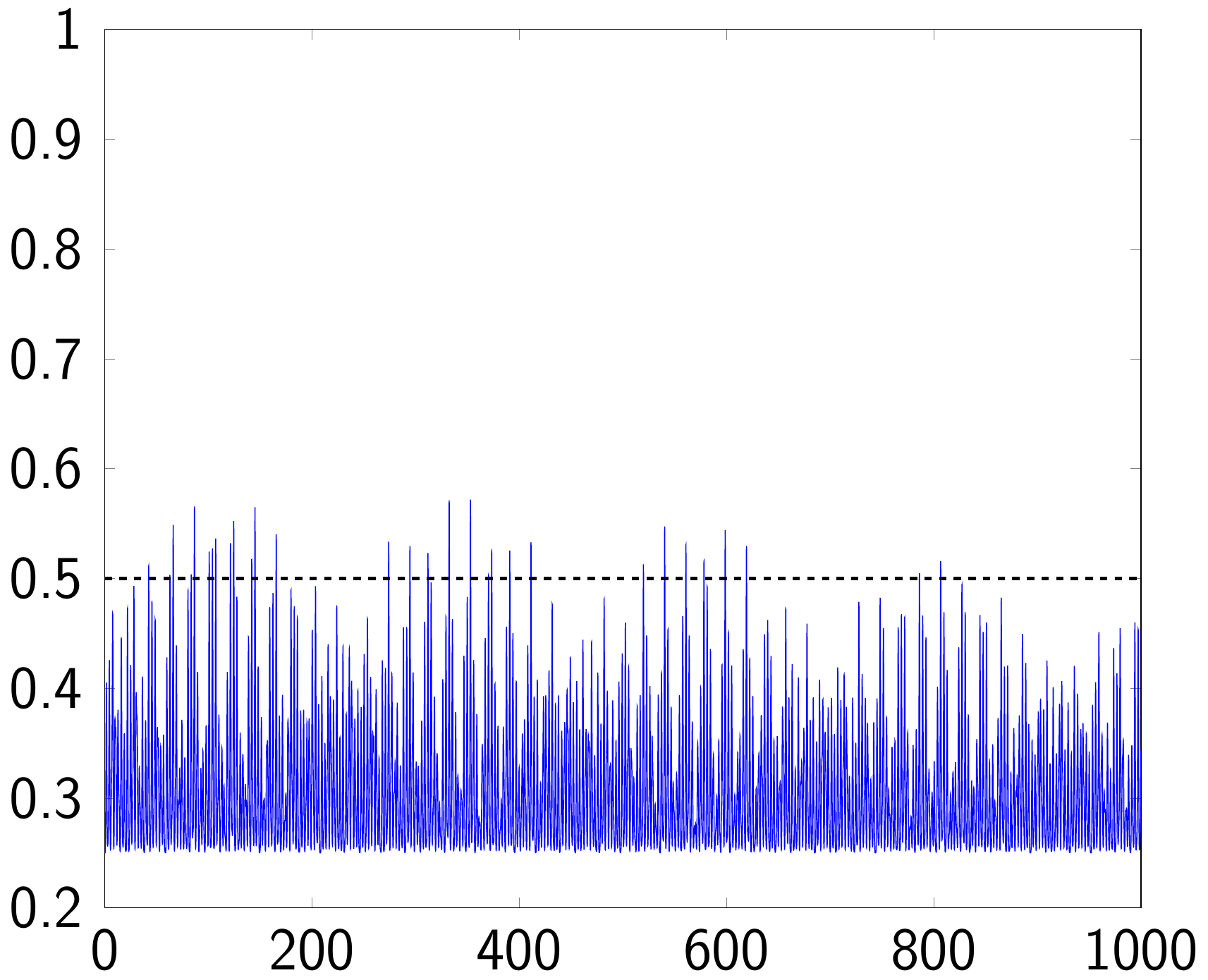} & \vspace{0.2cm}
		\includegraphics[ width=\linewidth, height=\linewidth, keepaspectratio]{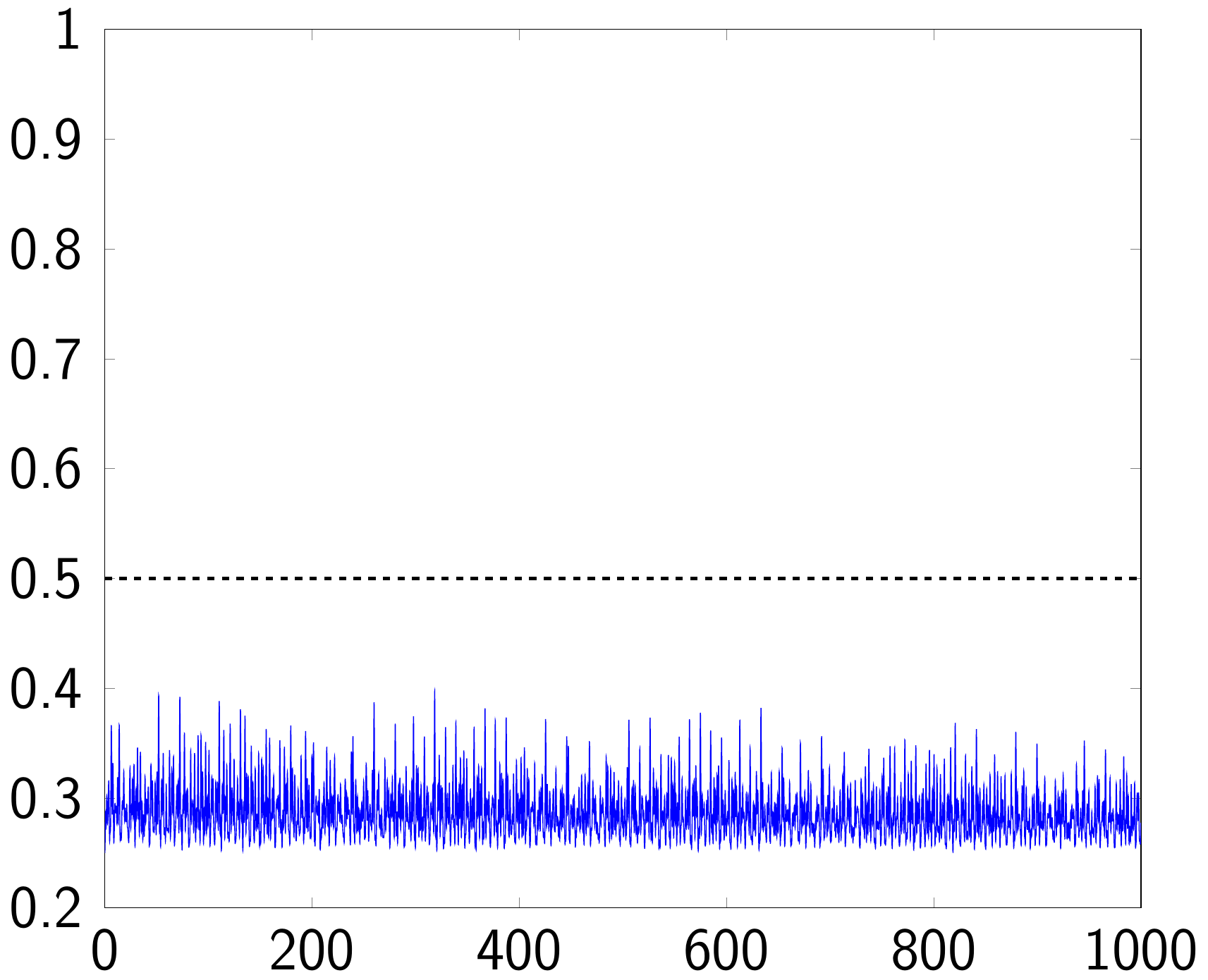} & \vspace{0.2cm}
		\includegraphics[ width=\linewidth, height=\linewidth, keepaspectratio]{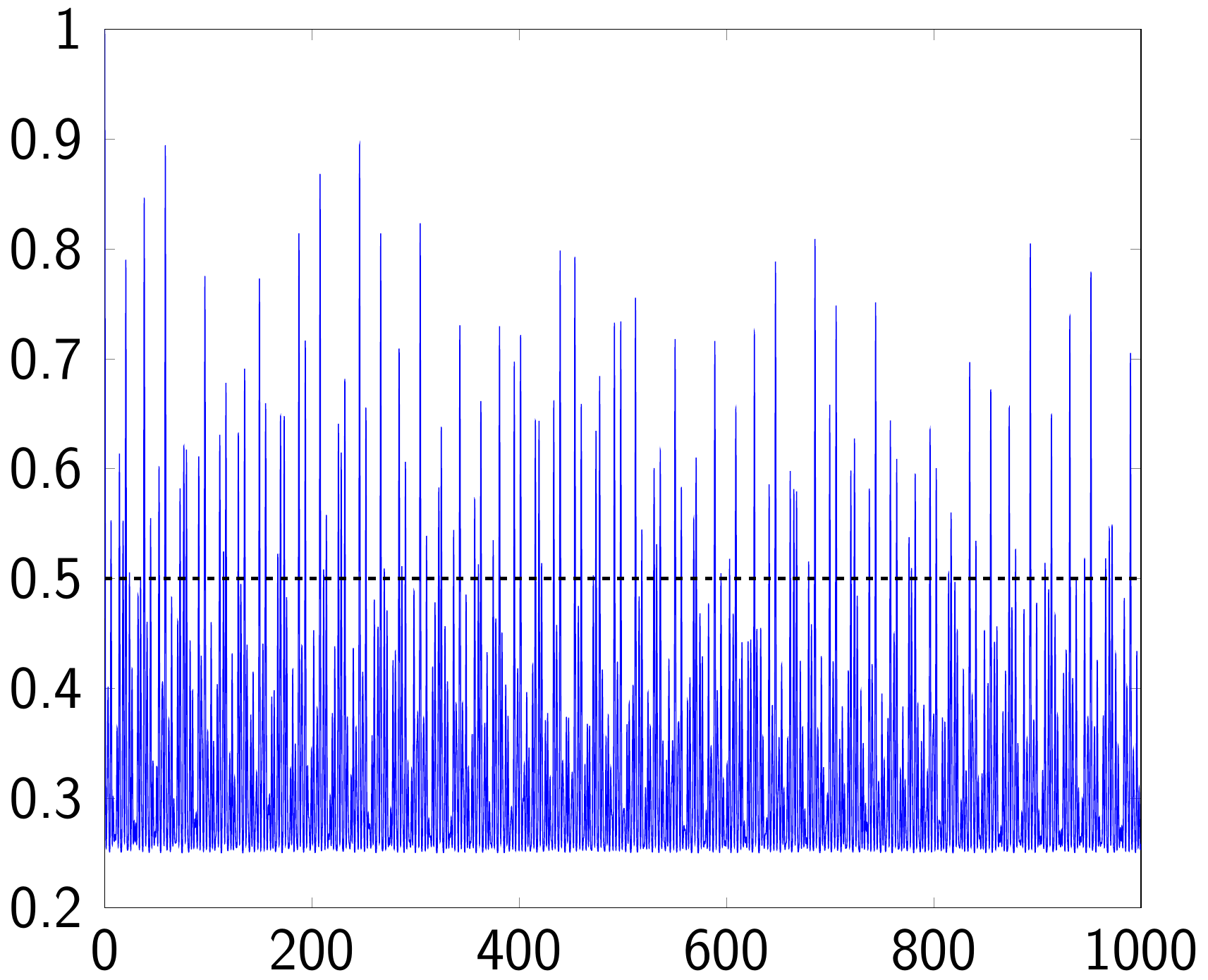} \\
		\hline
\end{tabularx}
\caption{Transfer of entanglement between nuclear spin pairs in a tricalcium biphosphate dimer with a \ce{C2} point-group symmetry. The $4$ nuclear spins in each dimer are indexed from $0$ through $3$. The notation $ij$ corresponds to spin pair comprising of the $i^{th}$ spin in the first dimer molecule, and the $j^{th}$ spin in the second dimer molecule. Each row represents the spin pair that was initialized in the singlet state (all other spins were left uncorrelated). The columns represent the spin pair for which the singlet probability is being calculated. The \textit{y} axis (limits from $0.2$ to $1$) in each plot represents the singlet probability, and the \textit{x} axis denotes time (a period of $1000$ seconds is covered).}
\label{tab:transfer_coherence_dimer}
\end{figure}

As discussed in the main text, the dimer is expected to exist as a stable, independent entity without coalescing into Posner molecules, and, thus, could potentially be a fascinating case study  of nuclear spin entanglement in pairs of molecules. It is also interesting to study whether coherence is transferred from one \ce{^{31}P} nuclear spin pair to another as the singlet state evolves over time. Below, we show this for all nuclear spin pairs for the dimer structure with a \ce{C2} point-group symmetry (see Fig.~\ref{tab:transfer_coherence_dimer}). Even though the singlet state is long-lived for each individual pair of \ce{^{31}P} nuclear spins, we see that, in some cases, coherence is indeed transferred to other spins pairs, suggesting that more than one nuclear spin pair might have long entanglement lifetimes. We expect a similar behavior for all other dimer structures. We provide an explanation for the exceptional longevity of the singlet state in the dimer, over the Posner molecule, in the following section (see section \ref{sec:effect_of_nP}).


\newpage

\section[{The effect of the number of coupled nuclear spins on the spin dynamics of a system}]{The effect of the number of coupled nuclear spins on the spin dynamics of a system}
\label{sec:effect_of_nP}

Let there be $n$ distinct spin energy levels in a molecule. Then, $n = 2^{n_\text{P}}$ in a molecule with no symmetry, with $n_\text{P}$ being the number of \ce{^{31}P} atoms in the molecule. To get the total number of unique, positive frequencies contributing to coherences in one molecule, we count the number of pairs of energy levels, i.e. $\sum_{i=1}^{n-1}i=\frac{n(n-1)}{2}=N$.
    
Every frequency that could contribute to disentanglement of the singlet state by destructive interference in the singlet probability oscillations, is a result of a unique pair of frequencies from the above $N$ frequencies across both molecules. Adding them up, we get:
\begin{align}
z &= \frac{1}{2}(N(N-1)) + N + N\,. \nonumber
\end{align}
Here, $z$ is the total number of oscillation frequencies potentially contributing to the disentanglement of the spin singlet state. The first term on the right-hand side corresponds to combinations of unique pairs of different frequencies from the two molecules, the second term corresponds to oscillations within each molecule independent of the other, i.e.\ , pairing up frequencies in one molecule to zero frequencies in the other molecule, and the last term corresponds to the combination of identical frequencies from both molecules. While symmetry and accidental degeneracies associated with specific choices of coupling constants will inevitably introduce redundant frequencies close to one another, and some others close to zero, $z$ provides an upper limit to the number of oscillation frequencies. Written in terms of the number of distinct spin energy levels $n$, we have
\begin{align}
z = \frac{1}{8}(n^4 - 2n^3 + 7n^2 - 6n)\,. \nonumber
\end{align}

Below, we plot the entanglement yield for systems with different numbers of \ce{^{31}P} atoms. The entanglement yield has been calculated as $k\int_{0}^{T}\text{max}(p_{s}(t)-0.5,0) e^{-kt}dt$. Here, $k=1/300$ $\text{s}^{-1}$ is related to the sampling duration, $T=1000$s is the total duration over which $p_s(t)$ was evaluated, and $p_s(t)$ is the singlet probability over time. Note that these have been calculated in the absence of spin relaxation. Including spin relaxation would have required the structural information for each system, but we do not anticipate the trend to change appreciably even after its inclusion. We performed $50$ simulations for each system size, and the \textit{J}-coupling constants for each of the above simulation were chosen randomly from a normal distribution such that $\sqrt{\sum_{i=1}^{n_\text{P}}J_{ij}^{2}}=1$ Hz, where $n_\text{P}$ is the number of \ce{^{31}P} atoms.

\begin{figure}[h!]
      \centering
      \includegraphics[width=0.45\textwidth]{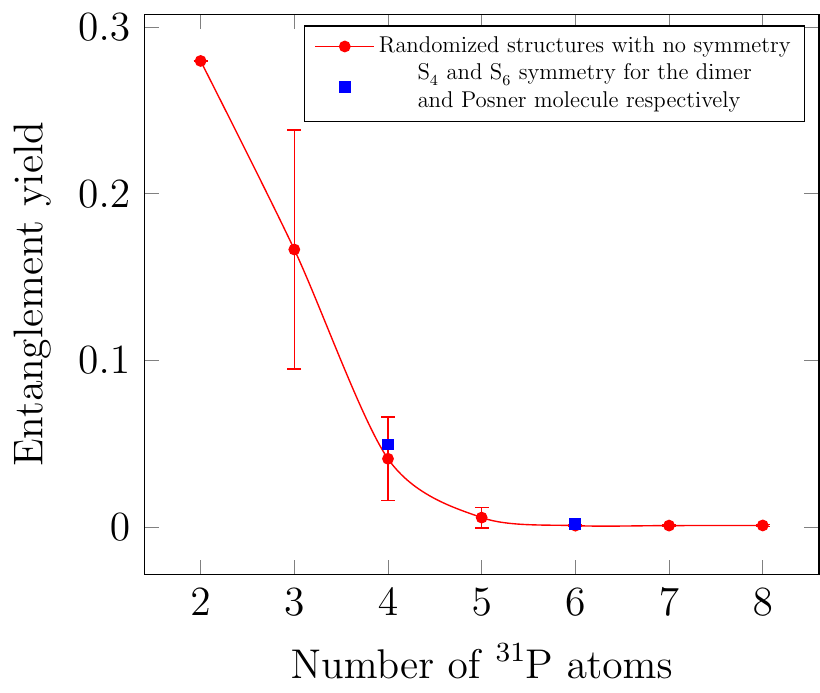}
      \caption{The entanglement yield monotonically decreases with the number of coupled nuclear spins. Note that the error bars and the data on the \ce{S4} and \ce{S6} -- symmetric dimer and Posner structures respectively show that for individual structures, a different set of coupling constants result in different yields for a given number of \ce{^{31}P} atoms.}
      \label{fig:singlet_entanglement_yield}
\end{figure}

Clearly, we see that a system with lesser number of coupled nuclear spins is much more adept at maintaining long-lived singlet states than a system with a larger number of coupled nuclear spins. This is in line with our results and our argument above. We reiterate that the accurate calculation of \textit{J}-coupling constants remains crucial because the above discussion only suggests that a smaller coupled nuclear spin system would be better at maintaining long-lived singlet states. To calculate the longevity of the singlet state, we still need accurate values of the coupling constants for a given structural configuration of the molecule.

Additionally, from Fig.\ \ref{fig:singlet_entanglement_yield}, it is also evident that, on average, symmetric configurations (represented as blue dots on the plot) are expected to have better entanglement yields. For the same structure, it is possible to move across the error bar by simply altering the coupling constants. Since the values of the coupling constants are linked to the symmetry of the molecule, the above observation confirms the importance of these features when studying the spin dynamics of a given system with a dynamic ensemble of structures.

\newpage

\section[Transfer of nuclear spin coherence in a pair of Posner molecules]{Transfer of coherence between \texorpdfstring{\ce{^{31}P}{ n}} nuclear spins in a pair of Posner molecules}

\begin{figure}
\begin{subfigure}[b]{0.3\textwidth}
\setlength\tabcolsep{2pt}
\centering
\raisebox{4.65cm}{
\begin{tabularx}{3cm}{|@{}c*{1}{|C|}@{}}
		\hline
		& $\textbf{00}$\\
		\hline 
		$\textbf{00}$ & \vspace{0.2cm}
		\includegraphics[ width=1.5cm, height=1.5cm, keepaspectratio]{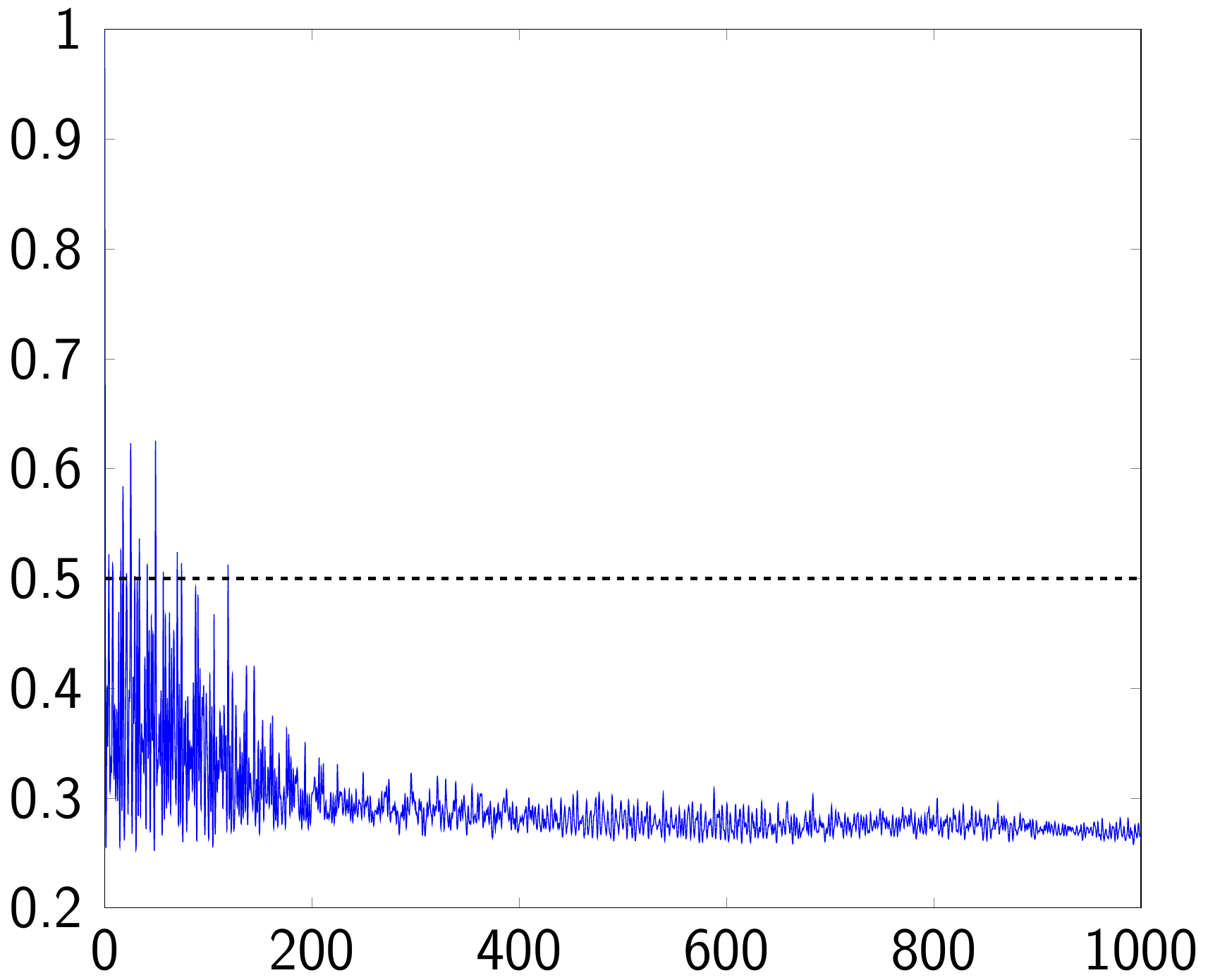} \\
		\hline
		$\textbf{01}$ & \vspace{0.2cm}
		\includegraphics[ width=1.5cm, height=1.5cm, keepaspectratio]{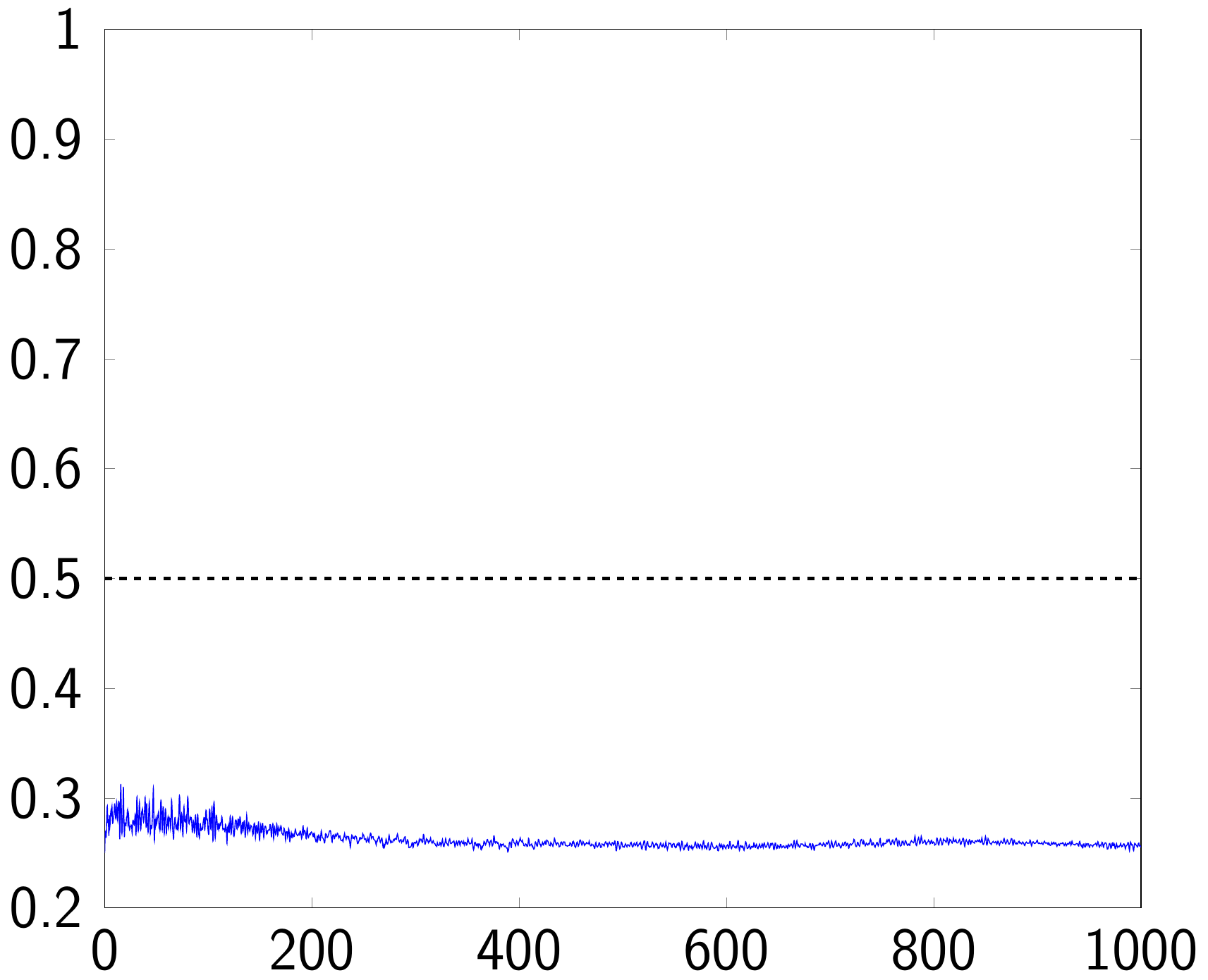} \\
		\hline
		$\textbf{02}$ & \vspace{0.2cm}
		\includegraphics[ width=1.5cm, height=1.5cm, keepaspectratio]{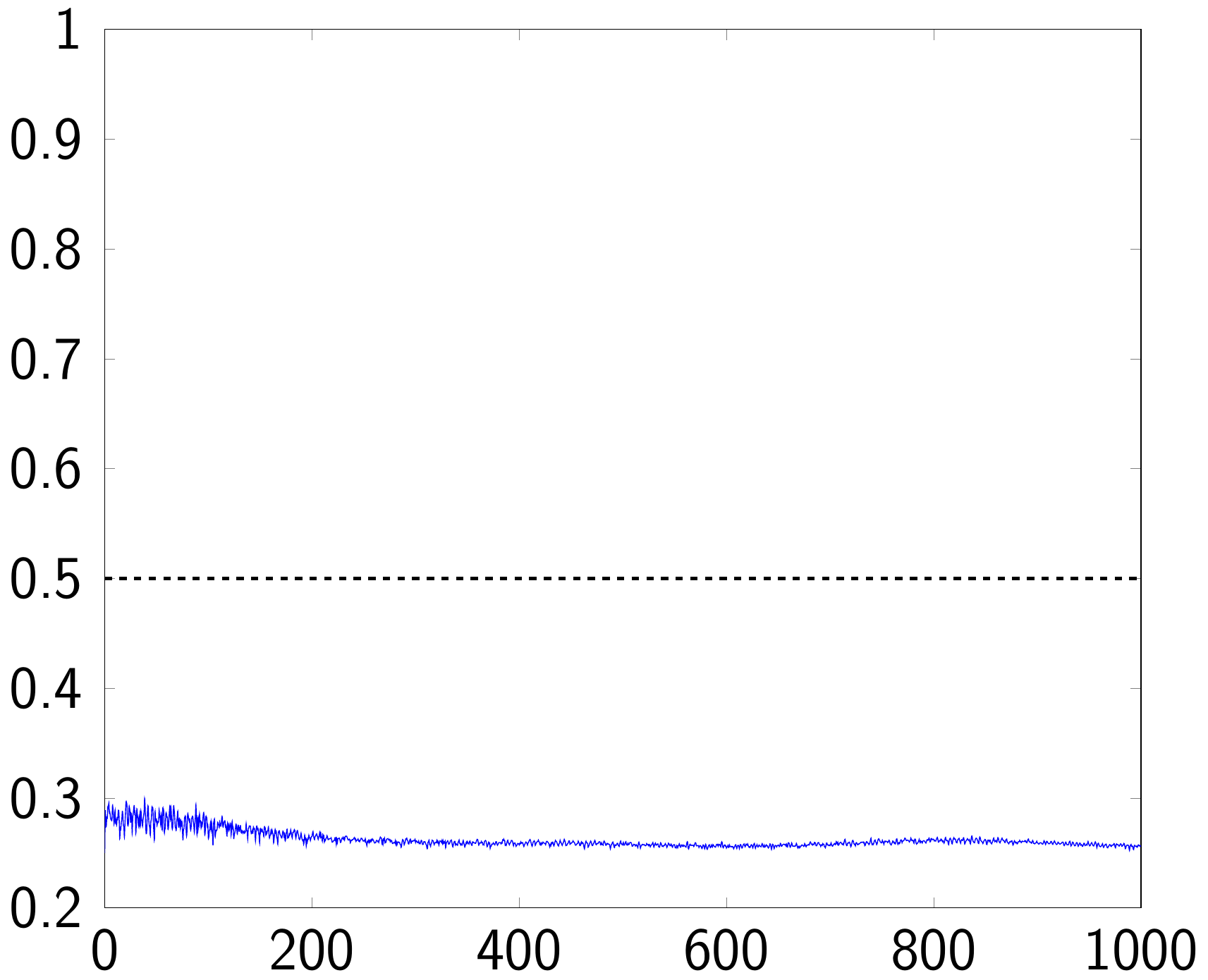} \\
		\hline
		$\textbf{03}$ & \vspace{0.2cm}
		\includegraphics[ width=1.5cm, height=1.5cm, keepaspectratio]{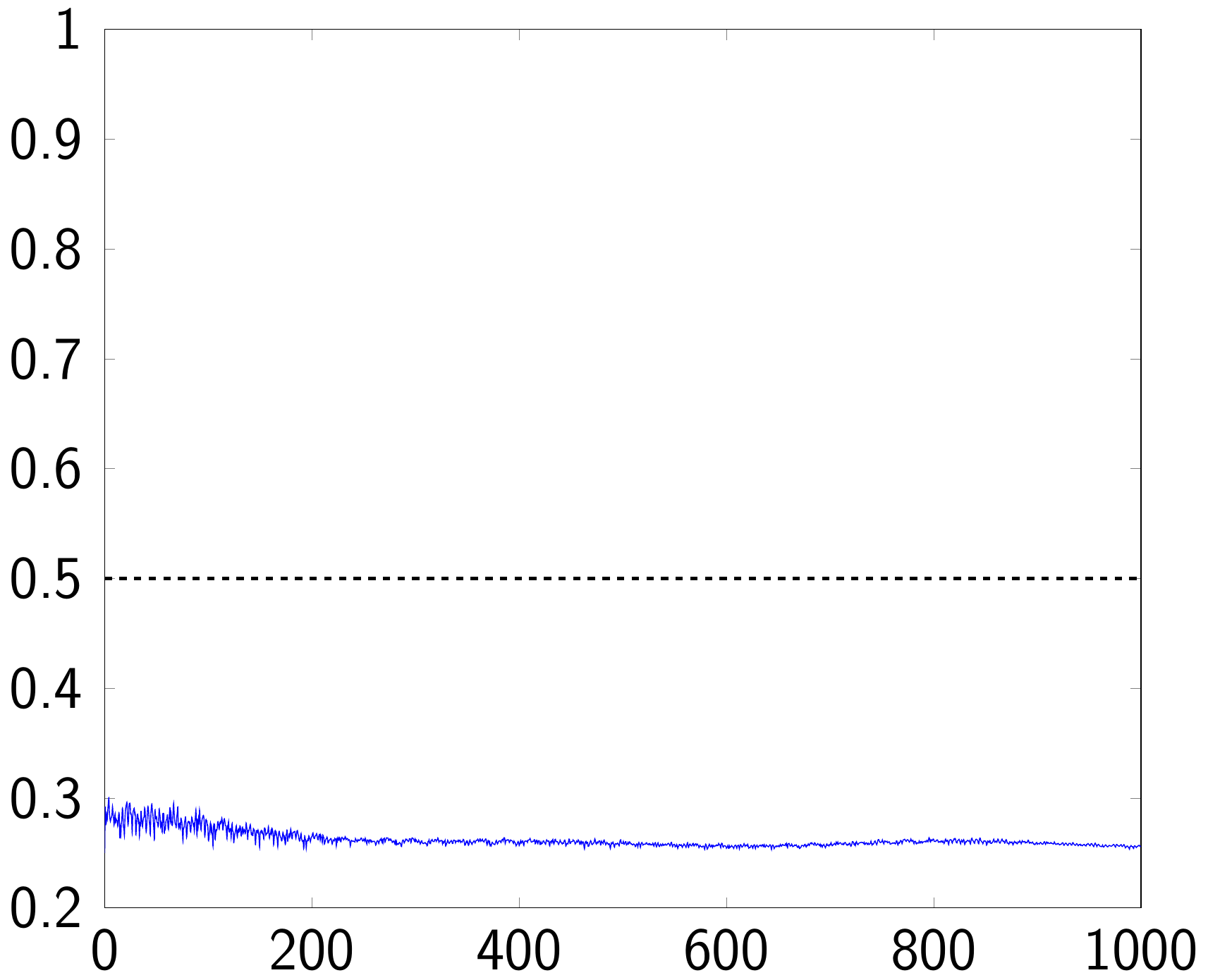} \\
		\hline
		$\textbf{04}$ & \vspace{0.2cm}
		\includegraphics[ width=1.5cm, height=1.5cm, keepaspectratio]{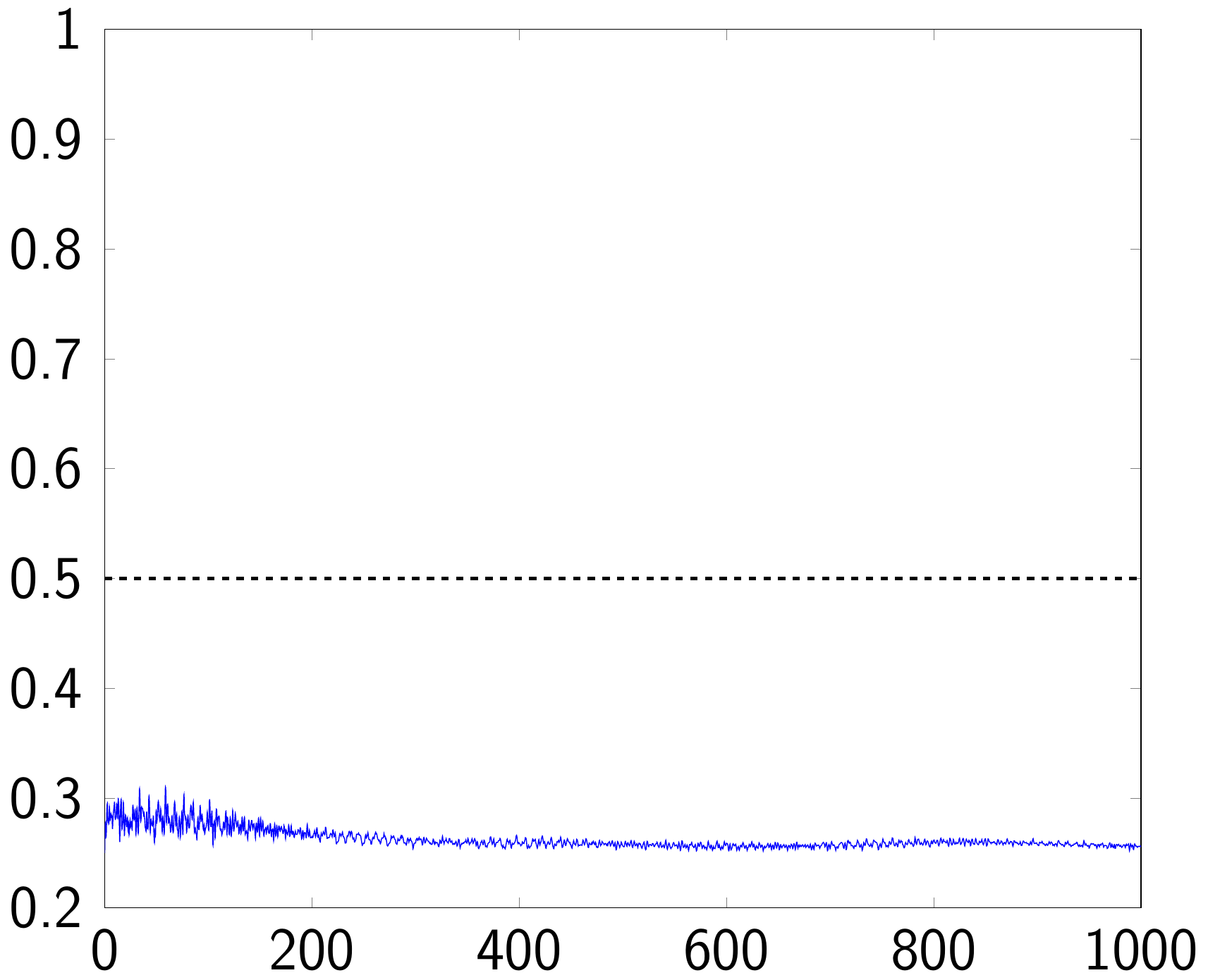} \\
		\hline
		$\textbf{05}$ & \vspace{0.2cm}
		\includegraphics[ width=1.5cm, height=1.5cm, keepaspectratio]{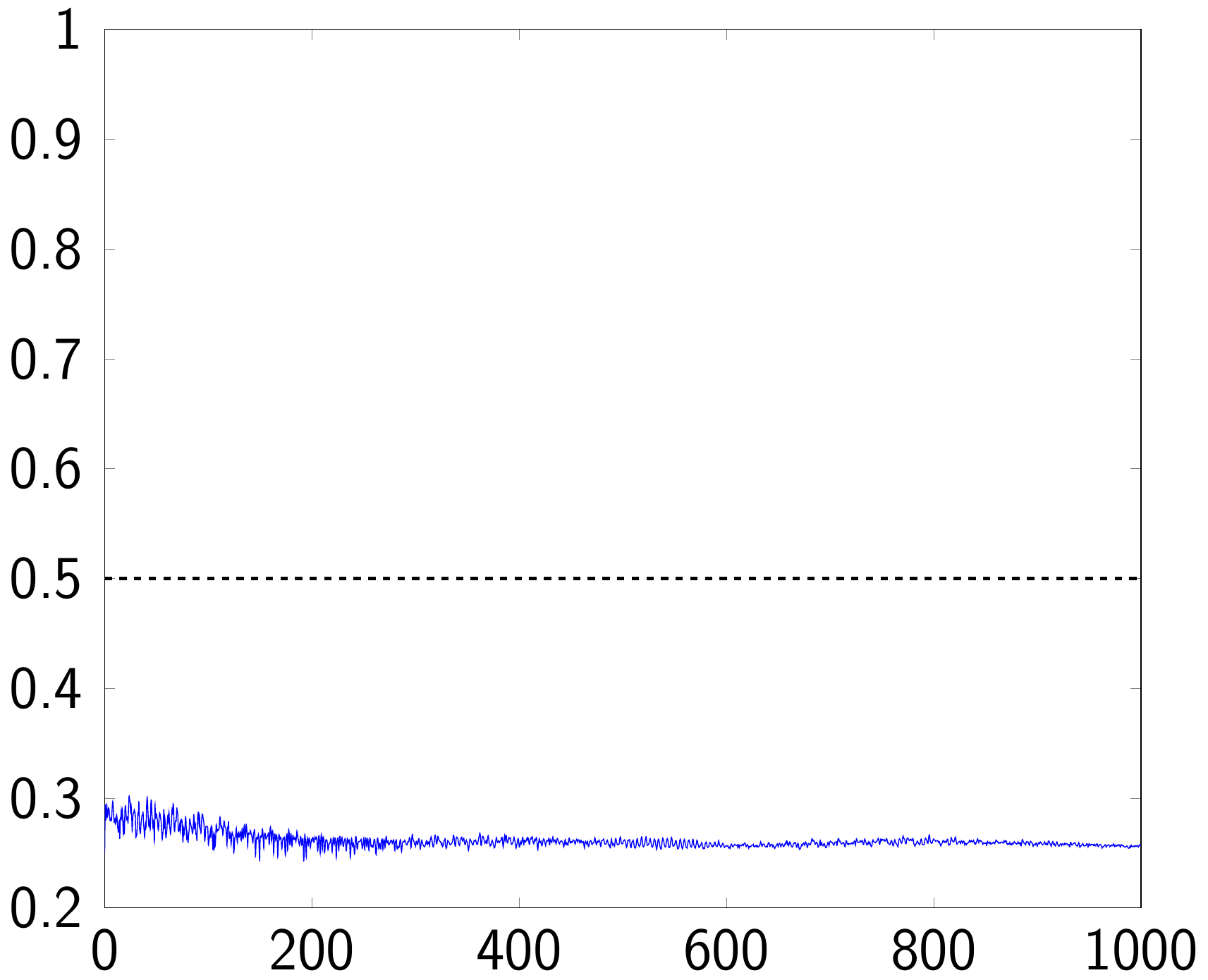} \\
		\hline
		$\textbf{11}$ & \vspace{0.2cm}
		\includegraphics[ width=1.5cm, height=1.5cm, keepaspectratio]{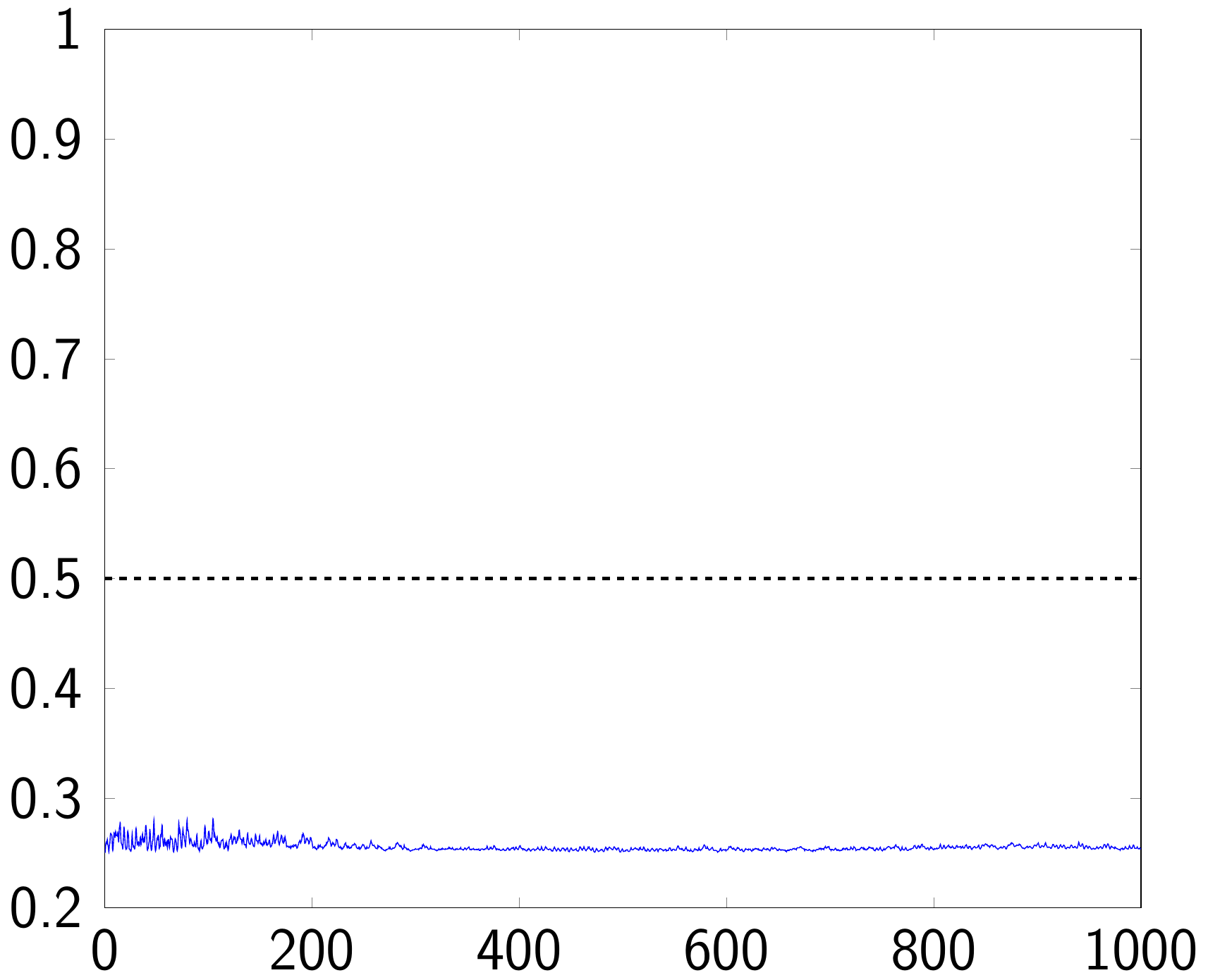} \\
		\hline
		$\textbf{12}$ & \vspace{0.2cm}
		\includegraphics[ width=1.5cm, height=1.5cm, keepaspectratio]{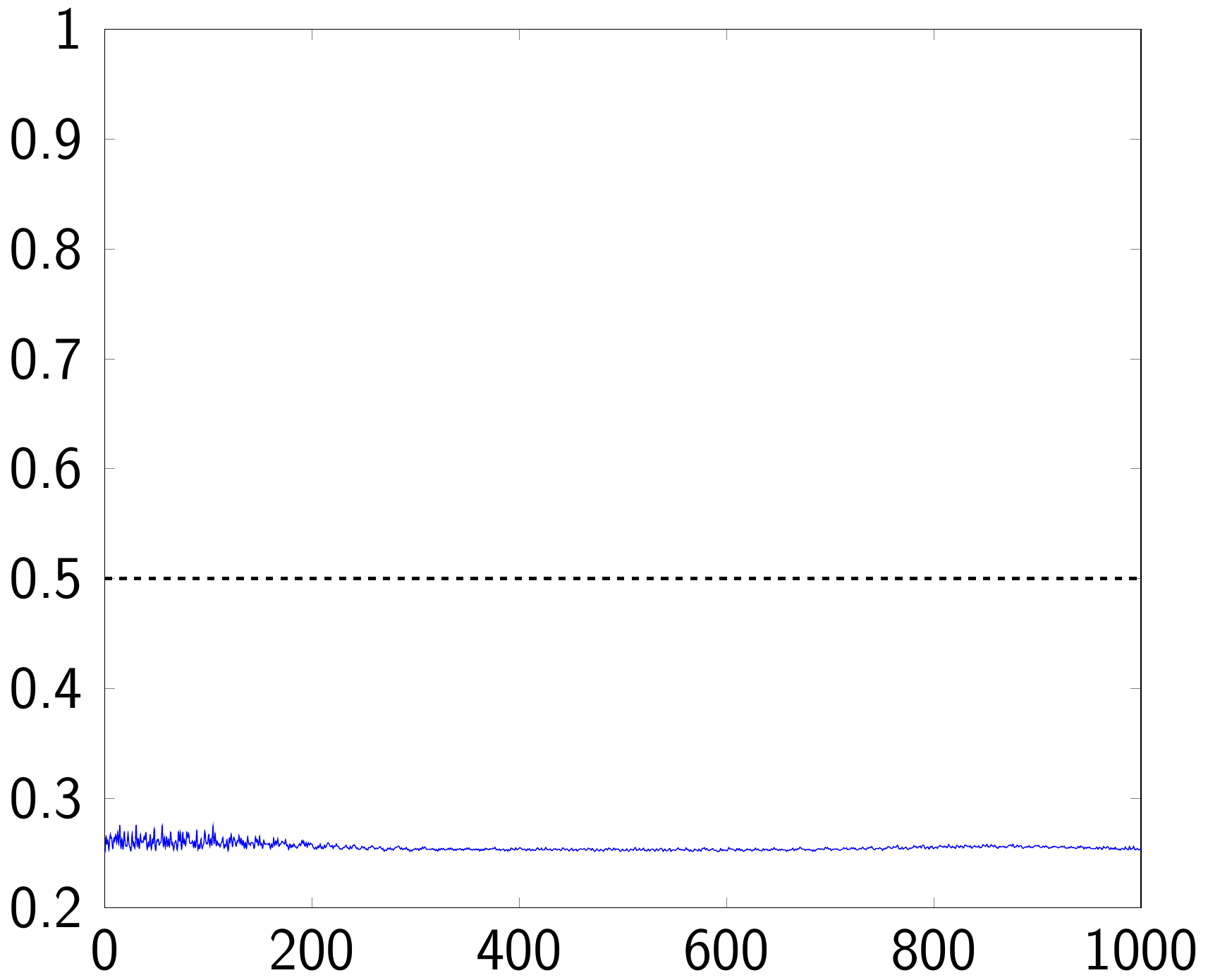} \\
		\hline
		$\textbf{13}$ & \vspace{0.2cm}
		\includegraphics[ width=1.5cm, height=1.5cm, keepaspectratio]{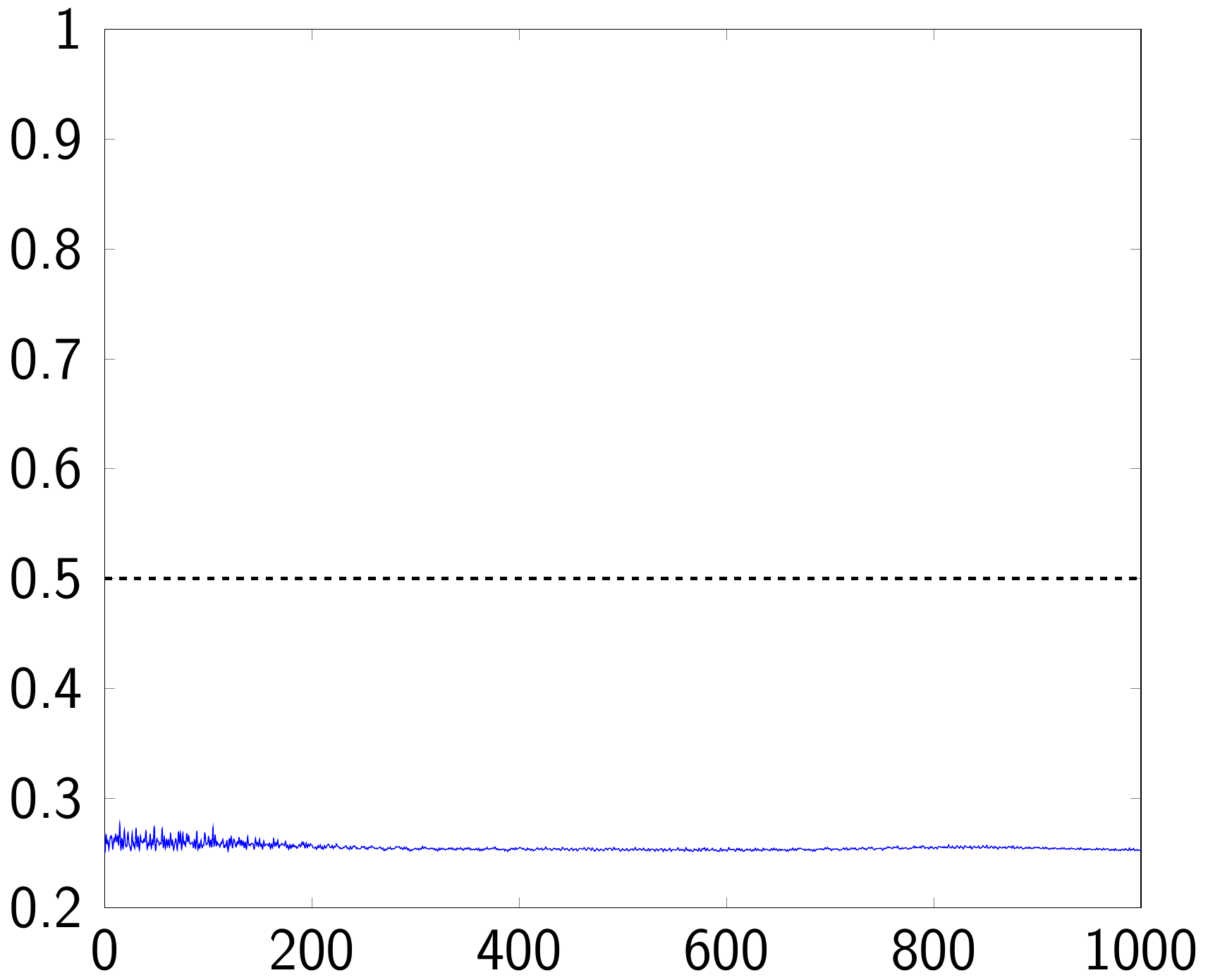} \\
		\hline
		$\textbf{14}$ & \vspace{0.2cm}
		\includegraphics[ width=1.5cm, height=1.5cm, keepaspectratio]{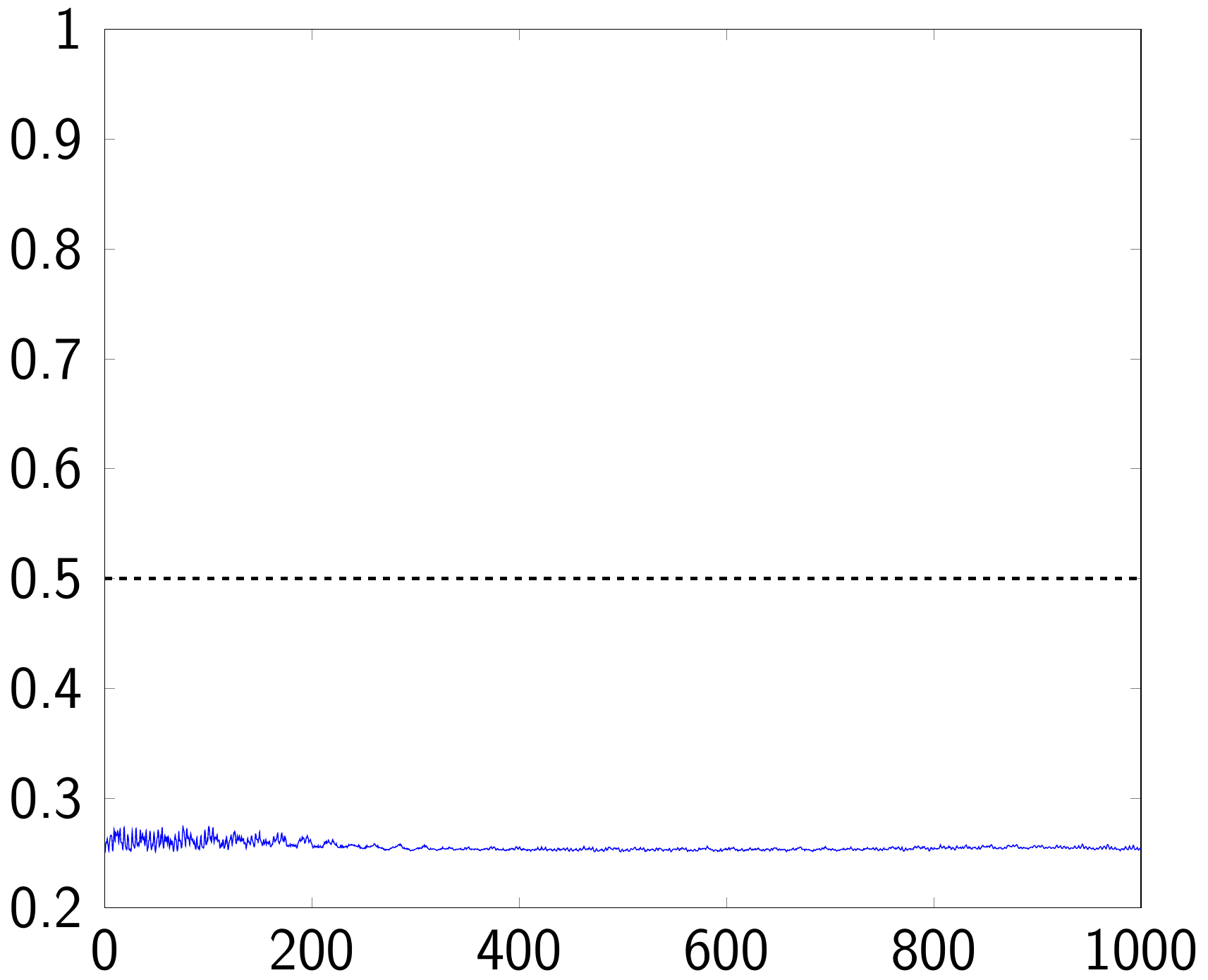} \\
		\hline
		$\textbf{15}$ & \vspace{0.2cm}
		\includegraphics[ width=1.5cm, height=1.5cm, keepaspectratio]{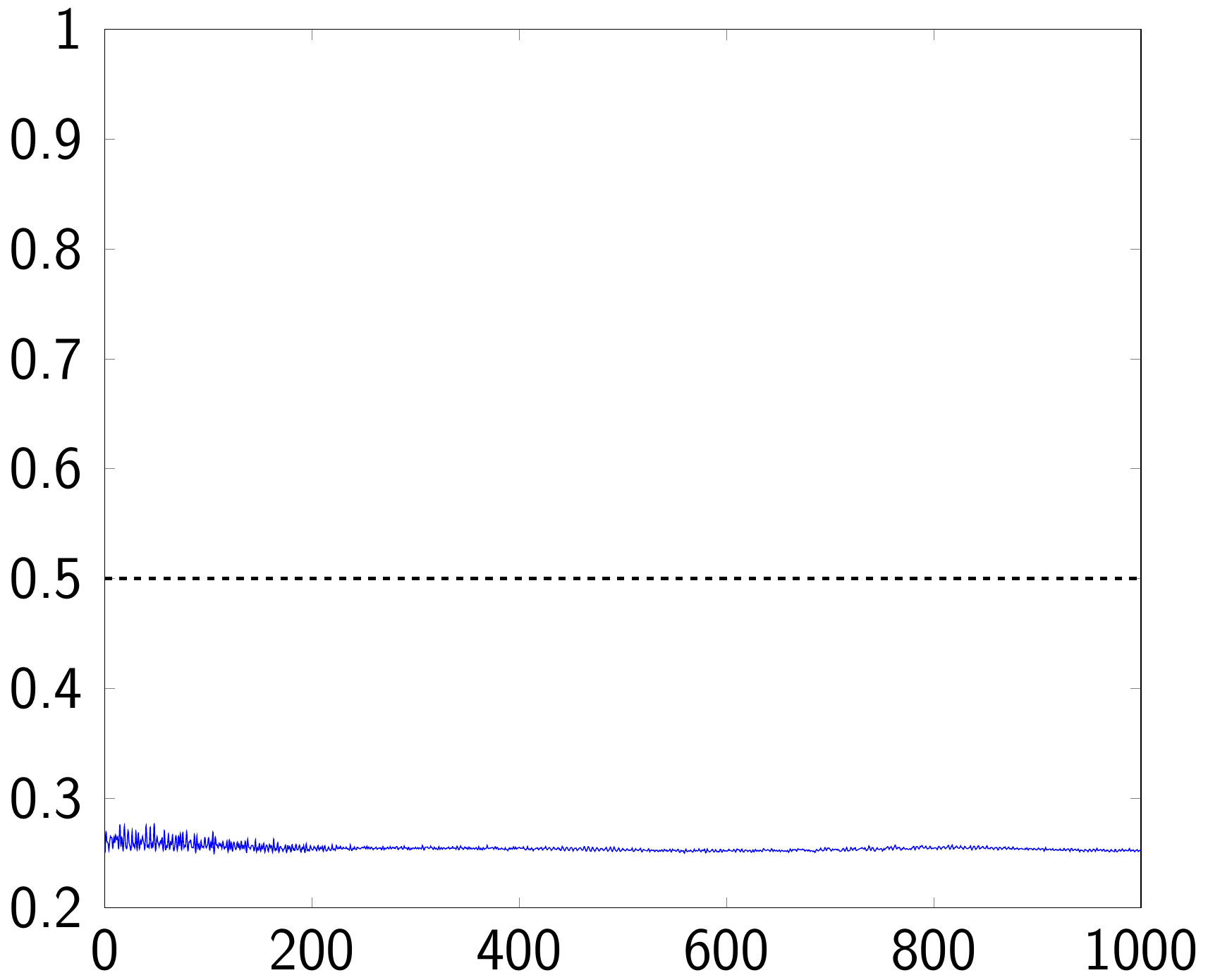} \\
		\hline
\end{tabularx}}
\end{subfigure}
\begin{subfigure}[b]{0.3\textwidth}
\setlength\tabcolsep{2pt}
\centering
\raisebox{5.5cm}{
\begin{tabularx}{3cm}{|@{}c*{1}{|C|}@{}}
		\hline
		& $\textbf{00}$\\
		\hline 
		$\textbf{22}$ & \vspace{0.2cm}
		\includegraphics[ width=1.5cm, height=1.5cm, keepaspectratio]{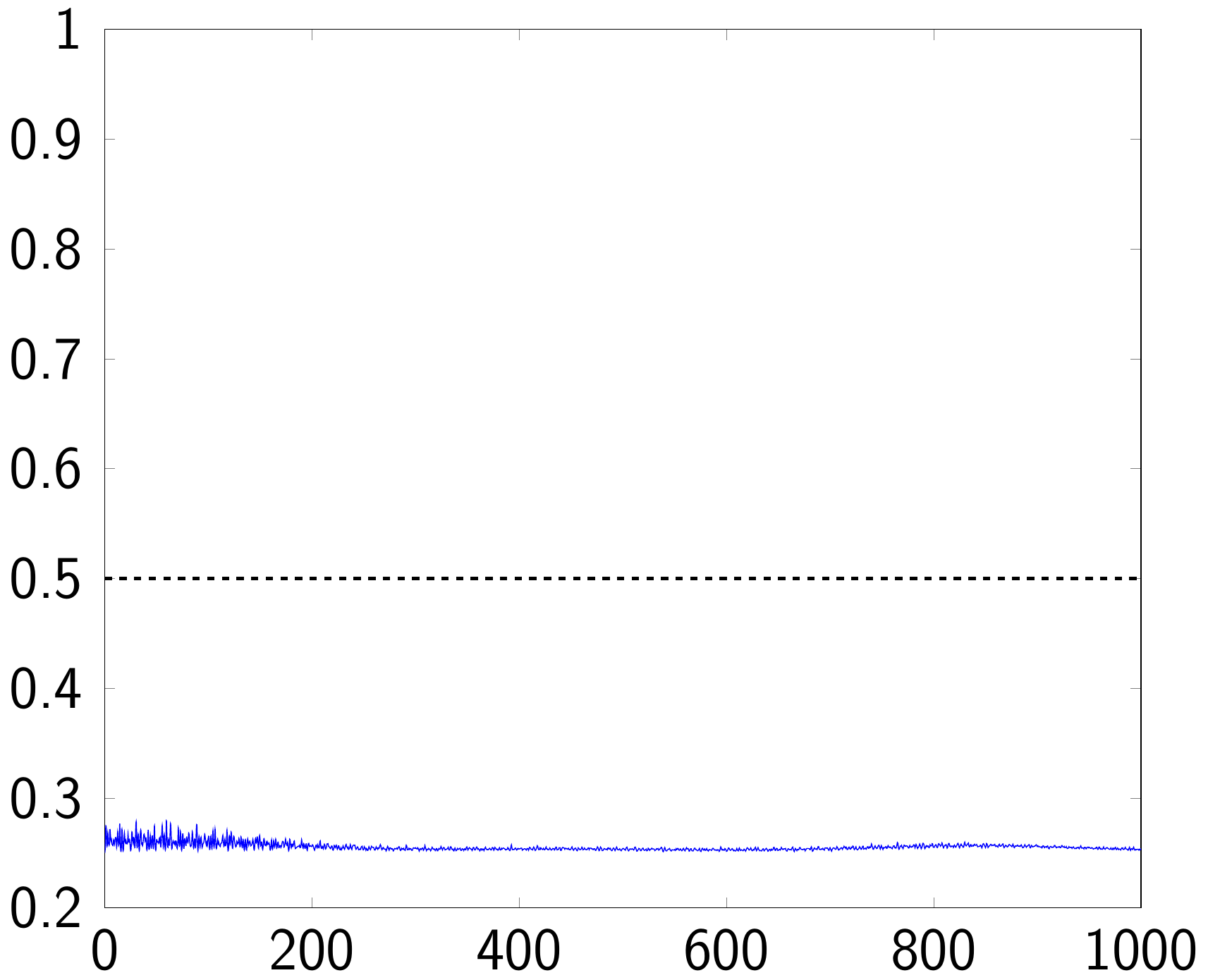} \\
		\hline
		$\textbf{23}$ & \vspace{0.2cm}
		\includegraphics[ width=1.5cm, height=1.5cm, keepaspectratio]{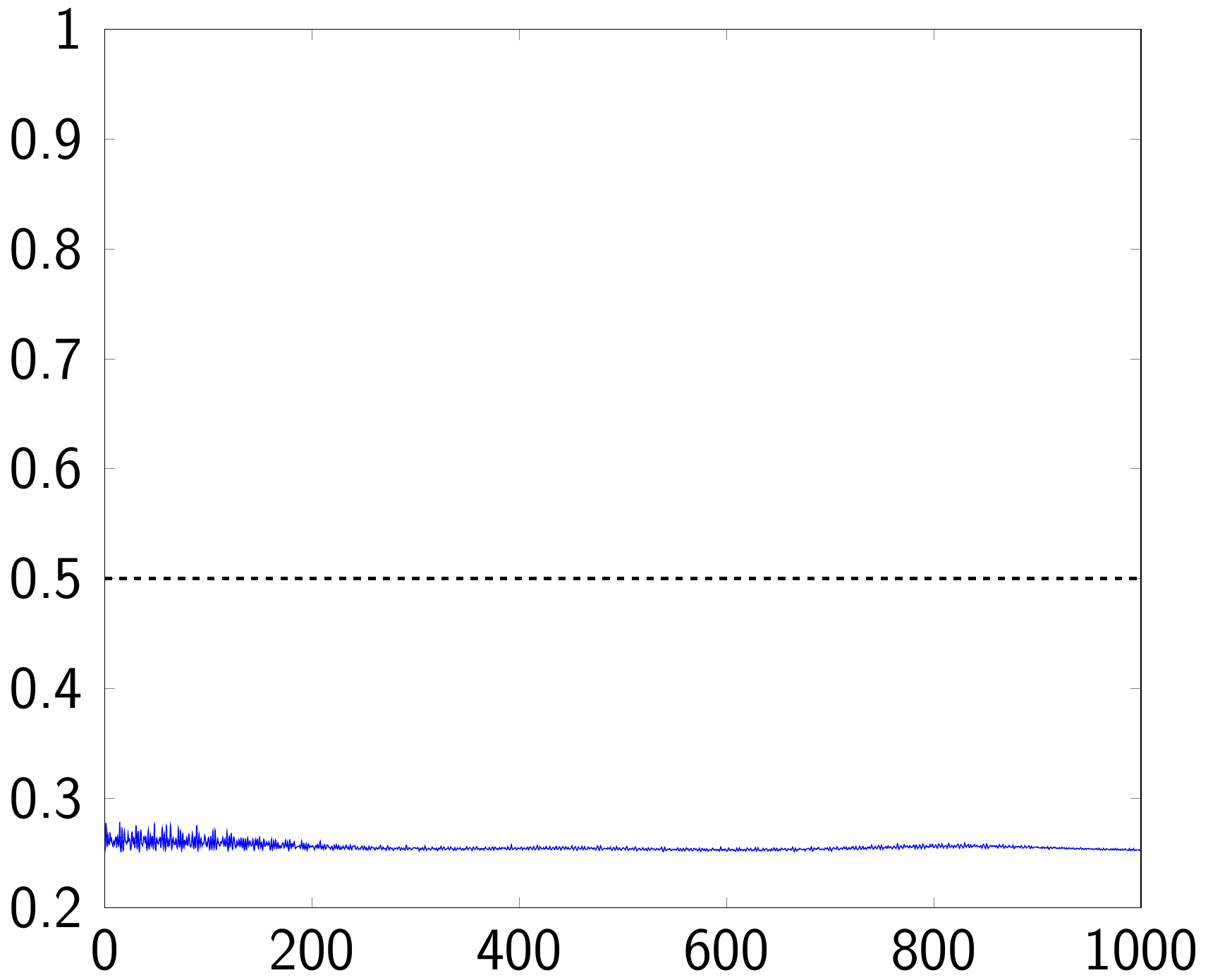} \\
		\hline
		$\textbf{24}$ & \vspace{0.2cm}
		\includegraphics[ width=1.5cm, height=1.5cm, keepaspectratio]{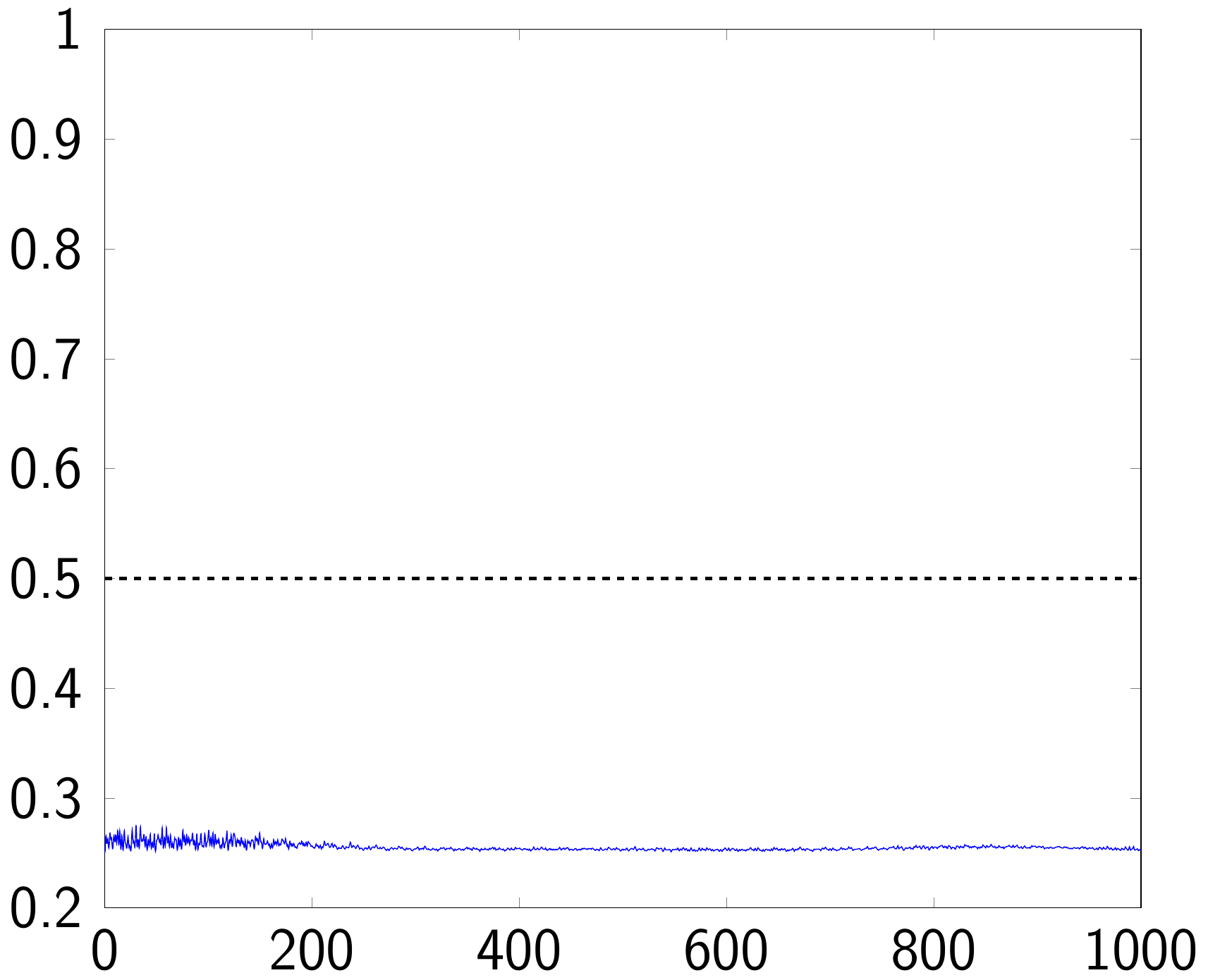} \\
		\hline
		$\textbf{25}$ & \vspace{0.2cm}
		\includegraphics[ width=1.5cm, height=1.5cm, keepaspectratio]{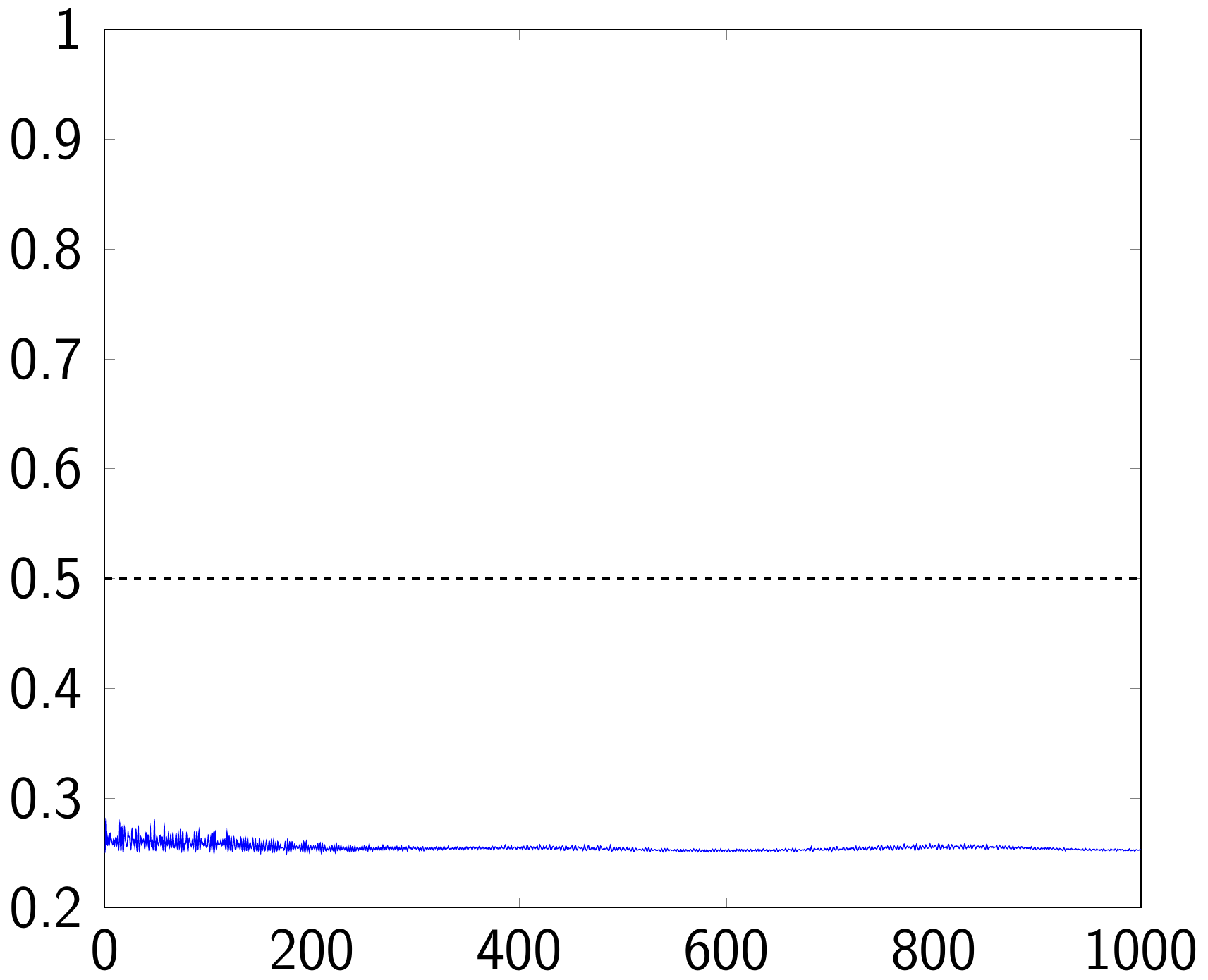} \\
		\hline
		$\textbf{33}$ & \vspace{0.2cm}
		\includegraphics[ width=1.5cm, height=1.5cm, keepaspectratio]{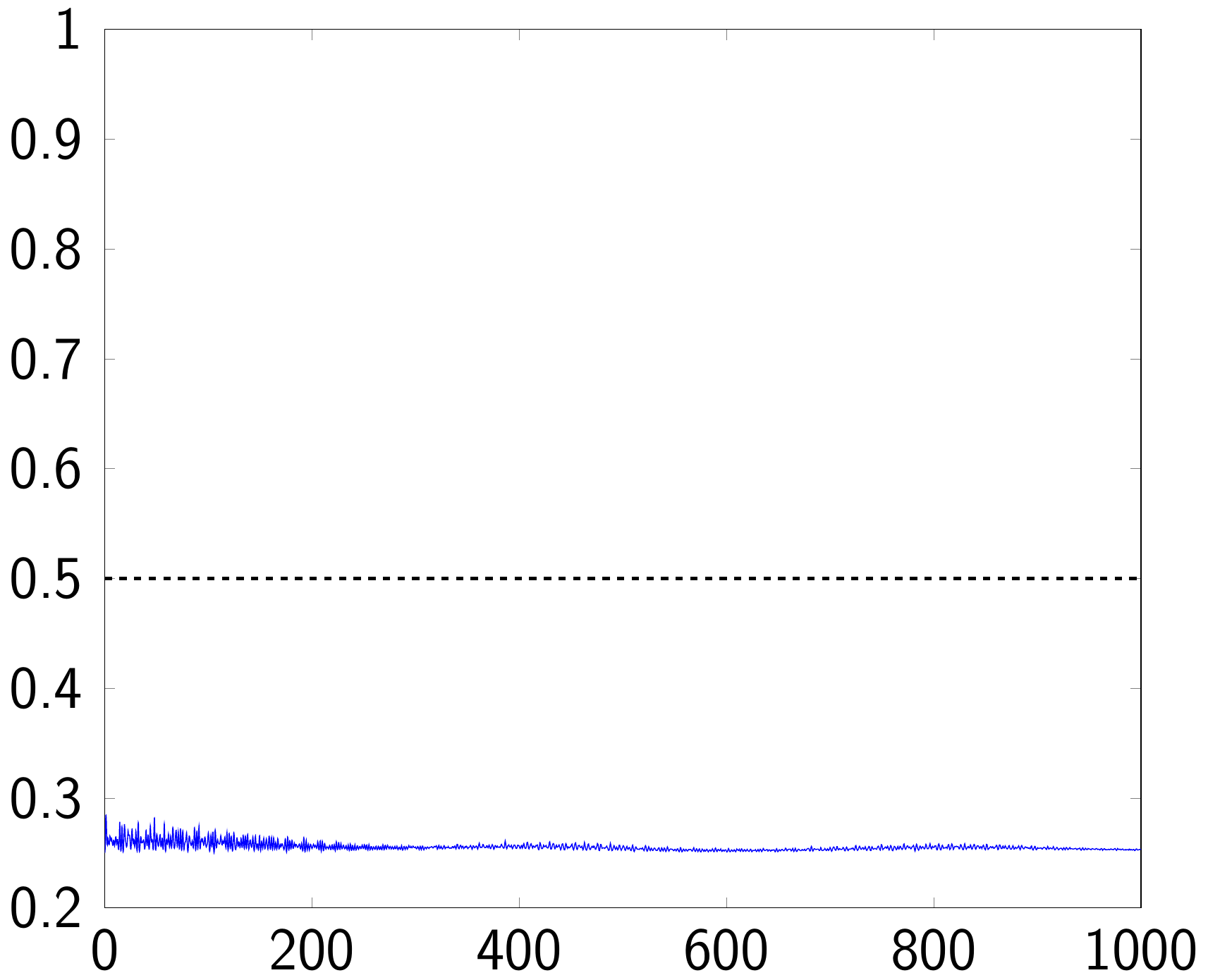} \\
		\hline
		$\textbf{34}$ & \vspace{0.2cm}
		\includegraphics[ width=1.5cm, height=1.5cm, keepaspectratio]{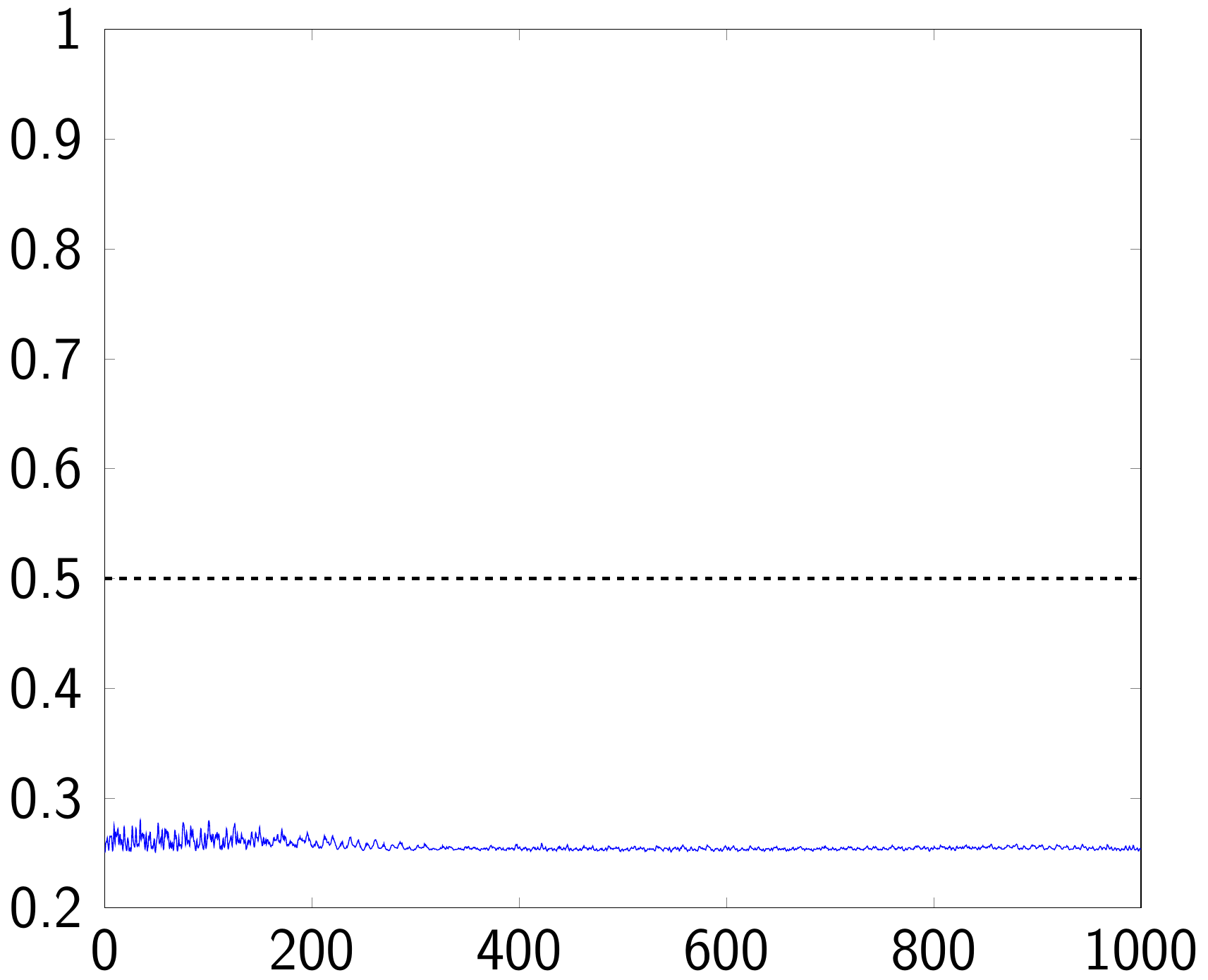} \\
		\hline
		$\textbf{35}$ & \vspace{0.2cm}
		\includegraphics[ width=1.5cm, height=1.5cm, keepaspectratio]{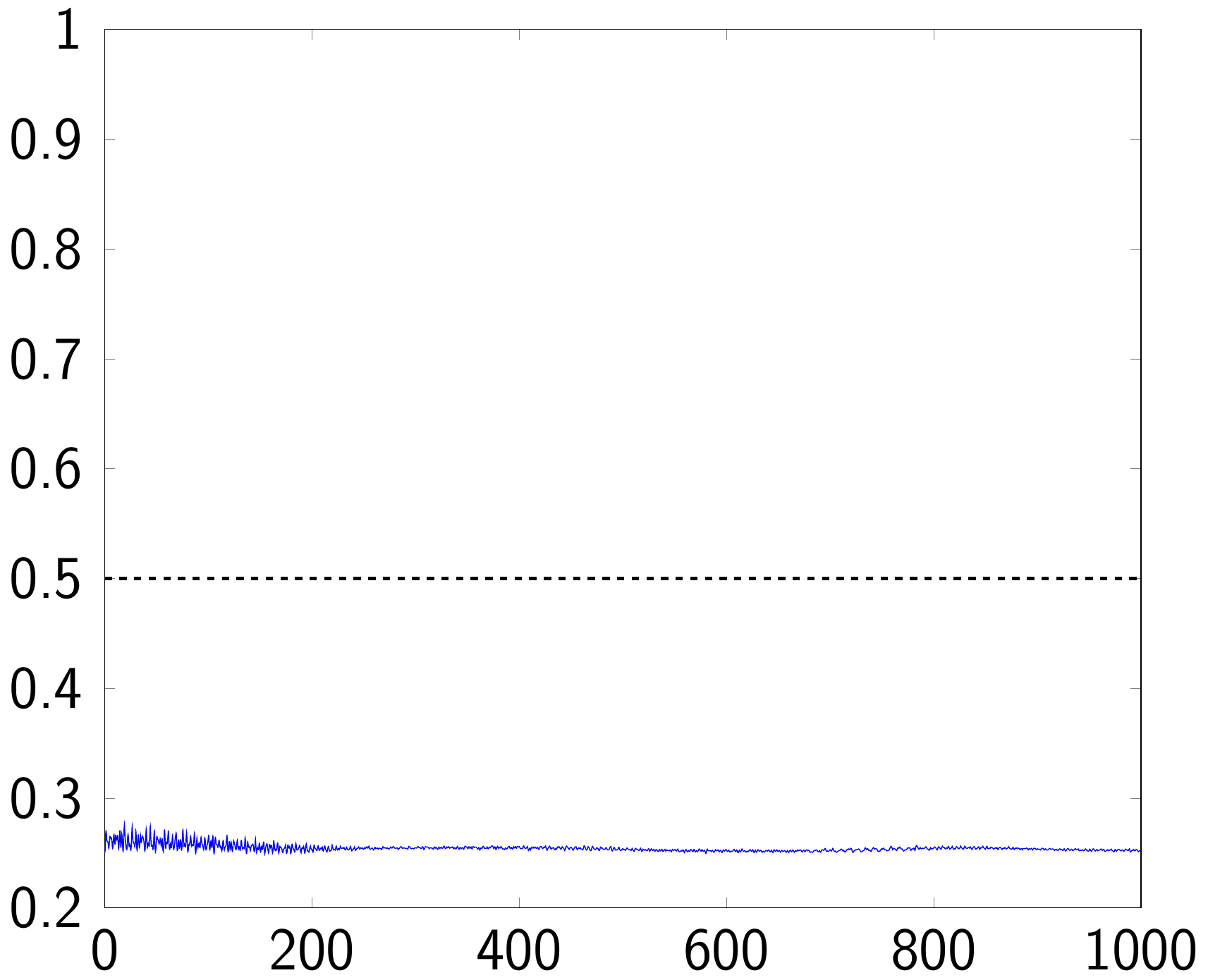} \\
		\hline
		$\textbf{44}$ & \vspace{0.2cm}
		\includegraphics[ width=1.5cm, height=1.5cm, keepaspectratio]{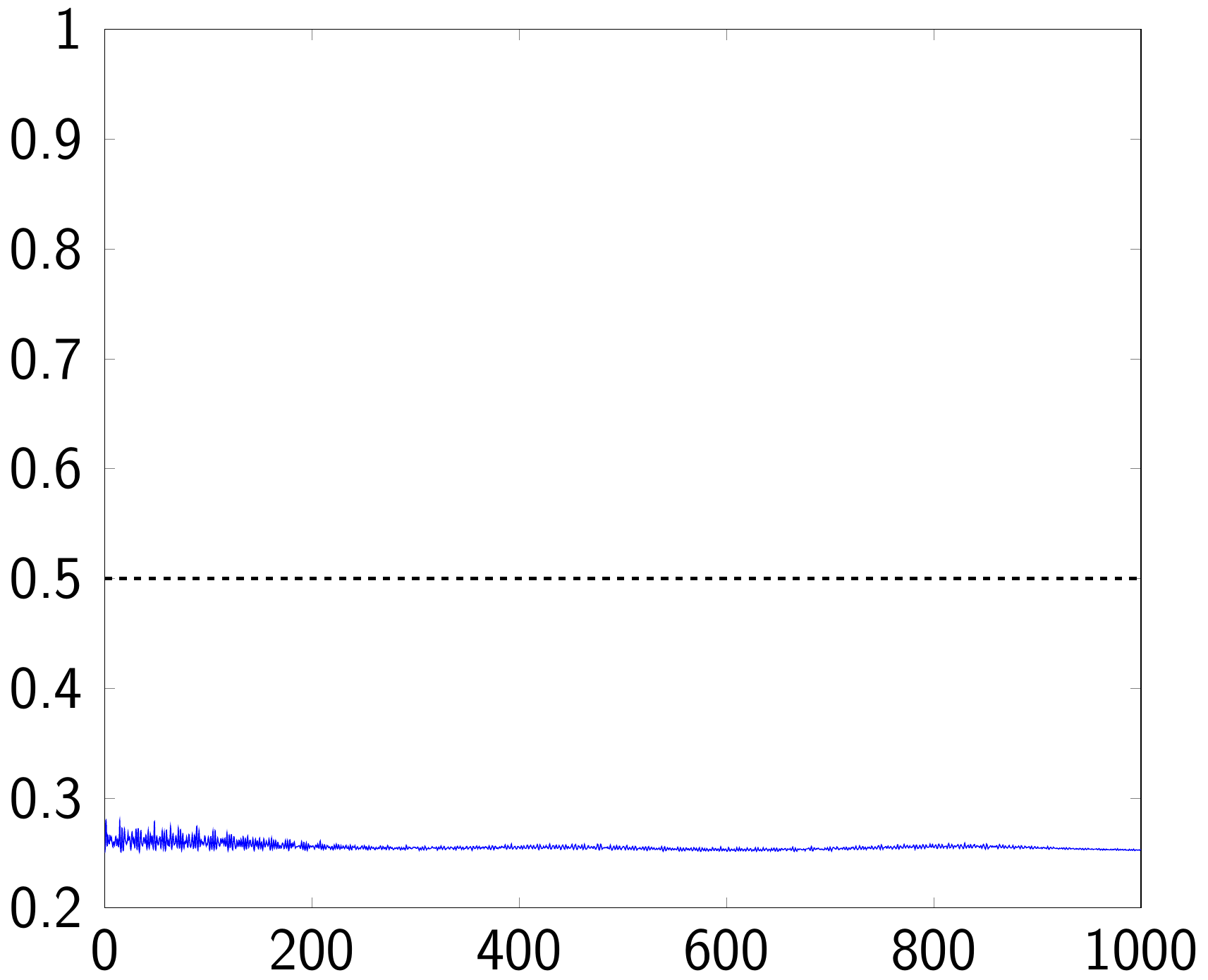} \\
		\hline
		$\textbf{45}$ & \vspace{0.2cm}
		\includegraphics[ width=1.5cm, height=1.5cm, keepaspectratio]{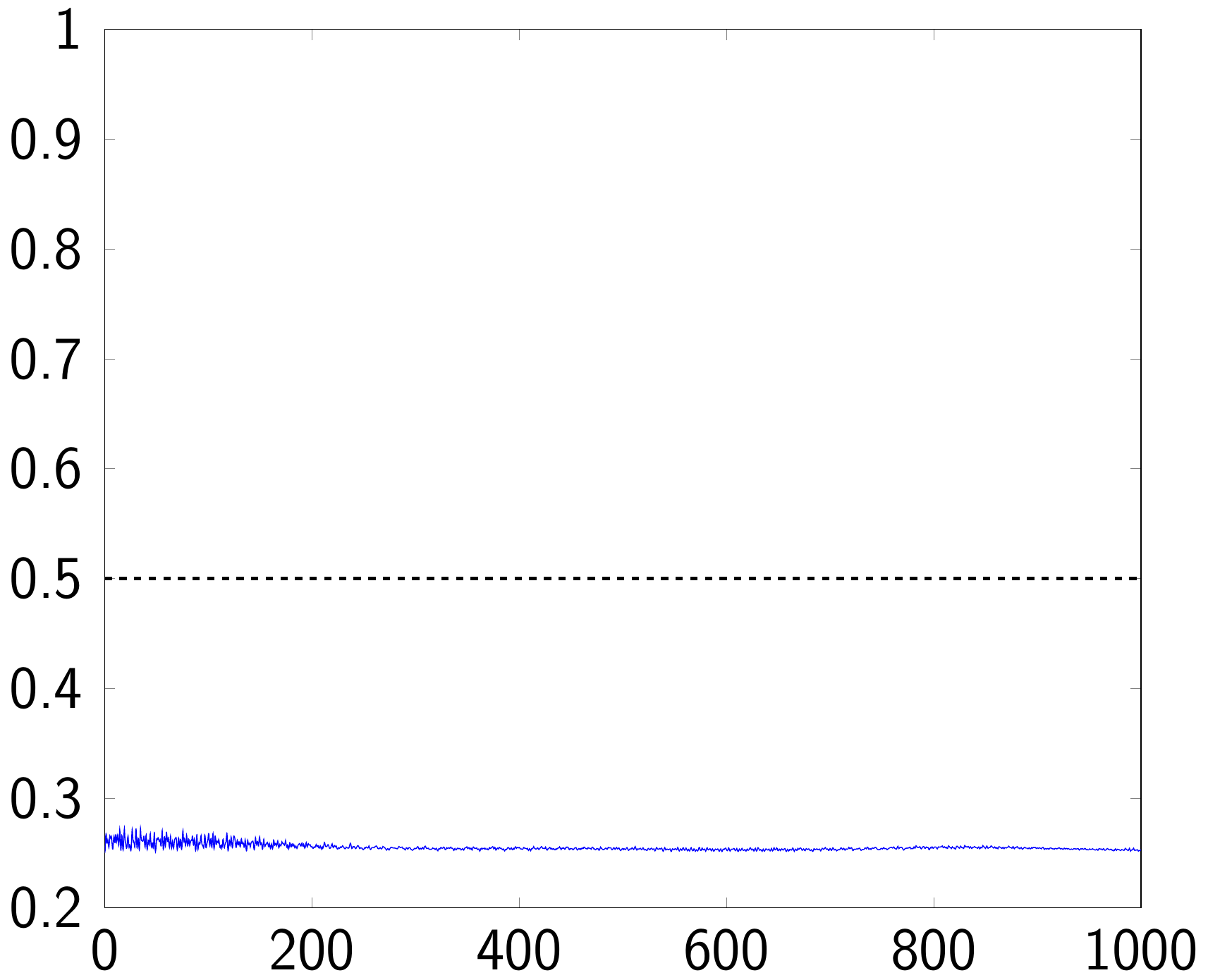} \\
		\hline
		$\textbf{55}$ & \vspace{0.2cm}
		\includegraphics[ width=1.5cm, height=1.5cm, keepaspectratio]{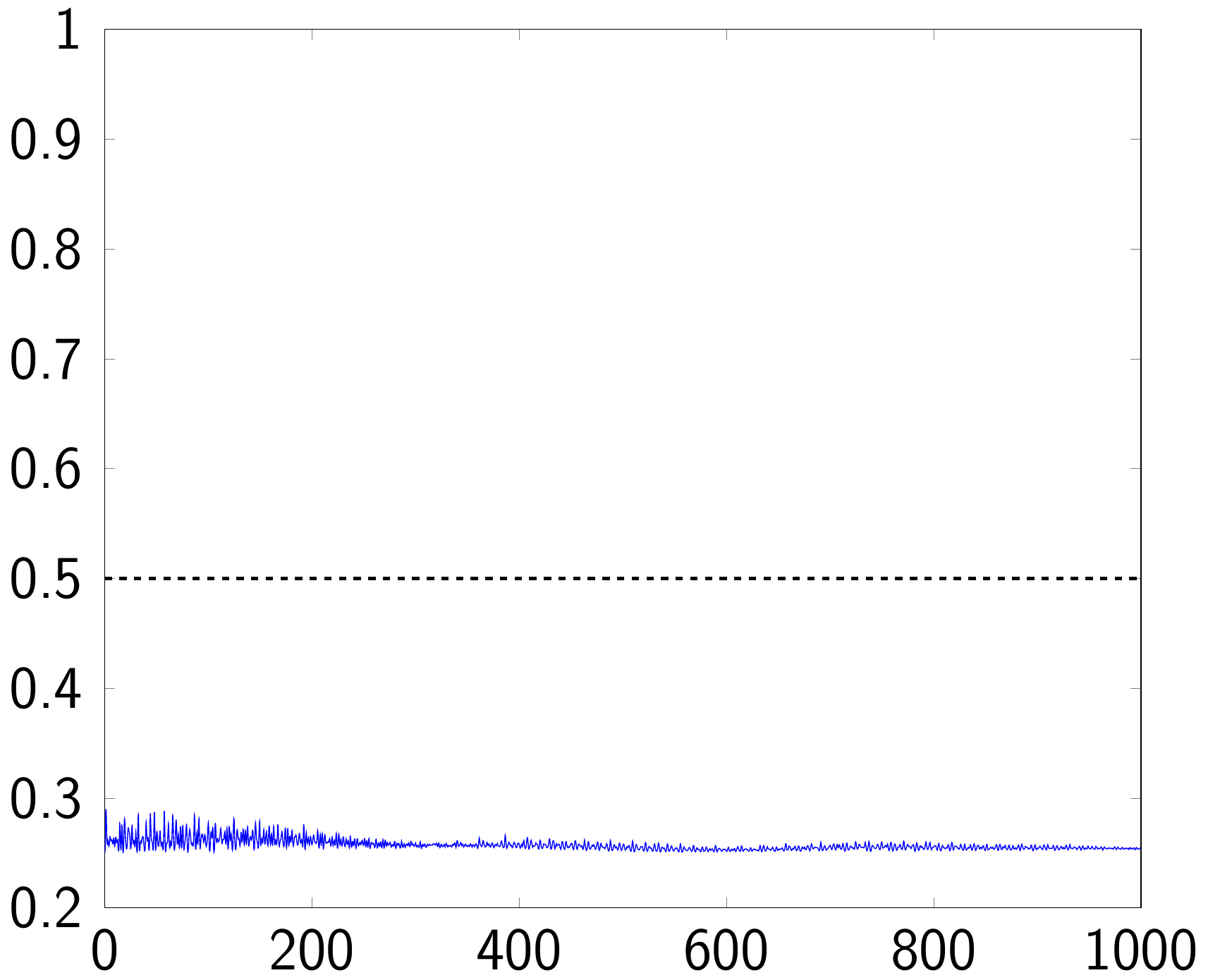} \\
		\hline
\end{tabularx}}
\end{subfigure}
\begin{subfigure}[b]{0.3\textwidth}
\setlength\tabcolsep{2pt}
\centering
\begin{tabularx}{3cm}{|@{}c*{1}{|C|}@{}}
		\hline
		& $\textbf{11}$\\
		\hline 
		$\textbf{11}$ & \vspace{0.2cm}
		\includegraphics[ width=1.5cm, height=1.5cm, keepaspectratio]{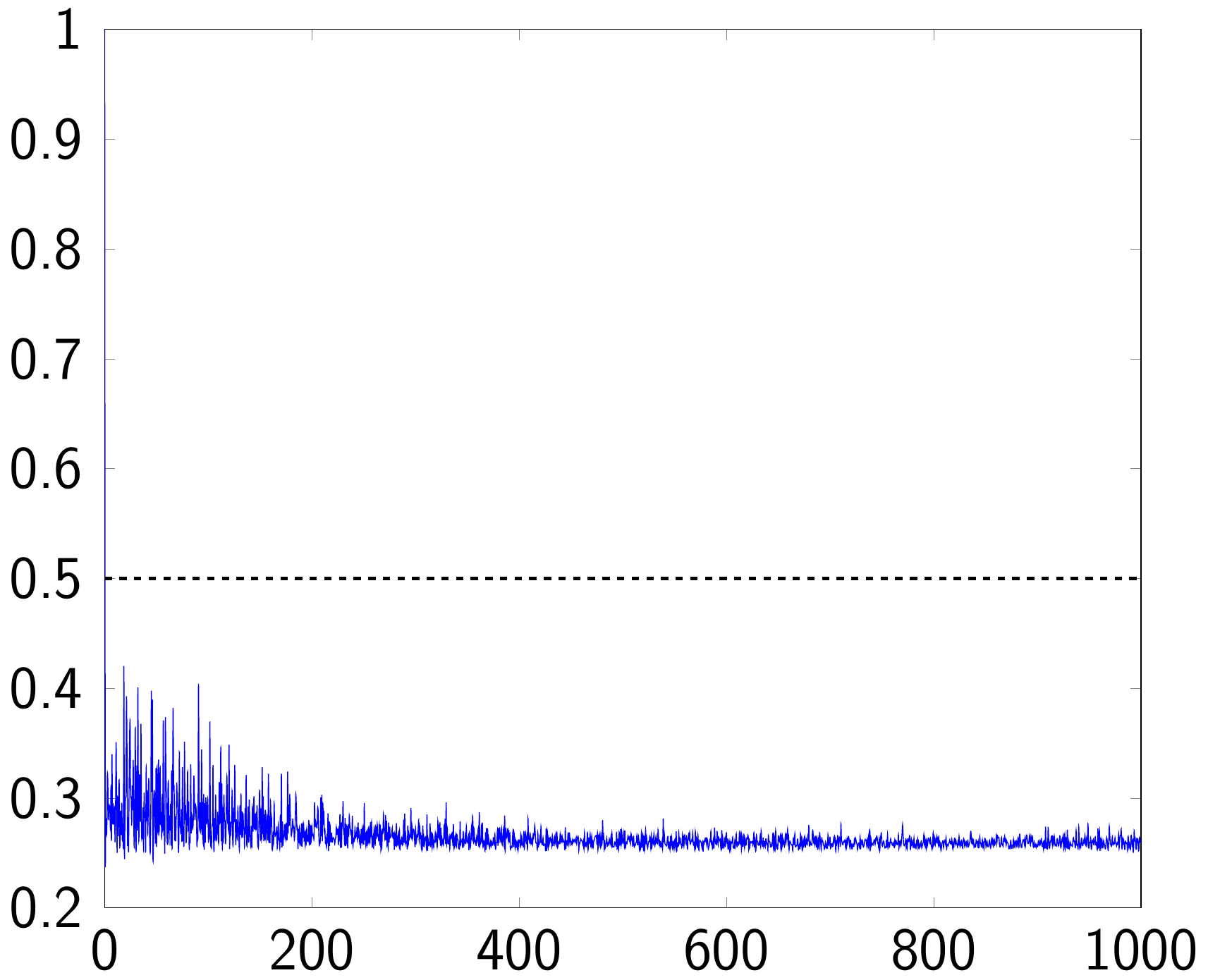} \\
		\hline
\end{tabularx} \\
\vspace{1cm}
\begin{tabularx}{3cm}{|@{}c*{1}{|C|}@{}}
		\hline
		& $\textbf{22}$\\
		\hline 
		$\textbf{22}$ & \vspace{0.2cm}
		\includegraphics[ width=1.5cm, height=1.5cm, keepaspectratio]{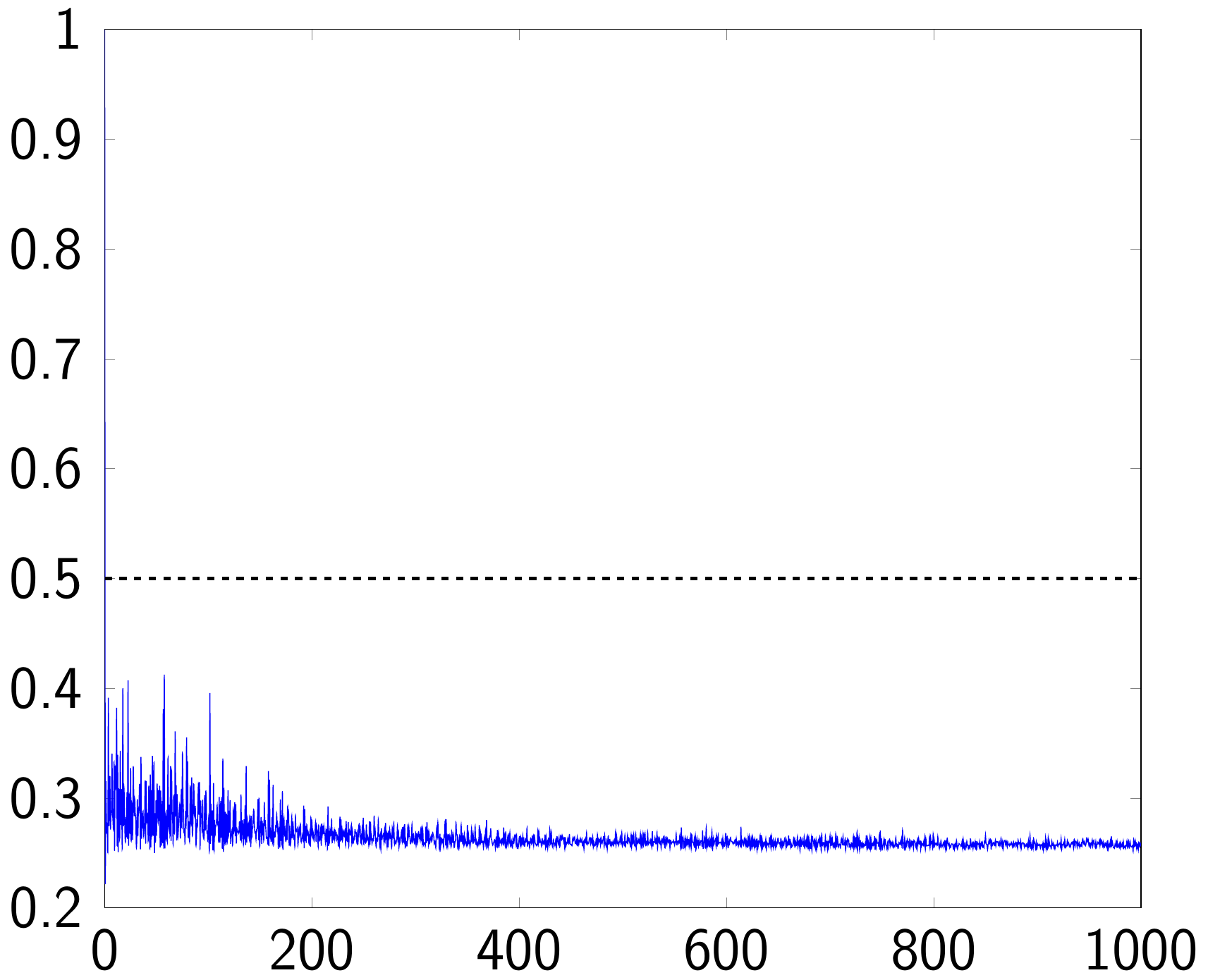} \\
		\hline
\end{tabularx} \\
\vspace{1cm}
\begin{tabularx}{3cm}{|@{}c*{1}{|C|}@{}}
		\hline
		& $\textbf{33}$\\
		\hline 
		$\textbf{33}$ & \vspace{0.2cm}
		\includegraphics[ width=1.5cm, height=1.5cm, keepaspectratio]{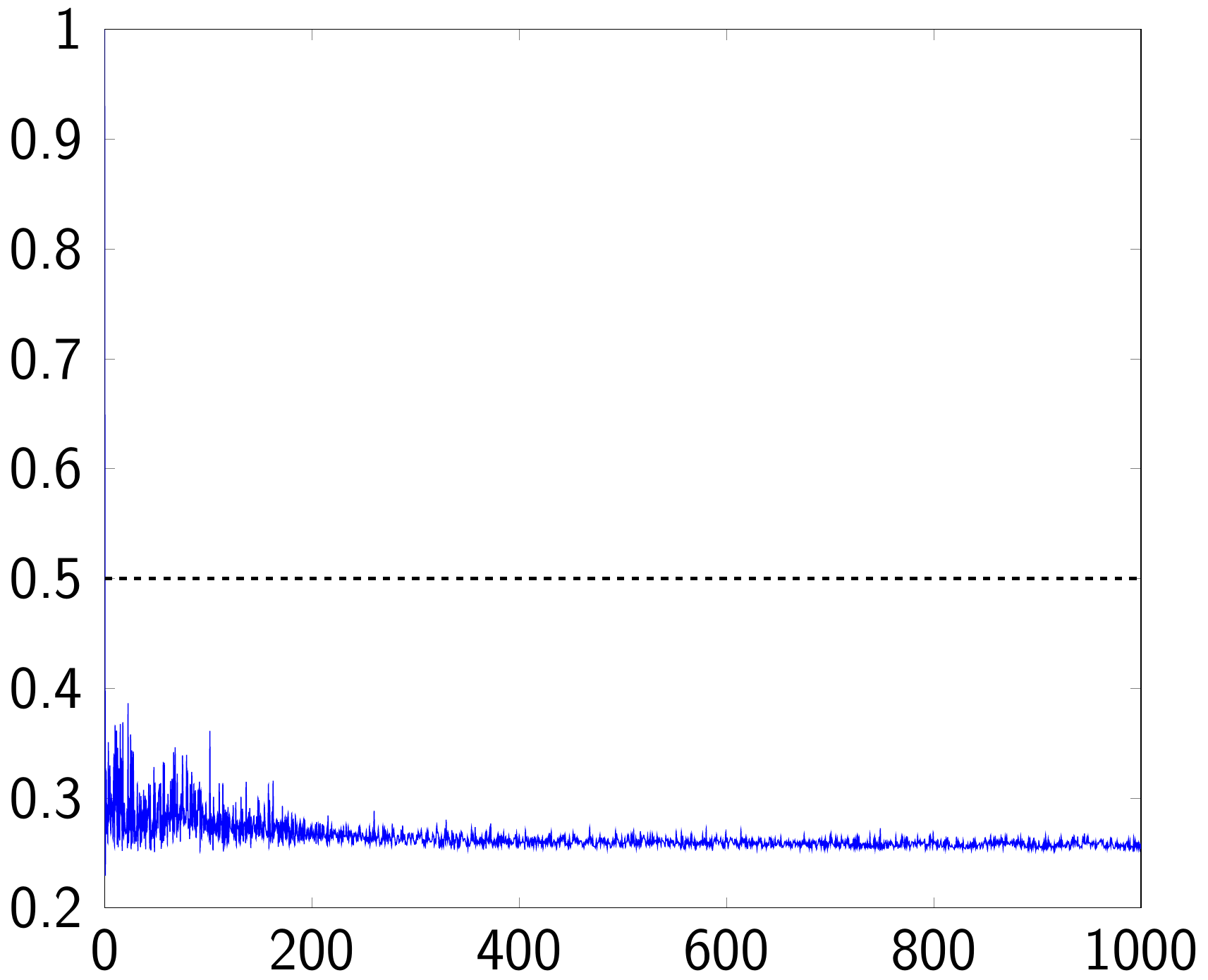} \\
		\hline
\end{tabularx} \\
\vspace{1cm}
\begin{tabularx}{3cm}{|@{}c*{1}{|C|}@{}}
		\hline
		& $\textbf{44}$\\
		\hline 
		$\textbf{44}$ & \vspace{0.2cm}
		\includegraphics[ width=1.5cm, height=1.5cm, keepaspectratio]{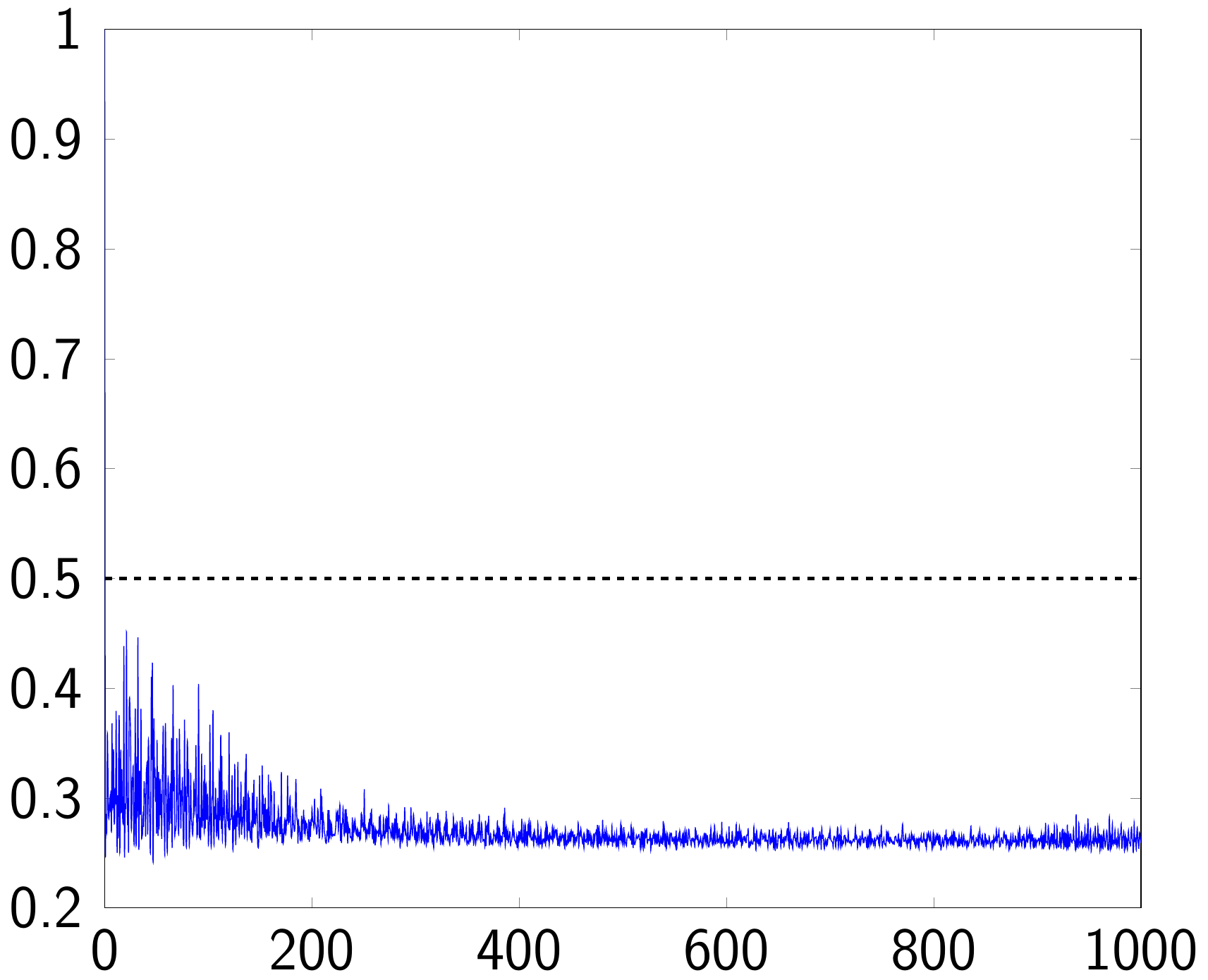} \\
		\hline
\end{tabularx} \\
\vspace{1cm}
\begin{tabularx}{3cm}{|@{}c*{1}{|C|}@{}}
		\hline
		& $\textbf{55}$\\
		\hline 
		$\textbf{55}$ & \vspace{0.2cm}
		\includegraphics[ width=1.5cm, height=1.5cm, keepaspectratio]{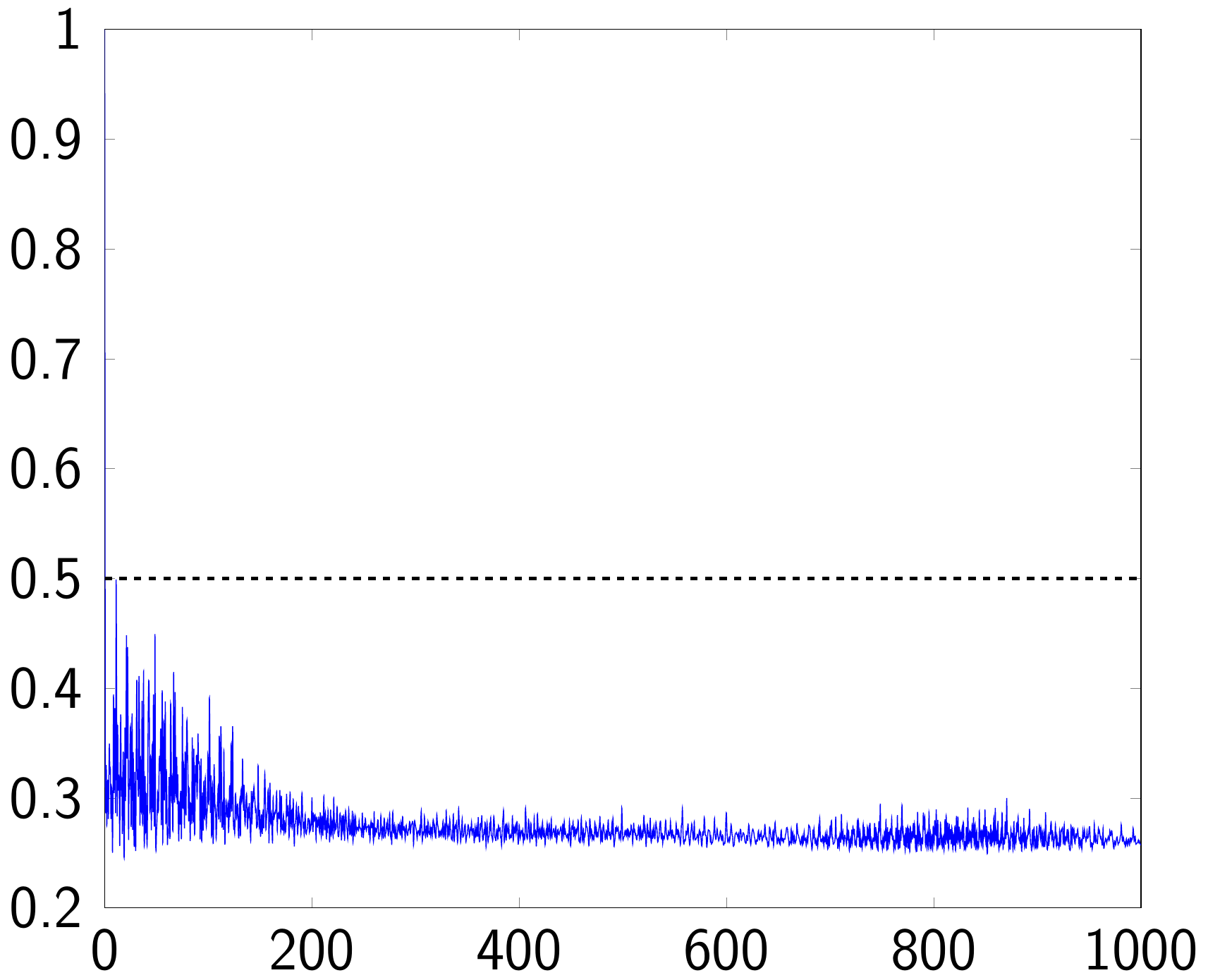} \\
		\hline
\end{tabularx}
\end{subfigure}
\caption{Transfer of entanglement between nuclear spin pairs in a Posner molecule. The $6$ nuclear spins in each Posner are indexed from $0$ through $5$. The notation $ij$ corresponds to spin pair comprising of the $i^{th}$ spin in the first Posner molecule, and the $j^{th}$ spin in the second Posner molecule. Each row represents the spin pair that was initialized in the singlet state (all other spins were left uncorrelated). The columns represent the spin pair for which the singlet probability is being calculated. The \textit{y} axis (limits from $0.2$ to $1$) in each plot represents the singlet probability, and the \textit{x} axis denotes time (a period of $1000$ seconds is covered).}
\label{fig:tranfer_coherence_posner}
\end{figure}

We calculate the transfer of coherence between \texorpdfstring{\ce{^{31}P}{ n}} nuclear spins in a pair of Posner molecules. Below, we show the results for the configuration of the molecule that has the longest-lived singlet state in our study (see Fig.~\ref{fig:tranfer_coherence_posner}). Due to the large number of nuclear spin pairs between two Posner molecules, we only show the data for a few representative \ce{^{31}P} nuclear spin pairs. Unsurprisingly, we observe that there is no marked transfer of the singlet probability, and no transfer of entanglement in any case. The singlet probabilities remain low ($\sim 0.25$) throughout, except for when the same nuclear spin pair is considered that was initialized in the singlet state. Note that this represents the best-case scenario because the structure considered had the longest-lived singlet state. We argue that for any other structural configuration of the Posner molecule considered in this study, the transfer of coherence between nuclear spin pairs will only be poorer. This further reinforces our claim that the Posner molecule, in any of its diverse structural configurations, fails to maintain long-lived entanglement amongst its \ce{^{31}P} nuclear spins.

\newpage

\providecommand{\latin}[1]{#1}
\makeatletter
\providecommand{\doi}
  {\begingroup\let\do\@makeother\dospecials
  \catcode`\{=1 \catcode`\}=2 \doi@aux}
\providecommand{\doi@aux}[1]{\endgroup\texttt{#1}}
\makeatother
\providecommand*\mcitethebibliography{\thebibliography}
\csname @ifundefined\endcsname{endmcitethebibliography}
  {\let\endmcitethebibliography\endthebibliography}{}

\end{document}